\documentclass[fleqn,10pt]{wlscirep}
\usepackage[utf8]{inputenc}
\usepackage[T1]{fontenc}
\usepackage{bm}
\usepackage{caption}
\usepackage{textcomp} 
\usepackage{pifont}
\usepackage{adjustbox}
\usepackage{graphicx}
\usepackage{multirow}
\usepackage{booktabs}
\usepackage{threeparttable}
\usepackage{xurl}
\makeatletter
\renewcommand\thefootnote{\textdaggerdbl}
\renewcommand\@makefnmark{\hbox{\textsuperscript{\thefootnote}}}
\makeatother
\title{A Clinically Validated Foundation Model for Comprehensive Lung Pathology Interpretation}

\author[1$\dagger$]{Zhengrui Guo}
\author[2,3,4$\dagger$]{Zhengyu Zhang}
\author[1]{Jiabo Ma}
\author[1]{Yihui Wang}
\author[1]{Fengtao Zhou}
\author[1]{Yingxue Xu}
\author[1]{Ling Liang}
\author[2,3,4,5]{Chenglong Zhao} 
\author[2,3,4]{Qi Xie}
\author[2,3,4,6]{Jinbang Li} 
\author[2,3,4]{Shujing Guo}
\author[2,3,4]{Fangyi Han}
\author[2,3,4]{Zhijian Cen}
\author[1]{Ziyi Liu}
\author[1]{Cheng Jin}
\author[1]{Junlin Hou}
\author[1]{Zhixuan Chen}
\author[1]{Yu Cai}
\author[7]{Lijuan Qu} 
\author[8]{Shifu Chen}
\author[9]{Yueping Liu}
\author[10,11]{Zhe Wang}
\author[12]{Xiuming Zhang}
\author[13 \ding{41}]{Muyan Cai}
\author[2,3,4,14 \ding{41}]{Li Liang}
\author[1,15,16,17,18 \ding{41}]{Hao Chen}
\affil[1]{Department of Computer Science and Engineering, Hong Kong University of Science and Technology, Hong Kong, China}
\affil[2]{Department of Pathology, Nanfang Hospital, Southern Medical University, Guangzhou, China.}
\affil[3]{Department of Pathology, School of Basic Medical Sciences, Southern Medical University, Guangzhou, China.}
\affil[4]{Guangdong Provincial Key Laboratory of Molecular Tumor Pathology, Guangzhou, China.}
\affil[5]{Department of Pathology, Shandong Provincial Qianfoshan Hospital, Jinan, Shandong, China}
\affil[6]{Department of Pathology, The Affiliated Qingyuan Hospital (Qingyuan People's Hospital), Guangzhou Medical University, Qingyuan, China}
\affil[7]{Department of Pathology, 900th Hospital of PLA Joint Logistic Support Force, Fuzhou, China}
\affil[8]{HaploX Biotechnology, Shenzhen, China}
\affil[9]{Department of Pathology, the Fourth Hospital of Hebei Medical University, Shijiazhuang, Hebei, China}
\affil[10]{State Key Laboratory of Holistic Integrative Management of Gastrointestinal Cancers, Department of Pathology, School of Basic Medicine and Xijing Hospital, Fourth Military Medical University, Xi'an, China}
\affil[11]{Department of Pathology, The First Affiliated Hospital of USTC, Division of Life Sciences and Medicine, University of Science and Technology of China, Hefei, China}
\affil[12]{Department of Pathology, The First Affiliated Hospital, School of Medicine, Zhejiang University, Hangzhou, China} 
\affil[13]{Department of Pathology, State Key Laboratory of Oncology in South China,  Guangdong Provincial Clinical Research Center for Cancer, Sun Yat-sen University  Cancer Center, Guangzhou, China.}
\affil[14]{Jinfeng Laboratory, Chongqing, China}
\affil[15]{Department of Chemical and Biological Engineering, Hong Kong University of Science and Technology, Hong Kong SAR, China}
\affil[16]{Division of Life Science, Hong Kong University of Science and Technology, Hong Kong SAR, China}
\affil[17]{State Key Laboratory of Nervous System Disorders, The Hong Kong University of Science and Technology, Hong Kong SAR, China}
\affil[18]{HKUST Shenzhen-Hong Kong Collaborative Innovation Research Institute, The Hong Kong University of Science
and Technology, Futian, Shenzhen, China}

\affil[$\dagger$]{\textbf{Equal Contribution}}
\affil[\ding{41}]{\textbf{Corresponding Authors}}
\makeatletter
\affil[ ]{\textbf{Lead Contact: Hao Chen (jhc@ust.hk)}}
\makeatother

\begin{abstract}
Pathological assessment is fundamental to lung cancer care, guiding diagnosis, treatment decisions, and prognostic evaluation. However, current computational pathology (CPath) approaches rely on task-specific models designed for isolated assessment objectives. While recent pan-cancer foundation models offer versatility, they lack subspecialty-level depth and have not undergone comprehensive evaluation across clinical workflows or prospective validation in real-world settings. We introduce PulmoFoundation, a multi-center, prospectively validated, randomized controlled trial (RCT)-evaluated foundation model for comprehensive lung pathology assessment across pre-operative, intra-operative, and post-operative care. Built upon Virchow2 via subspecialty-specific continual pretraining using more than 88 million tiles from around 40,000 diagnostic H\&E-stained whole slide images (WSIs) from 12 institutional and public sources, PulmoFoundation was systematically evaluated on over 26,000 WSIs across 32 clinically relevant tasks spanning diagnostic assessment, biomarker prediction, and prognostication, with rigorous validation across 32 internal cohorts and 21 external cohorts from 8 independent institutions. In addition to accurately predicting molecular biomarkers and patient survival, PulmoFoundation achieves clinical-grade performance in core diagnostic tasks across biopsy, frozen section, and surgical resection slides, outperforming four pan-cancer foundation models in retrospective benchmarking. In a registered prospective study of 1,357 consecutive patients across 11 diagnostic tasks in routine practice, PulmoFoundation achieved an average AUC of 92.3\%. Using pre-specified high-confidence triage thresholds, PulmoFoundation could reduce additional second-review burden for 68.8\% of biopsy cases and 83.0\% of frozen-section cases, and defer 44.5\% of IHC stain orders, with PPVs of 1.000, 0.991, and 0.966, respectively. Beyond prospective clinical evaluation, we conducted a crossover RCT with eight pathologists, in which AI assistance improved diagnostic accuracy across 5,264 matched case-reader pairs (91.7\% with AI vs. 83.2\% without AI; OR=2.31, 95\% CI: 2.13--2.51; p$<$0.001). AI assistance also reduced median diagnostic time by 18.3\%, increased diagnostic confidence by 9.0\%, and improved inter-rater agreement from moderate ($\kappa$=0.55) to substantial ($\kappa$=0.76). Together, these evaluations support PulmoFoundation as a clinically validated decision-support system for pulmonary pathology.
\end{abstract}
\begin{document}

\flushbottom
\maketitle

\thispagestyle{empty}

\section*{Introduction}
Lung cancer is the most common cancer worldwide in terms of both incidence and mortality, accounting for approximately 2.5 million new cases and 1.8 million deaths annually \cite{wild2020world,cao2024comparative,siegel2025cancer,kratzer2024lung}, making accurate pathological diagnosis critical for treatment selection and patient outcomes \cite{adams2023lung,song2023artificial,bera2019artificial,lipkova2022artificial,ma2025research}. Distinct diagnostic requirements emerge across different stages of pulmonary tissue sampling. For diagnostic biopsy specimens, pathologists first determine benign versus malignant status, subsequently establish definitive histologic classification for malignant lesions, and potentially utilize immunohistochemistry (IHC) to distinguish primary from metastatic tumors \cite{nooreldeen2021current,ning2021early}. Intra-operative frozen section diagnosis demands rapid cytomorphologic evaluation within stringent time constraints to guide surgical decision-making \cite{licker2009impact,zheng2025end}. Postoperative pathologic assessment necessitates the generation of standardized reports encompassing tumor classification, invasion depth, biological behavior, lymph node metastasis status, and molecular profiling \cite{jaklitsch2003use,coudray2018classification,kather2019deep,montagne2021role,chen2025development}.
These diagnostic tasks are frequently complicated by morphologic heterogeneity, overlapping histologic features, and substantial interobserver variability, with relatively low concordance rates for challenging cases such as frozen section assessment and adenocarcinoma subtyping \cite{yu2016predicting,kludt2024next,grilley2013validation,romagnoli2019poor}. Increasing specimen volumes further compound these challenges, straining both diagnostic efficiency and accuracy \cite{campanella2019clinical,van2021deep}.

\begin{figure}[htbp]
    \centering
    \includegraphics[width=\linewidth]{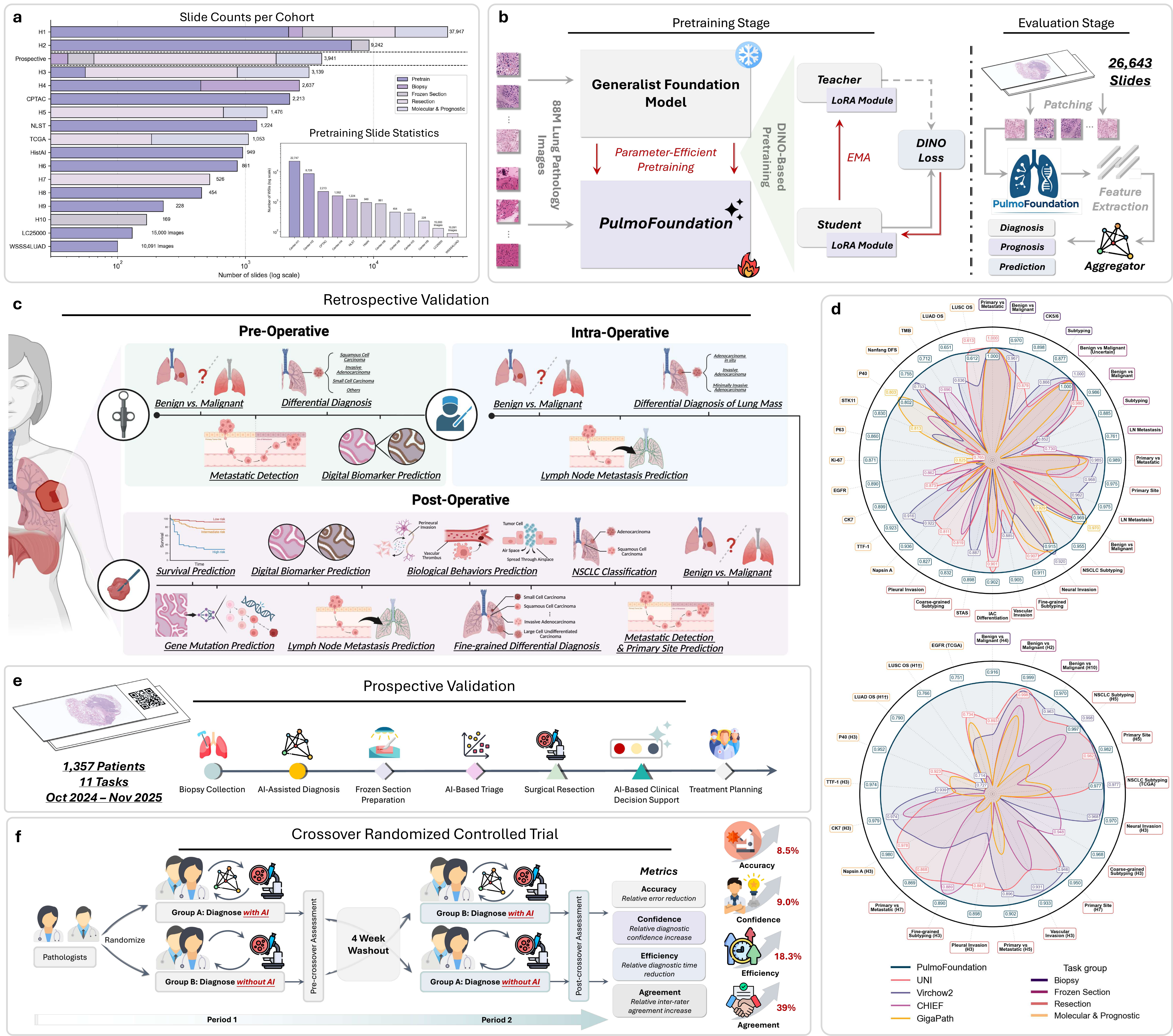}
    \caption{\textbf{Overview of PulmoFoundation study.} \textbf{a}, Distribution of WSIs across 16 medical centers and public data sources, encompassing 66,146 slides. Stacked horizontal bars show slide counts per cohort, colored by usage category, distinguishing slides used for self-supervised pretraining from those evaluated across biopsy, frozen section, resection, and molecular and prognostic tasks. Inset: pretraining WSI counts from 12 sources totaling 39,503 WSIs. \textbf{b}, Model architecture and evaluation workflow. Left: PulmoFoundation was developed through LoRA-based continual self-supervised pretraining on a generalist foundation model (Virchow2~\cite{zimmermann2024virchow2}) using a DINO teacher-student framework. Right: for downstream evaluation, WSIs are tiled into patches, encoded by PulmoFoundation, and aggregated to produce slide-level or case-level diagnostic, prognostic, and molecular predictions across 26,643 slides. \textbf{c}, Retrospective validation framework spanning 32 clinically relevant tasks across three diagnostic stages: biopsy diagnosis, intra-operative frozen section diagnosis, and post-operative surgical resection analysis, with representative tasks illustrated for each stage. \textbf{d}, Comparative performance of PulmoFoundation against four pan-cancer foundation model baselines (UNI~\cite{chen2024towards}, Virchow2~\cite{zimmermann2024virchow2}, CHIEF~\cite{wang2024pathology}, GigaPath~\cite{xu2024whole}) across retrospective tasks. Top: internal validation; bottom: external validation. Cohorts are grouped by task type (biopsy, frozen section, resection, and molecular and prognostic), with AUC or C-index values shown per task. 
    \textbf{e}, Prospective validation study design. 1,357 consecutive patients were enrolled across 11 diagnostic tasks at Nanfang Hospital (Center-H1) between October 2024 and November 2025, spanning biopsy, frozen-section, and surgical-resection pathology workflows.
    \textbf{f}, Crossover randomized controlled trial design. Eight pathologists evaluated 658 cases across four tasks under AI-assisted and unassisted conditions in two periods separated by a 4-week washout. AI assistance increased diagnostic accuracy by 8.5 percentage points, increased diagnostic confidence by 9.0\%, reduced diagnostic time by 18.3\%, and increased inter-rater agreement from 0.55 to 0.76 (Fleiss' $\kappa$).}
    \label{fig1}
\end{figure}

Artificial intelligence (AI)-empowered computational pathology (CPath) has demonstrated substantial progress in lung cancer diagnosis and prognosis. Deep learning models have shown promising results in classifying histologic subtypes~\cite{coudray2018classification,yang2021deep,davri2023deep,guo2025focus}, predicting patient survival~\cite{yu2016predicting,zhang2024histopathology,yang2025foundation,guo2024histgen}, and inferring molecular biomarkers~\cite{coudray2018classification,schmauch2020deep,diao2021human,guo2025context,jin2025genome} directly from hematoxylin and eosin (H\&E)-stained tissue.  However, these advances have predominantly relied on task-specific models designed for isolated diagnostic objectives, each requiring independent development, validation, and deployment~\cite{chen2024towards,lu2024visual}.  This fragmented approach limits clinical translation, as real-world pathology workflows demand integrated systems capable of addressing multiple interrelated diagnostic and prognostic questions simultaneously \cite{lu2024multimodal,serag2019translational,soenksen2022integrated,ma2025pathbench,campanella2025clinical,neidlinger2025benchmarking}.  Additionally, developing independent models for each clinical task faces fundamental scalability challenges, as expert pathologist annotations are resource-intensive to acquire and building task-specific datasets is challenging for rare histologic subtypes and staging features.
Furthermore, the inherent complexity and heterogeneity of lung pathology, spanning diverse histologic subtypes~\cite{solis2012histologic,ramos2025heterogeneity}, invasion patterns~\cite{kim2019patient}, and molecular features~\cite{campanella2025real,inamura2017lung,marino2019molecular,saller2022molecular}, necessitate models with deeper domain specialization than currently available general-purpose solutions can provide.

Recent pathology foundation models trained on large-scale, pan-cancer datasets have emerged as a promising paradigm, leveraging self-supervised learning to create versatile feature representations adaptable to diverse downstream tasks \cite{jin2026learning}, including patch-level~\cite{ma2025generalizable,chen2024towards,vorontsov2024foundation} and slide-level~\cite{wang2024pathology,xu2025multimodal,xu2024whole} approaches. While these models capture broad patterns across tissue types including cell morphology, tissue architecture, and staining characteristics \cite{vorontsov2024foundation}, existing foundation models excel at certain task types but struggle to effectively handle the diverse set of clinical objectives \cite{ma2025generalizable,maleki2025understanding}. Critically, their pan-cancer training strategy prioritizes breadth over depth, potentially limiting their ability to capture the subtle, disease-specific morphological features that distinguish diagnostically challenging cases within a single organ system \cite{moor2023foundation,lu2024visual,nori2023can}. Moreover, prior validation efforts have predominantly focused on isolated diagnostic and prognostic tasks rather than systematic evaluation across a clinically critical task spectrum \cite{campanella2025clinical,ma2025generalizable,neidlinger2025benchmarking,ochi2025pathology,xiong2025survey}. To our knowledge, no foundation model for lung pathology has been systematically validated across diagnostic biopsy interpretation, intra-operative frozen section assessment, and post-operative tumor characterization, molecular profiling, and prognostication. Furthermore, the absence of prospective validation in existing studies limits their clinical readiness and real-world deployment \cite{nagendran2020artificial,chen2024towards,lu2024multimodal,xu2024whole,ma2025generalizable,vorontsov2024foundation,da2026computational,liu2026advancing}. Beyond prospective validation establishing model accuracy in consecutive patients, understanding how AI systems integrate into actual diagnostic workflows requires direct evaluation of pathologist-AI interaction \cite{huang2025pathologist,bulten2022artificial,campanella2025real,skrede2020deep}.

Addressing these challenges requires a subspecialty-specific foundation model purpose-built for lung pathology, with comprehensive evaluation across the lung pathology task spectrum and rigorous assessment of pathologist-AI interaction in real-world diagnostic scenarios. This study introduces \textbf{PulmoFoundation}, a subspecialty-specific foundation model developed and evaluated using the largest lung pathology dataset assembled to date (Fig.~\ref{fig1}a), comprising $66,146$ whole-slide images (WSIs) across 16 medical centers and public data sources~\cite{weinstein2013cancer,edwards2015cptac,national2011nlst,nechaev2025histai,han2022wsss4luad,borkowski2019lung}, spanning biopsy, frozen section, and surgical resection specimens.
Leveraging this dataset, PulmoFoundation was built upon Virchow2 \cite{zimmermann2024virchow2}, a pan-cancer pathology foundation model, through subspecialty-specific continual self-supervised pretraining on $39,503$ WSIs ($88$ million image patches) from 12 sources (Fig.~\ref{fig1}a,b). This approach reflects the rationale that fine-grained morphological discrimination in lung pathology requires subspecialty depth beyond pan-cancer pretraining alone, and that continual self-supervised learning can build this depth while preserving general histopathological knowledge in the base model. We systematically evaluated PulmoFoundation across 32 clinically relevant tasks spanning the diagnostic workflow: pre-operative assessment, intra-operative decision support, and post-operative analysis (Fig.~\ref{fig1}c and Extended Data Fig.~\ref{ext_fig_cohort}).
As illustrated in Fig.~\ref{fig1}d, we benchmarked PulmoFoundation against four pan-cancer foundation-model baselines (patch-level: UNI~\cite{chen2024towards}, Virchow2~\cite{zimmermann2024virchow2}; slide-level: CHIEF~\cite{wang2024pathology}, GigaPath~\cite{xu2024whole}), and PulmoFoundation achieved consistently high performance across internal and external validations (21 external cohorts from 8 independent institutions). Critically, to establish real-world clinical utility, we conducted a registered prospective observational study across 11 diagnostic tasks, enrolling 1,357 consecutive patients in routine surgical workflows (Fig.~\ref{fig1}e and Fig.~\ref{fig6}, $\text{average AUC}=92.3\%$). To directly assess pathologist-AI interaction, we further conducted a crossover randomized controlled trial (RCT) in which eight pathologists evaluated 658 cases across four diagnostic tasks under both AI-assisted and unassisted conditions (Fig.~\ref{fig1}f and Fig.~\ref{fig7}), showing that AI assistance improved diagnostic accuracy, reduced diagnostic time, increased diagnostic confidence, and strengthened inter-rater agreement. Beyond performance gains, PulmoFoundation provides interpretable, clinically actionable insights. The model generates high-resolution attention maps that localize diagnostically critical features and accurately distinguishes subtle histologic states that guide treatment decisions (Figs.~\ref{fig4}d,e, \ref{fig5}c, and Extended Data Fig.~\ref{ext_fig_heatmap}). Together, these findings position PulmoFoundation as a clinically validated decision-support system for lung pathology and show how subspecialty-specific foundation models can extend computational pathology from isolated endpoints toward integrated clinical care.

\section*{Results}
\subsection*{Cohort characteristics and study design}
Building a foundation model capable of supporting clinical-grade decision-making across the lung pathology workflow requires capturing the breadth of morphological diversity encountered at each diagnostic juncture~\cite{chen2024towards,vorontsov2024foundation,campanella2019clinical,tham2025building}. Moreover, robust clinical translation necessitates validation not only across institutions and specimen types, but also through prospective evaluation in routine practice \cite{de2023perspectives,han2024randomised,you2025clinical}. To this end, we assembled a large-scale dataset of 66,146 H\&E-stained WSIs from 16 medical centers and public data sources spanning three specimen types: diagnostic biopsy specimens ($n=4,098$), intra-operative frozen sections ($n=12,789$), and post-operative surgical resections ($n=49,259$) (Fig.~\ref{fig1}a, Extended Data Fig.~\ref{ext_fig_cohort}, and Extended Data Tables~\ref{tab:pretrain}-\ref{tab:prospective}). This dataset covers a wide range of lung pathology encountered in surgical pathology practice, including benign lesions, primary lung cancers across major histologic subtypes, and metastatic tumors from diverse primary sites (Methods: \hyperref[sec:methods-dataset-curation]{Dataset Curation and Cohorts}).

PulmoFoundation was pretrained using self-supervised learning on 39,503 WSIs (88 million image patches) from 12 geographically diverse sources, to capture lung-specific morphological patterns (Fig.~\ref{fig1}a,b and Extended Data Table~\ref{tab:pretrain}, Methods: \hyperref[sec:methods-model-development]{PulmoFoundation Model Development}). We systematically evaluated PulmoFoundation across 32 clinically relevant tasks designed to mirror real-world diagnostic scenarios throughout lung cancer care (Fig.~\ref{fig1}c, Extended Data Fig.~\ref{ext_fig_cohort}, and Extended Data Table~\ref{tab:downstream}). These tasks spanned diagnostic biopsy assessment, intra-operative frozen section diagnosis, and post-operative analysis encompassing tumor classification and subtyping, staging and grading, molecular profiling, and survival prediction. Evaluation was conducted on 26,643 WSIs from 13,481 patients with rigorous validation across 32 internal cohorts and 21 external cohorts from 8 independent institutions (Methods: \hyperref[sec:clinical-downstream-tasks]{Clinical Downstream Tasks}).

\begin{figure}[htbp]
    \centering
    \includegraphics[width=\linewidth]{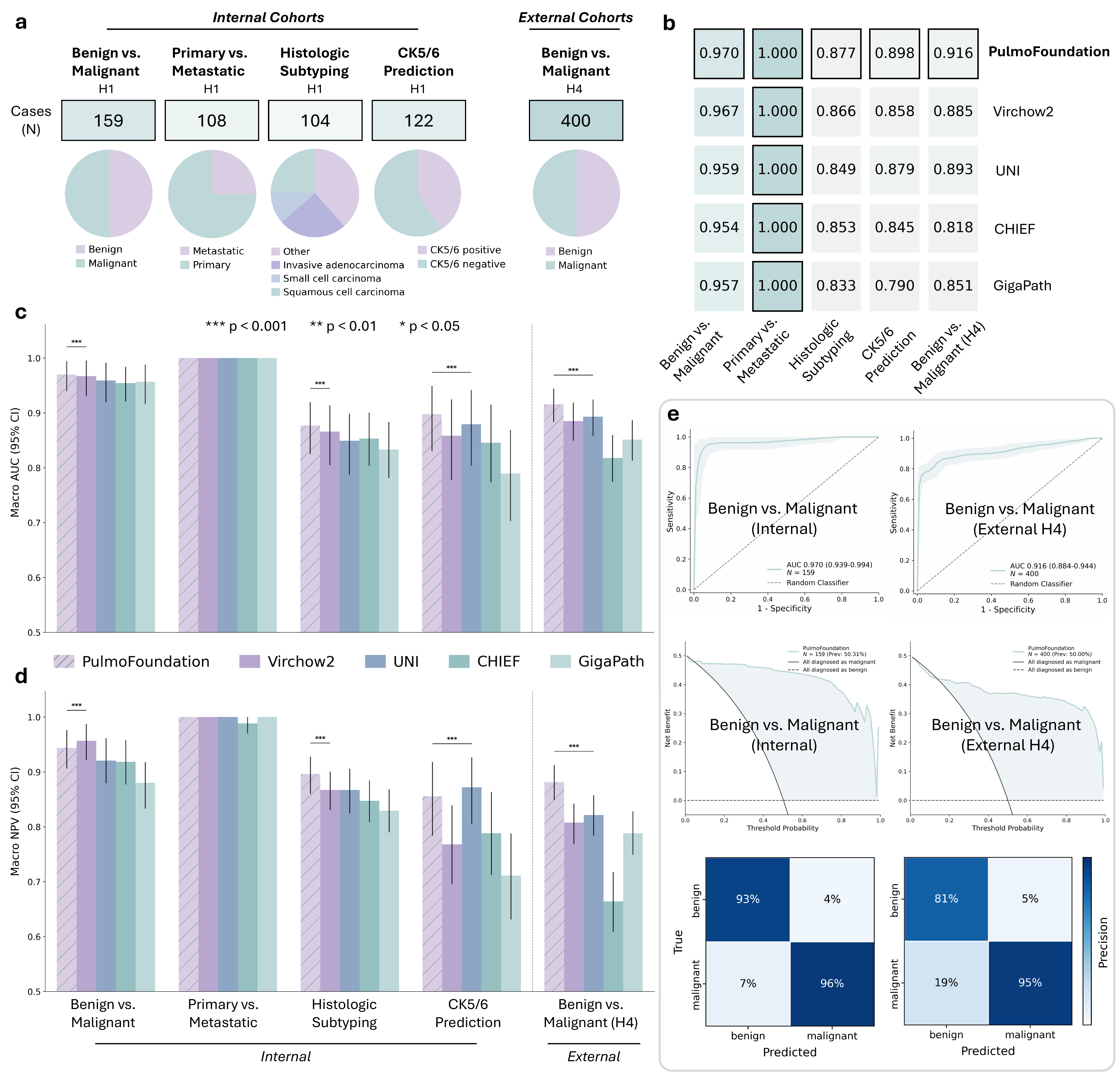}
    \caption{\textbf{Diagnostic biopsy assessment with PulmoFoundation across four clinically essential tasks.} \textbf{a}, Cohort overview for four internal task cohorts from Center-H1 and one external cohort from Center-H4: benign versus malignant, primary versus metastatic, histologic subtyping, and CK5/6 prediction. Class composition is shown beneath each cohort. \textbf{b}, Macro AUC heatmap for five models across the five task-cohort combinations; black outlines mark the best-performing model per task-cohort, with ties (e.g.\ primary versus metastatic, where all five models reach 1.000) shown in every tied cell. \textbf{c}, Macro AUC and \textbf{d}, Macro NPV at per-class Youden thresholds, comparing PulmoFoundation (lavender, hatched) with four pan-cancer pathology foundation models. Bars show the bootstrap mean and 95\% percentile confidence interval from paired bootstrap resampling ($n=1{,}000$). Brackets indicate paired Wilcoxon signed-rank tests comparing the best and second-best model for each task-cohort. \textbf{e}, Detailed evaluation on the benign versus malignant task at H1 internal (left) and H4 external (right). Top, ROC curves with shaded 95\% bootstrap confidence band; the AUC mean and 95\% CI are quoted from the same Macro AUC bootstrap replicates plotted in panel c. Middle, decision curve analysis with treat-all (solid) and treat-none (dashed) reference strategies. Bottom, confusion matrices normalized by predicted column.}
    \label{fig2}
\end{figure}

To establish real-world clinical utility, we conducted a registered prospective observational study enrolling 1,357 consecutive patients at Nanfang Hospital (Center-H1) between October 2024 and November 2025 (Fig.~\ref{fig1}e, Fig.~\ref{fig6}, and Extended Data Table~\ref{tab:prospective}). PulmoFoundation was evaluated across 11 clinically critical tasks, including biopsy and frozen-section benign-versus-malignant diagnosis, resection-based tumor subtyping, lymph-node-metastasis prediction, and IHC biomarker panel prediction (Methods: \hyperref[sec:methods-prospective-validation]{Prospective Validation Study Protocol}).

To directly assess the impact of AI assistance on pathologist decision-making, we conducted a crossover RCT (ClinicalTrials.gov ID: NCT07157618) with eight board-certified pathologists (4 senior, 4 junior) evaluating 658 cases across four diagnostic tasks: biopsy primary-versus-metastatic classification, frozen-section benign-versus-malignant diagnosis, resection-based NSCLC subtyping, and metastatic-tumor primary-site prediction (Extended Data Table~\ref{tab:RCT_pathologist}; Extended Data Table~\ref{tab:RCT_cases}; Fig.~\ref{fig1}f; Fig.~\ref{fig7}a). Each pathologist reviewed all cases under both AI-assisted and unassisted conditions separated by a 4-week washout period, generating 10,528 total diagnostic observations and 5,264 matched AI-assisted versus unassisted case-reader comparisons for within-subject comparison of accuracy, diagnostic time, confidence, and inter-rater agreement (Methods: \hyperref[sec:methods-rct]{Crossover Randomized Controlled Trial}).




\subsection*{PulmoFoundation supports diagnostic biopsy interpretation}
Biopsy represents the critical first diagnostic encounter with suspected lung malignancies, where pathologists must render definitive diagnoses from limited tissue samples \cite{mukhopadhyay2012utility,hofman2019challenges}. As pulmonary nodule detection through screening computed tomography (CT) increases \cite{ardila2019end,adams2023lung,bhamani2025low}, the volume and complexity of biopsy interpretation continues to grow, underscoring the need for robust computational decision support. We evaluated PulmoFoundation on four clinically essential biopsy tasks (Fig.~\ref{fig2}a), achieving averaged macro AUCs of 0.936 (internal) and 0.916 (external) (Fig.~\ref{fig2}b). Across all five task-cohort combinations, PulmoFoundation ranked first or tied first against four pan-cancer foundation model baselines (Fig.~\ref{fig2}b).

\textbf{Diagnostic gating: benign-versus-malignant classification and primary-versus-metastatic distinction.} The two binary decisions that gate all downstream workup are whether a lesion is malignant and, if so, whether it is a primary lung cancer or a metastasis. For benign-versus-malignant classification, PulmoFoundation achieved an internal AUC of 0.970 (95\% CI: 0.939--0.994) and an external AUC of 0.916 (95\% CI: 0.884--0.944) at Center-H4, outperforming the next-best baseline by 2.3 percentage points externally (UNI, 0.893) and by 9.8 percentage points over the lowest-ranked model (CHIEF, 0.818) (Fig.~\ref{fig2}b,c). Because a missed malignancy carries greater clinical consequence than a false positive in biopsy triage, negative predictive value (NPV) is a key safety metric; PulmoFoundation maintained high macro NPVs among evaluated models across validation cohorts (Fig.~\ref{fig2}d). Meanwhile, decision curve analysis showed positive net benefit across clinically relevant threshold probabilities in both the internal and external cohorts (Fig.~\ref{fig2}e). For primary-versus-metastatic classification, PulmoFoundation and all four baselines achieved perfect discrimination (AUC = 1.000) in the internal cohort spanning metastases from colon, breast, liver, and kidney \cite{herbst2018biology,wang2021toward}, indicating that this distinction is well captured by foundation-model representations.

\textbf{Tumor characterization: histologic subtyping and biomarker prediction.} Once malignancy and primary origin are established, accurate histologic classification guides treatment selection. Following IASLC criteria \cite{travis2011international,travis2013diagnosis}, we evaluated a four-class classification task (squamous cell carcinoma, invasive adenocarcinoma, small cell carcinoma, and other malignant neoplasms). PulmoFoundation achieved a macro AUC of 0.877 (95\% CI: 0.825--0.920), outperforming the next-best model (Virchow2, 0.866) (Fig.~\ref{fig2}b,c). To further support the clinically important distinction between squamous cell carcinoma and adenocarcinoma, we evaluated prediction of CK5/6 expression directly from H\&E-stained tissue. On 122 biopsy specimens with available CK5/6 IHC, PulmoFoundation achieved an AUC of 0.898 (95\% CI: 0.830--0.949), outperforming UNI (0.879), Virchow2 (0.858), CHIEF (0.845), and GigaPath (0.790) (Fig.~\ref{fig2}b,c), demonstrating the model's capacity to infer protein expression patterns from morphology alone.

Across biopsy tasks, PulmoFoundation ranked first or tied first by macro AUC in all five task-cohort combinations, with significant advantages in the non-tied comparisons confirmed by paired Wilcoxon signed-rank tests on bootstrap replicates (Fig.~\ref{fig2}c). These results position PulmoFoundation as an integrated decision-support tool for biopsy interpretation in surgical pathology practice. Detailed performance metrics for each task are provided in Extended Data Tables \ref{tab:biopsy_benign_malignant}--\ref{tab:biopsy_ck56}.


\begin{figure}[htbp]
    \centering
    \includegraphics[width=\linewidth]{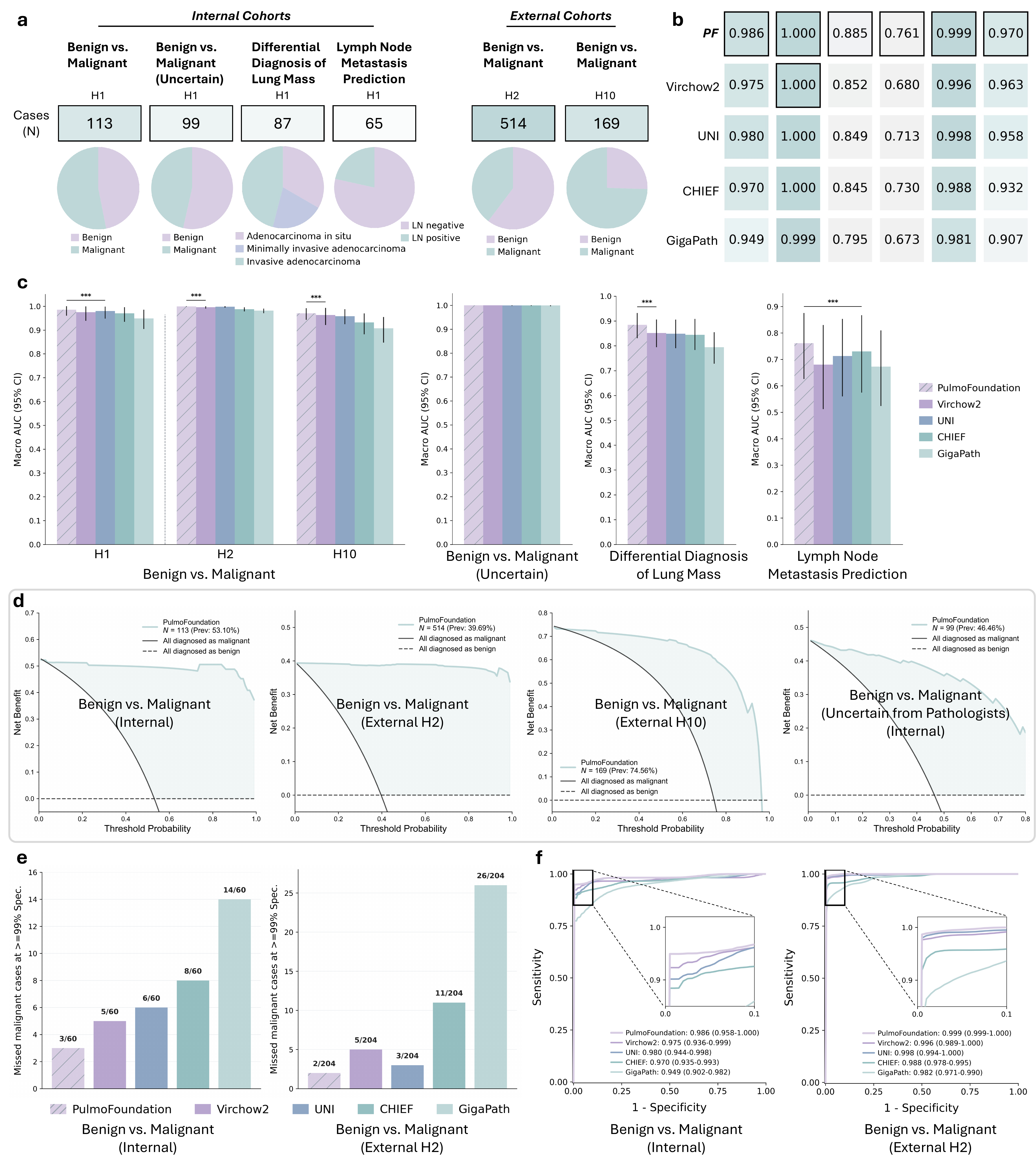}
    \caption{\textbf{Intra-operative frozen-section diagnosis with PulmoFoundation across four clinically essential tasks.} 
    \textbf{a}, Cohort overview for four internal task-cohorts at Center-H1 and two external benign-versus-malignant cohorts. Class composition is shown beneath each cohort. 
    \textbf{b}, Macro AUC heatmap for five models across the six task-cohort combinations; black outlines mark the best-performing model per task-cohort, with ties shown in every tied cell. 
    \textbf{c}, Macro AUC bars comparing PulmoFoundation with four pan-cancer foundation model baselines. Bars show the bootstrap mean and 95\% percentile confidence interval. 
    \textbf{d}, PulmoFoundation decision curve analysis for benign versus malignant at Center-H1, Center-H2, and Center-H10, and for pathologist-uncertain cases at Center-H1. 
    \textbf{e}, Missed malignant slides at the operating point yielding specificity at least 0.99, for the benign versus malignant task at Center-H1 (left) and Center-H2 (right). Bars are labelled as the count of false negatives over the total malignant count in each cohort. 
    \textbf{f}, ROC curves for benign versus malignant at Center-H1 (left) and Center-H2 (right), with the upper-left region magnified in an inset (zoom region marked on the main axes).}
    \label{fig3}
\end{figure}

\subsection*{PulmoFoundation enables time-critical frozen-section diagnosis}
Intra-operative frozen section diagnosis requires rapid histologic assessment (typically within 15--20 minutes) to guide immediate surgical decisions \cite{zhao2025clinical,novis1997interinstitutional}. For lung lesions, frozen section evaluation determines resection extent, classifies early-stage adenocarcinomas to guide surgical approach, and assesses lymph nodes for metastatic disease \cite{sienko2005frozen,li2016intraoperative}. However, interpretation is complicated by tissue freezing artifacts, limited sampling, and time pressure, resulting in higher rates of diagnostic uncertainty compared to permanent sections \cite{gupta2008lessons}. We evaluated PulmoFoundation across four frozen section diagnostic tasks (Fig.~\ref{fig3}a), achieving averaged macro AUCs of 0.908 (internal) and 0.985 (external). PulmoFoundation ranked first or tied first against four pan-cancer baselines on all six task-cohort combinations (Fig.~\ref{fig3}b).

\textbf{Malignancy assessment in frozen sections.} Determining whether a frozen section specimen contains malignancy is the primary intra-operative diagnostic question, directly influencing surgical extent \cite{liu2016precise,li2016intraoperative}. PulmoFoundation achieved the highest macro AUC across internal and external cohorts: 0.986 (95\% CI: 0.960--1.000) internally, 0.999 (95\% CI: 0.997--1.000) at Center-H2, and 0.970 (95\% CI: 0.943--0.992) at Center-H10 (Fig.~\ref{fig3}b,c,f). At an operating point yielding at least 99\% specificity, PulmoFoundation missed the fewest malignant cases among all evaluated models: 3 of 60 malignant slides at Center-H1 (vs.\ Virchow2 5, UNI 6, CHIEF 8, GigaPath 14) and 2 of 204 malignant slides at Center-H2 (vs.\ UNI 3, Virchow2 5, CHIEF 11, GigaPath 26) (Fig.~\ref{fig3}e). In the frozen-section setting, each missed malignancy can delay definitive management or require additional surgery; the reduction from 14 to 3 missed cases (Center-H1) and from 26 to 2 (Center-H2) relative to the lowest-ranked baseline illustrates how subspecialty pretraining can translate into clinically meaningful differences under high-specificity operating constraints. Moreover, decision curve analysis showed substantial positive net benefit across broad clinically relevant threshold ranges (Fig.~\ref{fig3}d). We further identified frozen section cases where pathologists expressed explicit diagnostic uncertainty during real-time evaluation (qualifying language such as ``likely malignant'', ``suspicious for'', or ``recommend permanent section correlation''), using the final diagnosis from permanent resection as ground truth \cite{gupta2008lessons}. On this subset, PulmoFoundation and all four baselines achieved near-perfect discrimination (AUC $\geq$ 0.999; Fig.~\ref{fig3}b), with decision curve analysis confirming positive net benefit across clinically relevant probability thresholds (Fig.~\ref{fig3}d).

\textbf{Surgical extent guidance: adenocarcinoma subtyping and lymph node metastasis prediction.} Two frozen section tasks inform surgical extent and nodal management. First, distinguishing adenocarcinoma in situ (AIS), minimally invasive adenocarcinoma (MIA), and invasive adenocarcinoma (IAC) is critical because AIS and MIA with a size $<$ 2\,cm may be adequately treated with wedge resection, whereas IAC typically requires lobectomy \cite{saji2022segmentectomy,altorki2023lobar}. PulmoFoundation achieved a macro AUC of 0.885 (95\% CI: 0.831--0.932) on this three-class task, exceeding the next-best baseline (Virchow2, 0.852) by 3.3 percentage points (Fig.~\ref{fig3}b,c). Second, we evaluated prediction of lymph node metastasis status from primary tumor morphology in frozen sections, a binary task using H\&E-stained frozen sections of the primary tumor as input with ground truth established by post-operative pathological examination of all resected regional lymph nodes \cite{li2016intraoperative,ortiz2025diagnostic}. PulmoFoundation achieved an AUC of 0.761 (95\% CI: 0.626--0.875) (Fig.~\ref{fig3}b,c). This task requires inference of metastatic behavior from primary tumor histology alone, and the moderate performance reflects this inherent difficulty; nonetheless, the prediction may complement conventional nodal assessment when comprehensive intra-operative sampling is impractical.

Beyond discrimination accuracy, practical deployment in the frozen-section workflow demands that inference completes in the narrow time window available for intra-operative consultation. Computational workflow analysis showed that frozen-section WSIs could be processed from slide to prediction in a median of 83.6\,s (IQR, 59.4--98.3) on a single NVIDIA GeForce RTX 3090 GPU (Extended Data Tables~\ref{tab:runtime_summary} and \ref{tab:runtime_components}), well within the typical 15--20 minute consultation window. Across frozen-section tasks, PulmoFoundation ranked first or tied first by macro AUC in all six task-cohort combinations, with the largest advantages on adenocarcinoma subtyping and missed-malignant analyses (Fig.~\ref{fig3}b,c,e). Detailed performance metrics for each task are provided in Extended Data Tables \ref{tab:frozen_benign_malignant}--\ref{tab:frozen_lymph_node_metastasis}.

\begin{figure}[htbp]
    \centering
    \includegraphics[width=0.95\linewidth]{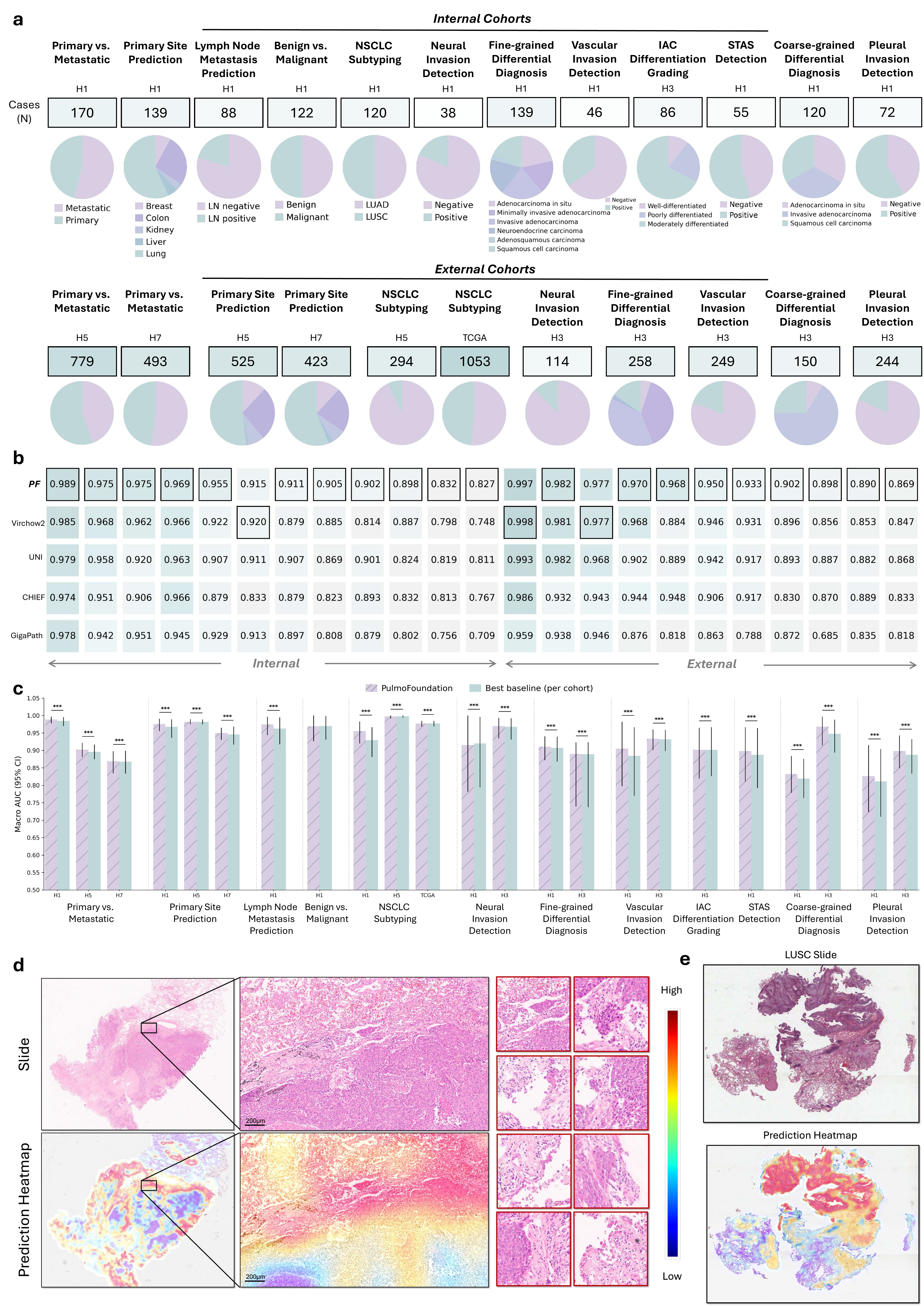}
\end{figure}
\begin{figure}[t]
\caption{\textbf{Post-operative resection: diagnostic classification, staging, and grading task evaluation with PulmoFoundation.}
\textbf{a}, Overview of resection tasks and validation cohorts. Twelve internal task-cohort columns and eleven external columns, spanning diagnostic classification (benign versus malignant discrimination, primary versus metastatic classification, primary site prediction, NSCLC subtyping, coarse-grained subtyping, fine-grained subtyping) and staging/grading (vascular invasion detection, neural invasion detection, pleural invasion detection, spread through air spaces detection, lymph node metastasis prediction, IAC differentiation grading), with cohort sizes and class composition shown.
\textbf{b}, Macro AUC heatmap of five foundation models (PulmoFoundation, Virchow2, UNI, GigaPath, CHIEF) across all 23 task-cohort columns, illustrating per-task performance distribution. 
\textbf{c}, Per-cohort macro AUC of PulmoFoundation versus the best-performing pan-cancer baseline of each cohort, with 95\% bootstrap percentile confidence intervals. Brackets indicate paired Wilcoxon signed-rank tests comparing PulmoFoundation with the best-performing baseline for each task-cohort column after Holm correction ($^{***}$: $p < 0.001$).
\textbf{d}, Interpretability analysis for vascular tumor thrombus detection. Top panels display the original WSI (left) and magnified view of the thrombus region (right). Bottom panels show the corresponding attention heatmap (left) and magnified attention map (right). Right panel displays top-scoring image patches ranked by the model's confidence.
\textbf{e}, Representative attention heatmap overlaid on a lung squamous cell carcinoma (LUSC) WSI, showing PulmoFoundation's localization of keratinization, intercellular bridges, and characteristic nuclear features distinctive of squamous differentiation.}
    \label{fig4}
\end{figure}

\subsection*{PulmoFoundation facilitates tumor classification, staging, and grading in surgical resection specimens}
Post-operative pathologic examination of surgical resection specimens guides adjuvant therapy decisions, prognostic stratification, and clinical trial eligibility \cite{nicholson20222021}. Unlike limited biopsy samples, resection specimens enable complete evaluation of tumor architecture, growth patterns, and invasion characteristics, and additionally require assessment of histologic features that independently predict outcomes, including vascular invasion, perineural spread, pleural involvement, and lymph node status \cite{nicholson20222021,thunnissen2012reproducibility,goldstraw2016iaslc}. We evaluated PulmoFoundation across twelve resection tasks internally and externally, spanning six diagnostic classification tasks and six staging/grading tasks (Fig.~\ref{fig4}a,b), achieving averaged macro AUCs of 0.921 (internal) and 0.940 (external). PulmoFoundation ranked first in 20 of 23 task-cohort combinations against four pan-cancer foundation model baselines; in the remaining three, performance was within 0.5 percentage points of the best-performing model (Fig.~\ref{fig4}b).

\textbf{Diagnostic classification:} Six tasks address the core diagnostic questions for resection specimens. While most surgical resections are performed for suspected malignancy, distinguishing benign from malignant lesions remains essential for excluding reactive, inflammatory, or infectious processes that may radiologically mimic malignancy; PulmoFoundation achieved an AUC of 0.969 (95\% CI: 0.926--1.000) internally. Determining whether a malignancy represents primary lung cancer or metastasis from an extrapulmonary site fundamentally influences treatment strategy and prognostic assessment; PulmoFoundation yielded AUCs of 0.989 (95\% CI: 0.978--0.997) internally, 0.902 (95\% CI: 0.881--0.922) externally at Center-H5, and 0.869 (95\% CI: 0.834--0.898) externally at Center-H7. For confirmed metastases from colorectal, breast, kidney, and liver primaries, identifying the primary site of origin is critical for selecting appropriate systemic therapy \cite{lu2021ai,moon2023machine}; PulmoFoundation achieved 0.975 (95\% CI: 0.955--0.991) internally, ranking first across all three cohorts (0.982 at H5, 0.950 at H7). Distinguishing adenocarcinoma from squamous cell carcinoma is among the most clinically consequential classification decisions, as these subtypes exhibit distinct molecular profiles and therapeutic vulnerabilities (e.g., pemetrexed and bevacizumab are contraindicated in squamous histology) \cite{hendriks2024non,huang2025advances}; PulmoFoundation yielded an internal AUC of 0.955 (95\% CI: 0.920--0.983), exceeding the next-best baseline internally (Gigapath, 0.929) by 2.6 percentage points, and external AUCs of 0.997 (95\% CI: 0.991--1.000) at Center-H5 and 0.977 (95\% CI: 0.969--0.985) in TCGA, with representative attention patterns localizing keratinization, intercellular bridges, and characteristic nuclear features (Fig.~\ref{fig4}e). At finer diagnostic resolution, coarse-grained three-class subtyping (squamous cell carcinoma, invasive adenocarcinoma, adenocarcinoma in situ) is relevant because AIS designation carries distinct prognostic and treatment implications and generally precludes adjuvant therapy \cite{travis2011international,travis2013diagnosis}; PulmoFoundation yielded 0.832 (95\% CI: 0.779--0.884) internally and 0.968 (95\% CI: 0.914--0.997) at Center-H3, outperforming the next-best external baseline (CHIEF, 0.948) by 2.1 percentage points. Six-class fine-grained classification following WHO criteria \cite{nicholson20222021}, encompassing rare and morphologically overlapping subtypes with known inter-observer variability \cite{thunnissen2012reproducibility}, achieved 0.911 (95\% CI: 0.872--0.940) internally and 0.890 (95\% CI: 0.739--0.923) at Center-H3 (Fig.~\ref{fig4}b,c).

\textbf{Staging and grading:} Six tasks assess histologic features that independently predict patient outcomes and inform staging, prognosis, adjuvant therapy decisions, and surveillance. Vascular tumor thrombus (tumor cells within blood or lymphatic vessels) is an independent adverse prognostic factor associated with increased recurrence risk \cite{shimada2010extratumoral}; however, the lung's abundant and intricately arranged vascular spaces make identification time-consuming. PulmoFoundation achieved an AUC of 0.905 (95\% CI: 0.798--0.982) internally and 0.933 (95\% CI: 0.901--0.960) at Center-H3, ranking first on both cohorts (next-best internal: Virchow2, 0.885; next-best external: Virchow2, 0.931). Attention patterns highlighted vessels containing tumor thrombi, with top-scoring patches highlighting tumor cell clusters within vessel lumens and the interface between intravascular tumor and vessel walls (Fig.~\ref{fig4}d). Perineural invasion (tumor cells infiltrating along nerve sheaths) is similarly prognostic, but identifying scattered tumor cells along small nerve bundles that are morphologically similar to surrounding stroma represents a recognized diagnostic challenge \cite{nicholson20222021}; PulmoFoundation yielded an internal AUC of 0.915 (95\% CI: 0.782--1.000) and an external AUC of 0.970 (95\% CI: 0.934--0.993) at Center-H3, with attention concentrated on nerve bundle regions with adjacent tumor infiltration (Extended Data Fig.~\ref{ext_fig_heatmap}). Spread through air spaces (STAS) has emerged as a critical prognostic feature in lung adenocarcinoma: its presence upgrades AIS and MIA to invasive adenocarcinoma under current classification criteria and is associated with recurrence risk after resection \cite{nicholson20222021,han2021tumor}. PulmoFoundation achieved an AUC of 0.898 (95\% CI: 0.811--0.966), exceeding the next-best model (Virchow2, 0.888) and the lowest-ranked baseline (GigaPath, 0.802) by 9.6 percentage points, demonstrating capability to identify this subtle pattern that requires careful distinction from sectioning artifacts; representative attention heatmaps are shown in Extended Data Fig.~\ref{ext_fig_heatmap}. Visceral pleural invasion upstages tumors in the TNM classification system and influences adjuvant therapy recommendations, but identification is complicated by the need to distinguish true pleural involvement from indentation or elastic fiber thickening \cite{goldstraw2016iaslc}; PulmoFoundation yielded 0.827 (95\% CI: 0.724--0.915) internally and 0.898 (95\% CI: 0.849--0.943) at Center-H3. Prediction of lymph node metastasis from primary tumor morphology, the single most important prognostic factor in resectable lung cancer that directly determines N stage and the need for adjuvant chemotherapy \cite{goldstraw2016iaslc}, achieved the highest internal AUC among staging tasks at 0.975 (95\% CI: 0.944--0.997), outperforming Virchow2 (0.962), GigaPath (0.951), UNI (0.920), and CHIEF (0.906). Finally, histologic grading of invasive adenocarcinomas based on predominant growth patterns provides independent prognostic information that influences surveillance intensity \cite{nicholson20222021}; PulmoFoundation yielded a macro AUC of 0.902 (95\% CI: 0.820--0.965) for three-tier differentiation grading (Fig.~\ref{fig4}b,c).

PulmoFoundation's consistent performance across resection tasks (Fig.~\ref{fig4}b,c, Extended Data Fig.~\ref{ext_fig_overall_performance}), combined with interpretable attention visualizations that localize diagnostically critical regions (Fig.~\ref{fig4}d,e), supports its utility as a decision-support tool for comprehensive pathologic assessment of surgical specimens. Resection WSIs were processed from digital slide to prediction in a median of 121.7\,s (IQR, 96.0--148.4) on a single NVIDIA GeForce RTX 3090 GPU (Extended Data Tables~\ref{tab:runtime_summary} and \ref{tab:runtime_components}), enabling same-day computational assessment alongside routine specimen processing. Detailed performance metrics for each task are provided in Extended Data Tables \ref{tab:resection_benign_malignant}--\ref{tab:post_iac_grading}.


\subsection*{PulmoFoundation extends H\&E morphology to molecular prediction and prognostic inference}
Beyond histologic diagnosis, comprehensive lung cancer characterization requires assessment of protein expression patterns and molecular alterations that guide treatment selection and predict therapeutic response \cite{liam2020tissue,travis2011pathological}. The ability to predict biomarker status directly from routine H\&E-stained sections could streamline diagnostic workflows and provide rapid preliminary assessments while awaiting ancillary test results \cite{kather2020pan,fu2020pan}. We evaluated PulmoFoundation on nine classification tasks (five IHC markers, one proliferation index, two gene mutations, and tumor mutational burden) and three survival endpoints (Fig.~\ref{fig5}a), achieving averaged macro AUCs of 0.858 (internal) and 0.927 (external) for classification tasks, and averaged C-indices of 0.673 (internal) and 0.778 (external) for survival. PulmoFoundation ranked first in 17 of 19 task-cohort combinations against four pan-cancer baselines; the two exceptions were within 0.2 percentage points of the best-performing model (Fig.~\ref{fig5}b).

\textbf{Diagnostic IHC panel:} IHC panels are routinely employed to distinguish adenocarcinoma from squamous cell carcinoma in diagnostically challenging cases \cite{nicholson20222021,mukhopadhyay2011subclassification,yatabe2019best,bishop2012p40}. The standard diagnostic panel includes adenocarcinoma markers (TTF-1, Napsin-A, CK-7) and squamous markers (P40, P63). For adenocarcinoma-associated markers, PulmoFoundation achieved internal AUCs of 0.923 (95\% CI: 0.868--0.965) for TTF-1, 0.936 (95\% CI: 0.893--0.972) for Napsin-A, and 0.899 (95\% CI: 0.825--0.956) for CK-7, with external validation at Center-H3 showing consistently higher performance (0.974, 0.980, and 0.979, respectively), outperforming the next-best baseline on each marker externally (Figs.~\ref{fig5}b). TTF-1 showed the largest internal baseline gap, exceeding UNI (0.840) by 8.2 percentage points. For squamous cell carcinoma markers, P63 achieved an internal AUC of 0.860 (95\% CI: 0.746--0.950), exceeding the next-best model (UNI, 0.765) by 9.5 percentage points, reflecting PulmoFoundation's capture of lung-specific morphological correlates of squamous differentiation. P40 yielded an internal AUC of 0.802 (95\% CI: 0.660--0.917) with markedly improved external performance (0.952, 95\% CI: 0.888--0.995), outperforming the next-best external baseline (UNI, 0.923) (Fig.~\ref{fig5}b). The consistent performance improvement from internal to external cohorts for squamous markers suggests that PulmoFoundation captures robust morphological correlates of squamous differentiation that generalize well across institutions. At a threshold of $\geq$90\% specificity on external IHC marker cohorts, PulmoFoundation missed fewer positive cases on average than all four pan-cancer baselines (3.8 versus 8.3 for Virchow2, 15.3 for UNI, 27.8 for CHIEF, and 35.3 for GigaPath), demonstrating higher positive-case sensitivity under stringent specificity constraints (Fig.~\ref{fig5}d).

\textbf{Proliferation and molecular biomarkers:} Beyond the diagnostic IHC panel, we evaluated PulmoFoundation's capacity to predict proliferative activity, actionable driver mutations, and tumor mutational burden. Ki-67, a nuclear protein expressed during active phases of the cell cycle, serves as a proliferation index with prognostic significance across multiple tumor types \cite{de2007ki,warth2014tumour}. PulmoFoundation achieved an averaged macro AUC of 0.871 (95\% CI: 0.786--0.943) for predicting Ki-67 expression categories (low $<$10\%, medium 10--50\%, high $>$50\%, following established clinical cutoffs \cite{li2021tumor,xiong2019ki}), exceeding the next-best baseline (GigaPath, 0.825) by 4.7 percentage points and the lowest-ranked model (CHIEF, 0.452) by a wide margin, consistent with lung-specific pretraining capturing morphological correlates of proliferative activity that were less well represented by pan-cancer baselines. For actionable driver mutations, EGFR mutations, which predict response to tyrosine kinase inhibitors and fundamentally alter first-line treatment strategies \cite{soria2018osimertinib,zhou2025changing}, were predicted with an internal AUC of 0.890 (95\% CI: 0.813--0.953), outperforming all four baselines (next-best: CHIEF, 0.862), and an external AUC of 0.751 (95\% CI: 0.682--0.822) in TCGA-LUAD, where PulmoFoundation again ranked first (next-best: UNI, 0.734). STK11, a tumor suppressor gene whose loss is associated with aggressive biology and reduced immunotherapy response \cite{skoulidis2018stk11,pons2021stk11}, was predicted with an AUC of 0.830 (95\% CI: 0.723--0.920), exceeding GigaPath (0.813), Virchow2 (0.803), and UNI (0.782). Tumor mutational burden (TMB), a predictive biomarker for immunotherapy response \cite{samstein2019tumor,chan2019development}, yielded an AUC of 0.712 (95\% CI: 0.618--0.800) in TCGA-LUAD/LUSC (Fig.~\ref{fig5}b); while modest in absolute terms, this exceeded all four baselines (next-best: CHIEF, 0.696), and the moderate performance aligns with prior observations that TMB represents a complex genomic phenotype with subtle morphologic manifestations \cite{jardim2021challenges,anagnostou2022status}.

These H\&E-based biomarker predictions are not intended to replace definitive ancillary testing, which remains the standard for treatment selection. Rather, they offer complementary clinical value by enabling rapid preliminary risk stratification, prioritizing cases for expedited molecular analysis, and providing prognostic insight when tissue is insufficient for comprehensive genomic profiling \cite{schmauch2020deep,kather2020pan}. Detailed performance metrics for each biomarker are provided in Extended Data Tables \ref{tab:post_TTF-1}--\ref{tab:post_STK11}.

\begin{figure}[htbp]
    \centering
    \includegraphics[width=\linewidth]{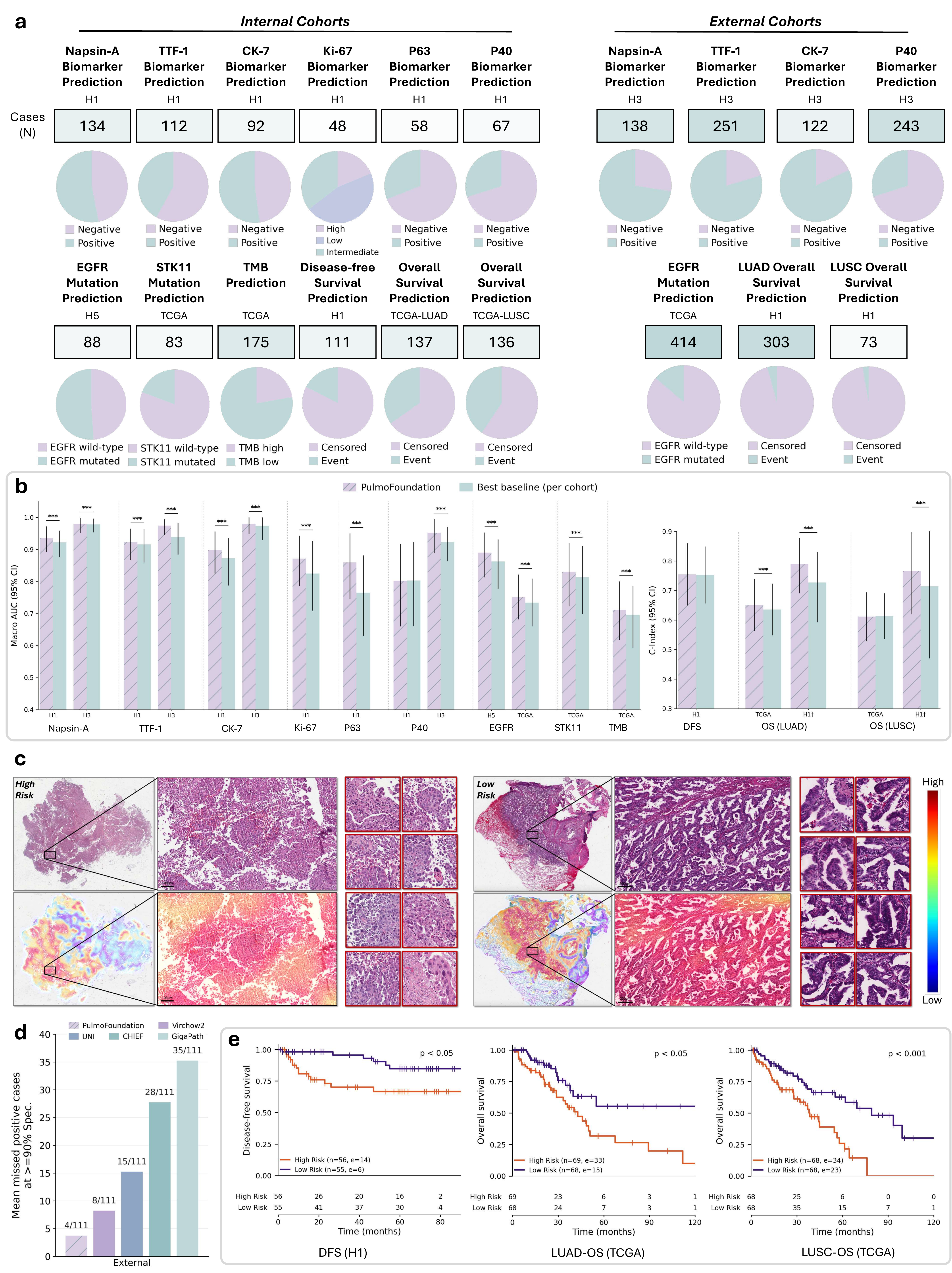}
\end{figure}
\begin{figure}[t]
    \caption{\textbf{PulmoFoundation enables molecular biomarker prediction and survival risk stratification directly from H\&E-stained tissue.}
    \textbf{a}, Cohort overview for all molecular and survival prediction tasks: 12 internal cohorts spanning IHC biomarkers (Napsin-A, TTF-1, CK-7, Ki-67, P63, P40), gene mutations (EGFR, STK11), tumor mutational burden (TMB), and survival (disease-free and overall survival), and 7 external cohorts. Pie charts show class distribution; $N$ denotes patient count per cohort.
    \textbf{b}, Macro AUC for nine classification tasks (left block) and C-index for three survival tasks (right block), comparing PulmoFoundation against the best-performing baseline across internal and external cohorts. Error bars: 95\% bootstrap CI (1,000 resamples). Brackets indicate paired Wilcoxon signed-rank tests comparing PulmoFoundation with the best-performing baseline for each task-cohort column after Holm correction ($^{***}$: $p < 0.001$).
    \textbf{c}, Attention heatmap visualization for survival risk stratification in lung adenocarcinoma. A representative high-risk case (left) and low-risk case (right) are shown, each displaying the original whole-slide image with magnified insets (top row) and the corresponding attention heatmap (bottom row); highest-attention patches are shown at right.
    \textbf{d}, Averaged number of missed positive cases at $\geq$90\% specificity on external IHC marker cohorts, comparing PulmoFoundation against four pan-cancer baselines.
    \textbf{e}, Kaplan-Meier survival curves showing patient stratification by PulmoFoundation risk score for disease-free survival (DFS, Center-H1), LUAD overall survival (TCGA-LUAD), and LUSC overall survival (TCGA-LUSC). At-risk tables and log-rank $p$-values are shown.
    }
    \label{fig5}
\end{figure}

\textbf{Survival risk stratification:} Accurate prognostic assessment is fundamental to lung cancer management, informing treatment intensity and surveillance strategies \cite{goldstraw2016iaslc}. While TNM staging remains the cornerstone of prognosis, it inadequately captures biological heterogeneity within stage-matched tumors \cite{detterbeck2017eighth,lim20188th}. Emerging evidence suggests that tumor morphology contains prognostically relevant information beyond traditional grading, with architectural patterns and stromal features reflecting underlying tumor aggressiveness \cite{yu2016predicting}. We evaluated PulmoFoundation's capacity to predict patient overall survival (OS) and disease-free survival (DFS) in LUAD and LUSC across multiple cohorts (Fig.~\ref{fig5}a).

For overall survival prediction in LUAD, PulmoFoundation achieved C-indices of 0.651 (95\% CI: 0.563--0.739) internally (48 events among 137 patients; median follow-up 23.5 months) and 0.790 (95\% CI: 0.691--0.878) externally at Center-H1 (13 events among 303 patients; median follow-up 61.0 months), outperforming the next-best baseline externally (Virchow2, 0.727) by 6.3 percentage points. For LUSC, PulmoFoundation achieved an internal C-index of 0.612 (95\% CI: 0.529--0.695; 57 events among 136 patients; median follow-up 23.6 months), effectively tied with UNI (0.613); the external LUSC cohort (2 events among 73 patients) yielded a C-index of 0.766 (95\% CI: 0.620--0.897), though this estimate is of limited precision given the low event count (Fig.~\ref{fig5}b). For disease-free survival prediction in surgically resected patients (20 events among 111 patients; median follow-up 46.0 months), PulmoFoundation achieved a C-index of 0.755 (95\% CI: 0.649--0.860), narrowly exceeding Virchow2 (0.753) (Fig.~\ref{fig5}b). Median-split risk stratification yielded clear survival separation for overall survival and disease-free survival, supporting PulmoFoundation's ability to stratify postoperative prognostic risk from H\&E morphology (Fig.~\ref{fig5}e).

To understand the morphological basis of prognostic predictions, we visualized PulmoFoundation's attention patterns on representative LUAD cases stratified by risk (Fig.~\ref{fig5}c). Representative high-risk cases showed high-attention regions overlapping solid growth patterns, micropapillary architecture, and necrotic components, which are histologic features known to portend poor prognosis \cite{travis2011international,warth2012novel,nitadori2013impact}. Conversely, low-risk cases showed attention focused on lepidic and acinar patterns with well-differentiated tumor areas, features associated with favorable outcomes \cite{yoshizawa2011impact,russell2011does}. The convergence between computational attention patterns and established pathology knowledge provides spatial localization of high-risk morphological features that could inform clinical decision-making. Detailed performance metrics are provided in Extended Data Tables \ref{tab:post_survival_luad}--\ref{tab:post_dfs}.

\begin{figure}[htbp]
    \centering
    \includegraphics[width=\linewidth]{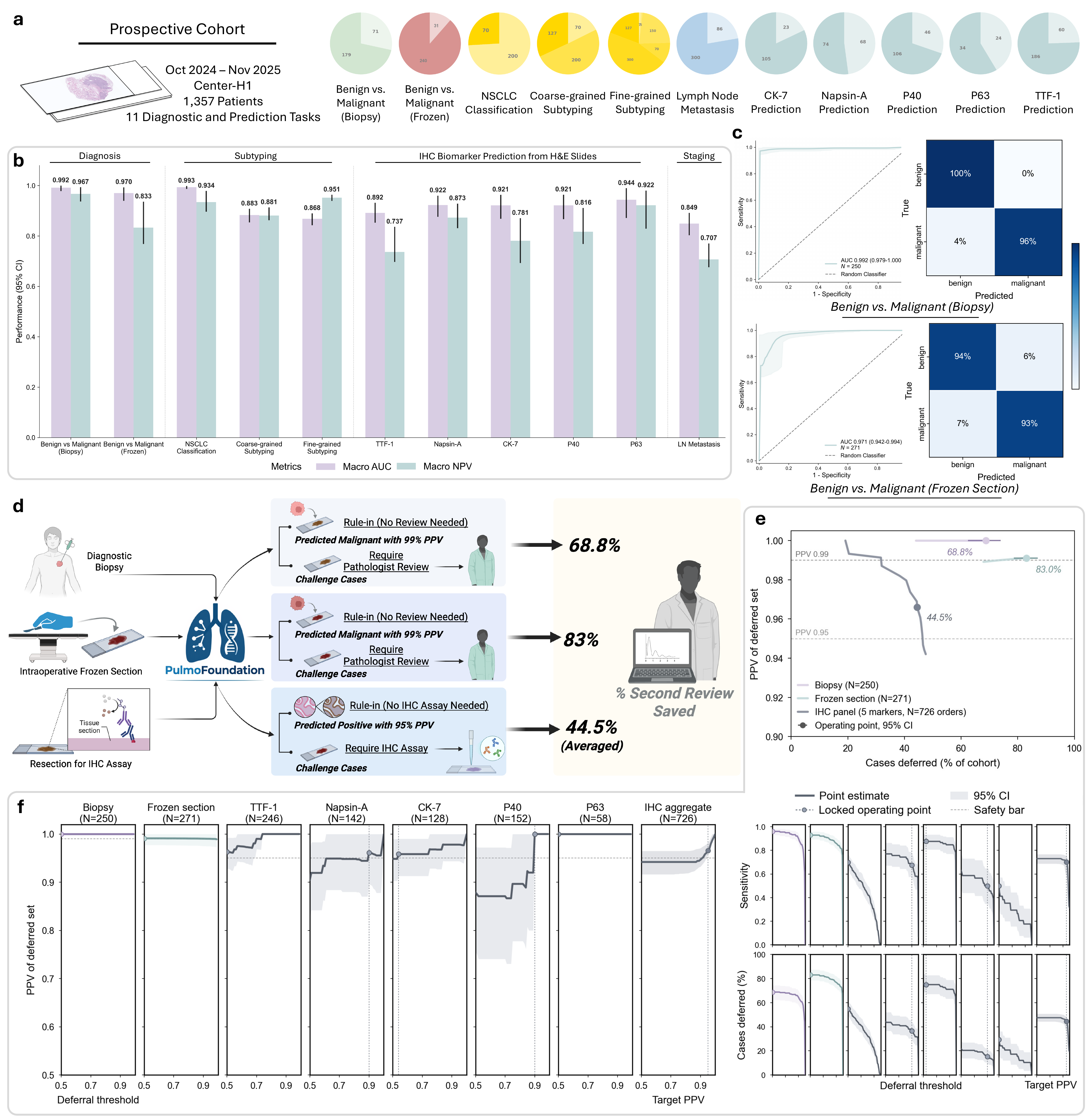}
    \caption{
    \textbf{Prospective validation and clinical triage feasibility of PulmoFoundation in consecutive patients.}
    \textbf{a}, Prospective observational cohort (Center-H1, October 2024 to November 2025; 1,357 consecutive patients) and class distribution across the 11 prospectively evaluated tasks; pie charts show case counts per class.
    \textbf{b}, Macro AUC and Macro NPV across all 11 prospective tasks, grouped by clinical category (Diagnosis, Subtyping, IHC biomarker prediction from H\&E slides, Staging); Macro NPV was computed at class-specific Youden-optimal operating points. Error bars: 95\% bootstrap CI (1,000 non-stratified resamples).
    \textbf{c}, ROC curves (shaded 95\% bootstrap CI) and confusion matrices for the two binary diagnostic tasks: benign versus malignant in diagnostic biopsies (top) and intra-operative frozen sections (bottom); $N$ and AUC annotated in the panel.
    \textbf{d}, Schematic of three prospective triage workflows (diagnostic biopsy, frozen section, and IHC ancillary-test ordering for resection specimens) with pre-specified safety bars (PPV $\geq$ 0.99 for binary diagnostic decisions; PPV $\geq$ 0.95 for each IHC marker before pooling).
    \textbf{e}, Aggregate PPV versus cases-deferred trade-off curves for the three workflows, with locked operating points shown as filled markers. Sample sizes: biopsy ($N=250$), frozen section ($N=271$), pooled IHC panel ($N=726$ stain orders).
    \textbf{f}, Per-task triage decomposition across eight workflow components showing PPV, sensitivity (triage coverage among reference-positive cases or stain orders, not overall diagnostic sensitivity), and deferral fraction. Locked operating points are marked with circles; dashed lines indicate safety bars. Shaded bands show 95\% bootstrap CIs from 1,000 resamples.
    }
    \label{fig6}
\end{figure}

\subsection*{Real-world validation in prospective clinical cohorts}
Prospective validation in consecutive patients represents the definitive benchmark for establishing clinical readiness of AI diagnostic systems \cite{nagendran2020artificial,de2023perspectives,han2024randomised}. Compared with retrospective studies, prospective evaluation in consecutive patients reduces case-selection bias, captures a broader range of diagnostic challenges encountered in routine practice, and tests whether model predictions maintain accuracy in real-world clinical settings \cite{vasey2022decide}. We conducted a registered prospective observational study at Center-H1, enrolling 1,357 consecutive patients undergoing routine surgical evaluation between October 2024 and November 2025 (Fig.~\ref{fig6}a; Extended Data Fig.~\ref{ext_fig_consort_prospective}, CONSORT-style case-flow diagram). All patients meeting predefined eligibility criteria were systematically enrolled without case selection, and all predictions were generated in a blinded manner prior to final pathological diagnosis. Importantly, no model retraining or parameter adjustment was performed during prospective evaluation.

\subsubsection*{PulmoFoundation maintains diagnostic performance across 11 prospective tasks}
We evaluated PulmoFoundation across 11 clinically critical tasks spanning pre-operative, intra-operative, and post-operative stages, organized into four categories: diagnostic classification, histologic subtyping, IHC biomarker prediction, and staging (Fig.~\ref{fig6}a,b). The class distributions reflected the natural prevalence of diagnostic entities in routine clinical practice (Fig.~\ref{fig6}a, Extended Data Table~\ref{tab:prospective}). Across all 11 prospective tasks, PulmoFoundation achieved an average AUC of 0.923, demonstrating broadly consistent performance in consecutive patients (Fig.~\ref{fig6}b; Extended Data Table~\ref{tab:perf_delta}).

\textbf{Performance on diagnostic classification, histologic subtyping, and staging.} Across the six tasks spanning the lung pathology diagnostic workflow, PulmoFoundation maintained discrimination under the natural case mix of consecutive patients. For the two binary diagnostic gates that drive immediate clinical action, PulmoFoundation distinguished benign from malignant lesions in diagnostic biopsies with an AUC of 0.992 (95\% CI: 0.978, 1.000), and in intra-operative frozen sections with an AUC of 0.970 (95\% CI: 0.940, 0.993) (Fig.~\ref{fig6}b,c). Adenocarcinoma versus squamous cell carcinoma classification, the principal therapy-selection branch in NSCLC, reached an AUC of 0.993 (95\% CI: 0.986, 0.998); coarse- and fine-grained histologic subtyping AUCs were 0.883 (95\% CI: 0.854, 0.908) and 0.868 (95\% CI: 0.843, 0.890), respectively, despite inclusion of rare entities at their natural prevalence (Fig.~\ref{fig6}b). For lymph node metastasis prediction from primary tumor tissue, an AUC of 0.849 (95\% CI: 0.803, 0.892) was achieved on consecutive surgical specimens (Fig.~\ref{fig6}b). Across the five non-staging diagnostic classification tasks, the average AUC was 0.941.

To characterize the gap between retrospective internal (AUC 0.975) and prospective (AUC 0.849) lymph-node-metastasis prediction, we performed a case-level failure-mode analysis (Extended Data Table~\ref{tab:prospective_lnm_failure_modes}). False-positive nodal upstaging in pathology-confirmed N0 cases was the dominant error pattern (95/300), whereas 8/86 N+ cases were false negatives. Invasive adenocarcinoma contributed the largest absolute number of false positives (62/95 overall; 62/151 within the subgroup). The false-positive fractions were 18/28 in squamous cell carcinoma, 5/7 in neuroendocrine neoplasm, 4/4 in other specified carcinoma, and 2/3 among cases with histologic subtype unavailable, although estimates for the smaller subgroups were descriptive. AIS and MIA had few errors, with 4/107 false positives among N0 cases and no false negatives in the single N+ case. False positives were also frequent in N0 strata with high-risk primary-tumor morphologies. This pattern is consistent with model sensitivity to aggressive histologic features that do not invariably correspond to actual nodal spread. These results associate the prospective performance gap with identifiable case-composition features rather than a uniform error pattern.

\textbf{Performance on IHC panel prediction from H\&E.} Direct prediction of the diagnostic IHC panel from H\&E morphology is clinically attractive because it could prioritize ancillary-test ordering in surgical specimens. It may also shorten the time to a working diagnosis. PulmoFoundation predicted the complete five-marker NSCLC panel with an averaged AUC of 0.920, comprising TTF-1 (0.892), Napsin-A (0.922), CK-7 (0.921), P40 (0.921), and P63 (0.944) (Fig.~\ref{fig6}b). The prospective five-marker average exceeded the corresponding internal five-marker average (0.920 vs 0.884), although TTF-1 and Napsin-A were lower than their internal estimates. Because the panel is interpreted jointly for adenocarcinoma-versus-squamous differentiation rather than as five independent decisions, single-marker errors are partially absorbed by the rest of the panel, a property we leverage when defining the IHC triage safety bar below.

\subsubsection*{Pre-specified triage workflows for clinical deployment}
Translating discrimination performance into clinical value requires defining workflows where confident model predictions can reduce low-yield diagnostic steps without compromising patient safety. In high-volume surgical pathology laboratories, second-pathologist review of straightforward cases, redundant IHC staining when morphology is diagnostic, and delayed triage of urgent biopsies represent workflow bottlenecks amenable to AI-assisted routing \cite{campanella2019clinical,van2024clinical,gehrung2021triage}. We designed three prospective triage workflows with pre-specified precision thresholds, evaluated entirely on the consecutive-patient cohort described above (Fig.~\ref{fig6}d--f).

\textbf{Triage workflow design.} Moving beyond aggregate discrimination metrics, we defined pre-specified PPV safety bars and selected the lowest score cutoff satisfying each bar for each prospective task, mirroring the AI-assisted screening paradigm established~\cite{campanella2025real}. We defined three triage workflows a priori on the prospective cohort (Fig.~\ref{fig6}d): (i) diagnostic biopsy and (ii) intra-operative frozen section triage, in which a confident malignant prediction fast-tracks the case under primary pathologist sign-out without an additional second-pathologist review; and (iii) surgical-resection IHC ancillary-test triage, in which a confident marker-positive prediction helps pathologists decide whether the corresponding stain needs to be ordered across the five-marker panel. Per-workflow safety bars were locked at PPV $\geq$ 0.99 for binary diagnostic decisions and PPV $\geq$ 0.95 for each IHC marker; the moderate IHC bar reflects the panel-buffering property described above, in which single-marker errors are absorbed by joint interpretation across the differential adenocarcinoma-versus-squamous panel. The triage signal is positioned as a parallel routing recommendation rather than an autonomous diagnostic decision: every case retains the conventional sign-out, and challenge cases proceed under the standard workup.
\textbf{Triage performance and projected clinical impact.} At the locked operating points, the proposed triage workflows would have identified 68.8\% of biopsy cases and 83.0\% of frozen-section cases as eligible for fast-track sign-out, with PPVs of 1.000 and 0.991, respectively. The IHC workflow would have identified 44.5\% of stain orders as eligible for deferral, with a pooled PPV of 0.966 (Fig.~\ref{fig6}e; Extended Data Table~\ref{tab:triage_supp}). Figure~\ref{fig6}e shows the aggregate PPV-deferral trade-off for the three workflows, with each locked operating point marking the highest observed deferral fraction that satisfied its pre-specified PPV safety bar. Figure~\ref{fig6}f decomposes this trade-off by task into PPV, sensitivity, and deferral fraction across the eight workflow components. Here, sensitivity denotes triage coverage among reference-positive cases or stain orders rather than overall diagnostic sensitivity. Bootstrap confidence intervals quantify uncertainty in the operating-point PPV estimates. For the IHC aggregate, per-marker thresholds achieving the target PPV are selected independently and the resulting deferred stain orders are pooled across all five markers. Translated to service-volume terms at Center-H1, the frozen-section operating point corresponds to approximately 1{,}350 second-pathologist reviews per year that need not be staffed; the biopsy operating point corresponds to approximately 1{,}240 expedited oncology referrals enabled without delaying confirmatory review; and the IHC operating point applies to approximately 470 surgical resection cases per year that currently undergo IHC assay, deferring 44.5\% of stain orders across the five-marker panel, with corresponding reagent-cost and turnaround-time savings. Because the PPV safety bars were pre-specified, the resulting score cutoffs can be reported as candidate operating points for future prospective deployment and external recalibration.

Together, the prospective performance figures and the locked triage operating points suggest that PulmoFoundation has the potential to support real-world diagnostic workflows in lung pathology (Fig.~\ref{fig6}a,b,d--f). All results were obtained without model retraining, parameter adjustment, or case selection. Prospective cohort characteristics, retrospective-to-prospective performance comparisons, and triage operating-point details are provided in Extended Data Tables~\ref{tab:prospective}, \ref{tab:perf_delta}, and \ref{tab:triage_supp}. Per-task prospective metrics are provided in Extended Data Tables~\ref{tab:cancer_vs_benign_biopsy_prospective}--\ref{tab:nanfang_lung_ttf1_prospective}.


\begin{figure}[htbp]
    \centering
    \includegraphics[width=\linewidth]{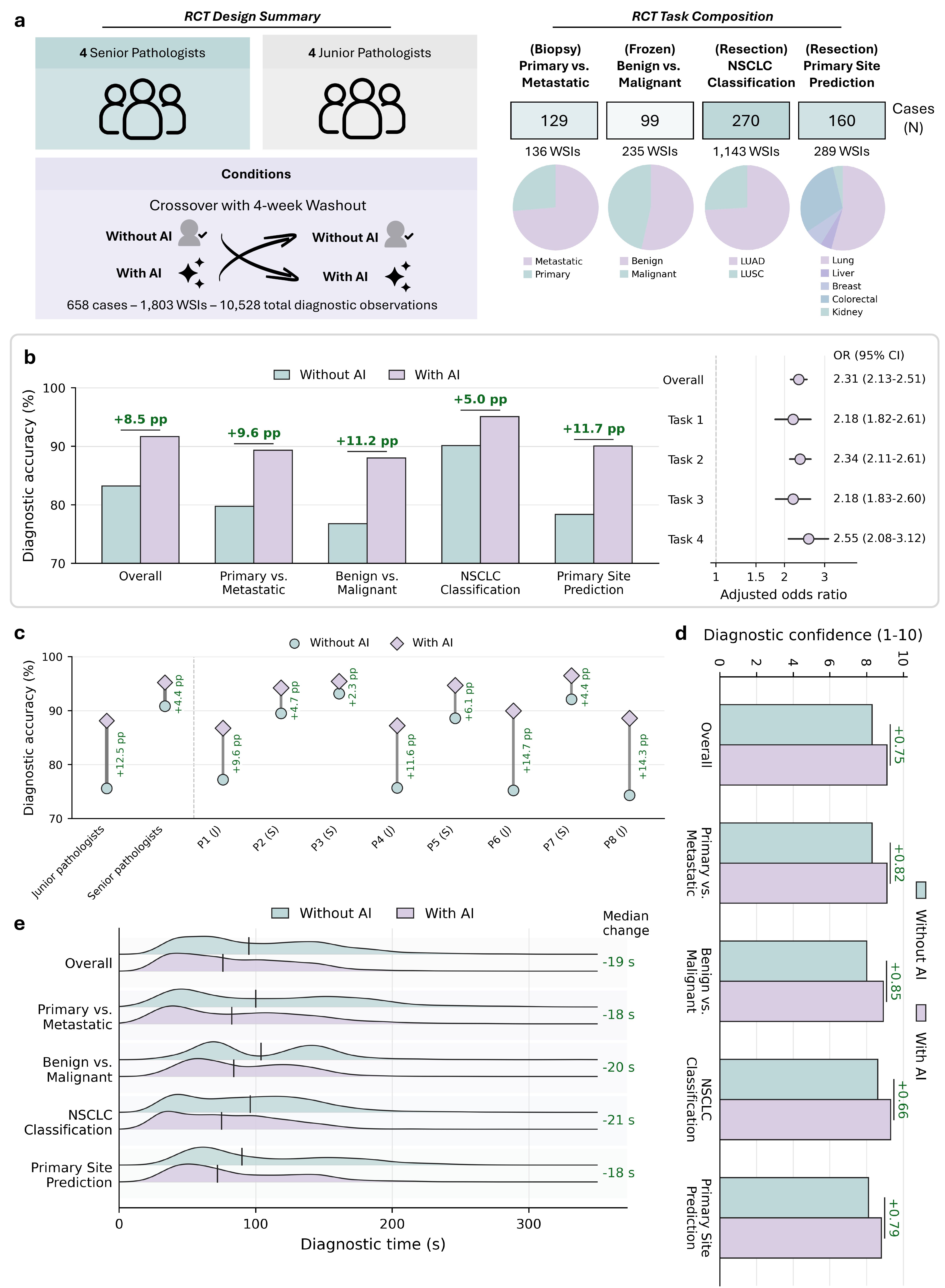}
\end{figure}
\begin{figure}[t]
    \caption{\textbf{Crossover randomized controlled trial of PulmoFoundation-assisted pathology review.}
    \textbf{a}, RCT design and task composition. Eight pathologists (4 senior and 4 junior) reviewed the same case sets with and without AI assistance in a two-period crossover design separated by a 4-week washout. The four RCT tasks comprised primary versus metastatic classification in biopsy specimens (129 cases, 136 WSIs), benign versus malignant classification in diagnostically uncertain frozen-section cases (99 cases, 235 WSIs), NSCLC subtype classification in resection specimens (270 cases, 1,143 WSIs), and primary-site prediction for metastatic tumors in resection specimens (160 cases, 289 WSIs), yielding 658 cases, 1,803 WSIs, and 10,528 total diagnostic observations. 
    \textbf{b}, Primary diagnostic accuracy with and without AI assistance across the overall RCT and each task. Bars show observed diagnostic accuracy. Green labels show absolute accuracy differences in percentage points. The right forest plot shows adjusted odds ratios (ORs) for correct diagnosis with AI assistance and 95\% CIs from generalized estimating equation (GEE) models adjusting for period, task, and experience level; the overall model used pathologist-case pair as the clustering unit, and task-specific models used pathologist as the clustering unit.
    \textbf{c}, Reader-level diagnostic accuracy by experience group and individual pathologist. Green labels show absolute differences in percentage points. The dashed vertical line separates pooled experience-level summaries from individual pathologists.
    \textbf{d}, Diagnostic confidence with and without AI assistance. Horizontal bars show mean confidence on a 1--10 scale, and green labels show adjusted GEE mean differences.
    \textbf{e}, Observation-level diagnostic-time distributions with and without AI assistance. Ridgelines show raw diagnostic-time distributions for observations; vertical ticks mark median times, and green labels show median time changes.}
    \label{fig7}
\end{figure}

\subsection*{Crossover RCT: pathologist-AI interaction evaluation}
To quantify the effect of AI assistance on pathologist decision-making, we conducted a crossover randomized controlled trial, a rigorous design for assessing diagnostic interventions, with eight board-certified pathologists (4 senior with $>$5 years subspecialty experience, 4 junior with $<$5 years, see Extended Data Table~\ref{tab:RCT_pathologist} for details) evaluating 658 cases across four diagnostic tasks spanning biopsy, frozen section, and surgical resection specimens (Extended Data Table~\ref{tab:RCT_cases}; Fig.~\ref{fig7}a; Extended Data Fig.~\ref{ext_fig_consort_rct}, CONSORT-style flow diagram). Pathologists were randomized to the order in which they received AI assistance, and case order was randomized within each period. Each pathologist reviewed the same case sets under AI-assisted and unassisted conditions, separated by a 4-week washout, generating 10,528 total diagnostic observations from 5,264 matched case-reader comparisons. This within-pathologist comparison reduced confounding from case difficulty and differences in pathologist experience.

\subsubsection*{AI assistance improves diagnostic accuracy across tasks and experience levels}
AI assistance significantly improved diagnostic accuracy across all pathologists and all tasks (91.7\% vs 83.2\%, adjusted OR=2.31, 95\% CI: 2.13--2.51, p$<$0.001; Fig.~\ref{fig7}b; Extended Data Table~\ref{tab:primary}). This 8.5 percentage point absolute improvement translates to approximately 85 additional correct diagnoses per 1,000 cases, indicating a clinically meaningful reduction in diagnostic errors with direct implications for patient management. Notably, all eight participating pathologists demonstrated individual improvement with AI assistance (Fig.~\ref{fig7}c; Extended Data Table~\ref{tab:rct_individual_reader_performance}), indicating benefit across varying expertise levels rather than isolated effects in select individuals.

Junior pathologists showed greater pooled absolute gains than senior pathologists (+12.5 vs +4.4 percentage points), suggesting that AI assistance may help bridge the expertise gap and reduce experience-related disparities in diagnostic accuracy (Fig.~\ref{fig7}c). This pattern suggests potential value for training programs and resource-limited settings where subspecialty expertise may be scarce.

The magnitude of AI benefit varied by clinical scenario. The largest task-level effect was observed for post-operative metastasis origin prediction (90.1\% vs 78.4\%, OR=2.55, 95\% CI: 2.08--3.12, p$<$0.001), a challenging differential diagnosis that requires primary-site attribution among metastatic tumors. AI assistance also improved intra-operative frozen-section diagnosis (88.0\% vs 76.8\%, OR=2.34, 95\% CI: 2.11--2.61, p$<$0.001), which had the lowest unassisted pathologist accuracy among the four tasks. In this time-sensitive setting, the final FFPE diagnosis provided the adjudicated endpoint. AI assistance further improved NSCLC subtyping (95.1\% vs 90.1\%, OR=2.18) and primary versus metastatic classification (89.3\% vs 79.7\%, OR=2.18), demonstrating consistent benefit across tasks (Fig.~\ref{fig7}b,c).

\subsubsection*{AI assistance reduces diagnostic time, enhances diagnosis confidence, and improves inter-rater agreement}

Beyond accuracy, AI assistance improved diagnostic efficiency and diagnostic confidence (Fig.~\ref{fig7}d,e; Extended Data Table~\ref{tab:secondary}). Median diagnostic time decreased by 18.3\% (76 seconds vs 95 seconds, time ratio=0.82, 95\% CI: 0.81--0.83, p$<$0.001), representing time savings that could translate to increased throughput or more time for complex cases in high-volume clinical practice. Diagnostic confidence increased with AI assistance (mean 9.1 vs 8.3 on a 10-point scale, difference=+0.75, 95\% CI: +0.71 to +0.79, Cohen's d=0.59), with the largest confidence gains observed for the diagnostically challenging frozen section task (+0.85 points). Diagnostic confidence increased alongside accuracy, with no evidence of a tradeoff between confidence and correctness.

AI assistance also improved diagnostic consistency among pathologists (Extended Data Table~\ref{tab:interrater}). Inter-rater agreement, measured by Fleiss' $\kappa$, increased from 0.55 (95\% CI: 0.51--0.58; moderate agreement) without AI to 0.76 (95\% CI: 0.73--0.79; substantial agreement) with AI assistance ($\Delta\kappa$=+0.21, p$<$0.001). The largest improvement was observed for intra-operative frozen section diagnosis, where agreement improved from fair ($\kappa$=0.37) to substantial ($\kappa$=0.62), representing a shift across two interpretive categories. For NSCLC subtyping and metastasis origin prediction, AI assistance elevated agreement to $\kappa$=0.83 and $\kappa$=0.81, respectively. This increase in agreement suggests that AI assistance may reduce inter-observer variability in diagnostic settings where reproducible classification is important.

\subsubsection*{Automation-bias risk is quantified across AI-assisted observations}

A critical concern with AI-assisted diagnosis is the potential for automation bias, \textit{i.e.}, uncritical acceptance of AI recommendations leading to errors when the model prediction is incorrect. We quantified this risk through three complementary analyses. First, outcome-based classification of all 5,264 AI-assisted observations showed that 4,327 observations (82.2\%) confirmed a correct diagnosis and 477 (9.1\%) changed from an incorrect unassisted diagnosis to a correct assisted diagnosis (Extended Data Table~\ref{tab:ai_utility}). Accuracy loss after AI occurred in 32 observations (0.6\%), including 12 observations (0.2\%) that met the definition of strict erroneous-AI adoption harm. Second, the displayed model prediction was incorrect in 120 of 5,264 AI-assisted observations (2.3\%; Extended Data Table~\ref{tab:rct_wrong_model_behavior}). Within this subset, pathologists produced a correct final diagnosis in 26 observations (21.7\%), while the final diagnosis matched the incorrect displayed prediction in 93 observations (77.5\%). Of these 93 matches, 12 represented a correct-to-incorrect change after AI assistance, whereas the unassisted diagnosis was already incorrect in the other 81 observations. Third, we examined accuracy changes by model-confidence and case-difficulty strata (Extended Data Tables~\ref{tab:rct_confidence_sensitivity} and \ref{tab:rct_difficulty_proxy_sensitivity}). Because independent case-difficulty annotations were unavailable, the proxy was defined as 1 minus unassisted accuracy across the eight pathologists for each task-case pair. Accuracy increased across all model-confidence strata. The largest gain occurred among cases classified as difficult using this proxy. Across all observations, diagnostic improvement was more frequent than accuracy loss under the tested assistance format.

The robustness of these findings was evaluated through multiple sensitivity analyses (Extended Data Table~\ref{tab:rct_carryover_robustness}). Testing for carryover effects revealed no significant Condition $\times$ Period interaction (p=0.18), indicating that AI benefit was consistent across evaluation periods. To address potential memory effects from reviewing the same cases twice, we performed a Period 1-only robustness analysis. This analysis used only the first read of each case, before any pathologist crossed over to the alternate condition. In Period 1, randomized sequence assignment compared Group A with AI assistance against Group B without AI assistance. The Period 1-only AI effect remained similar to the full crossover estimate (OR=2.48, 95\% CI: 2.19--2.81, p$<$0.001), supporting that repeat case exposure did not substantially explain the observed AI-associated accuracy gain.


\section*{Discussion}
In this study, we present PulmoFoundation, a subspecialty-specific foundation model for systematic lung pathology assessment, rigorously validated through retrospective evaluation across 32 internal tasks, external validation in 21 independent cohorts from 8 institutions, and prospective assessment in 1,357 consecutive patients. By pretraining on 88 million lung-specific image patches, PulmoFoundation demonstrates strong performance across key tasks in diagnostic biopsy diagnosis, intra-operative frozen section assessment, and post-operative tumor characterization, molecular profiling, and prognostic stratification, with prospective validation in routine practice supporting its translational potential for surgical pathology workflows.

\noindent\textbf{Prospective validation supports clinical translation beyond retrospective benchmarking.} The prospective validation of PulmoFoundation across 1,357 consecutive patients represents a critical advance beyond the retrospective and external validation paradigms that have characterized prior foundation model evaluations \cite{vorontsov2024foundation,chen2024towards,ma2025generalizable,xu2024whole}. Recent expert consensus has identified this disconnect between benchmark performance and clinical adoption as a fundamental ``reality gap'' in computational pathology \cite{da2026computational,liu2026advancing}. Prospective validation captures the natural case-mix encountered in routine practice and tests generalization to temporal distribution shifts between model development and clinical deployment \cite{vasey2022decide}. Across 11 clinically critical tasks spanning pre-operative, intra-operative, and post-operative stages, PulmoFoundation achieved an average AUC of 0.923. Strong prospective performance was observed for classification of benign and malignant lesions in biopsy (AUC 0.992 prospective vs 0.970 internal), frozen section benign versus malignant classification (0.970 prospective vs 0.986 internal), NSCLC subtyping (0.993 vs 0.955), and the complete IHC biomarker panel (average 0.920 prospective vs 0.884 internal). This performance stability in prospective evaluation is noteworthy, as prospective validation often reveals performance degradation compared to retrospective evaluation, suggesting PulmoFoundation captured robust morphological patterns that generalize across temporal cohorts. Lymph-node-metastasis prediction showed lower prospective performance than internal validation (AUC 0.849 vs 0.975), and failure-mode analysis indicated that this decrease was mainly associated with false-positive node-positive predictions in pathology-confirmed N0 cases with high-risk primary-tumor features (Extended Data Table~\ref{tab:prospective_lnm_failure_modes}). This performance variation reflects the natural spectrum of diagnostic difficulty in consecutive clinical practice. Critically, the overall performance achieved without model retraining demonstrates consistent generalization from development to prospective evaluation cohorts.

\noindent\textbf{Human-AI evaluation shows measurable benefit for pathologist decision support.} The crossover randomized controlled trial provides direct evidence that PulmoFoundation can support pathologist decision-making across key diagnostic decision points in the lung pathology workflow, with an 8.5 percentage point accuracy improvement (adjusted OR=2.31), an 18.3\% time reduction, and improved inter-rater agreement from moderate to substantial levels ($\kappa$: from 0.55 to 0.76) across 10,528 total diagnostic observations. Junior pathologists showed larger absolute gains than senior pathologists (+12.5 vs +4.4 percentage points), suggesting AI may help reduce experience-related disparities. The outcome-based analysis showed that diagnostic improvement after AI (9.1\%) exceeded accuracy loss after AI (0.6\%), with strict erroneous-AI adoption harm occurring in 0.2\% of observations. However, when the model prediction was incorrect (2.3\% of AI-assisted observations), pathologists adopted the erroneous label in 77.5\% of those cases. Overall harm remained low because model accuracy was high, although the high adoption rate among incorrect model predictions highlights the need for future interfaces that help pathologists calibrate trust in AI outputs. Future interface studies should test whether calibrated confidence, uncertainty estimates, or localization cues can further improve pathologist calibration and adoption behavior. Task-level gains were consistent across the four diagnostic scenarios, with the largest effect in post-operative metastasis origin prediction, a differential diagnosis requiring primary-site attribution among metastatic tumors. AI assistance also produced a substantial gain in intra-operative frozen-section diagnosis, a time-sensitive task with the lowest unassisted pathologist accuracy. While this single-institution study demonstrates proof-of-concept, multi-center studies with live workflow deployment across more diagnostic tasks are needed to confirm generalizability across diverse practice settings.

\noindent\textbf{Lung-specific continual pretraining extends capability beyond pan-cancer foundation models.} Direct comparison between PulmoFoundation and four contemporary pan-cancer foundation models (UNI \cite{chen2024towards}, Virchow2 \cite{zimmermann2024virchow2}, GigaPath \cite{xu2024whole}, and CHIEF \cite{wang2024pathology}) across 53 retrospective task cohorts reveals that while all four models achieve adequate performance on fundamental diagnostic distinctions such as benign versus malignant classification and adenocarcinoma versus squamous cell carcinoma discrimination, they do not provide the same breadth of performance as PulmoFoundation across the clinically critical lung pathology assessment continuum (Extended Data Fig.~\ref{ext_fig_overall_performance}). The comparison with Virchow2 serves as a controlled ablation: PulmoFoundation is built upon Virchow2 through lung-specific continual pretraining, directly isolating the contribution of subspecialty-specific adaptation. Because the adaptation introduces only lightweight LoRA parameters atop the frozen Virchow2 backbone, the observed gains cannot be attributed to increased model capacity or architectural differences, but specifically to lung-domain self-supervision. Performance gains concentrated in two capability domains that pan-cancer pretraining does not adequately capture. First, predicting immunohistochemical and molecular biomarkers directly from H\&E morphology, where PulmoFoundation improved over Virchow2 by 12.2\% for P63, 9.1\% for tumor mutational burden, and 8.5\% for Ki-67; these tasks require recognition of subtle morphological correlates specific to lung tumor subtypes. Second, fine-grained morphological grading and invasion assessment, including invasive adenocarcinoma differentiation (8.8\% improvement), lymph node metastasis prediction in frozen sections (8.1\%), and pleural invasion (7.9\%), which demand discrimination of diagnostically critical architectural patterns unique to lung pathology. These findings establish that subspecialty-specific pretraining unlocks clinically important capabilities, particularly biomarker prediction and nuanced morphological grading, that lie beyond the reach of pan-cancer feature representations, while preserving competitive performance on fundamental diagnostic tasks including cross-tissue discrimination.

\noindent\textbf{PulmoFoundation addresses interrelated diagnostic questions across the lung pathology workflow.} The comprehensive evaluation of PulmoFoundation across pre-operative, intra-operative, and post-operative stages addresses a fundamental limitation of prior computational pathology research: task-specific models designed for isolated diagnostic objectives are difficult to scale across real-world pathology workflows that demand integrated assessment of multiple interrelated diagnostic questions. Pathologists do not evaluate specimens for single tasks in isolation; rather, each specimen requires systematic assessment of malignancy status, histologic subtype, invasion patterns, biomarker profiles, staging features, and prognostic characteristics in an integrated diagnostic process. The consistent performance across biopsy, frozen section, and surgical resection specimens demonstrates that a single subspecialty-specific foundation model can address a broad set of diagnostic questions without requiring separate, independently deployed systems for each task. This integrated approach aligns with clinical workflow realities and practical deployment considerations, and positions itself within the broader shift from task-specific models to generalist AI systems \cite{moor2023foundation}. Moreover, the model's ability to adapt to fundamentally different specimen types and clinical contexts suggests that subspecialty-specific pretraining creates versatile feature representations that transfer effectively across the diverse diagnostic scenarios within a single organ system.

\noindent\textbf{Limitations and future directions.} Our study has several limitations that suggest directions for future investigation. First, prospective validation was conducted at a single institution, and while external validation across 21 cohorts from 8 institutions demonstrates geographic generalization, a fully prospective multi-center interventional study comparing pathologist performance with and without AI assistance is necessary to assess real-world clinical impact and workflow integration \cite{rivera2020guidelines}. Second, the dataset composition reflects the specimen-type distribution of a tertiary surgical pathology practice, with surgical resections constituting the majority of WSIs; performance characteristics in settings with different specimen-type distributions remain to be determined. Third, this study evaluated 32 representative tasks across biopsy, frozen section, and surgical resection assessment, but additional clinically important tasks (e.g., PD-L1 scoring, ALK rearrangement detection, surgical margin evaluation) represent natural extensions. PulmoFoundation currently operates on H\&E morphology alone; integration with immunohistochemical, molecular, and clinical data that pathologists routinely synthesize would enable more comprehensive diagnostic support. Fourth, the crossover RCT evaluated a defined AI-assisted review format in which pathologists saw the task-specific displayed prediction during digital WSI review. Future interface studies should test whether calibrated confidence, uncertainty estimates, localization cues, or differential diagnostic context further improve pathologist calibration and adoption behavior. Lastly, the crossover RCT used repeated review of the same cases, consistent with prior randomized crossover evaluation of AI-assisted digital pathology \cite{abdurrachim2025utility}. A 4-week washout and randomized case order were used to reduce recognition effects. The AI-associated effect persisted in the Period 1-only analysis, which excluded all second-period reads and therefore removed within-trial repeat exposure (OR=2.48, 95\% CI: 2.19--2.81, compared with OR=2.31, 95\% CI: 2.13--2.51 in the full crossover analysis). The Condition $\times$ Period interaction was not statistically significant (p=0.18), providing no evidence that the AI-associated effect differed detectably between periods. Together, these analyses did not indicate that repeat exposure materially accounted for the observed accuracy gain.
Additionally, this trial was conducted at a single institution with eight pathologists, and generalizability to other practice settings with different pathologist training backgrounds warrants further investigation.

This study establishes that subspecialty-specific foundation models offer a practical path to achieve consistently high diagnostic performance across extensive lung pathology assessment, as pan-cancer models prove insufficient for tasks requiring fine-grained morphological discrimination. Through rigorous retrospective validation, registered prospective validation, and crossover randomized controlled trial evaluation, this study outlines an evidence path from model development to clinical validation, a gap that remains a central challenge for the field \cite{da2026computational,liu2026advancing}, by demonstrating how pathology foundation models can progress from research tools to decision support systems that augment pathologist expertise across cancer care pathways.


\section*{Methods}\label{methods}
\subsection*{Ethics declarations}
This study was approved by the Ethics Committee of Nanfang Hospital (approval numbers NFEC-2024-535, NFEC-2025-403, and NFEC-2025-419) and conducted in accordance with the Declaration of Helsinki. Informed consent was obtained from all patients enrolled in the prospective validation cohort. All pathologists participating in the crossover RCT provided informed consent. The requirement for informed consent was waived by the institutional review board for retrospective cohorts involving analysis of deidentified, routinely collected clinical data. The retrospective study was registered in the ClinicalTrials.gov (NCT07239297). The prospective validation studies were prospectively registered in the ClinicalTrials.gov for pre-operative and post-operative cohorts (NCT07239063) and intra-operative cohort (NCT07157618). The crossover RCT evaluating pathologist-AI interaction was registered in the ClinicalTrials.gov (NCT07157618).

\subsection*{Dataset curation and cohorts}
\phantomsection\label{sec:methods-dataset-curation}
\noindent\textbf{Overview.} We assembled a large-scale dataset of 66,146 H\&E-stained WSIs from 16 medical centers and public repositories to capture a broad range of lung pathology across biopsy, frozen-section, and surgical-resection specimens. The dataset includes benign lesions, primary lung cancers across major histologic subtypes, and metastatic tumors from diverse primary sites, including colorectal, breast, kidney, and liver cancers. All slides were scanned at $20\times$, $40\times$, or $80\times$ magnification and de-identified before analysis. Per-source slide counts are provided in Extended Data Tables~\ref{tab:pretrain} and \ref{tab:downstream}.

\noindent\textbf{Definitions.} Extended Data Table~\ref{tab:pretrain} lists sources used for lung-specific continual self-supervised pretraining, whereas Extended Data Table~\ref{tab:downstream} lists sources used for downstream task evaluation. In these tables, \textit{Data Source} denotes an anonymized institution or public repository. \textit{Data Cohort} denotes the collection design and specimen type for that source, including retrospective biopsy, retrospective frozen section, retrospective surgical slide, prospective biopsy, prospective frozen section, or prospective surgical slide cohorts. Internal validation denotes a retrospective held-out test set from the same institution as the downstream task development cohort. External validation denotes retrospective evaluation on data from independent institutions that were not used to train the corresponding downstream task head. Prospective validation denotes evaluation on consecutive patients enrolled at Center-H1 after retrospective model development, without model retraining or parameter adjustment. N/S in the \textit{Time Range} column of Extended Data Table~\ref{tab:downstream} indicates that collection dates were not specified in the de-identified retrospective exports; these retrospective cohorts were collected before the prospective study period.

\noindent\textbf{Pretraining cohort.} The pretraining cohort encompassed biopsy ($n=1,592$), frozen-section ($n=8,728$), and surgical-resection ($n=29,183$) WSIs, enabling representation learning across the evaluated specimen types from pre-, intra-, and post-operative stages. PulmoFoundation was pretrained using self-supervised learning on 39,503 WSIs comprising 88 million image patches from 12 geographically diverse sources, including seven medical centers (Center-H1, Center-H2, Center-H3, Center-H4, Center-H6, Center-H8, and Center-H9) and five public repositories: Clinical Proteomic Tumor Analysis Consortium (CPTAC) \cite{edwards2015cptac}, National Lung Screening Trial (NLST) \cite{national2011nlst}, HistAI-Lung \cite{nechaev2025histai}, LC25000 \cite{borkowski2019lung}, and WSSS4LUAD \cite{han2022wsss4luad}. The distribution of WSIs and image patches across sources is shown in Fig.~\ref{fig1}a and Extended Data Table~\ref{tab:pretrain}.

\noindent\textbf{Retrospective training and validation cohorts.} Retrospective downstream evaluation used 26,643 WSIs organized into 32 internal validation cohorts and 21 external validation cohorts from eight independent institutions. Cohort definitions, source mappings, and task-level cohort assignments are provided in Extended Data Table~\ref{tab:downstream}.

\noindent\textbf{Prospective validation cohort.} We conducted a registered prospective observational study at Nanfang Hospital (Center-H1) between October 1, 2024, and November 25, 2025. The study was conducted under the ethics approvals and ClinicalTrials.gov registrations listed above. The final prospective cohort comprised 1,357 patients, including 250 pre-operative biopsy cases, 271 intra-operative frozen-section cases, and 836 post-operative surgical-resection cases. PulmoFoundation was evaluated across 11 clinically critical tasks. 
All predictions were generated in a blinded manner before final pathological diagnosis, without model retraining or parameter adjustment. Detailed eligibility criteria, enrollment procedures, blinding methods, and ground-truth procedures are described in \hyperref[sec:methods-prospective-validation]{Prospective Validation Study Protocol}. Patient-flow accounting is provided in Extended Data Fig.~\ref{ext_fig_consort_prospective}.

\noindent\textbf{Crossover RCT reader-study cohort.} To evaluate pathologist-AI interaction, we assembled a reader-study cohort comprising 658 cases and 1,803 WSIs across four diagnostic tasks: biopsy primary-versus-metastatic classification, frozen-section benign-versus-malignant diagnosis, resection-based NSCLC subtyping, and metastatic-tumor primary-site prediction. Eight board-certified pathologists from Center-H1 participated, including four senior pathologists and four junior pathologists. Each pathologist reviewed the same case set under AI-assisted and unassisted conditions, yielding 10,528 total diagnostic observations and 5,264 matched case-reader comparisons. Reader characteristics, case composition, and trial flow are provided in Extended Data Tables~\ref{tab:RCT_pathologist}, \ref{tab:RCT_cases} and Extended Data Fig.~\ref{ext_fig_consort_rct}; the full randomized crossover design is described in \hyperref[sec:methods-rct]{Crossover Randomized Controlled Trial}.

\subsection*{PulmoFoundation model development}
\phantomsection\label{sec:methods-model-development}
\textbf{Base model architecture.} PulmoFoundation is built upon Virchow2, a state-of-the-art pathology foundation model pretrained on 3.1 million H\&E-stained whole-slide images spanning diverse cancer types \cite{zimmermann2024virchow2}. Virchow2 employs a vision transformer (ViT) architecture with ViT-H/14 consisting of 632 million parameters. Each input patch of
$224\times224$ pixels is divided into $14\times14$ pixel tokens, which are then processed through 32 transformer layers with 1,280 hidden dimensions and 16 attention heads. The model outputs a 1,280-dimensional feature vector from the [CLS] token and individual patch token embeddings.

\noindent To specialize this general-purpose pan-cancer model for lung pathology while preserving its learned representations and generalizability, we employed low-rank adaptation (LoRA) \cite{hu2021lora}, a parameter-efficient fine-tuning (PEFT) method. LoRA introduces trainable low-rank decomposition matrices into specific layers of the pre-trained model while keeping the original weights frozen, thereby minimizing computational costs and catastrophic forgetting of the original pan-cancer knowledge. In PulmoFoundation, LoRA modules were integrated into the fused query-key-value projection and attention output projection modules of each attention layer in all transformer blocks. This design isolates lung-domain learning within the LoRA parameters while leaving the pan-cancer backbone intact, so that performance differences between PulmoFoundation and Virchow2 primarily reflect subspecialty-specific continual pretraining rather than architectural or capacity changes. We configured the rank dimension to 8 with a scaling coefficient ($\alpha$) of 16, yielding approximately 2 million trainable parameters, corresponding to 0.3\% of Virchow2's 632 million total parameters, enabling efficient subspecialty-specific adaptation while maintaining the base model's representation learning capability.

\noindent\textbf{Lung-specific continual pretraining.} PulmoFoundation was pretrained on 39,503 H\&E-stained WSIs comprising 88 million image patches from 12 geographically diverse sources using self-supervised learning (Fig.~\ref{fig1}a). We employed DINO (self-DIstillation with NO labels) \cite{caron2021emerging}, a self-supervised learning strategy which enables representation learning without manual annotations through a teacher-student architecture. In this paradigm, differentially augmented views of each image tile are processed independently by student and teacher networks. The student network is trained to match the teacher's output distribution, driving the emergence of semantically coherent feature representations. The teacher network's weights are updated through an exponential moving average (EMA) of the student weights, providing stable learning targets. During continued pretraining, only foreground tissue patches were used, excluding background regions. Foreground tissue was tessellated at the base level (level 0) using non-overlapping square patches with magnification-adaptive dimensions: $256 \times 256$ pixels for $20\times$ WSIs, $512 \times 512$ pixels for $40\times$ WSIs, and $1024 \times 1024$ pixels for $80\times$ WSIs. The step size was set equal to the patch size. This protocol preserved an approximately matched tissue field of view across scanned magnifications, corresponding to a $40\times$-equivalent reference scale of approximately $0.25~\mu$m/pixel. For each training iteration, two differently augmented views of the same patch were generated and processed by the student and teacher networks. The teacher network's weights were updated via an exponential moving average (EMA) of the student weights, with a momentum coefficient set to 0.9995. Only the LoRA parameters in the student network were trainable, while all other Virchow2 weights remained frozen.  The model was trained using the AdamW optimizer \cite{loshchilov2017decoupled} with $\beta_1=0.9$, $\beta_2=0.999$. The learning rate followed a cosine annealing schedule with linear warmup, starting from 1e-6 with a warmup period of 10,000 iterations. Mixed-precision (FP16) arithmetic was employed to accelerate training and reduce memory consumption. Continued pretraining was conducted using 8 NVIDIA H800 GPUs with a global batch size of 384 for 687,500 iterations. Training took approximately 1,300 GPU-hours.

\noindent\textbf{Fine-tuning for downstream Tasks.} For downstream clinical tasks, we implemented a slide-level prediction pipeline following standard computational pathology workflows (Fig.~\ref{fig1}b). Since WSIs are typically too large to be processed directly by deep learning models (often >100,000 pixels per dimension), the analysis pipeline was divided into three steps: patch-level feature extraction, feature aggregation, and slide-level prediction.

\noindent\textbf{Patch-level feature extraction.} From each WSI in the downstream datasets, we extracted non-overlapping patches at the base resolution (level 0) using the same magnification-adaptive preprocessing protocol: $256 \times 256$ pixel patches from $20\times$ WSIs, $512 \times 512$ pixel patches from $40\times$ WSIs, and $1024 \times 1024$ pixel patches from $80\times$ WSIs. The step size was set equal to the patch size. These settings preserved an approximately matched tissue field of view across scanned magnifications, corresponding to a $40\times$-equivalent reference scale of approximately $0.25~\mu$m/pixel. Each patch was processed through the frozen PulmoFoundation encoder after applying the Virchow2-compatible preprocessing transform, which resized the extracted tile to the fixed encoder input size of $224 \times 224$ pixels. From the final transformer layer, the [CLS] token embedding (1,280 dimensions) and the mean-pooled embeddings of all patch tokens (1,280 dimensions) were concatenated, forming a 2,560-dimensional feature vector for each patch. The encoder remained frozen during downstream task training to leverage the pretrained lung-specific representations.

\noindent\textbf{Feature aggregation.} To aggregate patch-level features into slide-level representations, we used an attention-based multiple instance learning (ABMIL) model \cite{ilse2018attention}. Multiple instance learning treats each WSI as a "bag" of patch instances, where the slide-level label applies to the entire bag but individual patch labels are unknown. The ABMIL module consists of two components: (1) an attention mechanism that assigns an attention weight to each patch feature vector, and (2) a weighted aggregation step that computes a slide-level feature representation as the attention-weighted sum of patch features. Specifically, attention weights were computed as:
\begin{equation*}
a_i = \frac{\exp(w^T \tanh(V h_i))}{\sum_{j=1}^{N} \exp(w^T \tanh(V h_j))}
\end{equation*}
where $h_i$ is the feature vector for patch $i$, $V$ is a learnable weight matrix, $w$ is a learnable attention vector, and $N$ is the total number of patches in the WSI. The slide-level representation was computed as $z = \sum_{i=1}^{N} a_i h_i$. The ABMIL attention mechanism used a hidden dimension of 512.

\noindent\textbf{Task-specific prediction heads and training.} Aggregated feature representations were passed through a single fully connected prediction layer with dropout rate 0.25. For classification tasks, the prediction head produced logits over the task-specific label set; binary tasks were implemented as two-logit softmax classifiers and multiclass tasks as K-logit softmax classifiers, and all classification heads were optimized with cross-entropy loss. For survival prediction tasks, discrete-time survival negative log-likelihood loss was used. Only the ABMIL aggregator and prediction layer were trained for each downstream task, while the PulmoFoundation encoder remained frozen. For each task, data were split at the patient level into training (70\%), validation (10\%), and test (20\%) sets, with stratification by class label where applicable. Models were optimized using Adam \cite{kingma2014adam} with a learning rate of $2\times10^{-4}$ and weight decay of $1\times10^{-5}$. Model checkpoints for classification tasks were selected by validation macro-AUC, with early stopping after 10 consecutive epochs without validation macro-AUC improvement. The maximum number of training epochs was task-specific and set to 25 for the released public TCGA training example. Training was performed on one NVIDIA H800 GPU with a batch size of 1.

\noindent\textbf{Baseline foundation-model comparison protocol.} PulmoFoundation, UNI~\cite{chen2024towards}, and Virchow2~\cite{zimmermann2024virchow2} are patch-level foundation models. For these models, frozen patch embeddings were extracted from each WSI, and the same ABMIL feature aggregator and task-specific prediction head described above were trained for each downstream task. CHIEF~\cite{wang2024pathology} and GigaPath~\cite{xu2024whole} are slide-level foundation models with native slide-level aggregation modules. For these models, we loaded the released pretrained slide-level components and fine-tuned the slide-level aggregation module together with the task-specific prediction head for each downstream task. Across all baseline comparisons, the same task splits, loss functions, optimizer settings, early-stopping rule, and model-selection rule were applied wherever model architecture allowed; the comparison protocol differed only in whether the model natively provided patch-level embeddings or slide-level aggregation.

\subsection*{Clinical Downstream Tasks}
\phantomsection\label{sec:clinical-downstream-tasks}
We designed 32 clinically relevant tasks to systematically evaluate PulmoFoundation across diagnostic assessment, biomarker prediction, and prognostication. Tasks were organized into three clinical stages, \textit{i.e.}, diagnostic biopsy assessment, intra-operative frozen section diagnosis, and post-operative surgical resection analysis, each reflecting real-world diagnostic scenarios encountered in routine pathology practice. All tasks utilized H\&E-stained WSIs as input, with ground truth labels established through clinical pathology reports, expert pathologist review, and ancillary testing (IHC and molecular analysis) where applicable.
Detailed results are reported in Extended Data Tables \ref{tab:biopsy_benign_malignant}-\ref{tab:post_dfs}. Task cohorts were constructed separately for each diagnostic objective; therefore, a patient or case could contribute to more than one task when the corresponding ground-truth label was available, and per-task counts should not be summed across tasks.
For each eligible case, task-relevant digitized H\&E-stained WSIs available for the corresponding diagnostic objective were included. For surgical resection tasks, this meant tumor-containing H\&E slides relevant to the endpoint. For biopsy and frozen-section material, cases typically contributed one digitized H\&E WSI; when additional task-relevant digitized sections were available, they were included for the corresponding diagnostic objective rather than manually selecting a single representative slide.
For tasks where the diagnostic question is inherently per-patient (e.g., histologic subtyping, tumor grading), task-relevant WSIs from the same patient were concatenated and the model produced a single patient-level prediction; for tasks where each slide carries its own independent diagnostic label (e.g., benign versus malignant classification), evaluation was performed at the slide level.

\subsubsection*{Biopsy assessment tasks}
\noindent\textbf{Classification of benign and malignant lesions (Biopsy).} Accurate distinction between benign and malignant lesions in biopsy specimens is the most fundamental diagnostic decision that determines whether patients proceed to surgical resection, undergo further molecular testing, or continue surveillance. This binary classification task distinguishes benign pulmonary lesions from malignant tumors in biopsy specimens. The internal cohort was curated from Center-H1 and comprised 797 cases (797 WSIs): 397 benign cases (including inflammatory lesions, hamartomas, organizing pneumonia, and benign epithelial proliferations) and 400 malignant cases (including all primary and metastatic lung malignancies). Data were split into training (70\%), validation (10\%), and test (20\%) sets using slide-level stratification. External validation was performed on an independent cohort from Center-H4 consisting of 400 WSIs (200 benign, 200 malignant). Prospective validation enrolled 250 consecutive biopsy cases from Center-H1 (October 2024 - August 2025), comprising 71 benign and 179 malignant lesions. Ground truth was established by senior pathologist consensus diagnosis combined with clinical follow-up and subsequent surgical resection confirmation where available.

\noindent\textbf{Classification of primary and metastatic tumors (Biopsy).} Distinguishing primary lung cancer from metastatic disease fundamentally alters treatment strategy, as metastatic tumors require systemic therapy directed at the primary site while primary lung cancers follow lung-specific treatment algorithms. This binary classification task distinguishes primary lung cancers from metastatic tumors to the lung in malignant biopsy specimens. The internal cohort from Center-H1 comprised 537 cases (568 WSIs): 402 primary lung cancers (adenocarcinoma, squamous cell carcinoma, small cell carcinoma, and other primary lung malignancies) and 135 metastatic tumors from diverse primary sites, including colorectal, breast, liver, and kidney. Data were split at a 7:1:2 ratio using patient-level stratification. Ground truth was established through combined clinical history, IHC panel results, and molecular profiling where available.

\noindent\textbf{Histologic classification of lung biopsy specimens.} Accurate histologic subtyping guides selection of appropriate molecular testing panels and informs first-line therapy decisions, as treatment recommendations differ substantially between adenocarcinoma, squamous cell carcinoma, and small cell carcinoma. This multi-class classification task categorizes malignant lung biopsies into four clinically relevant subtypes following IASLC diagnostic criteria: (1) invasive adenocarcinoma, (2) squamous cell carcinoma, (3) small cell carcinoma, and (4) other malignant neoplasms (including large cell neuroendocrine carcinoma, adenosquamous carcinoma, and poorly differentiated carcinomas not otherwise specified). The internal cohort from Center-H1 comprised 517 cases (541 WSIs): 129 invasive adenocarcinoma cases, 124 squamous cell carcinoma cases, 64 small cell carcinoma cases, and 200 cases of other malignancies. Data were split at a 7:1:2 ratio using patient-level label-stratified sampling. We specifically excluded adenocarcinoma in situ and minimally invasive adenocarcinoma from this task, as these entities are rarely definitively diagnosed on small biopsy specimens in clinical practice and their presence typically requires assessment of the entire tumor-stromal interface visible only in resection specimens. Ground truth was established by expert pathologist diagnosis supplemented with IHC when clinically indicated.

\noindent\textbf{Prediction of CK5/6 expression (Biopsy).} CK5/6 immunohistochemistry is a key ancillary test for distinguishing squamous cell carcinoma from adenocarcinoma when morphology is ambiguous, helping to guide appropriate molecular testing and therapy selection while conserving limited biopsy tissue. This binary classification task predicts CK5/6 IHC expression status directly from H\&E-stained biopsy morphology. CK5/6 is a basal/squamous differentiation marker routinely used to distinguish squamous cell carcinoma (typically CK5/6-positive) from adenocarcinoma (typically CK5/6-negative) in diagnostically challenging cases. The internal cohort comprised 606 cases (618 biopsy WSIs) with corresponding CK5/6 IHC results: 250 CK5/6-positive cases and 356 CK5/6-negative cases. CK5/6 positivity was defined as cytoplasmic staining in $\geq10\%$ of tumor cells, following standard clinical laboratory thresholds. Data were split at a 7:1:2 ratio with patient-level stratification. Ground truth was established by institutional clinical IHC results reviewed and validated by study pathologists.

\subsubsection*{Intra-operative frozen section diagnosis tasks}
\noindent\textbf{Classification of benign and malignant lesions in frozen sections.} Intra-operative frozen section diagnosis of malignancy directly determines surgical extent within minutes, influencing decisions regarding lobectomy versus wedge resection and the need for lymph node dissection. This binary classification task distinguishes benign pulmonary lesions from malignant tumors in intra-operative frozen section specimens. The internal cohort from Center-H1 comprised 562 cases (1,179 WSIs): 262 benign cases (including reactive lymph nodes, inflammatory lesions, and benign proliferations) and 300 malignant cases (primary lung cancers and metastases). Data were split at a 7:1:2 ratio using patient-level stratification. External validation was performed on two independent cohorts: Center-H2 with 449 cases (514 WSIs) (310 benign, 204 malignant) and Center-H10 with 169 cases (169 WSIs) (43 benign, 126 malignant). Prospective validation enrolled 271 consecutive cases (380 WSIs) from Center-H1 (August 2025 - November 2025), comprising 31 benign (44 WSIs) and 240 malignant (336 WSIs) cases. Ground truth was established by final diagnosis from permanent surgical resection specimens reviewed by senior pathologists, which serves as the clinical reference standard for frozen section accuracy.

\noindent\textbf{Classification of diagnostically uncertain frozen section cases.} Diagnostic uncertainty in frozen sections leads to deferred diagnosis and suboptimal intra-operative surgical decisions, with uncertain cases requiring repeat surgery or compromising oncologic outcomes. This binary classification task evaluates model performance on frozen section cases where pathologists expressed explicit diagnostic uncertainty during real-time intra-operative evaluation. From the Center-H1 frozen section cohort, we identified 494 cases (1,188 WSIs) where the original intra-operative frozen section report contained qualifying language indicating diagnostic uncertainty, including phrases such as: ``likely malignant", ``suspicious for malignancy", ``cannot exclude malignancy", ``favor benign/malignant", ``defer to permanent sections", or ``recommend permanent section correlation for definitive diagnosis." These cases represent scenarios where tissue artifacts (ice crystal formation, compression artifacts, cautery effect), limited sampling, time pressure, or ambiguous morphology prevented pathologists from rendering confident diagnoses. The cohort comprised 262 cases ultimately diagnosed as benign and 232 as malignant on permanent sections. Ground truth was established by the final diagnosis from permanent surgical resection specimens, which underwent standard formalin fixation, paraffin embedding, and expert pathologist review, supplemented with IHC when clinically indicated. This task was designed retrospectively, with data split at a 7:1:2 ratio using patient-level stratification.

\noindent\textbf{Classification of early-stage adenocarcinoma subtypes.} Distinguishing adenocarcinoma in situ and minimally invasive adenocarcinoma from invasive adenocarcinoma during frozen section evaluation determines whether patients can safely undergo limited sublobar resection or require anatomic lobectomy. This three-class classification task categorizes small pulmonary nodules suspicious for adenocarcinoma into: (1) adenocarcinoma in situ (AIS), (2) minimally invasive adenocarcinoma (MIA), and (3) invasive adenocarcinoma (IAC). These diagnostic categories follow the 2021 WHO classification and the IASLC/ATS/ERS international multidisciplinary classification criteria. The internal cohort from Center-H1 comprised 432 cases (919 frozen section WSIs): 141 AIS cases, 91 MIA cases, and 200 IAC cases. Data were split at 7:1:2 ratio with patient-level label-stratified sampling. Ground truth was established by final diagnosis on permanent sections, where complete tumor evaluation allows definitive assessment of invasion extent. For frozen sections, pathologists evaluate limited tissue samples and must infer invasion status, which can be particularly challenging given freezing artifacts that obscure cytologic detail. This distinction critically guides surgical decision-making during frozen section evaluation.

\noindent\textbf{Prediction of lymph node metastasis from primary tumor frozen sections.} Assessment of regional lymph nodes during frozen section examination informs the extent of intra-operative lymphadenectomy and influences staging-based treatment decisions \cite{li2016intraoperative,ortiz2025diagnostic}. This binary prediction task evaluates model ability to predict lymph node metastasis status by analyzing primary tumor morphology in frozen section specimens. The input consists of H\&E-stained frozen sections of the primary tumor. Ground truth was established by final pathological assessment of all resected regional lymph nodes from the surgical specimen. The internal cohort from Center-H1 comprised 321 cases (738 frozen section WSIs of primary tumor tissue): 250 node-negative (N0) and 71 node-positive (N+) cases. Data were split at a 7:1:2 ratio using patient-level stratification. The cohort included lymph nodes from hilar, interlobar, and mediastinal stations 2-14 according to the International Association for the Study of Lung Cancer lymph node map. Ground truth reflects actual nodal status determined by pathologist examination of lymph node specimens.

\subsubsection*{Post-operative surgical resection tasks}
\noindent\textbf{Classification of benign and malignant lesions (Surgical resection).} While most surgical resections are performed for suspected malignancy, definitive pathologic confirmation remains essential for treatment planning and accurate clinical documentation. This binary classification task distinguishes benign pulmonary lesions from malignant tumors in post-operative surgical resection specimens. The internal cohort from Center-H1 comprised 600 cases (600 WSIs): 300 benign cases (including organizing pneumonia, hamartomas, inflammatory pseudotumors, sclerosing pneumocytoma, and reactive lymphoid hyperplasia) and 300 malignant cases (primary lung cancers and metastases). Data were split at a 7:1:2 ratio using slide-level stratification. Ground truth was established by senior pathologist consensus diagnosis based on comprehensive histologic evaluation and clinical-pathologic correlation.

\noindent\textbf{Classification of primary and metastatic tumors (Surgical resection).} Determining primary versus metastatic origin in surgical specimens establishes appropriate staging systems (TNM for primary lung cancer versus stage IV disease for metastases) and guides adjuvant therapy recommendations. This binary classification task distinguishes primary lung cancers from metastatic tumors to the lung in malignant surgical resection specimens. The internal cohort from Center-H1 comprised 846 cases (1,422 WSIs): 389 primary lung cancers (all histologic subtypes) and 457 metastatic tumors from diverse primary sites including colorectal, breast, kidney, and liver. Data were split at 7:1:2 ratio with patient-level stratification. External validation was performed on two independent cohorts: Center-H5 with 779 cases / 1,307 WSIs (434 primary, 345 metastatic) and Center-H7 with 493 cases / 493 WSIs (237 primary, 256 metastatic). Ground truth was established through combined clinical history, imaging correlation, IHC panel results, and molecular profiling where available.

\noindent\textbf{Prediction of primary site in metastatic tumors (Surgical resection).} Identifying the tissue of origin for metastatic tumors is critical for selecting site-specific systemic therapy regimens and determining appropriate multidisciplinary management. This four-class classification task identifies the primary site of origin for metastatic tumors found in the lung, categorizing them as metastases from: (1) colorectal cancer, (2) breast cancer, (3) renal cell carcinoma, or (4) hepatocellular carcinoma. The internal cohort from Center-H1 comprised 693 cases (1,183 WSIs) of primary and metastatic tumors: 393 primary lung malignancies, 186 colorectal, 55 breast, 25 renal cell, and 34 hepatocellular. Data were split at a 7:1:2 ratio with label-stratified sampling. External validation cohorts included: Center-H5 with 525 cases / 967 WSIs (273 lung, 141 colorectal, 63 breast, 43 renal, 5 hepatocellular) and Center-H7 with 423 cases / 423 WSIs (237 lung, 96 colorectal, 50 breast, 30 renal, 10 hepatocellular). Ground truth was established through integrated review of clinical history, imaging findings, pathology reports, IHC panels, and molecular results where available.

\noindent\textbf{Classification of NSCLC histologic subtypes (Surgical resection).} The distinction between adenocarcinoma and squamous cell carcinoma is one of the most consequential diagnostic decisions in lung cancer, as these subtypes exhibit distinct molecular profiles, treatment responses, and contraindications to specific therapies. This binary classification task distinguishes the two major histologic subtypes of NSCLC: adenocarcinoma versus squamous cell carcinoma. The internal cohort from Center-H1 comprised 599 cases (599 WSIs): 300 adenocarcinomas and 299 squamous cell carcinomas. Data were split at a 7:1:2 ratio with patient-level stratification. External validation was performed on two cohorts: Center-H5 with 294 cases / 517 WSIs (273 adenocarcinoma, 21 squamous) and TCGA-LUAD/LUSC with 1,053 WSIs (541 adenocarcinoma from TCGA-LUAD, 512 squamous from TCGA-LUSC). Prospective validation enrolled 270 consecutive NSCLC cases (1,143 WSIs) from Center-H1 (200 adenocarcinoma, 70 squamous). Ground truth was established by expert pathologist diagnosis following the WHO 2021 classification criteria, supplemented with IHC in morphologically ambiguous cases.

\noindent\textbf{Differential diagnosis: coarse-grained classification (Surgical resection).} Distinguishing invasive adenocarcinoma from early-stage adenocarcinomas and squamous cell carcinoma informs surgical adequacy assessment, prognosis, and follow-up intensity. This three-class classification task distinguishes: (1) squamous cell carcinoma, (2) invasive adenocarcinoma, and (3) adenocarcinoma in situ (AIS). The internal cohort from Center-H1 comprised 599 cases / 599 WSIs: 199 squamous cell carcinoma, 200 invasive adenocarcinoma, and 200 AIS. Data were split at a 7:1:2 ratio with label-stratified sampling. External validation was performed on a cohort from Center-H3 with 150 cases / 511 WSIs (37 squamous, 100 invasive adenocarcinoma, 13 AIS). Prospective validation enrolled 397 consecutive cases / 1,673 WSIs from Center-H1 (70 squamous, 200 invasive adenocarcinoma, 127 AIS). Ground truth was established by expert pathologist evaluation following the WHO 2021 classification with comprehensive tumor sampling to confirm absence of invasion in AIS cases.

\noindent\textbf{Differential diagnosis: fine-grained classification (Surgical resection).} Comprehensive classification of all major histologic subtypes reflects the full diagnostic complexity pathologists face and determines subtype-specific treatment algorithms and prognostic counseling. This six-class classification task encompasses the complete spectrum of common lung malignancies following WHO 2021 classification: (1) squamous cell carcinoma, (2) invasive adenocarcinoma, (3) minimally invasive adenocarcinoma (MIA), (4) adenocarcinoma in situ (AIS), (5) neuroendocrine carcinomas, and (6) adenosquamous carcinoma. The internal cohort from Center-H1 comprised 694 cases (1,078 WSIs): 123 squamous cell carcinoma, 150 invasive adenocarcinoma, 121 MIA, 150 AIS, 131 neuroendocrine carcinomas, and 19 adenosquamous carcinoma. Data were split at a 7:1:2 ratio with label-stratified sampling. External validation was performed on a cohort from Center-H3 with 258 cases / 767 WSIs (37 squamous cell carcinoma, 100 invasive adenocarcinoma, 100 MIA, 13 AIS, 5 neuroendocrine carcinomas, and 3 adenosquamous carcinoma). Prospective validation enrolled 509 consecutive cases (1,802 WSIs) from Center-H1 (70 squamous cell carcinoma, 150 invasive adenocarcinoma, 150 MIA, 127 AIS, 5 neuroendocrine carcinomas, and 7 adenosquamous carcinoma). Ground truth was established by expert pathologist evaluation following the WHO 2021 classification, with IHC for neuroendocrine markers and proliferation markers used for neuroendocrine tumor subtyping.

\noindent\textbf{Vascular tumor thrombus detection.} The presence of vascular tumor thrombus is an independent adverse prognostic factor that upstages tumors and identifies patients at high risk for hematogenous metastasis who may benefit from more aggressive adjuvant therapy. This binary classification task detects the presence of vascular tumor thrombus, defined as tumor cells within blood vessel or lymphatic vessel lumens. The internal cohort from Center-H1 comprised 226 cases (517 WSIs): 76 cases with vascular tumor thrombus and 150 cases without. Vascular tumor thrombus was identified on H\&E sections as tumor cell clusters within endothelium-lined spaces, distinct from tumor cells compressing or surrounding vessels. Cases with ambiguous vascular involvement were adjudicated by two senior pathologists with IHC staining for endothelial markers when necessary. Data were split at a 7:1:2 ratio with patient-level stratification. External validation was performed on a cohort from Center-H3 with 249 cases / 807 WSIs (49 positive, 200 negative). Ground truth was established by systematic pathologist review of all tumor-containing slides with focused examination of peritumoral vessels.

\noindent\textbf{Perineural invasion detection.} Perineural invasion indicates aggressive tumor biology with increased risk of local recurrence and reduced survival, influencing decisions regarding adjuvant therapy and surveillance intensity. This binary classification task detects perineural invasion, defined as tumor cells infiltrating along or surrounding nerve structures. The internal cohort from Center-H1 comprised 186 cases (410 WSIs): 36 cases with perineural invasion and 150 cases without. Perineural invasion was identified as tumor cells in contact with nerves, forming a crescent or circumferential pattern around nerve bundles, or tumor cells tracking along nerve sheaths. Nerves were identified by their characteristic morphology (ovoid bundles with wavy spindle cells and central axons) and confirmed with S-100 immunostaining in ambiguous cases. Data were split at a 7:1:2 ratio with patient-level stratification. External validation was performed on a cohort from Center-H3 with 114 cases / 366 WSIs (14 positive, 100 negative). Ground truth was established by systematic pathologist review with particular attention to peritumoral stroma, as perineural invasion typically occurs at the invasive front.

\noindent\textbf{Spread through air spaces (STAS) detection.} Detection of STAS is critical for accurate classification (upgrading non-invasive lesions to invasive adenocarcinoma), prognosis (predicting recurrence risk), and surgical planning (influencing margin adequacy assessment). This binary classification task detects spread through air spaces (STAS), defined as tumor cells spreading within air spaces in the lung parenchyma beyond the edge of the main tumor mass. The internal cohort from Center-H1 comprised 274 cases (569 WSIs) of adenocarcinomas: 150 cases with STAS and 124 cases without. STAS was identified as micropapillary clusters, solid nests, or single cells within alveolar spaces beyond the tumor edge, requiring distinction from sectioning artifacts (artificially detached tumor fragments), atelectatic alveoli with entrapped tumor, or cells within lymphatic spaces (which have endothelial lining). Multiple sections from the tumor periphery were examined to confirm true STAS versus artifact. Data were split at a 7:1:2 ratio with patient-level stratification. Ground truth was established by consensus review of two pathologists examining all slides at the tumor-lung interface.

\noindent\textbf{Pleural invasion detection.} Visceral pleural invasion upstages tumors in the TNM classification system (affecting T stage), independently predicts worse survival, and influences adjuvant therapy recommendations. This binary classification task detects visceral pleural invasion, defined as tumor extension to or through the visceral pleura. The internal cohort from Center-H1 comprised 356 cases (1,192 WSIs): 206 cases with pleural invasion and 150 cases without. Pleural invasion was classified according to IASLC criteria: PL0 (tumor not reaching pleural surface), PL1 (tumor invading beyond elastic layer but not reaching pleural surface), PL2 (tumor reaching pleural surface), and PL3 (tumor invading parietal pleura or chest wall). For this task, we combined PL1, PL2, and PL3 as positive cases. Elastic layer staining (Verhoeff-van Gieson or Movat pentachrome) was performed on representative sections to confirm elastic layer invasion in ambiguous cases. Data were split at a 7:1:2 ratio with patient-level stratification. External validation was performed on a cohort from Center-H3 with 244 cases / 789 WSIs (44 positive, 200 negative). Ground truth was established by pathologist evaluation of tumor-pleura relationship on H\&E with elastic stain confirmation.

\noindent\textbf{Prediction of lymph node metastasis from primary tumor morphology (surgical resection).} Lymph node metastasis status is the most important prognostic factor in resectable lung cancer, directly determining pathologic N stage (N0 vs N1 vs N2) and the need for adjuvant chemotherapy~\cite{goldstraw2016iaslc}. This binary prediction task evaluates model ability to predict pathological N staging by analyzing primary tumor morphology in surgical resection specimens. The input consists of H\&E-stained slides of the primary tumor. Ground truth was established by systematic pathological examination of all resected regional lymph nodes with cytokeratin immunohistochemistry performed selectively to identify small metastatic deposits. The internal cohort from Center-H1 comprised 440 cases (737 WSIs of primary tumor tissue): 352 node-negative (N0) and 88 node-positive (N+). Data were split at a 7:1:2 ratio with patient-level stratification. Prospective validation enrolled 386 consecutive cases (1,770 WSIs of primary tumor tissue) from Center-H1 surgical resections (300 N0, 86 N+). The cohort included patients with lymph nodes examined from all regional stations (hilar, interlobar, mediastinal stations 2-14 per IASLC lymph node map). Ground truth reflects actual nodal status determined by pathologist examination of lymph node specimens, not the primary tumor slides used as model input.

\noindent\textbf{Invasive adenocarcinoma differentiation grading.} Histologic grade provides independent prognostic information and helps stratify patients for adjuvant therapy decisions and surveillance strategies. This three-class classification task grades invasive adenocarcinomas as well differentiated, moderately differentiated, or poorly differentiated based on predominant architectural patterns and cytologic features following the WHO 2021 classification. The internal cohort comprised 429 cases (1,425 invasive adenocarcinoma WSIs): 44 well differentiated, 289 moderately differentiated, and 96 poorly differentiated. Grading was based on a comprehensive evaluation of predominant growth patterns: lepidic and acinar patterns are considered low-grade; papillary patterns are intermediate-grade; solid and micropapillary patterns are high-grade. Data were split at a 7:1:2 ratio with label-stratified sampling. Ground truth was established by consensus grading by two pathologists following the WHO criteria, with semiquantitative assessment of the percentage of each histologic pattern.

\noindent\textbf{TTF-1 biomarker prediction.} Thyroid transcription factor 1 (TTF-1) is the most specific marker for pulmonary adenocarcinoma and is essential for distinguishing primary lung adenocarcinoma from metastatic adenocarcinomas of other origins in morphologically ambiguous cases. This binary classification task predicts TTF-1 IHC expression status directly from H\&E-stained morphology. Concretely, TTF-1 is a nuclear transcription factor highly specific for pulmonary adenocarcinoma and thyroid carcinoma, expressed in most lung adenocarcinomas but rarely in squamous cell carcinomas. The internal cohort from Center-H1 comprised 558 cases (774 surgical resection WSIs) with available TTF-1 IHC: 232 TTF-1-positive cases ($\geq$1\% nuclear staining) and 326 TTF-1-negative cases. External validation was performed on a cohort from Center-H3 with 251 cases / 774 WSIs (200 positive, 51 negative). Prospective validation enrolled 246 consecutive cases (1,056 WSIs) from Center-H1 (186 positive, 60 negative). Data were split at a 7:1:2 ratio for internal cohorts. Ground truth was established by institutional clinical IHC results.

\noindent\textbf{Napsin-A biomarker prediction.} Napsin-A serves as a highly specific marker for lung adenocarcinoma (particularly lepidic and acinar patterns) and renal cell carcinoma, supporting diagnostic classification when combined with other markers. This binary classification task predicts Napsin-A (aspartic proteinase) immunohistochemical expression. Napsin-A is a cytoplasmic enzyme expressed in a large fraction of lung adenocarcinomas, particularly those with lepidic or acinar patterns, and in renal cell carcinomas. The internal cohort from Center-H1 comprised 668 cases (821 WSIs): 354 Napsin-A-positive and 314 Napsin-A-negative. External validation included 138 cases (446 WSIs) from Center-H3 (100 positive, 38 negative). Prospective validation enrolled 142 consecutive cases / 700 WSIs (74 positive, 68 negative). Ground truth was established by clinical IHC using polyclonal antibodies or monoclonal clones.

\noindent\textbf{CK-7 biomarker prediction.} CK-7 expression patterns help distinguish lung adenocarcinomas from squamous cell carcinomas and colorectal metastases, aiding in determination of tumor origin. This binary classification task predicts cytokeratin 7 (CK-7) expression. CK-7 is an intermediate filament protein expressed in most adenocarcinomas of lung, breast, ovary, and endometrium, but not in most squamous cell carcinomas or colorectal adenocarcinomas. The internal cohort comprised 460 cases (635 WSIs) from Center-H1: 242 CK-7-positive and 218 CK-7-negative. External validation included 122 cases (377 WSIs) from Center-H3 (100 positive, 22 negative). Prospective validation enrolled 128 consecutive cases / 614 WSIs (105 positive, 23 negative). Ground truth was established by clinical IHC using OV-TL 12/30 or other validated clones.

\noindent\textbf{P40 biomarker prediction.} P40 is highly specific for squamous differentiation and is critical for confirming squamous cell carcinoma diagnosis, particularly in poorly differentiated cases where morphology alone is insufficient. This binary classification task predicts P40 immunohistochemical expression. P40 is a truncated isoform of p63 that is highly specific for squamous differentiation, expressed in most of squamous cell carcinomas but rarely in adenocarcinomas. The internal cohort from Center-H1 comprised 335 cases (442 WSIs): 99 P40-positive and 236 P40-negative. External validation included 243 cases (780 WSIs) from Center-H3 (43 positive, 200 negative). Prospective validation enrolled 152 consecutive cases / 737 WSIs (46 positive, 106 negative). Ground truth was established by clinical IHC using clone BC28 or polyclonal antibodies.

\noindent\textbf{P63 biomarker prediction.} P63 serves as a sensitive marker for squamous differentiation and is routinely used to support squamous cell carcinoma diagnosis in diagnostic panels. This binary classification task predicts p63 immunohistochemical expression. P63 is a nuclear transcription factor expressed in basal/squamous cells, positive in nearly 100\% of squamous cell carcinomas but also in some adenocarcinomas and neuroendocrine tumors. The internal cohort from Center-H1 comprised 287 cases (354 WSIs): 90 p63-positive and 197 p63-negative. Prospective validation enrolled 58 consecutive cases / 297 WSIs (34 positive, 24 negative). Ground truth was established by clinical IHC.

\noindent\textbf{Ki-67 proliferation index prediction.} Ki-67 proliferation index provides prognostic stratification and helps identify rapidly proliferating tumors that may benefit from more aggressive treatment approaches. This three-class classification task predicts Ki-67 proliferation index categories. Ki-67 is a nuclear protein expressed in all active phases of the cell cycle (G1, S, G2, and M) but absent in G0, serving as a proliferation marker with prognostic significance. The internal cohort from Center-H1 comprised 239 cases (630 WSIs): 112 low Ki-67 (<10\% positive nuclei), 84 medium Ki-67 (10-50\% positive nuclei), and 43 high Ki-67 (>50\% positive nuclei). Data were split at a 7:1:2 ratio with patient-level label-stratified sampling. Ground truth was established by clinical IHC using MIB-1 or other validated clones, with proliferation index calculated as percentage of positive tumor cell nuclei, typically assessed in hotspots of highest proliferation.

\noindent\textbf{Tumor mutational burden (TMB) prediction.} TMB is a predictive biomarker for immunotherapy response, with TMB-high tumors showing greater benefit from immune checkpoint inhibitors, making rapid H\&E-based prediction valuable for treatment planning. This binary classification task predicts tumor mutational burden (TMB) status, defined as the total number of somatic coding mutations per megabase of genome examined. TMB-high status was defined as $\geq$10 mutations/Mb following FDA-approved cutoffs for pembrolizumab. The cohort from TCGA-LUAD/LUSC comprised 875 cases (970 WSIs): 195 TMB-high and 680 TMB-low. Data were split at a 7:1:2 ratio. Ground truth was established by whole-exome sequencing from TCGA, with TMB calculated as total somatic mutations (synonymous and non-synonymous) divided by exome size.

\noindent\textbf{EGFR mutation prediction.} EGFR mutations are the most common actionable driver mutations in lung adenocarcinoma and predict dramatic response to EGFR tyrosine kinase inhibitors, fundamentally altering first-line treatment selection. This binary classification task predicts the presence of epidermal growth factor receptor (EGFR) activating mutations, including exon 19 deletions and L858R point mutation in exon 21, which account for most EGFR mutations. The internal cohort from Center-H5 comprised 438 cases (706 lung adenocarcinoma WSIs): 222 EGFR-mutant and 216 EGFR wild-type. External validation used TCGA-LUAD with 414 cases / 469 WSIs (57 EGFR-mutant, 357 wild-type). Data were split at a 7:1:2 ratio for internal cohorts. Ground truth was established by clinical molecular testing for internal cohorts.

\noindent\textbf{STK11/LKB1 mutation prediction.} STK11 loss is associated with aggressive tumor biology, reduced response to immunotherapy, and distinct metabolic vulnerabilities, providing prognostic information and informing treatment selection. This binary classification task predicts the presence of STK11 (serine/threonine kinase 11, also known as LKB1) loss-of-function mutations or deletions. The internal cohort from TCGA-LUAD comprised 414 cases (469 lung adenocarcinoma WSIs): 80 STK11-mutant and 334 STK11 wild-type. Data were split at a 7:1:2 ratio. Ground truth was established by targeted NGS panels or whole-exome sequencing.

\noindent\textbf{Overall survival prediction in lung adenocarcinoma (LUAD).} Morphology-based survival prediction enables personalized risk stratification beyond TNM staging, helping identify high-risk patients who may benefit from more intensive surveillance or adjuvant therapy. This continuous risk prediction task estimates overall survival (OS) risk scores for patients with surgically resected lung adenocarcinoma. Overall survival was defined as time from surgery to death from any cause, with patients censored at last follow-up or administrative censoring. The internal cohort from TCGA-LUAD comprised 455 LUAD cases (518 WSIs) with complete follow-up data. It contains 296 patients censored alive and 159 deceased patients. Data were split into training and held-out evaluation subsets at an approximately 7:3 ratio with patient-level stratification, ensuring balanced stage distribution across subsets. External validation was performed on an independent cohort from Center-H1 comprising 303 LUAD cases (1,387 WSIs). It contains 290 living patients and 13 deceased patients. No prospective validation was performed for survival tasks due to insufficient follow-up duration. The model was trained to predict a continuous risk score using discrete-time survival negative log-likelihood loss, with performance evaluated using the concordance index (C-index). Ground truth overall survival data for TCGA-LUAD were obtained from TCGA clinical annotations, whereas overall survival data for the Center-H1 external cohort were obtained from institutional tumor registries, medical records, and death certificates.

\noindent\textbf{Overall survival prediction in lung squamous cell carcinoma (LUSC).} Prognostic stratification within squamous cell carcinoma identifies patients at elevated risk for disease-specific mortality who may require intensified treatment or surveillance approaches. This continuous risk prediction task estimates overall survival risk scores for patients with surgically resected lung squamous cell carcinoma. The internal cohort from TCGA-LUSC comprised 452 LUSC cases (484 WSIs) with complete follow-up data. It contains 262 patients censored alive and 190 deceased patients. Data were split into training and held-out evaluation subsets at an approximately 7:3 ratio with patient-level stratification, ensuring balanced stage distribution across subsets. External validation was performed on an independent cohort from Center-H1 comprising 73 LUSC cases (382 WSIs). It contains 71 living patients and 2 deceased patients. No prospective validation was performed for survival tasks due to insufficient follow-up duration. Training and evaluation methodology were identical to LUAD OS prediction. Ground truth overall survival data for TCGA-LUSC were obtained from TCGA clinical annotations, whereas overall survival data for the Center-H1 external cohort were obtained using the institutional follow-up procedure described for LUAD overall survival.

\noindent\textbf{Disease-free survival prediction.} Prediction of recurrence risk enables personalized surveillance strategies and helps identify patients who may benefit from adjuvant therapy or enrollment in clinical trials. This continuous risk prediction task estimates disease-free survival (DFS) risk scores for patients with surgically resected lung cancer (both LUAD and LUSC). Disease-free survival was defined as time from surgery to first recurrence (local, regional, or distant), with patients censored at last follow-up without recurrence. The cohort from Center-H1 comprised 367 surgically resected lung cancer cases (1,715 WSIs) with 301 living patients and 66 recurrences observed. Data were split into training and held-out evaluation subsets at an approximately 7:3 ratio with patient-level stratification. Training used discrete-time survival negative log-likelihood loss, with performance evaluated using C-index. Ground truth disease-free survival data were obtained from institutional follow-up records and medical records, with recurrence status determined through clinical follow-up.


\subsection*{Prospective Validation Study Protocol}
\phantomsection\label{sec:methods-prospective-validation}
\noindent\textbf{Study design and ethics approval.} We conducted a prospective, single-center, observational study to evaluate PulmoFoundation's performance on consecutive patients undergoing routine surgical evaluation at Nanfang Hospital (Center-H1). The study was approved by the Institutional Review Board of Nanfang Hospital (approval numbers: NFEC-2024-535 and NFEC-2025-403). The study adhered to DECIDE-AI reporting guidelines \cite{vasey2022decide} for early-stage clinical evaluation of AI decision support systems.

\noindent\textbf{Study period and patient enrollment.} Consecutive patients were screened between October 1, 2024, and November 25, 2025. All patients meeting predefined eligibility criteria were systematically enrolled without case selection to ensure unbiased representation of routine clinical practice. Inclusion criteria were: (1) adult patients ($\geq$18 years) undergoing surgical biopsy or resection for suspected or confirmed lung lesions, (2) availability of H\&E-stained slides suitable for digitization (adequate tissue, acceptable staining quality), and (3) availability of clinical follow-up and final pathological diagnosis. Of 1,558 patients screened, 201 were excluded as ineligible. Among these, 136 pre-operative biopsy cases were excluded because they lacked a definitive diagnosis by H\&E or IHC and had no curative resection specimen available for supplementary diagnosis, or because the final diagnosis remained undetermined owing to loss to follow-up. The remaining 65 post-operative surgical-resection cases were excluded because of prior neoadjuvant therapy (n=20), absence of IHC-assisted testing when H\&E alone was insufficient to establish a definitive diagnosis (n=21), or failure of H\&E combined with IHC to establish a definitive diagnosis (n=24). The complete patient flow is shown in Extended Data Fig.~\ref{ext_fig_consort_prospective}.

\noindent The final prospective cohort comprised 1,357 patients: 250 pre-operative biopsy cases, 271 intra-operative frozen-section cases, and 836 post-operative surgical-resection cases. The cohort reflected the natural case distribution in routine clinical practice, including common histologic subtypes (adenocarcinoma and squamous cell carcinoma), rare entities (neuroendocrine carcinomas and adenosquamous carcinoma), and challenging diagnostic scenarios (poorly differentiated tumors and mixed histologic patterns).

\noindent\textbf{Blinding and prediction workflow.} The prospective evaluation followed a strictly observational design: model predictions did not influence clinical decision-making at any stage. To ensure unbiased assessment, the prediction and diagnostic workflows were conducted independently. H\&E-stained slides were digitized following standard institutional protocols, and PulmoFoundation predictions were generated automatically upon completion of digitization. Predictions were stored on a research server separate from the clinical pathology information system and were not accessible to the diagnosing pathologists or study investigators involved in ground truth adjudication. Clinical pathologists completed their diagnostic reports following standard institutional workflows, blinded to all model outputs. Final pathological diagnoses were established and locked in the hospital information system before model predictions were unblinded for comparative analysis. No model retraining or hyperparameter adjustment was performed between retrospective development and prospective evaluation; the identical fixed model weights were applied to all prospective cases. Baseline foundation models were not deployed in the prospective observational workflow; head-to-head model comparison was restricted to retrospective benchmarking under the matched evaluation protocol described above.

\noindent\textbf{Ground truth establishment.} Ground truth for prospective cases was established through routine clinical workflows following institutional standards. Board-certified pathologists with thoracic pathology expertise rendered the initial diagnosis. Diagnostically challenging cases underwent consensus review by two senior pathologists, each with more than 10 years of thoracic pathology experience. IHC staining was performed when clinically indicated according to standard diagnostic algorithms. Cases with initial diagnostic discordance were adjudicated by a third senior pathologist, and the final diagnosis was determined by majority consensus. All ground truth labels were finalized and locked in the hospital information system before model predictions were unblinded. For IHC biomarker prediction tasks, ground truth was established by institutional clinical laboratory IHC results reviewed by study pathologists, with positivity thresholds following the marker-specific clinical criteria described for each task.

\noindent \textbf{Tasks evaluated prospectively.} The prospective study evaluated PulmoFoundation across 11 clinically critical tasks: (1) Pre-operative classification of benign and malignant lesions in biopsies (n=250 WSIs), (2) Intra-operative classification of benign and malignant lesions in frozen sections (n=380 WSIs), (3) Post-operative NSCLC classification (n=1,143 WSIs), (4) Post-operative coarse-grained differential diagnosis: squamous cell carcinoma vs. invasive adenocarcinoma vs. adenocarcinoma in situ (n=1,673 WSIs), (5) Post-operative fine-grained differential diagnosis: six-class classification including rare subtypes (n=1,802 WSIs), (6) prediction of lymph node metastasis from primary-tumor surgical resection slides (n=1,770 WSIs), (7-11) Five IHC biomarker predictions: TTF-1 (n=1,056 WSIs), Napsin-A (n=700 WSIs), CK-7 (n=614 WSIs), P40 (n=737 WSIs), P63 (n=297 WSIs). These tasks were selected based on clinical relevance, availability of ground truth in routine practice, and coverage of the diagnostic workflow spanning biopsy, frozen section, and surgical resection specimens including diagnostic classification, histologic subtyping, lymph-node-metastasis prediction, and IHC biomarker prediction. Task-specific sample sizes and class distributions are provided in Extended Data Table~\ref{tab:prospective}.

\noindent\textbf{Model deployment and computational workflow.} All predictions were fully automated without manual intervention using one NVIDIA GeForce RTX 3090 GPU, and the identical computational pipeline was applied uniformly across all 11 prospective tasks. We measured the computational workflow from digitized WSI files to model prediction in 90 native-format WSIs, including 30 biopsy (.sdpc), 30 frozen-section (.svs), and 30 resection (.kfb) WSIs. The workflow included coordinate extraction, patch reading/cropping, feature extraction, feature loading, device transfer, and classifier prediction, and excluded physical slide scanning. Median computational workflow time was 394.8 s for biopsy WSIs (IQR, 284.8--515.8), 83.6 s for frozen-section WSIs (IQR, 59.4--98.3), 121.7 s for resection WSIs (IQR, 96.0--148.4), and 127.7 s across all 90 WSIs (IQR, 90.7--266.7; Extended Data Tables~\ref{tab:runtime_summary} and \ref{tab:runtime_components}).

\subsection*{Crossover Randomized Controlled Trial}
\phantomsection\label{sec:methods-rct}
\textbf{Study design and rationale.} To evaluate the clinical utility of AI-assisted diagnosis in a controlled digital pathology setting, we conducted a crossover randomized controlled trial (ClinicalTrials.gov ID: NCT07157618) assessing diagnostic performance with and without PulmoFoundation assistance. 
This within-subject design enabled direct comparison of pathologist performance under both conditions while controlling for individual expertise variation and case difficulty, consistent with prior randomized clinical-AI and digital-pathology reader studies that evaluated clinician performance with and without AI assistance \cite{goh2025gpt,tao2026llm,wu2025eyecare,abdurrachim2025utility}. The crossover design was chosen to maximize statistical power while minimizing the number of pathologists required, as each pathologist serves as their own control. The full trial design, including inclusion/exclusion criteria and randomization scheme, is detailed in Extended Data Fig.~\ref{ext_fig_consort_rct}.

\noindent\textbf{Participants and randomization.} Eight board-certified pathologists from Center-H1 (Nanfang Hospital) participated in this study. Pathologists were stratified by experience level into two groups: senior pathologists ($>5$ years of subspecialty experience in thoracic pathology, $n=4$) and junior pathologists ($<5$ years of subspecialty experience, $n=4$). Within each experience stratum, pathologists were randomly assigned to one of two sequence groups using computer-generated randomization: Group A (AI-first sequence: Period 1 with AI assistance, Period 2 without AI) or Group B (no-AI-first sequence: Period 1 without AI, Period 2 with AI assistance). This yielded balanced allocation with 2 senior and 2 junior pathologists in each sequence group. All participating pathologists provided informed consent. Pathologists were blinded to the study hypotheses and sequence assignments of other participants.

\noindent\textbf{Case selection and tasks.} We evaluated pathologist performance across four clinically critical diagnostic tasks spanning biopsy, frozen-section, and surgical-resection workflows: (1) \textit{Pre-operative primary versus metastatic tumor classification} in biopsy specimens (n=129 cases, 136 WSIs), distinguishing primary lung cancers from metastatic tumors to the lung; (2) \textit{Intra-operative benign versus malignant classification} in frozen-section cases (n=99 cases, 235 WSIs), distinguishing benign from malignant lesions during intra-operative consultation; (3) \textit{Post-operative NSCLC histologic subtyping} in surgical resection specimens (n=270 cases, 1,143 WSIs), distinguishing adenocarcinoma from squamous cell carcinoma; and (4) \textit{Post-operative primary-site prediction} for metastatic tumors in surgical resection specimens (n=160 cases, 289 WSIs), identifying tissue of origin among lung, colorectal, breast, renal, and hepatocellular carcinomas. For diagnostically challenging cases from biopsy primary-versus-metastatic, frozen-section benign-versus-malignant, and metastatic primary-site prediction tasks, definitive reference labels were established by review of available final diagnostic materials, including final FFPE diagnoses, ancillary IHC results, and other clinically available confirmatory information, with additional consultation by thoracic subspecialty pathologists when required. The study comprised 658 cases across 1,803 WSIs. All analyses were conducted at the case level; for cases with multiple slides, pathologists reviewed all available WSIs before rendering a single case-level diagnosis. Cases or slides were excluded only when a definitive diagnosis could not be established from available diagnostic material. Each pathologist reviewed identical cases in both periods, with randomized presentation order within each period to minimize order effects.

\noindent\textbf{Washout period.} A mandatory 4-week washout period separated Period 1 and Period 2 to minimize carryover effects and reduce the likelihood that pathologists would recall specific cases from the first period. During the washout period, pathologists were instructed to avoid reviewing study cases and to refrain from discussing study content with other participants.

\noindent\textbf{Evaluation procedure.} All case review was performed using digitized H\&E WSIs. In the unassisted condition, pathologists reviewed the WSIs without computational assistance. In the AI-assisted condition, pathologists reviewed the same WSIs with PulmoFoundation's task-specific predicted diagnostic label displayed as the model output. In both conditions, pathologists had unlimited time for case review and access to standard diagnostic tools (zooming, panning, annotation). No per-case time limit was imposed, but each evaluation period was completed within 2 weeks. Pathologists were instructed to render diagnoses as they would in routine clinical practice. Pathologists were necessarily unblinded to AI predictions in the AI-assisted condition but remained blinded to model development details, training data composition, and overall study hypotheses. Ground truth diagnoses were withheld from pathologists during evaluation and revealed only after completion of both periods.

\noindent For each case, pathologists completed a standardized electronic case report form capturing: (1) final diagnosis using task-specific diagnostic categories, (2) diagnostic time automatically recorded from case presentation to submission, and (3) diagnostic confidence on a 1--10 Likert scale (1=completely uncertain, 10=extremely confident).

\noindent\textbf{Outcome measures.} The primary outcome was case-level diagnostic accuracy per case-reader observation, defined as the proportion of diagnoses matching the ground-truth label. Secondary outcomes included: (1) diagnostic efficiency, measured as time per case in seconds, (2) diagnostic confidence on the 1--10 scale, and (3) inter-rater agreement among pathologists, measured using Fleiss' $\kappa$.

\noindent\textbf{Outcome-based AI impact classification.} To characterize human-AI collaboration dynamics beyond simple accuracy metrics, we developed a post-hoc outcome-based classification framework using the model prediction displayed during AI-assisted review as the AI recommendation for each observation. By comparing each pathologist's paired diagnoses with and without AI assistance against ground truth, each AI-assisted observation was assigned to one of four mutually exclusive categories: (1) \textit{Improved}, defined as an incorrect unassisted diagnosis followed by a correct assisted diagnosis; (2) \textit{Confirmed}, defined as a correct displayed model prediction with correct unassisted and assisted diagnoses; (3) \textit{Resilient}, defined as an incorrect displayed model prediction with correct unassisted and assisted diagnoses; and (4) \textit{Failed}, defined as an incorrect final assisted diagnosis. Failed cases were further decomposed into \textit{missed opportunity}, defined as a correct displayed model prediction with incorrect unassisted and assisted diagnoses; \textit{accuracy loss after AI}, defined as a correct unassisted diagnosis followed by an incorrect assisted diagnosis; and \textit{both failed}, defined as an incorrect displayed model prediction, incorrect unassisted diagnosis, and incorrect assisted diagnosis. We separately defined \textit{strict erroneous-AI adoption harm} as the subset of accuracy-loss observations in which the displayed model prediction was incorrect and the final assisted diagnosis matched that incorrect displayed model prediction.

\noindent\textbf{Model-error, confidence, and difficulty sensitivity analyses.} For the model-incorrect behavior analysis, the first row used all AI-assisted observations as the denominator, and subsequent rows used observations with an incorrect displayed model prediction as the denominator. Adoption of the displayed model prediction was defined as an assisted diagnosis that matched the task-specific prediction shown during AI-assisted review. The model-confidence sensitivity analysis included all 658 RCT cases and 5{,}264 case-reader observations. Model confidence was defined as the maximum predicted class probability at the case level. Cases were assigned within each task to low-, middle-, or high-confidence strata using task-specific confidence cutoffs. These probabilities were used only for post hoc confidence stratification and were not displayed to pathologists during RCT review. Case-difficulty sensitivity analysis used 1 minus the unassisted accuracy across the eight pathologists for each task-case pair as an exploratory post-hoc difficulty proxy. Difficulty strata were assigned within each task by ranked tertiles because independent per-case difficulty annotations were unavailable.


\subsection*{Statistical Analysis}
\noindent\phantomsection\label{sec:methods-performance-metrics}\textbf{Performance metrics.} For classification tasks, model discrimination was summarized using macro AUC. For each class \(c\), the predicted probability \(p_{ic}\) was evaluated in a one-versus-rest setting against all remaining classes. The class-specific AUC was interpreted as the probability that a randomly selected case from class \(c\) received a higher class-\(c\) score than a randomly selected case not from class \(c\), with ties assigned half credit:
\[
\mathrm{AUC}_c =
P(p_{ic}>p_{jc}\mid y_i=c,y_j\neq c)
+\frac{1}{2}P(p_{ic}=p_{jc}\mid y_i=c,y_j\neq c).
\]
Macro AUC was computed as the equal-weight mean of one-versus-rest AUCs across classes:
\[
\mathrm{Macro\ AUC}=\frac{1}{K}\sum_{c=1}^{K}\mathrm{AUC}_c.
\]
For class-label macro metrics in the retrospective and external model-comparison tables, the predicted class label was assigned by the argmax rule, \(\hat{y}_i=\arg\max_c p_{ic}\). For each class, sensitivity, specificity, PPV, and NPV were computed in a one-versus-rest manner from the corresponding true-positive, false-positive, true-negative, and false-negative counts:
\[
\mathrm{Sensitivity}_c=\frac{\mathrm{TP}_c}{\mathrm{TP}_c+\mathrm{FN}_c},\quad
\mathrm{Specificity}_c=\frac{\mathrm{TN}_c}{\mathrm{TN}_c+\mathrm{FP}_c},\quad
\mathrm{PPV}_c=\frac{\mathrm{TP}_c}{\mathrm{TP}_c+\mathrm{FP}_c},\quad
\mathrm{NPV}_c=\frac{\mathrm{TN}_c}{\mathrm{TN}_c+\mathrm{FN}_c}.
\]
Macro Sensitivity, macro Specificity, macro PPV, and macro NPV were computed by equal-weight averaging across classes:
\[
\mathrm{Macro\ }M=\frac{1}{K}\sum_{c=1}^{K}M_c,\quad
M\in\{\mathrm{Sensitivity},\mathrm{Specificity},\mathrm{PPV},\mathrm{NPV}\}.
\]
For classification metrics, 95\% CIs were estimated by non-parametric bootstrap resampling over cases with 1{,}000 resamples.

\noindent\textbf{Per-class sensitivity at Youden-optimal operating points.} In the Extended Data Tables, per-class sensitivity was summarized at class-specific one-versus-rest thresholds selected by maximizing Youden's statistic within each evaluation cohort. For each class \(c\), the predicted probability for that class was treated as a continuous score against all remaining classes, and the operating threshold was selected on the evaluation cohort by maximizing Youden's statistic:
\[
t_c=\arg\max_t\{\mathrm{Sensitivity}_c(t)+\mathrm{Specificity}_c(t)-1\}.
\]
The reported per-class sensitivity was \(\mathrm{Sensitivity}_c(t_c)\). Bootstrap CIs were computed by resampling cases with replacement 1{,}000 times and recomputing both the threshold and sensitivity within each bootstrap replicate. Prospective triage operating points were defined separately using workflow-specific PPV safety bars, as described in the triage analysis below.

\noindent\textbf{Survival analysis.} For overall survival (OS) and disease-free survival (DFS) prediction, we used the concordance index (C-index) as the primary discrimination metric. Survival models output discrete survival probabilities \(S_{im}\) over four time bins, and the case-level risk score was computed as
\[
r_i=-\sum_{m=1}^{4}S_{im}.
\]
Let \(T_i\) denote observed follow-up time and \(\delta_i\) indicate an observed event. Comparable patient pairs were defined as \(\mathcal{P}=\{(i,j):T_i<T_j,\delta_i=1\}\), meaning that patient \(i\) had an observed event before patient \(j\)'s follow-up time. The C-index was computed as
\[
\mathrm{C\mbox{-}index}
=
\frac{1}{|\mathcal{P}|}
\sum_{(i,j)\in\mathcal{P}}
\left[
\mathbf{1}\{r_i>r_j\}
+\frac{1}{2}\mathbf{1}\{r_i=r_j\}
\right].
\]
A C-index of 0.5 corresponds to random risk ordering in expectation, and a C-index of 1.0 indicates perfect concordance. CIs for C-index were estimated with 1{,}000 bootstrap resamples over cases. Patients were right-censored at the date of last follow-up or administrative censoring. For each survival cohort, we report the total number of cases, the number of observed events (deaths for OS; recurrences for DFS), and the median follow-up time with interquartile range (IQR), computed over the subset of patients used for metric evaluation.

\noindent\textbf{Decision curve analysis.} Following standard DCA methodology \cite{vickers2006decision,shao2025mri}, we performed DCA for key retrospective diagnostic tasks to quantify net benefit across threshold probabilities. DCA compares model-guided decisions with treat-all and treat-none reference strategies, which are standard methodological comparators for decision-curve interpretation. Net benefit is defined as:
\begin{equation*}
    \text{Net Benefit} = \frac{\text{True Positives}}{N} - \frac{\text{False Positives}}{N} \times \frac{p_t}{1-p_t},
\end{equation*}
\noindent where $N$ is the sample size, and $p_t$ is the probability threshold. A model has positive clinical net benefit at thresholds where its curve lies above both reference strategies. For prospective clinical-utility assessment, we complemented DCA with pre-specified PPV-constrained triage operating points for diagnostic biopsy, intra-operative frozen section, and IHC ancillary-test triage (Fig.~\ref{fig6}d--f; Extended Data Table~\ref{tab:triage_supp}).

\noindent\textbf{Statistical significance testing.}
For pairwise comparisons between PulmoFoundation and each baseline model (e.g., Virchow2), we used the paired Wilcoxon signed-rank test \cite{wilcoxon1945individual} applied to the bootstrap-replicate AUCs of the two models, paired by replicate index, and we report the resulting two-sided $P$-value.

\noindent\textbf{Cohort-pair performance differences.} For each PulmoFoundation task evaluated on more than one cohort (Extended Data Table~\ref{tab:perf_delta}), we report the difference between cohort-marginal AUC or C-Index point estimates together with a 95\% percentile bootstrap CI on the difference. CIs were obtained by paired-replicate non-parametric bootstrap on the underlying prediction probabilities: each cohort was independently resampled 1{,}000 times under a synchronized replicate index, and the 2.5th and 97.5th percentiles of the resulting delta distribution were taken as the lower and upper bounds. Cohorts contain disjoint patients, so the pairing reflects replicate alignment rather than case alignment, as is standard for the difference of two independent-sample estimates.


\noindent\textbf{Statistical analysis for prospective cohort.} Prospective discrimination was summarized using the same macro one-versus-rest AUC procedure described above. In Fig.~\ref{fig6}b, Macro NPV was computed at class-specific Youden-optimal operating points. In the per-task prospective Extended Data Tables, operating-point macro Sensitivity, Specificity, PPV, and NPV were also computed at class-specific Youden-optimal thresholds. Bootstrap 95\% CIs used 1{,}000 case-level resamples, with operating-point thresholds recomputed within each bootstrap replicate. Prospective triage operating points were analyzed separately under the pre-specified PPV safety bars described in Extended Data Table~\ref{tab:triage_supp}. Given the observational nature of this study and the lack of intervention, no formal sample size calculation for superiority testing was performed. Instead, we aimed to enroll sufficient patients to achieve stable performance estimates with narrow confidence intervals.

\noindent For the prospective lymph-node-metastasis failure-mode analysis, case-level model scores were linked to pathology-report annotations from the same prospective surgical-resection cohort. The analysis was performed at the Youden-optimal threshold used for the prospective lymph-node-metastasis performance table. False positives were defined as node-positive model predictions in pathology-confirmed N0 cases, and false negatives were defined as node-negative model predictions in N+ cases. Histologic subtype was obtained from the structured subtype field when available and recovered from the final diagnostic text otherwise. Cases were assigned to mutually exclusive categories comprising adenocarcinoma in situ, minimally invasive adenocarcinoma, invasive adenocarcinoma, squamous cell carcinoma, neuroendocrine neoplasm, adenosquamous carcinoma, other specified carcinoma, and histologic subtype unavailable. Other specified carcinoma included SMARCA4-deficient non-small-cell lung cancer, mucoepidermoid carcinoma, invasive carcinoma not otherwise specified, and pleomorphic or sarcomatoid carcinoma. Cases without an assignable subtype in either report source were categorized as histologic subtype unavailable. Nodal burden, tumor size, IASLC grade, lymphovascular invasion, STAS, and pleural invasion were derived from structured pathology-report fields. No post hoc review of individual slide images was performed.

\noindent\textbf{Statistical analysis for crossover RCT.} The statistical analysis plan was finalized before outcome data were analyzed. All analyses included all randomized pathologists and assigned case-reader observations. For diagnostic accuracy, we used generalized estimating equations (GEE) with binomial family, logit link, and exchangeable correlation structure:
\begin{equation*}
\mathrm{logit}\{\Pr(Y_{gr}=1)\}
=
\beta_0
+\beta_1 \mathrm{AI}_{gr}
+\beta_2 \mathrm{Period}_{gr}
+\beta_3 \mathrm{Experience}_{g}
+\boldsymbol{\beta}_4^{\top}\mathbf{Task}_{g}.
\end{equation*}
\noindent Here, \(Y_{gr}\) indicates whether read \(r\) within pathologist-case pair \(g\) was correct, and \(\mathbf{Task}_{g}\) denotes task indicator variables. The overall primary GEE model used an exchangeable working correlation within pathologist-case pairs to account for paired AI-assisted and unassisted reads of the same case by the same pathologist. Task-specific and stratified GEE models used pathologist as the clustering unit. The coefficient \(\beta_1\) quantifies the AI effect and is reported as an odds ratio with 95\% confidence interval calculated using robust sandwich variance estimators.

\noindent For the model-confidence and case-difficulty sensitivity analyses, the same GEE specification was fitted within each stratum. 

\noindent To assess whether the AI effect varied by evaluation period, we fitted a binomial GEE model for diagnostic correctness using all 10,528 observations. This model included AI condition, evaluation period, the Condition $\times$ Period interaction, and pathologist experience level. As a sensitivity analysis for repeat exposure, we analyzed Period 1 data only, before crossover and within-trial repeated review. This analysis compared the four pathologists receiving AI assistance in Period 1 with the four receiving unassisted review. The Period 1 model included AI condition and pathologist experience level as predictors. Both models used an exchangeable working correlation structure with pathologist as the clustering unit.

\noindent For diagnostic time, we used GEE with Gaussian family on log-transformed values and pathologist as the clustering unit; the exponentiated coefficient represents the time ratio. For confidence, we used GEE with Gaussian family and pathologist as the clustering unit; effect size was quantified using Cohen's d.

\noindent For inter-rater agreement, we computed Fleiss' kappa separately for each condition (with and without AI) across all eight pathologists. Bootstrap resampling with 1{,}000 iterations was used to calculate 95\% confidence intervals for each kappa estimate. A 10{,}000-iteration paired permutation test was used to test whether $\Delta\kappa$ differed from zero. Kappa values were interpreted as: $<$0.20 poor, 0.21--0.40 fair, 0.41--0.60 moderate, 0.61--0.80 substantial, $>$0.80 almost perfect.

\noindent All tests were two-sided with significance threshold $\alpha$=0.05. Subgroup analyses by task and experience level were pre-specified but considered hypothesis-generating; no multiple comparison adjustment was applied.

\noindent\textbf{Software.} Statistical analyses were performed using Python 3.10.9 with scikit-learn 1.5.2. Deep learning models were implemented in PyTorch 2.0.1. WSI preprocessing, including tissue segmentation and magnification-adaptive patch extraction, used PrePATH (\url{https://github.com/birkhoffkiki/PrePATH}). WSI-level downstream task evaluation used a modified mSTAR toolkit (\url{https://github.com/Innse/mSTAR/tree/main/downstream_task}). Crossover RCT analyses were performed using Python 3.10.9 with statsmodels 0.14 for GEE models and scipy 1.11 for statistical tests.

\section*{Data Availability}
Publicly available TCGA and CPTAC data, including WSIs and clinical metadata, can be obtained from the NIH Genomic Data Commons portal (\url{https://portal.gdc.cancer.gov}) and the Proteomic Data Commons (\url{https://proteomic.datacommons.cancer.gov}). Whole-slide images from the National Lung Screening Trial (NLST) can be accessed through The Cancer Imaging Archive \cite{clark2013cancer} (\url{https://www.cancerimagingarchive.net/collection/nlst/}). HistAI-Lung slides \cite{nechaev2025histai} can be obtained from HuggingFace (\url{https://huggingface.co/datasets/histai/HISTAI-metadata}). WSSS4LUAD images \cite{han2022wsss4luad} can be accessed from Grand Challenge (\url{https://wsss4luad.grand-challenge.org/}). LC25000 \cite{borkowski2019lung} is available at Kaggle (\url{https://www.kaggle.com/datasets/javaidahmadwani/lc25000}). Institutional datasets, including retrospective internal and external validation cohorts, prospective validation cohorts, and crossover RCT reader-study data, are not publicly released because of patient privacy and institutional data-protection requirements. Human-subject data were acquired under approved Institutional Review Board protocols. Requests for de-identified institutional datasets for non-commercial academic research should be directed to the corresponding authors and will be reviewed according to institutional data-access policies. Aggregated data supporting the study conclusions are provided in the manuscript, figures, and Extended Data tables.

\section*{Code Availability}
The source code developed for this work will be made publicly available upon acceptance. The WSI preprocessing pipeline used for tissue segmentation and magnification-adaptive patch extraction is available through PrePATH at \url{https://github.com/birkhoffkiki/PrePATH}.

\section*{Author Contribution}
Z.R.G and Z.Y.Z contributed equally to this work. Z.R.G, Z.Y.Z, J.B.M, Y.H.W, F.T.Z, Y.X.X, L.L, M.Y.C, L.L, and H.C conceived and designed the study. Z.R.G, Z.Y.Z, Z.J.C, Q.X, C.L.Z, J.B.L, S.J.G, Z.Y.L, L.J.Q, S.F.C, Y.P.L, Z.W, X.M.Z, M.Y.C, and L.L collected and curated the multi-center WSI dataset. Z.R.G, J.B.M, Y.H.W, and F.T.Z developed the PulmoFoundation model and conducted computational experiments for retrospective and prospective validations. Z.R.G, Z.Y.Z, L.L, and H.C coordinated the prospective study and enrolled patients. Z.Y.Z, C.L.Z, Z.J.C, Q.X, F.Y.H, and L.L participated as pathologists in the crossover RCT. Z.R.G and Z.Y.Z performed statistical analysis. Z.Y.Z, C.L.Z, Z.J.C, Q.X, J.B.L, S.J.G, F.Y.H, and L.L provided clinical annotation and pathological expertise. Z.R.G and Z.Y.Z wrote the original manuscript draft, while C.J, J.L.H, Z.X.C, and Y.C provided critical
revisions for important intellectual content. All authors approved the final version for publication and agree to be accountable for all aspects of the work. M.Y.C, L.L, and H.C supervised the study and acquired funding.

\section*{Acknowledgment}
This work was supported by Innovation and Technology Commission (Project No. MHP/002/22 and ITCPD/17-9), Research Grants Council of the Hong Kong Special Administrative Region, China (Project No. T45-401/22-N, R6003-22 and C4024-22GF), Frontier Technology Research for Joint Institutes with Industry Scheme (Project No. OKT24EG01), National Key R\&D Program of China (Project No. 2023YFE0204000), and Shenzhen Science and Technology Innovation Committee Fund (Project No. KCXFZ20230731094059008). We sincerely appreciate the pathologists, Dr. Huiting Ma and Dr. Feifei Wang, who participated in this study for their invaluable expertise and dedication during the execution of the crossover RCT.

\bibliography{ref.bib}

\newpage
\section*{Extended Data}
\setcounter{figure}{0}
\begin{figure}[!htbp]
    \centering
    \includegraphics[width=0.87\linewidth]{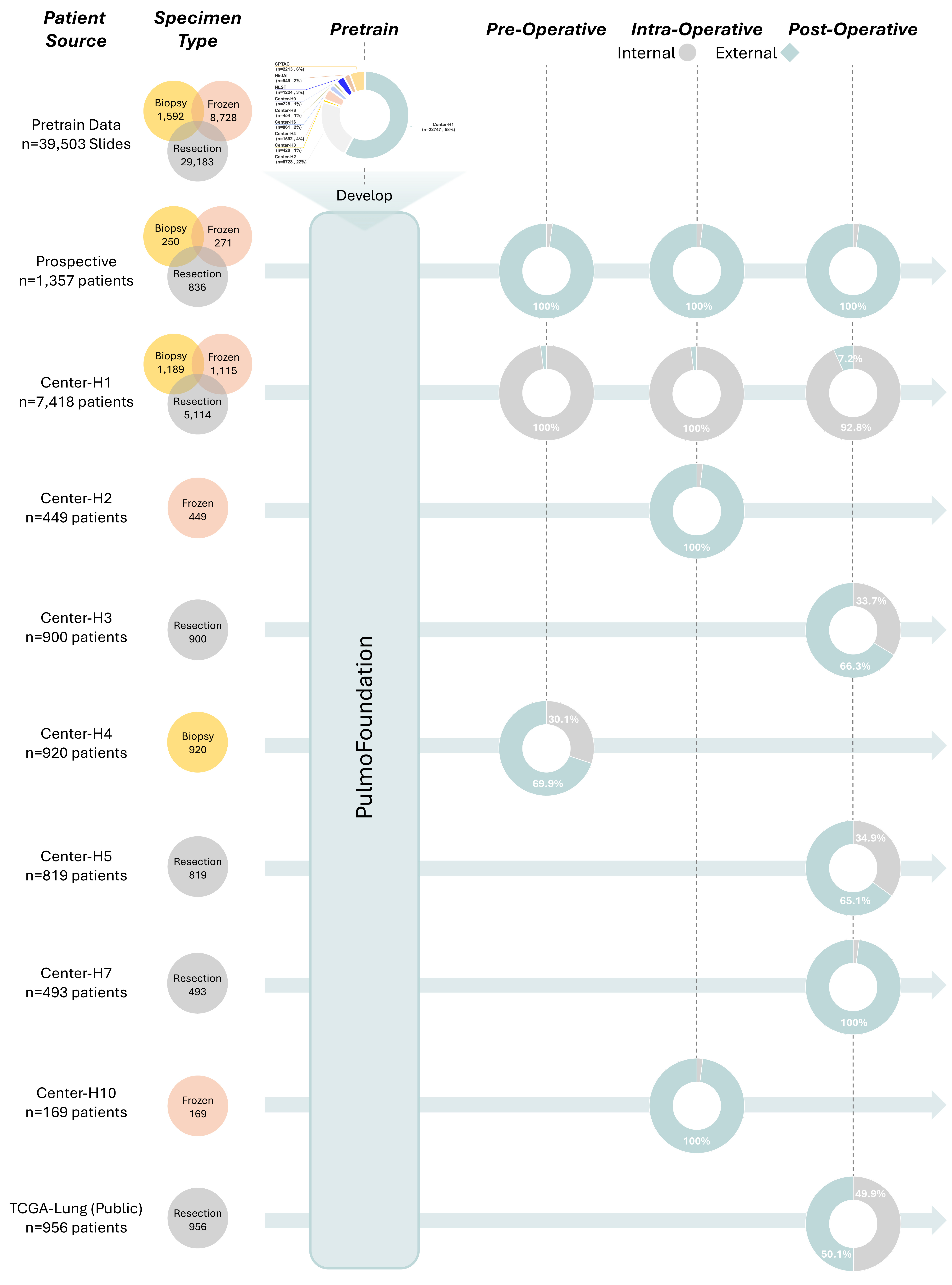}
    \captionsetup{name=Extended Data Figure}
    \caption{\textbf{Pipeline for development and validation of PulmoFoundation under multi-center, retrospective, and prospective collection spanning pre-operative, intra-operative, and post-operative clinical stages.}}
    \label{ext_fig_cohort}
\end{figure}


\begin{figure}[htbp]
    \centering
    \includegraphics[width=\linewidth]{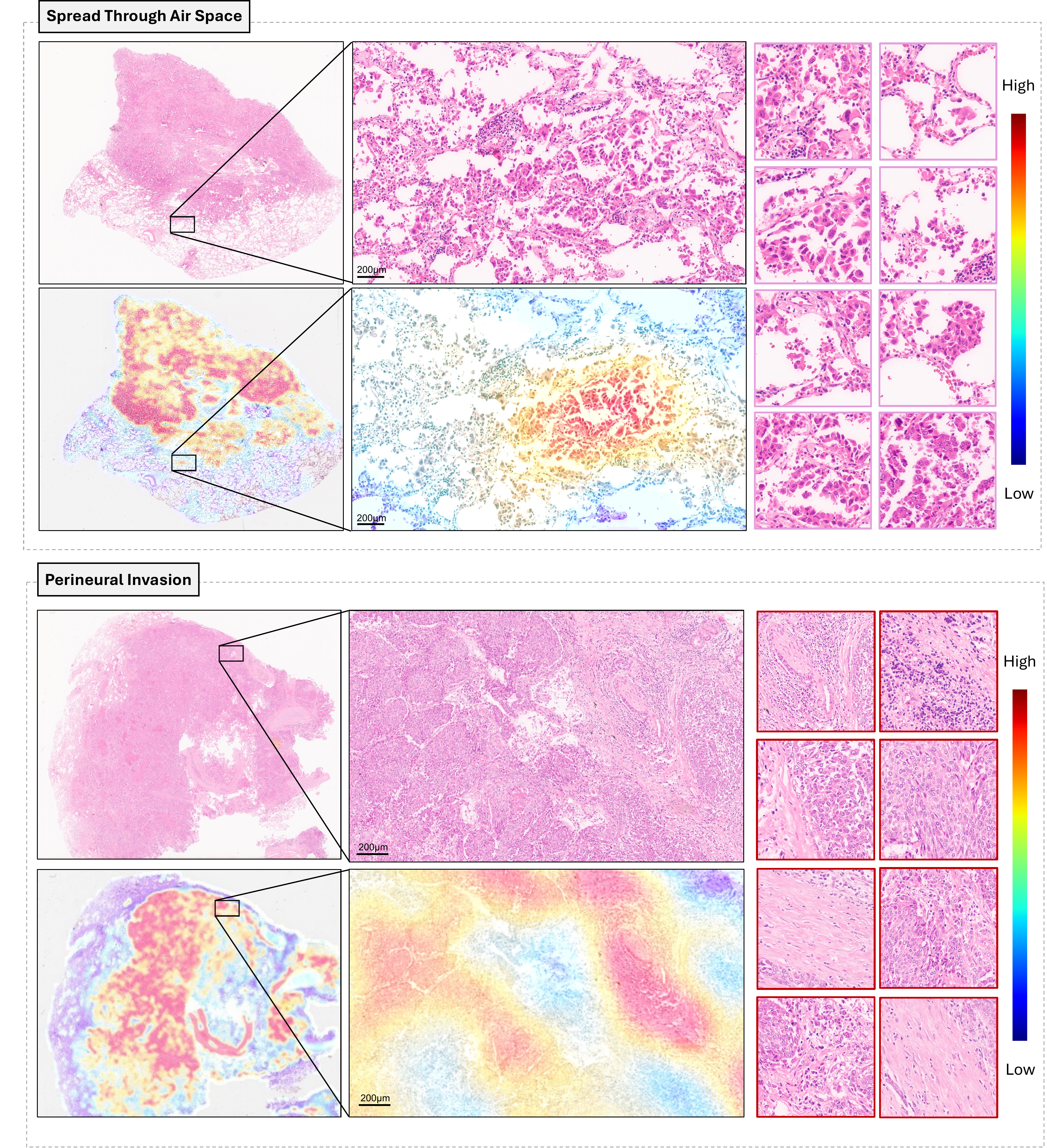}
    \captionsetup{name=Extended Data Figure}
    \caption{\textbf{Interpretability analysis of PulmoFoundation for detection of tumor spread through air spaces (STAS) and perineural invasion.}
    \textbf{Top panels, STAS detection:} Whole-slide image from lung adenocarcinoma, corresponding attention heatmap, magnified H\&E region showing STAS at the tumor periphery, zoomed attention heatmap, and top-ranked patches by attention score. Red indicates higher attention and blue indicates lower attention.
    \textbf{Bottom panels, perineural invasion detection:} Whole-slide image from a surgical resection specimen, corresponding attention heatmap, magnified H\&E region showing tumor cells infiltrating along nerve bundles, zoomed attention heatmap, and top-ranked patches by attention score. These examples illustrate that high-attention regions overlapped with diagnostically relevant STAS and perineural-invasion morphology.}
    \label{ext_fig_heatmap}
\end{figure}

\begin{figure}[htbp]
    \centering
    \includegraphics[width=\linewidth]{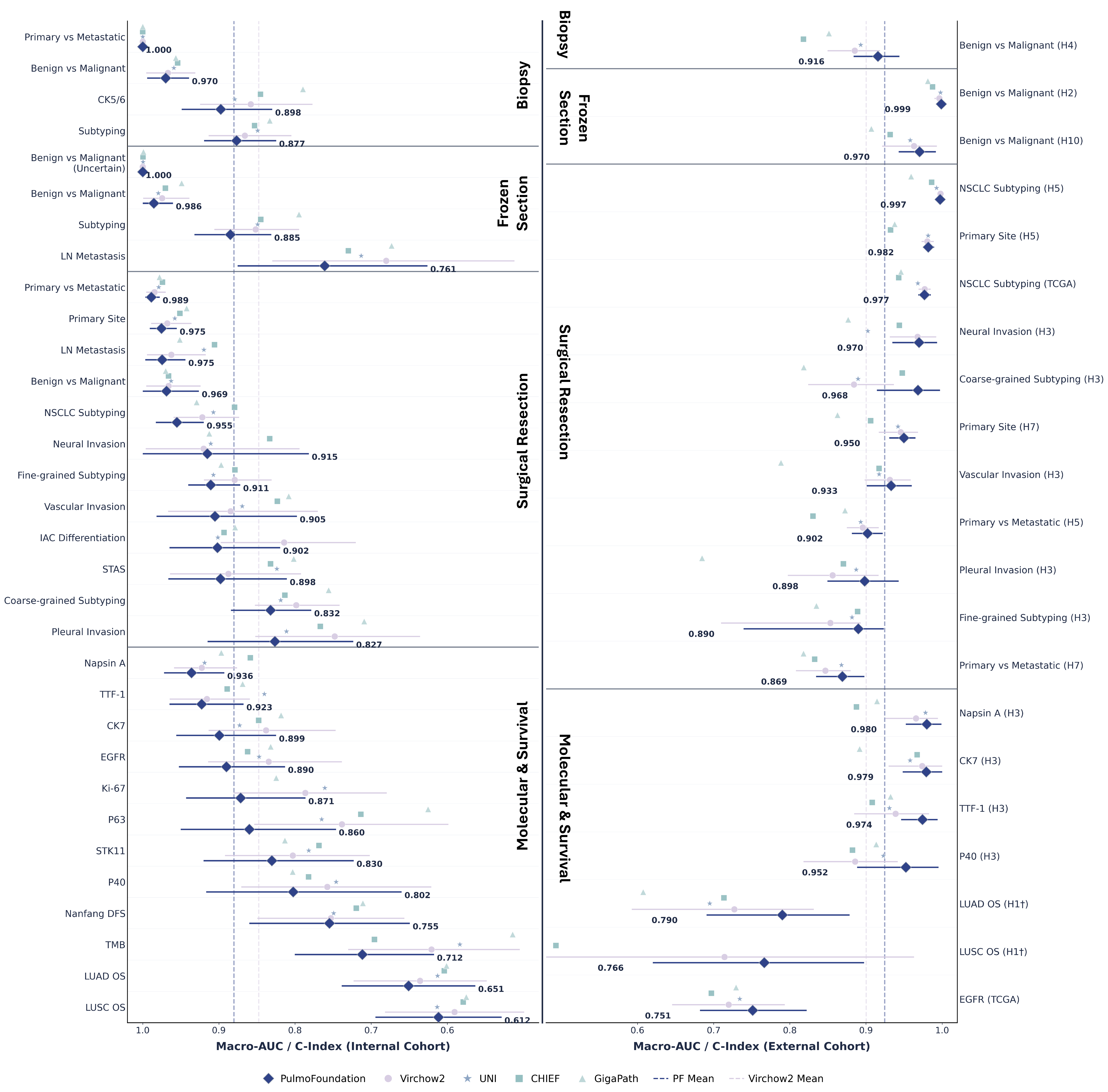}
    \captionsetup{name=Extended Data Figure}
    \caption{\textbf{Head-to-head retrospective comparison of PulmoFoundation against four pan-cancer foundation-model baselines.}
    Per-task performance is shown for PulmoFoundation and four baselines (UNI, Virchow2, CHIEF, and GigaPath) across 53 retrospective task cohorts, comprising 32 internal task cohorts (left) and 21 external task cohorts (right). Horizontal bars demonstrate PulmoFoundation and Virchow2 point estimates, with 95\% bootstrap confidence intervals shown for both models. UNI, CHIEF, and GigaPath are shown as individual markers. Tasks are grouped by clinical category: diagnostic biopsy, frozen section, surgical resection, molecular and IHC biomarker prediction, and survival analysis. Macro AUC is reported for classification tasks, and C-index is reported for survival tasks.}
    \label{ext_fig_overall_performance}
\end{figure}

\clearpage

\begin{figure}[htbp]
    \centering
    \includegraphics[width=\linewidth]{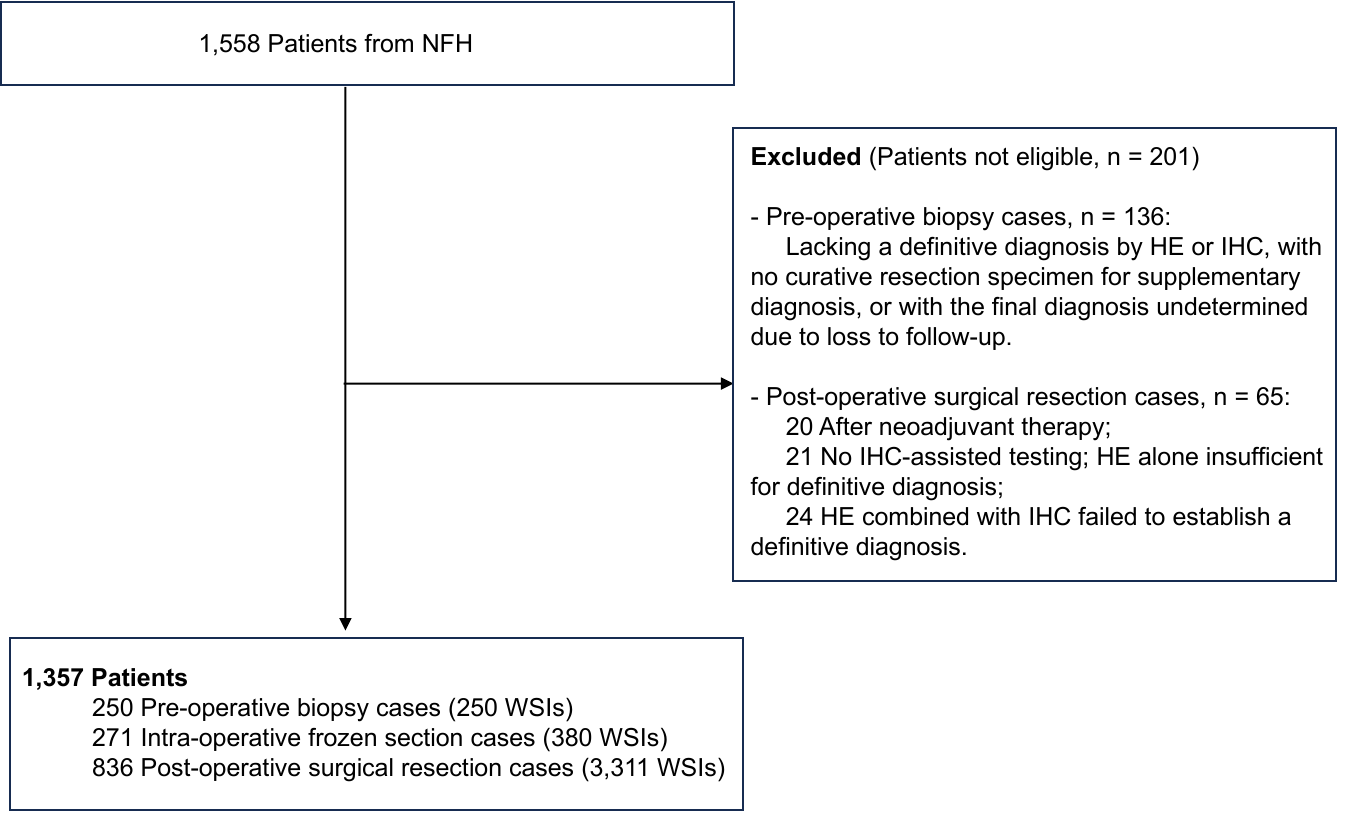}
    \captionsetup{name=Extended Data Figure}
    \caption{\textbf{CONSORT-style case-flow diagram for the prospective validation study.}
    Of 1,558 patients screened at Center-H1, 201 were excluded as ineligible. These exclusions comprised 136 pre-operative biopsy cases lacking a definitive diagnosis by H\&E or IHC, with no curative resection specimen available for supplementary diagnosis, or with the final diagnosis undetermined because of loss to follow-up, and 65 post-operative surgical-resection cases, including 20 cases after neoadjuvant therapy, 21 cases without IHC-assisted testing in which H\&E alone was insufficient for definitive diagnosis, and 24 cases in which H\&E combined with IHC failed to establish a definitive diagnosis. The final prospective cohort included 1,357 patients, comprising 250 pre-operative biopsy cases (250 WSIs), 271 intra-operative frozen-section cases (380 WSIs), and 836 post-operative surgical-resection cases (3,311 WSIs).}
    \label{ext_fig_consort_prospective}
\end{figure}

\begin{figure}[htbp]
    \centering
    \includegraphics[width=\linewidth]{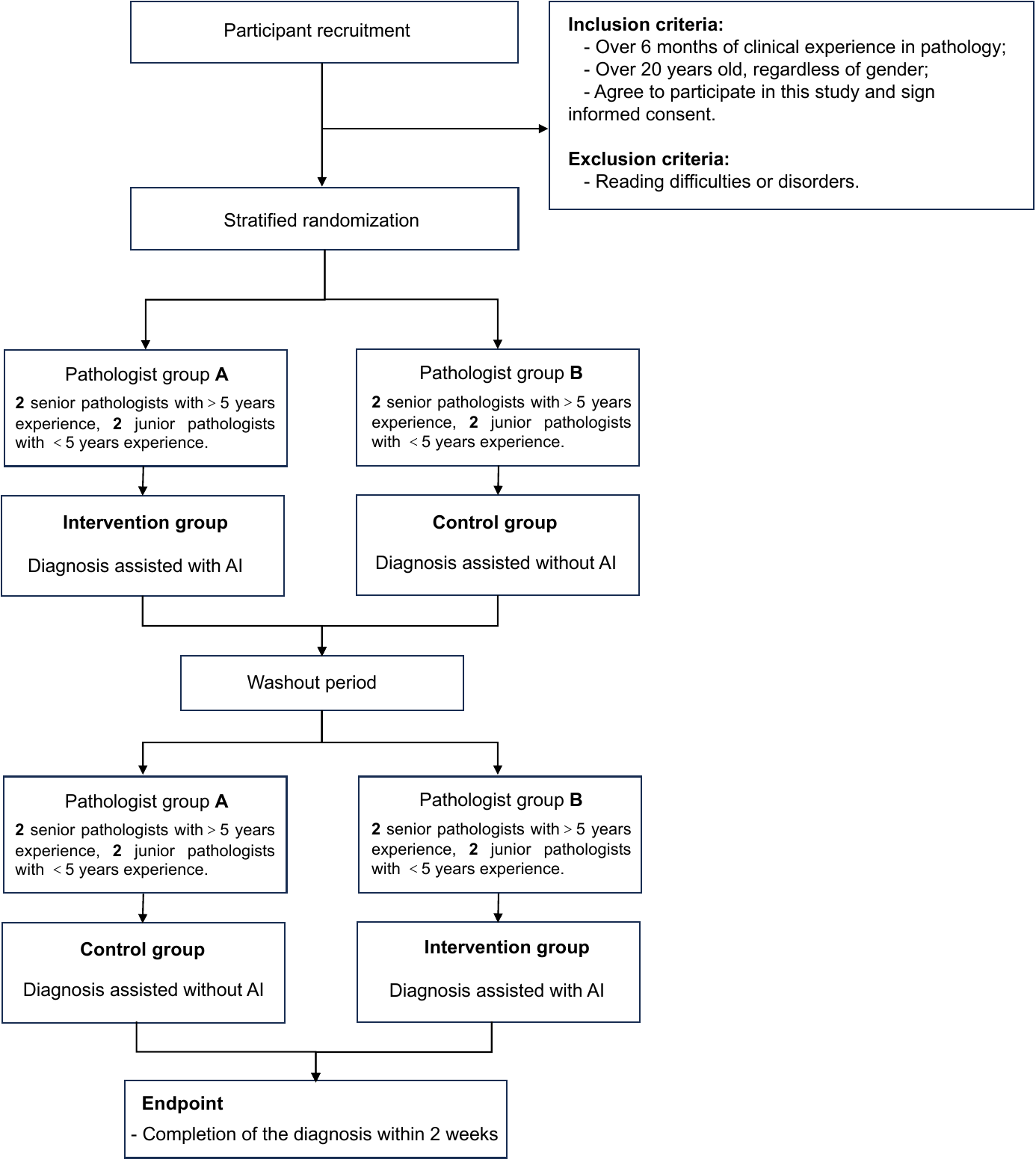}
    \captionsetup{name=Extended Data Figure}
    \caption{\textbf{CONSORT-style crossover RCT design for evaluating pathologist-AI interaction.}
    Eight board-certified pathologists from Center-H1 were enrolled, including four senior pathologists with $>$5 years of subspecialty experience and four junior pathologists with $<$5 years of subspecialty experience. Within each experience stratum, pathologists were randomized to Group A (AI-first sequence) or Group B (no-AI-first sequence), yielding 2 senior and 2 junior pathologists per sequence group. In Period 1, Group A reviewed cases with AI assistance and Group B reviewed cases without AI assistance; after a 4-week washout, groups crossed over in Period 2. Each pathologist reviewed the same case set in both periods, with randomized case order within each period. Inclusion criteria were $>$6 months of clinical pathology experience, age $>$20 years, and informed consent. The exclusion criterion was reading difficulty or disorder. Each evaluation period was completed within 2 weeks.}
    \label{ext_fig_consort_rct}
\end{figure}

\setcounter{table}{0}
\renewcommand{\tablename}{Extended Data Table}
\begin{table}[htbp]
\centering
\caption{Pretraining data sources used for lung-specific continual self-supervised pretraining. Slide Number reports available WSI counts by institutional or public source. LC25000 is a patch-level public dataset and therefore contributes 15{,}000 image patches but no WSI count. The pretraining cohort comprised 39{,}503 WSIs and approximately 88 million image patches.}
\label{tab:pretrain}
\resizebox{0.75\linewidth}{!}{%
\begin{tabular}{ccccc}
\hline
Usage & Data Source & Data Cohort & Slide Number & Patch Number \\
\hline
\multirow{12}{*}{Pretrain} & H1 & Surgical Slide Cohort & 22,747 & 63.5 Million \\
 & H2 & Frozen Section Cohort & 8,728 & 3 Million \\
 & H3 & Surgical Slide Cohort & 420 & 1.2 Million \\
 & H4 & Biopsy Cohort & 1,592 & 4.7 Million \\
 & H6 & Surgical Slide Cohort & 861 & 2.5 Million \\
 & H8 & Surgical Slide Cohort & 454 & 1.2 Million \\
 & H9 & Surgical Slide Cohort & 228 & 0.7 Million \\
 & CPTAC \cite{edwards2015cptac} & Surgical Slide Cohort & 2,213 & 5.7 Million \\
 & NLST \cite{national2011nlst} & Surgical Slide Cohort & 1,224 & 3.5 Million \\
 & HistAI \cite{nechaev2025histai} & Surgical Slide Cohort & 949 & 1.8 Million \\
 & WSSS4LUAD \cite{han2022wsss4luad} & Surgical Slide Cohort & 87 & 10,091 \\
 & LC25000 \cite{borkowski2019lung} & Surgical Slide Cohort & - & 15,000 \\
\hline
Total & - & - & 39,503 & 88 Million \\
\hline
\end{tabular}%
}
\end{table}

\begin{table}[htbp]
\centering
\caption{Source-level mapping of institutional and public data used for PulmoFoundation downstream task evaluation. Rows summarize source-level patient and WSI counts rather than task-specific denominators. \textit{Data Source} denotes an anonymized institution or public repository. \textit{Data Cohort} denotes the collection design and specimen type for that source. \textit{Time Range} denotes the available collection period; N/S indicates that collection dates were not specified in the de-identified retrospective exports, which were collected before the prospective study period.}
\label{tab:downstream}
\resizebox{\linewidth}{!}{%
\begin{tabular}{cccccc}
\hline
Usage & Data Source & Data Cohort & Patient Number & Slide Number & Time Range \\
\hline
\multirow{12}{*}{Evaluation} & H1 & Retrospective Biopsy & 1,189 & 1,211 & N/S \\
 & H1 & Retrospective Frozen Section & 1,115 & 2,998 & N/S \\
 & H1 & Retrospective Surgical Slide & 5,114 & 10,991 & N/S \\
 & H1 & Prospective Biopsy & 250 & 250 & Oct. 2024 - Aug. 2025 \\
 & H1 & Prospective Frozen Section & 271 & 380 & Aug. 2025 - Nov. 2025 \\
 & H1 & Prospective Surgical Slide & 836 & 3,311 & Oct. 2024 - Aug. 2025 \\
& H2 & Retrospective Frozen Section & 449 & 514 & N/S \\
 & H3 & Retrospective Surgical Slide & 900 & 2,719 & N/S \\
 & H4 & Retrospective Biopsy & 920 & 1,045 & N/S \\
 & H5 & Retrospective Surgical Slide & 819 & 1,476 & N/S \\
 & H7 & Retrospective Surgical Slide & 493 & 526 & N/S \\
 & H10 & Retrospective Frozen Section & 169 & 169 & N/S \\
 & TCGA \cite{weinstein2013cancer} & Retrospective Surgical Slide & 956 & 1,053 & N/S \\
\hline
Total & - & - & 13,481 & 26,643 & - \\
\hline
\end{tabular}%
}
\end{table}

\begin{table}[htbp]
\centering
\caption{Evaluation scope of PulmoFoundation across all 32 tasks. Each task is categorized by clinical stage and evaluated across four validation levels: internal retrospective evaluation, external retrospective evaluation at independent institutions, prospective evaluation of consecutive patients at Center-H1 from October 2024 to November 2025, and crossover RCT evaluation at Center-H1. Numeric entries are WSI counts. The Internal column reports the full retrospective internal task cohort used for downstream model development and held-out testing, rather than only the held-out test fold; per-task Extended Data Tables report held-out test-fold case counts, WSI counts, and class balance for internal performance evaluation. External validation sources are listed by anonymized center identifier or public repository with cohort size. The RCT column reports WSI counts used in the crossover RCT reader study. The Summary row reports the number of tasks or cohorts evaluated at each validation level. A dash (--) indicates that the task was not evaluated at that level. $^\dagger$Held-out Center-H1 cohort for survival validation.}
\label{tab:eval_scope}
\normalsize
\setlength{\tabcolsep}{4pt}
\renewcommand{\arraystretch}{0.85}
\begin{adjustbox}{max width=\textwidth}
\begin{tabular}{lrccc}
\toprule
\textbf{Task} & \textbf{Internal} & \textbf{External (center, $n$)} & \textbf{Prospective} & \textbf{RCT} \\
\midrule
\multicolumn{5}{l}{\textbf{Diagnostic biopsy}} \\
\multicolumn{5}{l}{\textit{\hspace{0.5em}Classification}} \\
\hspace{1em} Benign vs. malignant & 797 & H4 (400) & 250 & -- \\
\hspace{1em} Primary vs. metastatic & 568 & -- & -- & 136 \\
\hspace{1em} Histologic subtyping (4-class) & 541 & -- & -- & -- \\
\midrule
\multicolumn{5}{l}{\textit{\hspace{0.5em}Biomarker}} \\
\hspace{1em} CK5/6 prediction & 618 & -- & -- & -- \\
\midrule
\multicolumn{5}{l}{\textbf{Intra-operative frozen section}} \\
\multicolumn{5}{l}{\textit{\hspace{0.5em}Classification}} \\
\hspace{1em} Benign vs. malignant & 1,179 & H2 (514), H10 (169) & 380 & -- \\
\hspace{1em} Benign vs. malignant (uncertain cases) & 1,188 & -- & -- & 235 \\
\hspace{1em} AIS/MIA/IAC classification & 919 & -- & -- & -- \\
\midrule
\multicolumn{5}{l}{\textit{\hspace{0.5em}Staging}} \\
\hspace{1em} Lymph node metastasis from primary tumor & 738 & -- & -- & -- \\
\midrule
\multicolumn{5}{l}{\textbf{Post-operative surgical resection}} \\
\multicolumn{5}{l}{\textit{\hspace{0.5em}Classification}} \\
\hspace{1em} Benign vs. malignant & 600 & -- & -- & -- \\
\hspace{1em} NSCLC subtyping (ADC vs. SCC) & 599 & H5 (517), TCGA (1,053) & 1,143 & 1,143 \\
\hspace{1em} Coarse-grained (3-class) & 599 & H3 (511) & 1,673 & -- \\
\hspace{1em} Fine-grained (6-class) & 1,078 & H3 (767) & 1,802 & -- \\
\hspace{1em} Primary vs. metastatic & 1,422 & H5 (1,307), H7 (493) & -- & -- \\
\hspace{1em} Primary site prediction & 1,183 & H5 (967), H7 (423) & -- & 289 \\
\midrule
\multicolumn{5}{l}{\textit{\hspace{0.5em}Staging and grading}} \\
\hspace{1em} Lymph node metastasis & 737 & -- & 1,770 & -- \\
\hspace{1em} Spread through air spaces (STAS) & 569 & -- & -- & -- \\
\hspace{1em} Perineural invasion & 410 & H3 (366) & -- & -- \\
\hspace{1em} Pleural invasion & 1,192 & H3 (789) & -- & -- \\
\hspace{1em} Vascular tumor thrombus & 517 & H3 (807) & -- & -- \\
\hspace{1em} IAC differentiation grading & 1,425 & -- & -- & -- \\
\midrule
\multicolumn{5}{l}{\textbf{Molecular and IHC biomarker prediction}} \\
\multicolumn{5}{l}{\textit{\hspace{0.5em}Diagnostic IHC panel}} \\
\hspace{1em} TTF-1 & 774 & H3 (774) & 1,056 & -- \\
\hspace{1em} Napsin-A & 821 & H3 (446) & 700 & -- \\
\hspace{1em} CK-7 & 635 & H3 (377) & 614 & -- \\
\hspace{1em} P40 & 442 & H3 (780) & 737 & -- \\
\hspace{1em} P63 & 354 & -- & 297 & -- \\
\midrule
\multicolumn{5}{l}{\textit{\hspace{0.5em}Proliferation and molecular}} \\
\hspace{1em} Ki-67 & 630 & -- & -- & -- \\
\hspace{1em} EGFR mutation & 706 & TCGA (469) & -- & -- \\
\hspace{1em} STK11 mutation & 469 & -- & -- & -- \\
\hspace{1em} TMB & 970 & -- & -- & -- \\
\midrule
\multicolumn{5}{l}{\textbf{Survival prediction}} \\
\hspace{1em} LUAD overall survival & 518 & H1$^\dagger$ (1,387) & -- & -- \\
\hspace{1em} LUSC overall survival & 484 & H1$^\dagger$ (382) & -- & -- \\
\hspace{1em} Disease-free survival & 1,715 & -- & -- & -- \\
\midrule
\multicolumn{5}{l}{\textit{Summary}} \\
\hspace{1em} Total & 32 tasks & 21 cohorts & 11 tasks & 4 tasks \\
\bottomrule
\end{tabular}
\end{adjustbox}
\renewcommand{\arraystretch}{1.0}
\end{table}

\clearpage

\begin{table}[htbp]
  \centering 
  \caption{Patient characteristics of the prospective Center-H1 cohort. The table summarizes overall patient-level characteristics for the 1{,}357 enrolled prospective patients and 3{,}941 WSIs. These counts describe the prospective cohort overall and are not task-specific denominators; task-specific case counts, WSI counts, and class distributions are reported in the corresponding prospective performance tables.}
  \label{tab:prospective}
  \begin{tabular}{lr}
    \toprule 
    \textbf{Characteristics} &  \\
    \midrule 
    \textbf{Patient Counts} & 1,357 \\
    \textbf{Slide Counts} & 3,941 \\
    \textbf{Received Date} & Oct. 2024 - Nov. 2025 \\
    \textbf{Diagnosis for biopsy} & \\ 
    \hspace{1em} Malignant & 179 \\
    \hspace{1em} Benign & 71 \\
    \textbf{Diagnosis for frozen section} & \\ 
    \hspace{1em} Malignant & 240 \\
    \hspace{1em} Benign & 31 \\
    \textbf{Surgical subtype diagnosis} & \\ 
    \hspace{1em} Adenocarcinoma in situ & 127 \\
    \hspace{1em} Invasive adenocarcinoma & 425 \\
    \hspace{1em} Squamous cell carcinoma  & 70 \\
    \hspace{1em} Minimally invasive adenocarcinoma  & 219 \\
    \hspace{1em} Neuroendocrine carcinoma  & 5 \\
    \hspace{1em} Adenosquamous carcinoma  & 7 \\
    \textbf{NSCLC diagnosis for surgical slides} & \\ 
    \hspace{1em} Adenocarcinoma & 771 \\
    \hspace{1em} Squamous cell carcinoma & 70 \\
    \textbf{Tumor staging for surgical slides} & \\ 
    \hspace{1em} Positive Lymph Node Metastasis & 86 \\
    \hspace{1em} Negative Lymph Node Metastasis & 643 \\
    \textbf{CK-7 biomarker for surgical slides} & \\ 
    \hspace{1em} Positive & 105 \\
    \hspace{1em} Negative & 23 \\
    \textbf{Napsin-A biomarker for surgical slides} & \\ 
    \hspace{1em} Positive & 74\\
    \hspace{1em} Negative & 68 \\
    \textbf{P40 biomarker for surgical slides} & \\ 
    \hspace{1em} Positive & 46 \\
    \hspace{1em} Negative & 106 \\
    \textbf{P63 biomarker for surgical slides} & \\ 
    \hspace{1em} Positive & 34 \\
    \hspace{1em} Negative & 24 \\
    \textbf{TTF-1 biomarker for surgical slides} & \\ 
    \hspace{1em} Positive & 186 \\
    \hspace{1em} Negative & 60 \\
    \bottomrule 
  \end{tabular}
\end{table}

\begin{table}[htbp]
\centering
\caption{\textbf{Companion to Fig.~\ref{fig1}d and Extended Data Fig.~\ref{ext_fig_overall_performance}: per-task deltas of PulmoFoundation versus each foundation-model baseline.} $\Delta$ denotes PulmoFoundation (PF) minus baseline on the primary metric: macro AUC for classification tasks and C-index for survival tasks. Values in parentheses are paired bootstrap 95\% percentile confidence intervals for the delta from 1{,}000 bootstrap replicates. The table is organized by clinical stage across two table blocks: diagnostic biopsy, frozen-section, and post-operative resection tasks are shown here; molecular, biomarker, and survival tasks are shown in the continued block. Rows are grouped by task; horizontal rules separate tasks and clinical-stage groups. ``--'' indicates that the baseline was not evaluated for that task-cohort combination. H1$^\dagger$ denotes the held-out Center-H1 survival validation cohort.}
\label{tab:companion_fig8}
\renewcommand{\arraystretch}{0.95}
\setlength{\tabcolsep}{2pt}
\scriptsize
\begin{adjustbox}{max width=\textwidth,center}
\begin{tabular}{l l r p{3.0cm} p{3.0cm} p{3.0cm} p{3.0cm}}
\toprule
\textbf{Task} & \textbf{Cohort} & \textbf{PF} & \textbf{$\Delta$ vs UNI} & \textbf{$\Delta$ vs Virchow2} & \textbf{$\Delta$ vs CHIEF} & \textbf{$\Delta$ vs GigaPath} \\
\midrule
\multicolumn{7}{l}{\textit{Biopsy}}\\
\midrule
Subtyping & Internal & $0.877$ & $+0.028$ ($-0.007,\,+0.066$) & $+0.011$ ($-0.015,\,+0.038$) & $+0.024$ ($-0.020,\,+0.068$) & $+0.044$ ($-0.006,\,+0.091$) \\
\cmidrule(lr){1-7}
Primary vs Metastatic & Internal & $1.000$ & $+0.000$ ($+0.000,\,+0.000$) & $+0.000$ ($+0.000,\,+0.000$) & $-0.000$ ($+0.000,\,+0.000$) & $-0.000$ ($+0.000,\,+0.000$) \\
\cmidrule(lr){1-7}
CK5/6 & Internal & $0.898$ & $+0.018$ ($-0.039,\,+0.086$) & $+0.039$ ($-0.011,\,+0.097$) & $+0.052$ ($-0.016,\,+0.122$) & $+0.108$ ($+0.032,\,+0.186$) \\
\cmidrule(lr){1-7}
Benign vs Malignant & Internal & $0.970$ & $+0.011$ ($-0.002,\,+0.027$) & $+0.003$ ($-0.005,\,+0.013$) & $+0.016$ ($-0.008,\,+0.042$) & $+0.013$ ($-0.005,\,+0.034$) \\
 & H4 & $0.916$ & $+0.023$ ($-0.020,\,+0.065$) & $+0.030$ ($+0.009,\,+0.054$) & $+0.098$ ($+0.044,\,+0.149$) & $+0.064$ ($+0.016,\,+0.112$) \\
\midrule
\multicolumn{7}{l}{\textit{Frozen Section}}\\
\midrule
Subtyping & Internal & $0.885$ & $+0.036$ ($-0.001,\,+0.078$) & $+0.033$ ($-0.005,\,+0.072$) & $+0.040$ ($-0.014,\,+0.092$) & $+0.090$ ($+0.041,\,+0.145$) \\
\cmidrule(lr){1-7}
LN Metastasis & Internal & $0.761$ & $+0.048$ ($-0.092,\,+0.195$) & $+0.081$ ($-0.075,\,+0.262$) & $+0.031$ ($-0.108,\,+0.197$) & $+0.088$ ($-0.024,\,+0.194$) \\
\cmidrule(lr){1-7}
Benign vs Malignant & Internal & $0.986$ & $+0.006$ ($-0.002,\,+0.017$) & $+0.011$ ($-0.001,\,+0.028$) & $+0.015$ ($-0.003,\,+0.042$) & $+0.037$ ($+0.009,\,+0.073$) \\
 & H10 & $0.970$ & $+0.012$ ($-0.027,\,+0.054$) & $+0.007$ ($-0.014,\,+0.037$) & $+0.039$ ($-0.004,\,+0.088$) & $+0.063$ ($+0.000,\,+0.128$) \\
 & H2 & $0.999$ & $+0.001$ ($-0.002,\,+0.005$) & $+0.003$ ($+0.000,\,+0.007$) & $+0.011$ ($+0.003,\,+0.021$) & $+0.018$ ($+0.009,\,+0.028$) \\
\cmidrule(lr){1-7}
Benign vs Malignant (Uncertain) & Internal & $1.000$ & $+0.000$ ($+0.000,\,+0.002$) & $+0.000$ ($+0.000,\,+0.000$) & $+0.000$ ($+0.000,\,+0.002$) & $+0.001$ ($+0.000,\,+0.004$) \\
\midrule
\multicolumn{7}{l}{\textit{Resection}}\\
\midrule
LN Metastasis & Internal & $0.975$ & $+0.055$ ($+0.006,\,+0.121$) & $+0.012$ ($-0.002,\,+0.035$) & $+0.069$ ($+0.009,\,+0.161$) & $+0.023$ ($-0.016,\,+0.066$) \\
\cmidrule(lr){1-7}
Primary Site & Internal & $0.975$ & $+0.017$ ($+0.002,\,+0.038$) & $+0.008$ ($-0.007,\,+0.026$) & $+0.024$ ($+0.003,\,+0.052$) & $+0.033$ ($+0.013,\,+0.055$) \\
 & H5 & $0.982$ & $+0.000$ ($-0.011,\,+0.010$) & $+0.001$ ($-0.003,\,+0.006$) & $+0.049$ ($+0.029,\,+0.076$) & $+0.044$ ($+0.026,\,+0.064$) \\
 & H7 & $0.950$ & $+0.008$ ($-0.018,\,+0.034$) & $+0.004$ ($-0.008,\,+0.018$) & $+0.044$ ($+0.012,\,+0.080$) & $+0.087$ ($+0.035,\,+0.149$) \\
\cmidrule(lr){1-7}
Coarse-grained Subtyping & Internal & $0.832$ & $+0.013$ ($-0.028,\,+0.054$) & $+0.034$ ($-0.000,\,+0.072$) & $+0.019$ ($-0.020,\,+0.058$) & $+0.076$ ($+0.024,\,+0.134$) \\
 & H3 & $0.968$ & $+0.079$ ($+0.007,\,+0.146$) & $+0.084$ ($+0.047,\,+0.126$) & $+0.021$ ($-0.044,\,+0.088$) & $+0.150$ ($+0.062,\,+0.229$) \\
\cmidrule(lr){1-7}
STAS & Internal & $0.898$ & $+0.074$ ($+0.000,\,+0.163$) & $+0.010$ ($-0.037,\,+0.064$) & $+0.066$ ($-0.031,\,+0.166$) & $+0.096$ ($-0.007,\,+0.227$) \\
\cmidrule(lr){1-7}
NSCLC Subtyping & Internal & $0.955$ & $+0.048$ ($+0.017,\,+0.086$) & $+0.033$ ($+0.005,\,+0.066$) & $+0.076$ ($+0.033,\,+0.124$) & $+0.026$ ($-0.005,\,+0.060$) \\
 & H5 & $0.997$ & $+0.005$ ($-0.005,\,+0.018$) & $-0.000$ ($-0.005,\,+0.002$) & $+0.011$ ($-0.004,\,+0.039$) & $+0.038$ ($-0.002,\,+0.117$) \\
 & TCGA & $0.977$ & $+0.009$ ($-0.004,\,+0.023$) & $-0.000$ ($-0.003,\,+0.003$) & $+0.034$ ($+0.019,\,+0.049$) & $+0.031$ ($+0.015,\,+0.047$) \\
\cmidrule(lr){1-7}
Neural Invasion & Internal & $0.915$ & $+0.004$ ($-0.077,\,+0.088$) & $-0.005$ ($-0.036,\,+0.022$) & $+0.082$ ($-0.026,\,+0.206$) & $+0.002$ ($-0.103,\,+0.096$) \\
 & H3 & $0.970$ & $+0.067$ ($-0.018,\,+0.192$) & $+0.002$ ($-0.011,\,+0.014$) & $+0.026$ ($-0.024,\,+0.084$) & $+0.093$ ($+0.009,\,+0.193$) \\
\cmidrule(lr){1-7}
Pleural Invasion & Internal & $0.827$ & $+0.015$ ($-0.035,\,+0.072$) & $+0.079$ ($+0.002,\,+0.154$) & $+0.060$ ($-0.021,\,+0.144$) & $+0.117$ ($+0.011,\,+0.220$) \\
 & H3 & $0.898$ & $+0.011$ ($-0.059,\,+0.080$) & $+0.042$ ($+0.001,\,+0.082$) & $+0.028$ ($-0.046,\,+0.110$) & $+0.213$ ($+0.123,\,+0.301$) \\
\cmidrule(lr){1-7}
Benign vs Malignant & Internal & $0.969$ & $+0.006$ ($-0.007,\,+0.020$) & $+0.003$ ($-0.010,\,+0.015$) & $+0.003$ ($-0.025,\,+0.023$) & $-0.001$ ($-0.015,\,+0.011$) \\
\cmidrule(lr){1-7}
Fine-grained Subtyping & Internal & $0.911$ & $+0.003$ ($-0.014,\,+0.022$) & $+0.031$ ($+0.004,\,+0.059$) & $+0.032$ ($+0.009,\,+0.054$) & $+0.014$ ($-0.008,\,+0.035$) \\
 & H3 & $0.890$ & $+0.008$ ($-0.140,\,+0.157$) & $+0.037$ ($+0.008,\,+0.063$) & $+0.001$ ($-0.157,\,+0.163$) & $+0.055$ ($-0.102,\,+0.195$) \\
\cmidrule(lr){1-7}
Vascular Invasion & Internal & $0.905$ & $+0.036$ ($-0.035,\,+0.123$) & $+0.020$ ($-0.022,\,+0.069$) & $+0.082$ ($-0.012,\,+0.189$) & $+0.097$ ($-0.021,\,+0.237$) \\
 & H3 & $0.933$ & $+0.016$ ($-0.030,\,+0.062$) & $+0.002$ ($-0.008,\,+0.011$) & $+0.016$ ($-0.031,\,+0.058$) & $+0.145$ ($+0.074,\,+0.224$) \\
\cmidrule(lr){1-7}
Primary vs Metastatic & Internal & $0.989$ & $+0.010$ ($-0.000,\,+0.023$) & $+0.004$ ($-0.003,\,+0.013$) & $+0.015$ ($+0.003,\,+0.030$) & $+0.011$ ($+0.001,\,+0.024$) \\
 & H5 & $0.902$ & $+0.009$ ($-0.020,\,+0.037$) & $+0.006$ ($-0.005,\,+0.018$) & $+0.072$ ($+0.035,\,+0.106$) & $+0.030$ ($-0.002,\,+0.060$) \\
 & H7 & $0.869$ & $+0.001$ ($-0.046,\,+0.047$) & $+0.022$ ($+0.006,\,+0.039$) & $+0.036$ ($-0.010,\,+0.089$) & $+0.051$ ($+0.001,\,+0.105$) \\
\cmidrule(lr){1-7}
IAC Differentiation & Internal & $0.902$ & $+0.000$ ($-0.056,\,+0.050$) & $+0.088$ ($+0.013,\,+0.179$) & $+0.009$ ($-0.046,\,+0.075$) & $+0.023$ ($-0.031,\,+0.074$) \\
\bottomrule
\end{tabular}
\end{adjustbox}
\vspace{3pt}
\begin{flushleft}\footnotesize\textit{Note.} Values are rounded to three decimals; cells displayed as $+0.000$ or $-0.000$ reflect absolute deltas below $0.0005$ after rounding.\end{flushleft}
\end{table}
\clearpage

\begin{table}[htbp]
\ContinuedFloat
\centering
\caption[]{\textbf{(continued)} part B: molecular, biomarker, and survival tasks.}
\renewcommand{\arraystretch}{0.95}
\setlength{\tabcolsep}{2pt}
\scriptsize
\begin{adjustbox}{max width=\textwidth,center}
\begin{tabular}{l l r p{3.0cm} p{3.0cm} p{3.0cm} p{3.0cm}}
\toprule
\textbf{Task} & \textbf{Cohort} & \textbf{PF} & \textbf{$\Delta$ vs UNI} & \textbf{$\Delta$ vs Virchow2} & \textbf{$\Delta$ vs CHIEF} & \textbf{$\Delta$ vs GigaPath} \\
\midrule
\multicolumn{7}{l}{\textit{Molecular \& Biomarker Prediction}}\\
\midrule
EGFR & Internal & $0.890$ & $+0.043$ ($-0.009,\,+0.105$) & $+0.056$ ($+0.008,\,+0.114$) & $+0.028$ ($-0.031,\,+0.090$) & $+0.058$ ($+0.006,\,+0.111$) \\
 & TCGA & $0.751$ & $+0.017$ ($-0.077,\,+0.109$) & $+0.032$ ($-0.023,\,+0.088$) & $+0.054$ ($-0.042,\,+0.151$) & $+0.022$ ($-0.076,\,+0.124$) \\
\cmidrule(lr){1-7}
CK7 & Internal & $0.899$ & $+0.026$ ($-0.008,\,+0.063$) & $+0.061$ ($+0.013,\,+0.119$) & $+0.052$ ($+0.008,\,+0.105$) & $+0.081$ ($+0.027,\,+0.140$) \\
 & H3 & $0.979$ & $+0.021$ ($-0.036,\,+0.093$) & $+0.005$ ($-0.015,\,+0.032$) & $+0.012$ ($-0.028,\,+0.062$) & $+0.087$ ($+0.005,\,+0.188$) \\
\cmidrule(lr){1-7}
Ki-67 & Internal & $0.871$ & $+0.110$ ($+0.033,\,+0.190$) & $+0.085$ ($+0.003,\,+0.174$) & $+0.419$ ($+0.279,\,+0.562$) & $+0.047$ ($-0.020,\,+0.125$) \\
\cmidrule(lr){1-7}
Napsin A & Internal & $0.936$ & $+0.017$ ($-0.010,\,+0.047$) & $+0.014$ ($-0.007,\,+0.037$) & $+0.077$ ($+0.032,\,+0.129$) & $+0.039$ ($-0.002,\,+0.085$) \\
 & H3 & $0.980$ & $+0.002$ ($-0.030,\,+0.032$) & $+0.014$ ($+0.002,\,+0.032$) & $+0.092$ ($+0.034,\,+0.163$) & $+0.065$ ($+0.013,\,+0.128$) \\
\cmidrule(lr){1-7}
P40 & Internal & $0.802$ & $+0.056$ ($-0.021,\,+0.136$) & $+0.045$ ($-0.072,\,+0.155$) & $+0.020$ ($-0.163,\,+0.194$) & $-0.001$ ($-0.116,\,+0.117$) \\
 & H3 & $0.952$ & $+0.030$ ($-0.050,\,+0.107$) & $+0.067$ ($+0.024,\,+0.117$) & $+0.070$ ($-0.005,\,+0.145$) & $+0.039$ ($-0.041,\,+0.110$) \\
\cmidrule(lr){1-7}
P63 & Internal & $0.860$ & $+0.095$ ($-0.000,\,+0.205$) & $+0.122$ ($+0.031,\,+0.222$) & $+0.147$ ($+0.031,\,+0.283$) & $+0.235$ ($+0.116,\,+0.362$) \\
\cmidrule(lr){1-7}
TTF-1 & Internal & $0.923$ & $+0.082$ ($+0.028,\,+0.143$) & $+0.007$ ($-0.021,\,+0.036$) & $+0.034$ ($+0.000,\,+0.073$) & $+0.054$ ($+0.003,\,+0.109$) \\
 & H3 & $0.974$ & $+0.044$ ($-0.012,\,+0.107$) & $+0.035$ ($+0.008,\,+0.069$) & $+0.066$ ($+0.014,\,+0.124$) & $+0.042$ ($-0.005,\,+0.091$) \\
\cmidrule(lr){1-7}
STK11 & Internal & $0.830$ & $+0.048$ ($-0.047,\,+0.166$) & $+0.027$ ($-0.060,\,+0.122$) & $+0.062$ ($-0.036,\,+0.175$) & $+0.017$ ($-0.076,\,+0.117$) \\
\cmidrule(lr){1-7}
TMB & Internal & $0.712$ & $+0.128$ ($+0.044,\,+0.220$) & $+0.091$ ($+0.021,\,+0.172$) & $+0.016$ ($-0.041,\,+0.076$) & $+0.197$ ($+0.083,\,+0.312$) \\
\midrule
\multicolumn{7}{l}{\textit{Survival Prediction}}\\
\midrule
LUAD OS & Internal & $0.651$ & $+0.039$ ($-0.088,\,+0.168$) & $+0.015$ ($-0.111,\,+0.126$) & $+0.048$ ($-0.079,\,+0.174$) & $+0.048$ ($-0.088,\,+0.187$) \\
 & H1$^\dagger$ & $0.790$ & $+0.095$ ($-0.090,\,+0.282$) & $+0.063$ ($-0.007,\,+0.133$) & $+0.077$ ($-0.085,\,+0.248$) & $+0.183$ ($+0.002,\,+0.362$) \\
\cmidrule(lr){1-7}
LUSC OS & Internal & $0.612$ & $+0.001$ ($-0.109,\,+0.123$) & $+0.020$ ($-0.098,\,+0.131$) & $+0.036$ ($-0.093,\,+0.166$) & $+0.036$ ($-0.068,\,+0.142$) \\
 & H1$^\dagger$ & $0.766$ & $+0.386$ ($-0.134,\,+0.846$) & $+0.052$ ($-0.242,\,+0.294$) & $+0.273$ ($-0.030,\,+0.538$) & $+0.417$ ($+0.159,\,+0.630$) \\
\cmidrule(lr){1-7}
Nanfang DFS & Internal & $0.755$ & $+0.004$ ($-0.129,\,+0.149$) & $+0.003$ ($-0.134,\,+0.132$) & $+0.031$ ($-0.121,\,+0.175$) & $+0.042$ ($-0.118,\,+0.203$) \\
\bottomrule
\end{tabular}
\end{adjustbox}
\end{table}
\clearpage

\begin{table}[htbp]
\centering
\caption{\textbf{Benign vs Malignant (Biopsy)}: Per-task performance of PulmoFoundation and four baseline foundation models (UNI, Virchow2, CHIEF, GigaPath). Performance metrics are reported with 95\% bootstrap percentile CIs. Macro AUC is threshold-free. Macro Sensitivity, Specificity, PPV, and NPV are macro-averaged one-versus-rest metrics computed from predicted class labels; per-class sensitivity rows are computed at class-specific Youden-optimal operating points. CIs are bootstrap percentiles over cases ($N{=}1{,}000$ replicates). Best-performing model per metric and cohort is shown in bold. Cohort-composition rows report case counts, with slide counts in parentheses; class percentages are case-level percentages. All metrics are computed on the same cohort.}
\label{tab:biopsy_benign_malignant}
\renewcommand{\arraystretch}{1.15}
\setlength{\tabcolsep}{5pt}
\small
\begin{adjustbox}{max width=\textwidth,center}
\begin{tabular}{l l c c}
\toprule
\textbf{Metric} & \textbf{Model} & \textbf{Internal} & \textbf{External H4} \\
\midrule
Total cases (slides) & -- & 159 (159) & 400 (400) \\
Benign & -- & 79 (79; 49.7\%) & 200 (200; 50.0\%) \\
Malignant & -- & 80 (80; 50.3\%) & 200 (200; 50.0\%) \\
\midrule
\multirow{5}{*}{\textbf{Macro AUC}} & UNI & 0.959 (0.919, 0.992) & 0.893 (0.858, 0.924) \\
 & Virchow2 & 0.967 (0.931, 0.995) & 0.885 (0.849, 0.918) \\
 & CHIEF & 0.954 (0.921, 0.984) & 0.818 (0.774, 0.860) \\
 & GigaPath & 0.957 (0.916, 0.988) & 0.851 (0.813, 0.887) \\
 & PulmoFoundation & \textbf{0.970 (0.939, 0.994)} & \textbf{0.916 (0.884, 0.944)} \\
\midrule
\multirow{5}{*}{\textbf{Macro Sensitivity}} & UNI & 0.918 (0.873, 0.961) & 0.821 (0.783, 0.858) \\
 & Virchow2 & \textbf{0.956 (0.920, 0.987)} & 0.774 (0.735, 0.812) \\
 & CHIEF & 0.912 (0.867, 0.954) & 0.608 (0.568, 0.646) \\
 & GigaPath & 0.844 (0.794, 0.894) & 0.763 (0.724, 0.802) \\
 & PulmoFoundation & 0.943 (0.905, 0.976) & \textbf{0.869 (0.835, 0.901)} \\
\midrule
\multirow{5}{*}{\textbf{Macro Specificity}} & UNI & 0.918 (0.873, 0.961) & 0.821 (0.783, 0.858) \\
 & Virchow2 & \textbf{0.956 (0.920, 0.987)} & 0.774 (0.735, 0.812) \\
 & CHIEF & 0.912 (0.867, 0.954) & 0.608 (0.568, 0.646) \\
 & GigaPath & 0.844 (0.794, 0.894) & 0.763 (0.724, 0.802) \\
 & PulmoFoundation & 0.943 (0.905, 0.976) & \textbf{0.869 (0.835, 0.901)} \\
\midrule
\multirow{5}{*}{\textbf{Macro PPV}} & UNI & 0.921 (0.879, 0.962) & 0.822 (0.784, 0.858) \\
 & Virchow2 & \textbf{0.957 (0.921, 0.987)} & 0.808 (0.769, 0.842) \\
 & CHIEF & 0.918 (0.877, 0.958) & 0.664 (0.608, 0.717) \\
 & GigaPath & 0.880 (0.833, 0.918) & 0.788 (0.749, 0.828) \\
 & PulmoFoundation & 0.944 (0.906, 0.976) & \textbf{0.881 (0.849, 0.912)} \\
\midrule
\multirow{5}{*}{\textbf{Macro NPV}} & UNI & 0.921 (0.879, 0.962) & 0.822 (0.784, 0.858) \\
 & Virchow2 & \textbf{0.957 (0.921, 0.987)} & 0.808 (0.769, 0.842) \\
 & CHIEF & 0.918 (0.877, 0.958) & 0.664 (0.608, 0.717) \\
 & GigaPath & 0.880 (0.833, 0.918) & 0.788 (0.749, 0.828) \\
 & PulmoFoundation & 0.944 (0.906, 0.976) & \textbf{0.881 (0.849, 0.912)} \\
\midrule
\multirow{5}{*}{\textbf{Benign Sensitivity (Youden)}} & UNI & 0.962 (0.905, 1.000) & \textbf{0.935 (0.878, 0.985)} \\
 & Virchow2 & \textbf{0.975 (0.921, 1.000)} & 0.820 (0.780, 0.939) \\
 & CHIEF & 0.975 (0.894, 1.000) & 0.840 (0.779, 0.937) \\
 & GigaPath & 0.975 (0.917, 1.000) & 0.855 (0.686, 0.939) \\
 & PulmoFoundation & 0.962 (0.908, 1.000) & 0.890 (0.866, 0.995) \\
\midrule
\multirow{5}{*}{\textbf{Malignant Sensitivity (Youden)}} & UNI & 0.938 (0.898, 0.989) & 0.770 (0.690, 0.834) \\
 & Virchow2 & 0.938 (0.897, 0.989) & \textbf{0.825 (0.704, 0.882)} \\
 & CHIEF & 0.863 (0.800, 0.959) & 0.715 (0.603, 0.787) \\
 & GigaPath & 0.900 (0.853, 0.975) & 0.705 (0.617, 0.863) \\
 & PulmoFoundation & \textbf{0.950 (0.902, 0.989)} & 0.780 (0.737, 0.904) \\
\bottomrule
\end{tabular}
\end{adjustbox}
\end{table}

\clearpage

\begin{table}[htbp]
\centering
\caption{\textbf{Primary vs Metastatic (Biopsy)}: Per-task performance of PulmoFoundation and four baseline foundation models (UNI, Virchow2, CHIEF, GigaPath). Performance metrics are reported with 95\% bootstrap percentile CIs. Macro AUC is threshold-free. Macro Sensitivity, Specificity, PPV, and NPV are macro-averaged one-versus-rest metrics computed from predicted class labels; per-class sensitivity rows are computed at class-specific Youden-optimal operating points. CIs are bootstrap percentiles over cases ($N{=}1{,}000$ replicates). Best-performing model per metric and cohort is shown in bold. Cohort-composition rows report case counts, with slide counts in parentheses; class percentages are case-level percentages. All metrics are computed on the same cohort.}
\label{tab:biopsy_primary_metastatic}
\renewcommand{\arraystretch}{1.15}
\setlength{\tabcolsep}{5pt}
\small
\begin{adjustbox}{max width=\textwidth,center}
\begin{tabular}{l l c}
\toprule
\textbf{Metric} & \textbf{Model} & \textbf{Internal} \\
\midrule
Total cases (slides) & -- & 108 (114) \\
Metastatic & -- & 27 (28; 25.0\%) \\
Primary & -- & 81 (86; 75.0\%) \\
\midrule
\multirow{5}{*}{\textbf{Macro AUC}} & UNI & 1.000 (1.000, 1.000) \\
 & Virchow2 & 1.000 (1.000, 1.000) \\
 & CHIEF & 1.000 (1.000, 1.000) \\
 & GigaPath & 1.000 (1.000, 1.000) \\
 & PulmoFoundation & \textbf{1.000 (1.000, 1.000)} \\
\midrule
\multirow{5}{*}{\textbf{Macro Sensitivity}} & UNI & 1.000 (1.000, 1.000) \\
 & Virchow2 & 1.000 (1.000, 1.000) \\
 & CHIEF & 0.964 (0.913, 1.000) \\
 & GigaPath & 1.000 (1.000, 1.000) \\
 & PulmoFoundation & \textbf{1.000 (1.000, 1.000)} \\
\midrule
\multirow{5}{*}{\textbf{Macro Specificity}} & UNI & 1.000 (1.000, 1.000) \\
 & Virchow2 & 1.000 (1.000, 1.000) \\
 & CHIEF & 0.964 (0.913, 1.000) \\
 & GigaPath & 1.000 (1.000, 1.000) \\
 & PulmoFoundation & \textbf{1.000 (1.000, 1.000)} \\
\midrule
\multirow{5}{*}{\textbf{Macro PPV}} & UNI & 1.000 (1.000, 1.000) \\
 & Virchow2 & 1.000 (1.000, 1.000) \\
 & CHIEF & 0.988 (0.970, 1.000) \\
 & GigaPath & 1.000 (1.000, 1.000) \\
 & PulmoFoundation & \textbf{1.000 (1.000, 1.000)} \\
\midrule
\multirow{5}{*}{\textbf{Macro NPV}} & UNI & 1.000 (1.000, 1.000) \\
 & Virchow2 & 1.000 (1.000, 1.000) \\
 & CHIEF & 0.988 (0.970, 1.000) \\
 & GigaPath & 1.000 (1.000, 1.000) \\
 & PulmoFoundation & \textbf{1.000 (1.000, 1.000)} \\
\midrule
\multirow{5}{*}{\textbf{Metastatic Sensitivity (Youden)}} & UNI & 1.000 (1.000, 1.000) \\
 & Virchow2 & 1.000 (1.000, 1.000) \\
 & CHIEF & 1.000 (1.000, 1.000) \\
 & GigaPath & 1.000 (1.000, 1.000) \\
 & PulmoFoundation & \textbf{1.000 (1.000, 1.000)} \\
\midrule
\multirow{5}{*}{\textbf{Primary Sensitivity (Youden)}} & UNI & 1.000 (1.000, 1.000) \\
 & Virchow2 & 1.000 (1.000, 1.000) \\
 & CHIEF & 1.000 (1.000, 1.000) \\
 & GigaPath & 1.000 (1.000, 1.000) \\
 & PulmoFoundation & \textbf{1.000 (1.000, 1.000)} \\
\bottomrule
\end{tabular}
\end{adjustbox}
\end{table}

\clearpage

\begin{table}[htbp]
\centering
\caption{\textbf{Subtyping (Biopsy)}: Per-task performance of PulmoFoundation and four baseline foundation models (UNI, Virchow2, CHIEF, GigaPath). Performance metrics are reported with 95\% bootstrap percentile CIs. Macro AUC is threshold-free. Macro Sensitivity, Specificity, PPV, and NPV are macro-averaged one-versus-rest metrics computed from predicted class labels; per-class sensitivity rows are computed at class-specific Youden-optimal operating points. CIs are bootstrap percentiles over cases ($N{=}1{,}000$ replicates). Best-performing model per metric and cohort is shown in bold. Cohort-composition rows report case counts, with slide counts in parentheses; class percentages are case-level percentages. All metrics are computed on the same cohort.}
\label{tab:biopsy_subtyping}
\renewcommand{\arraystretch}{1.15}
\setlength{\tabcolsep}{5pt}
\small
\begin{adjustbox}{max width=\textwidth,center}
\begin{tabular}{l l c c c c c}
\toprule
\textbf{Metric} & \textbf{Cohort} & \textbf{UNI} & \textbf{Virchow2} & \textbf{CHIEF} & \textbf{GigaPath} & \textbf{PulmoFoundation} \\
\midrule
Total cases (slides) & \multicolumn{6}{l}{Internal: 104 (114)} \\
Other malignancies & \multicolumn{6}{l}{Internal: 40 (43; 38.5\%)} \\
Invasive adenocarcinoma & \multicolumn{6}{l}{Internal: 26 (28; 25.0\%)} \\
Small cell carcinoma & \multicolumn{6}{l}{Internal: 13 (14; 12.5\%)} \\
Squamous cell carcinoma & \multicolumn{6}{l}{Internal: 25 (29; 24.0\%)} \\
\midrule
\multirow{1}{*}{\textbf{Macro AUC}} & Internal & 0.849 (0.787, 0.898) & 0.866 (0.805, 0.913) & 0.853 (0.803, 0.901) & 0.833 (0.781, 0.883) & \textbf{0.877 (0.825, 0.919)} \\
\midrule
\multirow{1}{*}{\textbf{Macro Sensitivity}} & Internal & 0.623 (0.531, 0.708) & 0.651 (0.575, 0.720) & 0.566 (0.485, 0.644) & 0.566 (0.466, 0.659) & \textbf{0.695 (0.620, 0.760)} \\
\midrule
\multirow{1}{*}{\textbf{Macro Specificity}} & Internal & 0.855 (0.817, 0.889) & 0.859 (0.823, 0.890) & 0.834 (0.800, 0.869) & 0.829 (0.792, 0.863) & \textbf{0.879 (0.847, 0.910)} \\
\midrule
\multirow{1}{*}{\textbf{Macro PPV}} & Internal & 0.707 (0.608, 0.801) & 0.636 (0.529, 0.738) & 0.505 (0.432, 0.573) & 0.656 (0.569, 0.738) & \textbf{0.725 (0.600, 0.839)} \\
\midrule
\multirow{1}{*}{\textbf{Macro NPV}} & Internal & 0.867 (0.824, 0.906) & 0.867 (0.831, 0.900) & 0.847 (0.808, 0.885) & 0.829 (0.790, 0.868) & \textbf{0.896 (0.860, 0.927)} \\
\midrule
\multirow{1}{*}{\textbf{Other malignancies Sens. (Youden)}} & Internal & 0.750 (0.488, 0.976) & 0.875 (0.729, 1.000) & 0.625 (0.487, 1.000) & 0.700 (0.366, 0.946) & \textbf{0.875 (0.720, 0.978)} \\
\midrule
\multirow{1}{*}{\textbf{Invasive adenocarcinoma Sens. (Youden)}} & Internal & 0.615 (0.542, 1.000) & 0.769 (0.667, 1.000) & \textbf{0.923 (0.823, 1.000)} & 0.769 (0.652, 1.000) & 0.846 (0.724, 1.000) \\
\midrule
\multirow{1}{*}{\textbf{Small cell carcinoma Sens. (Youden)}} & Internal & \textbf{1.000 (0.900, 1.000)} & 0.923 (0.750, 1.000) & 0.923 (0.733, 1.000) & 0.846 (0.750, 1.000) & 0.923 (0.916, 1.000) \\
\midrule
\multirow{1}{*}{\textbf{Squamous cell carcinoma Sens. (Youden)}} & Internal & 0.720 (0.652, 1.000) & 0.800 (0.676, 1.000) & 0.760 (0.600, 1.000) & \textbf{0.920 (0.773, 1.000)} & 0.840 (0.680, 0.967) \\
\bottomrule
\end{tabular}
\end{adjustbox}
\end{table}

\clearpage

\begin{table}[htbp]
\centering
\caption{\textbf{CK5/6 (Biopsy)}: Per-task performance of PulmoFoundation and four baseline foundation models (UNI, Virchow2, CHIEF, GigaPath). Performance metrics are reported with 95\% bootstrap percentile CIs. Macro AUC is threshold-free. Macro Sensitivity, Specificity, PPV, and NPV are macro-averaged one-versus-rest metrics computed from predicted class labels; per-class sensitivity rows are computed at class-specific Youden-optimal operating points. CIs are bootstrap percentiles over cases ($N{=}1{,}000$ replicates). Best-performing model per metric and cohort is shown in bold. Cohort-composition rows report case counts, with slide counts in parentheses; class percentages are case-level percentages. All metrics are computed on the same cohort.}
\label{tab:biopsy_ck56}
\renewcommand{\arraystretch}{1.15}
\setlength{\tabcolsep}{5pt}
\small
\begin{adjustbox}{max width=\textwidth,center}
\begin{tabular}{l l c}
\toprule
\textbf{Metric} & \textbf{Model} & \textbf{Internal} \\
\midrule
Total cases (slides) & -- & 122 (126) \\
Positive & -- & 50 (52; 41.0\%) \\
Negative & -- & 72 (74; 59.0\%) \\
\midrule
\multirow{5}{*}{\textbf{Macro AUC}} & UNI & 0.879 (0.803, 0.942) \\
 & Virchow2 & 0.858 (0.777, 0.925) \\
 & CHIEF & 0.845 (0.773, 0.915) \\
 & GigaPath & 0.790 (0.703, 0.869) \\
 & PulmoFoundation & \textbf{0.898 (0.830, 0.949)} \\
\midrule
\multirow{5}{*}{\textbf{Macro Sensitivity}} & UNI & \textbf{0.844 (0.772, 0.910)} \\
 & Virchow2 & 0.775 (0.702, 0.847) \\
 & CHIEF & 0.786 (0.711, 0.860) \\
 & GigaPath & 0.711 (0.626, 0.787) \\
 & PulmoFoundation & 0.839 (0.767, 0.908) \\
\midrule
\multirow{5}{*}{\textbf{Macro Specificity}} & UNI & \textbf{0.844 (0.772, 0.910)} \\
 & Virchow2 & 0.775 (0.702, 0.847) \\
 & CHIEF & 0.786 (0.711, 0.860) \\
 & GigaPath & 0.711 (0.626, 0.787) \\
 & PulmoFoundation & 0.839 (0.767, 0.908) \\
\midrule
\multirow{5}{*}{\textbf{Macro PPV}} & UNI & \textbf{0.872 (0.805, 0.926)} \\
 & Virchow2 & 0.768 (0.696, 0.839) \\
 & CHIEF & 0.788 (0.712, 0.863) \\
 & GigaPath & 0.711 (0.631, 0.788) \\
 & PulmoFoundation & 0.856 (0.784, 0.918) \\
\midrule
\multirow{5}{*}{\textbf{Macro NPV}} & UNI & \textbf{0.872 (0.805, 0.926)} \\
 & Virchow2 & 0.768 (0.696, 0.839) \\
 & CHIEF & 0.788 (0.712, 0.863) \\
 & GigaPath & 0.711 (0.631, 0.788) \\
 & PulmoFoundation & 0.856 (0.784, 0.918) \\
\midrule
\multirow{5}{*}{\textbf{Positive Sensitivity (Youden)}} & UNI & 0.780 (0.652, 0.897) \\
 & Virchow2 & \textbf{0.860 (0.667, 0.942)} \\
 & CHIEF & 0.820 (0.651, 0.920) \\
 & GigaPath & 0.840 (0.583, 0.946) \\
 & PulmoFoundation & 0.760 (0.660, 0.917) \\
\midrule
\multirow{5}{*}{\textbf{Negative Sensitivity (Youden)}} & UNI & 0.931 (0.853, 1.000) \\
 & Virchow2 & 0.792 (0.724, 0.969) \\
 & CHIEF & 0.778 (0.685, 0.937) \\
 & GigaPath & 0.667 (0.557, 0.898) \\
 & PulmoFoundation & \textbf{0.944 (0.807, 0.987)} \\
\bottomrule
\end{tabular}
\end{adjustbox}
\end{table}

\clearpage

\begin{table}[htbp]
\centering
\caption{\textbf{Benign vs Malignant (Frozen)}: Per-task performance of PulmoFoundation and four baseline foundation models (UNI, Virchow2, CHIEF, GigaPath). Performance metrics are reported with 95\% bootstrap percentile CIs. Macro AUC is threshold-free. Macro Sensitivity, Specificity, PPV, and NPV are macro-averaged one-versus-rest metrics computed from predicted class labels; per-class sensitivity rows are computed at class-specific Youden-optimal operating points. CIs are bootstrap percentiles over cases ($N{=}1{,}000$ replicates). Best-performing model per metric and cohort is shown in bold. Cohort-composition rows report case counts, with slide counts in parentheses; class percentages are case-level percentages. All metrics are computed on the same cohort.}
\label{tab:frozen_benign_malignant}
\renewcommand{\arraystretch}{1.15}
\setlength{\tabcolsep}{5pt}
\small
\begin{adjustbox}{max width=\textwidth,center}
\begin{tabular}{l l c c c}
\toprule
\textbf{Metric} & \textbf{Model} & \textbf{Internal} & \textbf{External H10} & \textbf{External H2} \\
\midrule
Total cases (slides) & -- & 113 (220) & 169 (169) & 449 (514) \\
Benign & -- & 53 (112; 46.9\%) & 43 (43; 25.4\%) & -- (310; 60.3\%) \\
Malignant & -- & 60 (108; 53.1\%) & 126 (126; 74.6\%) & -- (204; 39.7\%) \\
\midrule
\multirow{5}{*}{\textbf{Macro AUC}} & UNI & 0.980 (0.948, 0.999) & 0.958 (0.924, 0.987) & 0.998 (0.994, 1.000) \\
 & Virchow2 & 0.975 (0.939, 0.999) & 0.963 (0.921, 0.993) & 0.996 (0.990, 1.000) \\
 & CHIEF & 0.970 (0.934, 0.996) & 0.932 (0.882, 0.969) & 0.988 (0.978, 0.995) \\
 & GigaPath & 0.949 (0.904, 0.985) & 0.907 (0.847, 0.954) & 0.981 (0.971, 0.990) \\
 & PulmoFoundation & \textbf{0.986 (0.960, 1.000)} & \textbf{0.970 (0.943, 0.992)} & \textbf{0.999 (0.997, 1.000)} \\
\midrule
\multirow{5}{*}{\textbf{Macro Sensitivity}} & UNI & 0.933 (0.887, 0.973) & 0.898 (0.842, 0.948) & 0.988 (0.976, 0.998) \\
 & Virchow2 & 0.927 (0.876, 0.972) & 0.900 (0.842, 0.957) & 0.924 (0.903, 0.942) \\
 & CHIEF & 0.920 (0.868, 0.965) & 0.839 (0.766, 0.905) & 0.959 (0.941, 0.975) \\
 & GigaPath & 0.824 (0.751, 0.890) & 0.812 (0.733, 0.881) & 0.795 (0.766, 0.822) \\
 & PulmoFoundation & \textbf{0.967 (0.933, 0.992)} & \textbf{0.927 (0.874, 0.972)} & \textbf{0.990 (0.979, 0.998)} \\
\midrule
\multirow{5}{*}{\textbf{Macro Specificity}} & UNI & 0.933 (0.887, 0.973) & 0.898 (0.842, 0.948) & 0.988 (0.976, 0.998) \\
 & Virchow2 & 0.927 (0.876, 0.972) & 0.900 (0.842, 0.957) & 0.924 (0.903, 0.942) \\
 & CHIEF & 0.920 (0.868, 0.965) & 0.839 (0.766, 0.905) & 0.959 (0.941, 0.975) \\
 & GigaPath & 0.824 (0.751, 0.890) & 0.812 (0.733, 0.881) & 0.795 (0.766, 0.822) \\
 & PulmoFoundation & \textbf{0.967 (0.933, 0.992)} & \textbf{0.927 (0.874, 0.972)} & \textbf{0.990 (0.979, 0.998)} \\
\midrule
\multirow{5}{*}{\textbf{Macro PPV}} & UNI & 0.934 (0.886, 0.973) & 0.851 (0.788, 0.913) & \textbf{0.992 (0.984, 0.998)} \\
 & Virchow2 & 0.933 (0.885, 0.974) & \textbf{0.943 (0.897, 0.981)} & 0.906 (0.881, 0.927) \\
 & CHIEF & 0.920 (0.867, 0.965) & 0.834 (0.766, 0.900) & 0.956 (0.937, 0.974) \\
 & GigaPath & 0.846 (0.777, 0.912) & 0.812 (0.732, 0.885) & 0.806 (0.779, 0.833) \\
 & PulmoFoundation & \textbf{0.966 (0.927, 0.992)} & 0.932 (0.879, 0.976) & 0.990 (0.981, 0.998) \\
\midrule
\multirow{5}{*}{\textbf{Macro NPV}} & UNI & 0.934 (0.886, 0.973) & 0.851 (0.788, 0.913) & \textbf{0.992 (0.984, 0.998)} \\
 & Virchow2 & 0.933 (0.885, 0.974) & \textbf{0.943 (0.897, 0.981)} & 0.906 (0.881, 0.927) \\
 & CHIEF & 0.920 (0.867, 0.965) & 0.834 (0.766, 0.900) & 0.956 (0.937, 0.974) \\
 & GigaPath & 0.846 (0.777, 0.912) & 0.812 (0.732, 0.885) & 0.806 (0.779, 0.833) \\
 & PulmoFoundation & \textbf{0.966 (0.927, 0.992)} & 0.932 (0.879, 0.976) & 0.990 (0.981, 0.998) \\
\midrule
\multirow{5}{*}{\textbf{Benign Sensitivity (Youden)}} & UNI & 1.000 (0.906, 1.000) & 0.884 (0.788, 0.976) & 0.994 (0.984, 1.000) \\
 & Virchow2 & 1.000 (0.936, 1.000) & \textbf{0.930 (0.837, 1.000)} & 1.000 (0.987, 1.000) \\
 & CHIEF & 0.981 (0.944, 1.000) & 0.860 (0.771, 0.977) & 0.987 (0.973, 1.000) \\
 & GigaPath & 0.943 (0.833, 1.000) & 0.837 (0.778, 0.978) & 0.984 (0.918, 0.997) \\
 & PulmoFoundation & \textbf{1.000 (0.956, 1.000)} & 0.884 (0.795, 0.976) & \textbf{1.000 (0.987, 1.000)} \\
\midrule
\multirow{5}{*}{\textbf{Malignant Sensitivity (Youden)}} & UNI & 0.900 (0.862, 1.000) & 0.960 (0.899, 0.992) & 0.985 (0.964, 1.000) \\
 & Virchow2 & 0.917 (0.877, 1.000) & 0.944 (0.915, 1.000) & 0.975 (0.956, 0.995) \\
 & CHIEF & 0.917 (0.838, 0.982) & 0.889 (0.793, 0.959) & 0.951 (0.923, 0.980) \\
 & GigaPath & 0.833 (0.743, 0.970) & 0.897 (0.758, 0.944) & 0.882 (0.852, 0.950) \\
 & PulmoFoundation & \textbf{0.950 (0.897, 1.000)} & \textbf{0.976 (0.877, 1.000)} & \textbf{0.985 (0.970, 1.000)} \\
\bottomrule
\end{tabular}
\end{adjustbox}
\end{table}

\clearpage

\begin{table}[htbp]
\centering
\caption{\textbf{Benign vs Malignant, Uncertain (Frozen)}: Per-task performance of PulmoFoundation and four baseline foundation models (UNI, Virchow2, CHIEF, GigaPath). Performance metrics are reported with 95\% bootstrap percentile CIs. Macro AUC is threshold-free. Macro Sensitivity, Specificity, PPV, and NPV are macro-averaged one-versus-rest metrics computed from predicted class labels; per-class sensitivity rows are computed at class-specific Youden-optimal operating points. CIs are bootstrap percentiles over cases ($N{=}1{,}000$ replicates). Best-performing model per metric and cohort is shown in bold. Cohort-composition rows report case counts, with slide counts in parentheses; class percentages are case-level percentages. All metrics are computed on the same cohort.}
\label{tab:frozen_benign_malignant_uncertain}
\renewcommand{\arraystretch}{1.15}
\setlength{\tabcolsep}{5pt}
\small
\begin{adjustbox}{max width=\textwidth,center}
\begin{tabular}{l l c}
\toprule
\textbf{Metric} & \textbf{Model} & \textbf{Internal} \\
\midrule
Total cases (slides) & -- & 99 (235) \\
Benign & -- & 53 (112; 53.5\%) \\
Malignant & -- & 46 (123; 46.5\%) \\
\midrule
\multirow{5}{*}{\textbf{Macro AUC}} & UNI & 1.000 (0.998, 1.000) \\
 & Virchow2 & 1.000 (1.000, 1.000) \\
 & CHIEF & 1.000 (0.998, 1.000) \\
 & GigaPath & 0.999 (0.996, 1.000) \\
 & PulmoFoundation & \textbf{1.000 (1.000, 1.000)} \\
\midrule
\multirow{5}{*}{\textbf{Macro Sensitivity}} & UNI & 0.981 (0.948, 1.000) \\
 & Virchow2 & \textbf{0.991 (0.968, 1.000)} \\
 & CHIEF & 0.980 (0.949, 1.000) \\
 & GigaPath & 0.868 (0.805, 0.927) \\
 & PulmoFoundation & 0.956 (0.908, 0.990) \\
\midrule
\multirow{5}{*}{\textbf{Macro Specificity}} & UNI & 0.981 (0.948, 1.000) \\
 & Virchow2 & \textbf{0.991 (0.968, 1.000)} \\
 & CHIEF & 0.980 (0.949, 1.000) \\
 & GigaPath & 0.868 (0.805, 0.927) \\
 & PulmoFoundation & 0.956 (0.908, 0.990) \\
\midrule
\multirow{5}{*}{\textbf{Macro PPV}} & UNI & 0.979 (0.943, 1.000) \\
 & Virchow2 & \textbf{0.989 (0.963, 1.000)} \\
 & CHIEF & 0.980 (0.948, 1.000) \\
 & GigaPath & 0.884 (0.831, 0.934) \\
 & PulmoFoundation & 0.964 (0.927, 0.992) \\
\midrule
\multirow{5}{*}{\textbf{Macro NPV}} & UNI & 0.979 (0.943, 1.000) \\
 & Virchow2 & \textbf{0.989 (0.963, 1.000)} \\
 & CHIEF & 0.980 (0.948, 1.000) \\
 & GigaPath & 0.884 (0.831, 0.934) \\
 & PulmoFoundation & 0.964 (0.927, 0.992) \\
\midrule
\multirow{5}{*}{\textbf{Benign Sensitivity (Youden)}} & UNI & 0.981 (0.964, 1.000) \\
 & Virchow2 & 1.000 (1.000, 1.000) \\
 & CHIEF & 0.981 (0.963, 1.000) \\
 & GigaPath & 1.000 (0.960, 1.000) \\
 & PulmoFoundation & \textbf{1.000 (1.000, 1.000)} \\
\midrule
\multirow{5}{*}{\textbf{Malignant Sensitivity (Youden)}} & UNI & 1.000 (0.963, 1.000) \\
 & Virchow2 & 1.000 (1.000, 1.000) \\
 & CHIEF & 1.000 (0.973, 1.000) \\
 & GigaPath & 0.978 (0.953, 1.000) \\
 & PulmoFoundation & \textbf{1.000 (1.000, 1.000)} \\
\bottomrule
\end{tabular}
\end{adjustbox}
\end{table}

\clearpage

\begin{table}[htbp]
\centering
\caption{\textbf{Subtyping (Frozen)}: Per-task performance of PulmoFoundation and four baseline foundation models (UNI, Virchow2, CHIEF, GigaPath). Performance metrics are reported with 95\% bootstrap percentile CIs. Macro AUC is threshold-free. Macro Sensitivity, Specificity, PPV, and NPV are macro-averaged one-versus-rest metrics computed from predicted class labels; per-class sensitivity rows are computed at class-specific Youden-optimal operating points. CIs are bootstrap percentiles over cases ($N{=}1{,}000$ replicates). Best-performing model per metric and cohort is shown in bold. Cohort-composition rows report case counts, with slide counts in parentheses; class percentages are case-level percentages. All metrics are computed on the same cohort.}
\label{tab:frozen_subtyping}
\renewcommand{\arraystretch}{1.15}
\setlength{\tabcolsep}{5pt}
\small
\begin{adjustbox}{max width=\textwidth,center}
\begin{tabular}{l l c c c c c}
\toprule
\textbf{Metric} & \textbf{Cohort} & \textbf{UNI} & \textbf{Virchow2} & \textbf{CHIEF} & \textbf{GigaPath} & \textbf{PulmoFoundation} \\
\midrule
Total cases (slides) & \multicolumn{6}{l}{Internal: 87 (169)} \\
Adenocarcinoma in situ & \multicolumn{6}{l}{Internal: 29 (57; 33.3\%)} \\
Minimally invasive adenocarcinoma & \multicolumn{6}{l}{Internal: 18 (33; 20.7\%)} \\
Invasive adenocarcinoma & \multicolumn{6}{l}{Internal: 40 (79; 46.0\%)} \\
\midrule
\multirow{1}{*}{\textbf{Macro AUC}} & Internal & 0.849 (0.791, 0.905) & 0.852 (0.795, 0.906) & 0.845 (0.783, 0.908) & 0.795 (0.728, 0.855) & \textbf{0.885 (0.831, 0.932)} \\
\midrule
\multirow{1}{*}{\textbf{Macro Sensitivity}} & Internal & 0.588 (0.521, 0.658) & 0.581 (0.521, 0.643) & \textbf{0.628 (0.536, 0.726)} & 0.541 (0.456, 0.631) & 0.583 (0.509, 0.660) \\
\midrule
\multirow{1}{*}{\textbf{Macro Specificity}} & Internal & 0.842 (0.795, 0.889) & 0.836 (0.792, 0.882) & \textbf{0.848 (0.799, 0.895)} & 0.808 (0.763, 0.855) & 0.837 (0.793, 0.885) \\
\midrule
\multirow{1}{*}{\textbf{Macro PPV}} & Internal & 0.544 (0.439, 0.681) & 0.548 (0.462, 0.667) & \textbf{0.637 (0.531, 0.754)} & 0.538 (0.445, 0.635) & 0.558 (0.470, 0.658) \\
\midrule
\multirow{1}{*}{\textbf{Macro NPV}} & Internal & \textbf{0.859 (0.812, 0.906)} & 0.855 (0.812, 0.897) & 0.850 (0.802, 0.895) & 0.807 (0.758, 0.858) & 0.843 (0.797, 0.889) \\
\midrule
\multirow{1}{*}{\textbf{Adenocarcinoma in situ Sens. (Youden)}} & Internal & 0.828 (0.706, 1.000) & 0.897 (0.750, 1.000) & 0.759 (0.630, 1.000) & \textbf{1.000 (0.750, 1.000)} & 0.724 (0.677, 1.000) \\
\midrule
\multirow{1}{*}{\textbf{Minimally invasive adenocarcinoma Sens. (Youden)}} & Internal & 0.833 (0.640, 1.000) & 0.944 (0.632, 1.000) & \textbf{1.000 (0.687, 1.000)} & 1.000 (0.550, 1.000) & 0.778 (0.700, 1.000) \\
\midrule
\multirow{1}{*}{\textbf{Invasive adenocarcinoma Sens. (Youden)}} & Internal & 0.900 (0.788, 1.000) & 0.850 (0.791, 1.000) & 0.900 (0.737, 0.978) & 0.925 (0.725, 1.000) & \textbf{0.950 (0.891, 1.000)} \\
\bottomrule
\end{tabular}
\end{adjustbox}
\end{table}

\begin{table}[htbp]
\centering
\caption{\textbf{LN Metastasis (Frozen)}: Per-task performance of PulmoFoundation and four baseline foundation models (UNI, Virchow2, CHIEF, GigaPath). Performance metrics are reported with 95\% bootstrap percentile CIs. Macro AUC is threshold-free. Macro Sensitivity, Specificity, PPV, and NPV are macro-averaged one-versus-rest metrics computed from predicted class labels; per-class sensitivity rows are computed at class-specific Youden-optimal operating points. CIs are bootstrap percentiles over cases ($N{=}1{,}000$ replicates). Best-performing model per metric and cohort is shown in bold. Cohort-composition rows report case counts, with slide counts in parentheses; class percentages are case-level percentages. All metrics are computed on the same cohort.}
\label{tab:frozen_lymph_node_metastasis}
\renewcommand{\arraystretch}{1.15}
\setlength{\tabcolsep}{5pt}
\small
\begin{adjustbox}{max width=\textwidth,center}
\begin{tabular}{l l c}
\toprule
\textbf{Metric} & \textbf{Model} & \textbf{Internal} \\
\midrule
Total cases (slides) & -- & 65 (134) \\
Negative & -- & 51 (102; 78.5\%) \\
Positive & -- & 14 (32; 21.5\%) \\
\midrule
\multirow{5}{*}{\textbf{Macro AUC}} & UNI & 0.713 (0.560, 0.853) \\
 & Virchow2 & 0.680 (0.512, 0.830) \\
 & CHIEF & 0.730 (0.575, 0.868) \\
 & GigaPath & 0.673 (0.524, 0.810) \\
 & PulmoFoundation & \textbf{0.761 (0.626, 0.875)} \\
\midrule
\multirow{5}{*}{\textbf{Macro Sensitivity}} & UNI & 0.512 (0.421, 0.626) \\
 & Virchow2 & 0.616 (0.474, 0.752) \\
 & CHIEF & \textbf{0.619 (0.494, 0.763)} \\
 & GigaPath & 0.500 (0.500, 0.500) \\
 & PulmoFoundation & 0.523 (0.435, 0.644) \\
\midrule
\multirow{5}{*}{\textbf{Macro Specificity}} & UNI & 0.512 (0.421, 0.626) \\
 & Virchow2 & 0.616 (0.474, 0.752) \\
 & CHIEF & \textbf{0.619 (0.494, 0.763)} \\
 & GigaPath & 0.500 (0.500, 0.500) \\
 & PulmoFoundation & 0.523 (0.435, 0.644) \\
\midrule
\multirow{5}{*}{\textbf{Macro PPV}} & UNI & 0.518 (0.371, 0.702) \\
 & Virchow2 & 0.605 (0.480, 0.730) \\
 & CHIEF & \textbf{0.642 (0.492, 0.800)} \\
 & GigaPath & 0.393 (0.339, 0.439) \\
 & PulmoFoundation & 0.540 (0.371, 0.753) \\
\midrule
\multirow{5}{*}{\textbf{Macro NPV}} & UNI & 0.518 (0.371, 0.702) \\
 & Virchow2 & 0.605 (0.480, 0.730) \\
 & CHIEF & \textbf{0.642 (0.492, 0.800)} \\
 & GigaPath & 0.393 (0.339, 0.439) \\
 & PulmoFoundation & 0.540 (0.371, 0.753) \\
\midrule
\multirow{5}{*}{\textbf{Negative Sensitivity (Youden)}} & UNI & 0.725 (0.377, 0.873) \\
 & Virchow2 & 0.745 (0.231, 0.913) \\
 & CHIEF & 0.667 (0.389, 0.943) \\
 & GigaPath & 0.725 (0.250, 0.852) \\
 & PulmoFoundation & \textbf{0.765 (0.380, 0.906)} \\
\midrule
\multirow{5}{*}{\textbf{Positive Sensitivity (Youden)}} & UNI & 0.714 (0.538, 1.000) \\
 & Virchow2 & 0.643 (0.435, 1.000) \\
 & CHIEF & \textbf{0.786 (0.462, 1.000)} \\
 & GigaPath & 0.643 (0.571, 1.000) \\
 & PulmoFoundation & 0.714 (0.625, 1.000) \\
\bottomrule
\end{tabular}
\end{adjustbox}
\end{table}

\clearpage

\begin{table}[htbp]
\centering
\caption{\textbf{Benign vs Malignant (Resection)}: Per-task performance of PulmoFoundation and four baseline foundation models (UNI, Virchow2, CHIEF, GigaPath). Performance metrics are reported with 95\% bootstrap percentile CIs. Macro AUC is threshold-free. Macro Sensitivity, Specificity, PPV, and NPV are macro-averaged one-versus-rest metrics computed from predicted class labels; per-class sensitivity rows are computed at class-specific Youden-optimal operating points. CIs are bootstrap percentiles over cases ($N{=}1{,}000$ replicates). Best-performing model per metric and cohort is shown in bold. Cohort-composition rows report case counts, with slide counts in parentheses; class percentages are case-level percentages. All metrics are computed on the same cohort.}
\label{tab:resection_benign_malignant}
\renewcommand{\arraystretch}{1.15}
\setlength{\tabcolsep}{5pt}
\small
\begin{adjustbox}{max width=\textwidth,center}
\begin{tabular}{l l c}
\toprule
\textbf{Metric} & \textbf{Model} & \textbf{Internal} \\
\midrule
Total cases (slides) & -- & 122 (122) \\
Benign & -- & 61 (61; 50.0\%) \\
Cancerous & -- & 61 (61; 50.0\%) \\
\midrule
\multirow{5}{*}{\textbf{Macro AUC}} & UNI & 0.963 (0.919, 0.995) \\
 & Virchow2 & 0.966 (0.924, 0.995) \\
 & CHIEF & 0.966 (0.922, 0.994) \\
 & GigaPath & \textbf{0.970 (0.931, 0.998)} \\
 & PulmoFoundation & 0.969 (0.926, 1.000) \\
\midrule
\multirow{5}{*}{\textbf{Macro Sensitivity}} & UNI & 0.934 (0.887, 0.975) \\
 & Virchow2 & 0.901 (0.848, 0.952) \\
 & CHIEF & 0.918 (0.867, 0.961) \\
 & GigaPath & 0.951 (0.908, 0.984) \\
 & PulmoFoundation & \textbf{0.959 (0.921, 0.992)} \\
\midrule
\multirow{5}{*}{\textbf{Macro Specificity}} & UNI & 0.934 (0.887, 0.975) \\
 & Virchow2 & 0.901 (0.848, 0.952) \\
 & CHIEF & 0.918 (0.867, 0.961) \\
 & GigaPath & 0.951 (0.908, 0.984) \\
 & PulmoFoundation & \textbf{0.959 (0.921, 0.992)} \\
\midrule
\multirow{5}{*}{\textbf{Macro PPV}} & UNI & 0.935 (0.885, 0.976) \\
 & Virchow2 & 0.909 (0.858, 0.955) \\
 & CHIEF & 0.926 (0.881, 0.966) \\
 & GigaPath & 0.952 (0.910, 0.984) \\
 & PulmoFoundation & \textbf{0.961 (0.925, 0.991)} \\
\midrule
\multirow{5}{*}{\textbf{Macro NPV}} & UNI & 0.935 (0.885, 0.976) \\
 & Virchow2 & 0.909 (0.858, 0.955) \\
 & CHIEF & 0.926 (0.881, 0.966) \\
 & GigaPath & 0.952 (0.910, 0.984) \\
 & PulmoFoundation & \textbf{0.961 (0.925, 0.991)} \\
\midrule
\multirow{5}{*}{\textbf{Benign Sensitivity (Youden)}} & UNI & 0.918 (0.873, 1.000) \\
 & Virchow2 & 0.951 (0.873, 1.000) \\
 & CHIEF & 0.918 (0.862, 1.000) \\
 & GigaPath & 0.967 (0.918, 1.000) \\
 & PulmoFoundation & \textbf{0.967 (0.935, 1.000)} \\
\midrule
\multirow{5}{*}{\textbf{Cancerous Sensitivity (Youden)}} & UNI & 0.934 (0.857, 1.000) \\
 & Virchow2 & 0.934 (0.871, 1.000) \\
 & CHIEF & 0.902 (0.828, 1.000) \\
 & GigaPath & 0.951 (0.891, 1.000) \\
 & PulmoFoundation & \textbf{0.951 (0.914, 1.000)} \\
\bottomrule
\end{tabular}
\end{adjustbox}
\end{table}

\clearpage

\begin{table}[htbp]
\centering
\caption{\textbf{Primary vs Metastatic (Resection)}: Per-task performance of PulmoFoundation and four baseline foundation models (UNI, Virchow2, CHIEF, GigaPath). Performance metrics are reported with 95\% bootstrap percentile CIs. Macro AUC is threshold-free. Macro Sensitivity, Specificity, PPV, and NPV are macro-averaged one-versus-rest metrics computed from predicted class labels; per-class sensitivity rows are computed at class-specific Youden-optimal operating points. CIs are bootstrap percentiles over cases ($N{=}1{,}000$ replicates). Best-performing model per metric and cohort is shown in bold. Cohort-composition rows report case counts, with slide counts in parentheses; class percentages are case-level percentages. All metrics are computed on the same cohort.}
\label{tab:post_primary_metastatic}
\renewcommand{\arraystretch}{1.15}
\setlength{\tabcolsep}{5pt}
\small
\begin{adjustbox}{max width=\textwidth,center}
\begin{tabular}{l l c c c}
\toprule
\textbf{Metric} & \textbf{Model} & \textbf{Internal} & \textbf{External H5} & \textbf{External H7} \\
\midrule
Total cases (slides) & -- & 170 (280) & 779 (1307) & 493 (493) \\
Metastatic & -- & 92 (145; 54.1\%) & 345 (633; 44.3\%) & 256 (256; 51.9\%) \\
Primary & -- & 78 (135; 45.9\%) & 434 (674; 55.7\%) & 237 (237; 48.1\%) \\
\midrule
\multirow{5}{*}{\textbf{Macro AUC}} & UNI & 0.979 (0.960, 0.993) & 0.893 (0.871, 0.913) & 0.868 (0.833, 0.899) \\
 & Virchow2 & 0.985 (0.970, 0.996) & 0.896 (0.875, 0.917) & 0.847 (0.808, 0.880) \\
 & CHIEF & 0.974 (0.952, 0.989) & 0.830 (0.802, 0.858) & 0.833 (0.789, 0.870) \\
 & GigaPath & 0.978 (0.959, 0.992) & 0.872 (0.849, 0.896) & 0.818 (0.778, 0.856) \\
 & PulmoFoundation & \textbf{0.989 (0.978, 0.997)} & \textbf{0.902 (0.881, 0.922)} & \textbf{0.869 (0.834, 0.898)} \\
\midrule
\multirow{5}{*}{\textbf{Macro Sensitivity}} & UNI & 0.916 (0.873, 0.955) & 0.718 (0.693, 0.745) & 0.816 (0.783, 0.846) \\
 & Virchow2 & 0.900 (0.856, 0.943) & 0.765 (0.739, 0.791) & \textbf{0.826 (0.793, 0.859)} \\
 & CHIEF & 0.899 (0.849, 0.941) & 0.742 (0.712, 0.773) & 0.781 (0.743, 0.817) \\
 & GigaPath & \textbf{0.926 (0.883, 0.961)} & \textbf{0.794 (0.766, 0.822)} & 0.738 (0.700, 0.777) \\
 & PulmoFoundation & 0.920 (0.876, 0.960) & 0.682 (0.659, 0.706) & 0.823 (0.791, 0.856) \\
\midrule
\multirow{5}{*}{\textbf{Macro Specificity}} & UNI & 0.916 (0.873, 0.955) & 0.718 (0.693, 0.745) & 0.816 (0.783, 0.846) \\
 & Virchow2 & 0.900 (0.856, 0.943) & 0.765 (0.739, 0.791) & \textbf{0.826 (0.793, 0.859)} \\
 & CHIEF & 0.899 (0.849, 0.941) & 0.742 (0.712, 0.773) & 0.781 (0.743, 0.817) \\
 & GigaPath & \textbf{0.926 (0.883, 0.961)} & \textbf{0.794 (0.766, 0.822)} & 0.738 (0.700, 0.777) \\
 & PulmoFoundation & 0.920 (0.876, 0.960) & 0.682 (0.659, 0.706) & 0.823 (0.791, 0.856) \\
\midrule
\multirow{5}{*}{\textbf{Macro PPV}} & UNI & 0.921 (0.879, 0.959) & 0.779 (0.754, 0.803) & 0.854 (0.824, 0.882) \\
 & Virchow2 & 0.918 (0.880, 0.955) & 0.792 (0.767, 0.820) & 0.861 (0.832, 0.888) \\
 & CHIEF & 0.900 (0.851, 0.941) & 0.741 (0.710, 0.771) & 0.785 (0.746, 0.821) \\
 & GigaPath & 0.924 (0.882, 0.959) & \textbf{0.793 (0.765, 0.820)} & 0.741 (0.702, 0.780) \\
 & PulmoFoundation & \textbf{0.929 (0.889, 0.965)} & 0.767 (0.743, 0.789) & \textbf{0.867 (0.838, 0.892)} \\
\midrule
\multirow{5}{*}{\textbf{Macro NPV}} & UNI & 0.921 (0.879, 0.959) & 0.779 (0.754, 0.803) & 0.854 (0.824, 0.882) \\
 & Virchow2 & 0.918 (0.880, 0.955) & 0.792 (0.767, 0.820) & 0.861 (0.832, 0.888) \\
 & CHIEF & 0.900 (0.851, 0.941) & 0.741 (0.710, 0.771) & 0.785 (0.746, 0.821) \\
 & GigaPath & 0.924 (0.882, 0.959) & \textbf{0.793 (0.765, 0.820)} & 0.741 (0.702, 0.780) \\
 & PulmoFoundation & \textbf{0.929 (0.889, 0.965)} & 0.767 (0.743, 0.789) & \textbf{0.867 (0.838, 0.892)} \\
\midrule
\multirow{5}{*}{\textbf{Metastatic Sensitivity (Youden)}} & UNI & 0.935 (0.860, 1.000) & 0.846 (0.810, 0.889) & 0.961 (0.938, 0.992) \\
 & Virchow2 & 0.946 (0.861, 0.989) & 0.858 (0.699, 0.899) & 0.965 (0.941, 0.989) \\
 & CHIEF & \textbf{0.967 (0.865, 1.000)} & 0.872 (0.623, 0.909) & \textbf{0.984 (0.950, 1.000)} \\
 & GigaPath & 0.913 (0.851, 0.967) & \textbf{0.893 (0.749, 0.928)} & 0.961 (0.929, 0.981) \\
 & PulmoFoundation & 0.913 (0.865, 1.000) & 0.783 (0.749, 0.905) & 0.973 (0.923, 0.993) \\
\midrule
\multirow{5}{*}{\textbf{Primary Sensitivity (Youden)}} & UNI & 0.910 (0.818, 0.987) & 0.823 (0.780, 0.854) & 0.684 (0.615, 0.744) \\
 & Virchow2 & 0.936 (0.901, 1.000) & 0.751 (0.715, 0.911) & \textbf{0.692 (0.631, 0.751)} \\
 & CHIEF & 0.872 (0.800, 0.974) & 0.624 (0.587, 0.867) & 0.667 (0.616, 0.739) \\
 & GigaPath & 0.974 (0.936, 1.000) & 0.703 (0.672, 0.852) & 0.633 (0.569, 0.697) \\
 & PulmoFoundation & \textbf{0.974 (0.883, 1.000)} & \textbf{0.862 (0.741, 0.895)} & 0.684 (0.626, 0.760) \\
\bottomrule
\end{tabular}
\end{adjustbox}
\end{table}

\clearpage

\begin{table}[htbp]
\centering
\caption{\textbf{Primary Site (Resection)}: Per-task performance of PulmoFoundation and four baseline foundation models (UNI, Virchow2, CHIEF, GigaPath). Performance metrics are reported with 95\% bootstrap percentile CIs. Macro AUC is threshold-free. Macro Sensitivity, Specificity, PPV, and NPV are macro-averaged one-versus-rest metrics computed from predicted class labels; per-class sensitivity rows are computed at class-specific Youden-optimal operating points. CIs are bootstrap percentiles over cases ($N{=}1{,}000$ replicates). Best-performing model per metric and cohort is shown in bold. Cohort-composition rows report case counts, with slide counts in parentheses; class percentages are case-level percentages. All metrics are computed on the same cohort.}
\label{tab:post_primary_site_prediction}
\renewcommand{\arraystretch}{1.15}
\setlength{\tabcolsep}{5pt}
\small
\begin{adjustbox}{max width=\textwidth,center}
\begin{tabular}{l l c c c c c}
\toprule
\textbf{Metric} & \textbf{Cohort} & \textbf{UNI} & \textbf{Virchow2} & \textbf{CHIEF} & \textbf{GigaPath} & \textbf{PulmoFoundation} \\
\midrule
Total cases (slides) & \multicolumn{6}{l}{Internal: 139 (237); H5: 525 (967); H7: 423 (423)} \\
Breast & \multicolumn{6}{l}{Internal: 11 (13; 7.9\%); H5: 63 (104; 12.0\%); H7: 50 (50; 11.8\%)} \\
Colorectal & \multicolumn{6}{l}{Internal: 37 (69; 26.6\%); H5: 141 (279; 26.9\%); H7: 96 (96; 22.7\%)} \\
Renal & \multicolumn{6}{l}{Internal: 5 (8; 3.6\%); H5: 43 (87; 8.2\%); H7: 30 (30; 7.1\%)} \\
Hepatocellular & \multicolumn{6}{l}{Internal: 7 (16; 5.0\%); H5: 5 (10; 1.0\%); H7: 10 (10; 2.4\%)} \\
Lung & \multicolumn{6}{l}{Internal: 79 (131; 56.8\%); H5: 273 (487; 52.0\%); H7: 237 (237; 56.0\%)} \\
\midrule
\multirow{3}{*}{\textbf{Macro AUC}} & Internal & 0.958 (0.921, 0.985) & 0.968 (0.936, 0.989) & 0.951 (0.919, 0.977) & 0.942 (0.914, 0.968) & \textbf{0.975 (0.955, 0.991)} \\
 & H5 & 0.982 (0.975, 0.989) & 0.981 (0.973, 0.989) & 0.932 (0.908, 0.951) & 0.938 (0.922, 0.955) & \textbf{0.982 (0.975, 0.990)} \\
 & H7 & 0.942 (0.921, 0.960) & 0.946 (0.917, 0.968) & 0.906 (0.875, 0.928) & 0.863 (0.804, 0.911) & \textbf{0.950 (0.930, 0.965)} \\
\midrule
\multirow{3}{*}{\textbf{Macro Sensitivity}} & Internal & 0.727 (0.577, 0.864) & \textbf{0.766 (0.627, 0.898)} & 0.618 (0.482, 0.761) & 0.611 (0.466, 0.762) & 0.645 (0.505, 0.795) \\
 & H5 & 0.850 (0.761, 0.909) & \textbf{0.907 (0.877, 0.933)} & 0.567 (0.470, 0.677) & 0.719 (0.597, 0.819) & 0.873 (0.768, 0.933) \\
 & H7 & 0.702 (0.621, 0.782) & \textbf{0.755 (0.670, 0.832)} & 0.543 (0.472, 0.615) & 0.569 (0.521, 0.619) & 0.747 (0.661, 0.821) \\
\midrule
\multirow{3}{*}{\textbf{Macro Specificity}} & Internal & 0.953 (0.928, 0.972) & \textbf{0.954 (0.932, 0.972)} & 0.949 (0.927, 0.969) & 0.946 (0.926, 0.965) & 0.951 (0.930, 0.970) \\
 & H5 & 0.963 (0.954, 0.972) & \textbf{0.967 (0.958, 0.975)} & 0.911 (0.899, 0.924) & 0.946 (0.935, 0.955) & 0.966 (0.958, 0.974) \\
 & H7 & 0.936 (0.923, 0.947) & \textbf{0.950 (0.938, 0.961)} & 0.910 (0.894, 0.924) & 0.921 (0.908, 0.933) & 0.944 (0.932, 0.954) \\
\midrule
\multirow{3}{*}{\textbf{Macro PPV}} & Internal & \textbf{0.864 (0.668, 0.930)} & 0.783 (0.628, 0.897) & 0.757 (0.536, 0.883) & 0.636 (0.497, 0.791) & 0.740 (0.548, 0.893) \\
 & H5 & \textbf{0.903 (0.874, 0.931)} & 0.867 (0.788, 0.918) & 0.808 (0.614, 0.874) & 0.698 (0.622, 0.784) & 0.821 (0.723, 0.898) \\
 & H7 & 0.718 (0.628, 0.796) & \textbf{0.795 (0.711, 0.863)} & 0.746 (0.559, 0.804) & 0.522 (0.475, 0.566) & 0.760 (0.673, 0.829) \\
\midrule
\multirow{3}{*}{\textbf{Macro NPV}} & Internal & \textbf{0.962 (0.942, 0.979)} & 0.952 (0.930, 0.972) & 0.958 (0.937, 0.975) & 0.956 (0.938, 0.974) & 0.959 (0.938, 0.976) \\
 & H5 & \textbf{0.965 (0.956, 0.973)} & 0.964 (0.955, 0.973) & 0.938 (0.927, 0.948) & 0.950 (0.939, 0.959) & 0.960 (0.951, 0.969) \\
 & H7 & 0.926 (0.914, 0.939) & \textbf{0.943 (0.931, 0.954)} & 0.910 (0.896, 0.924) & 0.910 (0.897, 0.923) & 0.932 (0.919, 0.944) \\
\midrule
\multirow{3}{*}{\textbf{Breast Sens. (Youden)}} & Internal & 0.909 (0.750, 1.000) & 0.909 (0.833, 1.000) & 1.000 (0.900, 1.000) & 0.909 (0.778, 1.000) & \textbf{1.000 (1.000, 1.000)} \\
 & H5 & 0.921 (0.871, 1.000) & 0.937 (0.867, 0.985) & 0.841 (0.677, 0.939) & 0.889 (0.667, 0.981) & \textbf{0.968 (0.921, 1.000)} \\
 & H7 & 0.860 (0.789, 1.000) & 0.880 (0.673, 1.000) & 0.920 (0.816, 1.000) & 0.900 (0.612, 0.978) & \textbf{1.000 (0.900, 1.000)} \\
\midrule
\multirow{3}{*}{\textbf{Colorectal Sens. (Youden)}} & Internal & 0.892 (0.788, 1.000) & \textbf{0.973 (0.889, 1.000)} & 0.892 (0.787, 0.976) & 0.811 (0.711, 0.969) & 0.946 (0.879, 1.000) \\
 & H5 & \textbf{0.936 (0.896, 0.971)} & 0.915 (0.886, 0.976) & 0.801 (0.739, 0.887) & 0.879 (0.824, 0.937) & 0.915 (0.866, 0.963) \\
 & H7 & \textbf{0.979 (0.859, 1.000)} & 0.938 (0.867, 1.000) & 0.906 (0.853, 0.966) & 0.885 (0.849, 0.977) & 0.969 (0.912, 1.000) \\
\midrule
\multirow{3}{*}{\textbf{Renal Sens. (Youden)}} & Internal & 1.000 (1.000, 1.000) & 1.000 (1.000, 1.000) & 0.800 (0.667, 1.000) & 1.000 (1.000, 1.000) & \textbf{1.000 (1.000, 1.000)} \\
 & H5 & \textbf{1.000 (0.974, 1.000)} & 0.953 (0.903, 1.000) & 0.907 (0.818, 1.000) & 0.907 (0.846, 1.000) & 0.953 (0.920, 1.000) \\
 & H7 & 0.967 (0.850, 1.000) & 0.800 (0.684, 0.967) & 0.767 (0.630, 0.939) & 0.867 (0.739, 0.972) & \textbf{0.967 (0.862, 1.000)} \\
\midrule
\multirow{3}{*}{\textbf{Hepatocellular Sens. (Youden)}} & Internal & 0.857 (0.714, 1.000) & 0.857 (0.750, 1.000) & 0.857 (0.750, 1.000) & \textbf{1.000 (1.000, 1.000)} & 0.857 (0.800, 1.000) \\
 & H5 & 1.000 (1.000, 1.000) & 1.000 (1.000, 1.000) & 0.600 (0.500, 1.000) & 1.000 (1.000, 1.000) & \textbf{1.000 (1.000, 1.000)} \\
 & H7 & 1.000 (0.800, 1.000) & 0.700 (0.545, 1.000) & 0.900 (0.583, 1.000) & 0.800 (0.571, 1.000) & \textbf{1.000 (0.778, 1.000)} \\
\midrule
\multirow{3}{*}{\textbf{Lung Sens. (Youden)}} & Internal & 0.899 (0.795, 0.987) & 0.886 (0.797, 0.974) & \textbf{0.975 (0.816, 1.000)} & 0.835 (0.787, 0.988) & 0.911 (0.833, 1.000) \\
 & H5 & 0.861 (0.829, 0.949) & 0.908 (0.866, 0.974) & 0.861 (0.767, 0.911) & 0.842 (0.818, 0.935) & \textbf{0.949 (0.860, 0.971)} \\
 & H7 & 0.743 (0.694, 0.881) & \textbf{0.911 (0.839, 0.974)} & 0.688 (0.634, 0.772) & 0.654 (0.604, 0.823) & 0.759 (0.704, 0.823) \\
\bottomrule
\end{tabular}
\end{adjustbox}
\end{table}

\clearpage

\begin{table}[htbp]
\centering
\caption{\textbf{NSCLC Subtyping (Resection)}: Per-task performance of PulmoFoundation and four baseline foundation models (UNI, Virchow2, CHIEF, GigaPath). Performance metrics are reported with 95\% bootstrap percentile CIs. Macro AUC is threshold-free. Macro Sensitivity, Specificity, PPV, and NPV are macro-averaged one-versus-rest metrics computed from predicted class labels; per-class sensitivity rows are computed at class-specific Youden-optimal operating points. CIs are bootstrap percentiles over cases ($N{=}1{,}000$ replicates). Best-performing model per metric and cohort is shown in bold. Cohort-composition rows report case counts, with slide counts in parentheses; class percentages are case-level percentages. All metrics are computed on the same cohort.}
\label{tab:post_nsclc}
\renewcommand{\arraystretch}{1.15}
\setlength{\tabcolsep}{5pt}
\small
\begin{adjustbox}{max width=\textwidth,center}
\begin{tabular}{l l c c c}
\toprule
\textbf{Metric} & \textbf{Model} & \textbf{Internal} & \textbf{External H5} & \textbf{External TCGA} \\
\midrule
Total cases (slides) & -- & 120 (120) & 294 (517) & 1053 (1053) \\
LUAD & -- & 60 (60; 50.0\%) & 273 (487; 92.9\%) & 541 (541; 51.4\%) \\
LUSC & -- & 60 (60; 50.0\%) & 21 (30; 7.1\%) & 512 (512; 48.6\%) \\
\midrule
\multirow{5}{*}{\textbf{Macro AUC}} & UNI & 0.907 (0.854, 0.950) & 0.993 (0.981, 1.000) & 0.968 (0.957, 0.977) \\
 & Virchow2 & 0.922 (0.873, 0.960) & \textbf{0.998 (0.994, 1.000)} & \textbf{0.977 (0.969, 0.985)} \\
 & CHIEF & 0.879 (0.816, 0.936) & 0.986 (0.959, 1.000) & 0.943 (0.929, 0.956) \\
 & GigaPath & 0.929 (0.881, 0.967) & 0.959 (0.879, 0.998) & 0.946 (0.932, 0.959) \\
 & PulmoFoundation & \textbf{0.955 (0.920, 0.983)} & 0.997 (0.991, 1.000) & 0.977 (0.969, 0.985) \\
\midrule
\multirow{5}{*}{\textbf{Macro Sensitivity}} & UNI & 0.783 (0.704, 0.852) & 0.960 (0.905, 0.991) & 0.899 (0.880, 0.917) \\
 & Virchow2 & 0.808 (0.739, 0.871) & 0.968 (0.916, 0.996) & 0.901 (0.883, 0.917) \\
 & CHIEF & 0.774 (0.700, 0.845) & 0.880 (0.815, 0.921) & 0.866 (0.845, 0.886) \\
 & GigaPath & 0.766 (0.690, 0.841) & 0.907 (0.843, 0.946) & 0.849 (0.828, 0.869) \\
 & PulmoFoundation & \textbf{0.883 (0.819, 0.934)} & \textbf{0.971 (0.919, 0.998)} & \textbf{0.929 (0.914, 0.945)} \\
\midrule
\multirow{5}{*}{\textbf{Macro Specificity}} & UNI & 0.783 (0.704, 0.852) & 0.960 (0.905, 0.991) & 0.899 (0.880, 0.917) \\
 & Virchow2 & 0.808 (0.739, 0.871) & 0.968 (0.916, 0.996) & 0.901 (0.883, 0.917) \\
 & CHIEF & 0.774 (0.700, 0.845) & 0.880 (0.815, 0.921) & 0.866 (0.845, 0.886) \\
 & GigaPath & 0.766 (0.690, 0.841) & 0.907 (0.843, 0.946) & 0.849 (0.828, 0.869) \\
 & PulmoFoundation & \textbf{0.883 (0.819, 0.934)} & \textbf{0.971 (0.919, 0.998)} & \textbf{0.929 (0.914, 0.945)} \\
\midrule
\multirow{5}{*}{\textbf{Macro PPV}} & UNI & 0.783 (0.704, 0.852) & 0.842 (0.756, 0.919) & 0.900 (0.882, 0.917) \\
 & Virchow2 & 0.822 (0.751, 0.882) & 0.899 (0.816, 0.976) & 0.906 (0.890, 0.922) \\
 & CHIEF & 0.778 (0.702, 0.850) & 0.637 (0.589, 0.690) & 0.868 (0.847, 0.887) \\
 & GigaPath & 0.792 (0.718, 0.862) & 0.669 (0.607, 0.735) & 0.862 (0.843, 0.880) \\
 & PulmoFoundation & \textbf{0.886 (0.823, 0.938)} & \textbf{0.934 (0.862, 0.998)} & \textbf{0.928 (0.913, 0.944)} \\
\midrule
\multirow{5}{*}{\textbf{Macro NPV}} & UNI & 0.783 (0.704, 0.852) & 0.842 (0.756, 0.919) & 0.900 (0.882, 0.917) \\
 & Virchow2 & 0.822 (0.751, 0.882) & 0.899 (0.816, 0.976) & 0.906 (0.890, 0.922) \\
 & CHIEF & 0.778 (0.702, 0.850) & 0.637 (0.589, 0.690) & 0.868 (0.847, 0.887) \\
 & GigaPath & 0.792 (0.718, 0.862) & 0.669 (0.607, 0.735) & 0.862 (0.843, 0.880) \\
 & PulmoFoundation & \textbf{0.886 (0.823, 0.938)} & \textbf{0.934 (0.862, 0.998)} & \textbf{0.928 (0.913, 0.944)} \\
\midrule
\multirow{5}{*}{\textbf{LUAD Sensitivity (Youden)}} & UNI & \textbf{0.917 (0.620, 0.984)} & \textbf{0.993 (0.893, 1.000)} & 0.885 (0.872, 0.971) \\
 & Virchow2 & 0.817 (0.656, 0.927) & 0.982 (0.967, 1.000) & \textbf{0.961 (0.900, 0.975)} \\
 & CHIEF & 0.733 (0.654, 0.959) & 0.978 (0.960, 1.000) & 0.915 (0.833, 0.941) \\
 & GigaPath & 0.900 (0.776, 0.985) & 0.978 (0.891, 0.996) & 0.902 (0.859, 0.927) \\
 & PulmoFoundation & 0.900 (0.786, 0.966) & 0.985 (0.974, 1.000) & 0.937 (0.917, 0.960) \\
\midrule
\multirow{5}{*}{\textbf{LUSC Sensitivity (Youden)}} & UNI & 0.733 (0.661, 1.000) & 0.952 (0.909, 1.000) & \textbf{0.938 (0.849, 0.956)} \\
 & Virchow2 & 0.867 (0.803, 1.000) & 1.000 (1.000, 1.000) & 0.891 (0.873, 0.953) \\
 & CHIEF & 0.883 (0.667, 0.968) & 0.952 (0.846, 1.000) & 0.848 (0.821, 0.928) \\
 & GigaPath & 0.767 (0.708, 0.955) & 0.905 (0.810, 1.000) & 0.898 (0.874, 0.940) \\
 & PulmoFoundation & \textbf{0.900 (0.833, 0.985)} & \textbf{1.000 (1.000, 1.000)} & 0.936 (0.912, 0.958) \\
\bottomrule
\end{tabular}
\end{adjustbox}
\end{table}

\clearpage

\begin{table}[htbp]
\centering
\caption{\textbf{Coarse-grained Subtyping (Resection)}: Per-task performance of PulmoFoundation and four baseline foundation models (UNI, Virchow2, CHIEF, GigaPath). Performance metrics are reported with 95\% bootstrap percentile CIs. Macro AUC is threshold-free. Macro Sensitivity, Specificity, PPV, and NPV are macro-averaged one-versus-rest metrics computed from predicted class labels; per-class sensitivity rows are computed at class-specific Youden-optimal operating points. CIs are bootstrap percentiles over cases ($N{=}1{,}000$ replicates). Best-performing model per metric and cohort is shown in bold. Cohort-composition rows report case counts, with slide counts in parentheses; class percentages are case-level percentages. All metrics are computed on the same cohort.}
\label{tab:post_3class}
\renewcommand{\arraystretch}{1.15}
\setlength{\tabcolsep}{5pt}
\small
\begin{adjustbox}{max width=\textwidth,center}
\begin{tabular}{l l c c c c c}
\toprule
\textbf{Metric} & \textbf{Cohort} & \textbf{UNI} & \textbf{Virchow2} & \textbf{CHIEF} & \textbf{GigaPath} & \textbf{PulmoFoundation} \\
\midrule
Total cases (slides) & \multicolumn{6}{l}{Internal: 120 (120); H3: 150 (511)} \\
Adenocarcinoma in situ & \multicolumn{6}{l}{Internal: 40 (40; 33.3\%); H3: 13 (31; 8.7\%)} \\
Invasive adenocarcinoma & \multicolumn{6}{l}{Internal: 40 (40; 33.3\%); H3: 100 (324; 66.7\%)} \\
Squamous cell carcinoma & \multicolumn{6}{l}{Internal: 40 (40; 33.3\%); H3: 37 (156; 24.7\%)} \\
\midrule
\multirow{2}{*}{\textbf{Macro AUC}} & Internal & 0.819 (0.763, 0.876) & 0.798 (0.742, 0.853) & 0.813 (0.753, 0.871) & 0.756 (0.692, 0.815) & \textbf{0.832 (0.779, 0.884)} \\
 & H3 & 0.889 (0.830, 0.948) & 0.884 (0.824, 0.937) & 0.948 (0.894, 0.988) & 0.818 (0.748, 0.887) & \textbf{0.968 (0.914, 0.997)} \\
\midrule
\multirow{2}{*}{\textbf{Macro Sensitivity}} & Internal & \textbf{0.659 (0.583, 0.739)} & 0.659 (0.579, 0.734) & 0.625 (0.534, 0.706) & 0.564 (0.486, 0.642) & 0.643 (0.565, 0.720) \\
 & H3 & 0.703 (0.619, 0.765) & 0.736 (0.627, 0.828) & \textbf{0.881 (0.795, 0.960)} & 0.624 (0.556, 0.705) & 0.611 (0.546, 0.681) \\
\midrule
\multirow{2}{*}{\textbf{Macro Specificity}} & Internal & \textbf{0.830 (0.789, 0.872)} & 0.830 (0.789, 0.868) & 0.812 (0.768, 0.856) & 0.782 (0.739, 0.823) & 0.822 (0.780, 0.861) \\
 & H3 & 0.811 (0.776, 0.843) & 0.861 (0.818, 0.898) & \textbf{0.953 (0.927, 0.975)} & 0.859 (0.811, 0.904) & 0.863 (0.818, 0.907) \\
\midrule
\multirow{2}{*}{\textbf{Macro PPV}} & Internal & \textbf{0.659 (0.577, 0.741)} & 0.652 (0.567, 0.735) & 0.626 (0.538, 0.709) & 0.579 (0.485, 0.676) & 0.646 (0.561, 0.730) \\
 & H3 & 0.689 (0.647, 0.725) & 0.686 (0.633, 0.738) & 0.802 (0.716, 0.881) & 0.611 (0.556, 0.667) & \textbf{0.813 (0.576, 0.958)} \\
\midrule
\multirow{2}{*}{\textbf{Macro NPV}} & Internal & 0.832 (0.792, 0.874) & \textbf{0.835 (0.792, 0.874)} & 0.813 (0.768, 0.856) & 0.791 (0.744, 0.833) & 0.827 (0.785, 0.869) \\
 & H3 & 0.789 (0.756, 0.822) & 0.818 (0.780, 0.855) & 0.926 (0.889, 0.959) & 0.837 (0.789, 0.881) & \textbf{0.948 (0.928, 0.967)} \\
\midrule
\multirow{2}{*}{\textbf{Adenocarcinoma in situ Sens. (Youden)}} & Internal & 0.625 (0.510, 0.932) & 0.725 (0.615, 0.974) & 0.750 (0.500, 0.977) & 0.500 (0.400, 1.000) & \textbf{0.800 (0.690, 0.974)} \\
 & H3 & 0.692 (0.529, 1.000) & 0.769 (0.533, 1.000) & 0.769 (0.615, 1.000) & 0.615 (0.455, 1.000) & \textbf{0.923 (0.727, 1.000)} \\
\midrule
\multirow{2}{*}{\textbf{Invasive adenocarcinoma Sens. (Youden)}} & Internal & \textbf{0.850 (0.471, 0.973)} & 0.600 (0.349, 0.939) & 0.750 (0.619, 0.906) & 0.800 (0.500, 0.941) & 0.450 (0.368, 0.950) \\
 & H3 & 0.940 (0.802, 0.990) & 0.900 (0.687, 0.970) & 0.960 (0.880, 0.990) & \textbf{1.000 (0.689, 1.000)} & 0.980 (0.890, 1.000) \\
\midrule
\multirow{2}{*}{\textbf{Squamous cell carcinoma Sens. (Youden)}} & Internal & \textbf{0.975 (0.846, 1.000)} & 0.925 (0.828, 1.000) & 0.925 (0.705, 0.978) & 0.875 (0.700, 0.971) & 0.875 (0.800, 1.000) \\
 & H3 & 1.000 (1.000, 1.000) & 0.973 (0.909, 1.000) & 1.000 (1.000, 1.000) & 1.000 (0.949, 1.000) & \textbf{1.000 (1.000, 1.000)} \\
\bottomrule
\end{tabular}
\end{adjustbox}
\end{table}

\clearpage

\begin{table}[htbp]
\centering
\caption{\textbf{Fine-grained Subtyping (Resection)}: Per-task performance of PulmoFoundation and four baseline foundation models (UNI, Virchow2, CHIEF, GigaPath). Performance metrics are reported with 95\% bootstrap percentile CIs. Macro AUC is threshold-free. Macro Sensitivity, Specificity, PPV, and NPV are macro-averaged one-versus-rest metrics computed from predicted class labels; per-class sensitivity rows are computed at class-specific Youden-optimal operating points. CIs are bootstrap percentiles over cases ($N{=}1{,}000$ replicates). Best-performing model per metric and cohort is shown in bold. Cohort-composition rows report case counts, with slide counts in parentheses; class percentages are case-level percentages. All metrics are computed on the same cohort.}
\label{tab:post_finegrained}
\renewcommand{\arraystretch}{1.15}
\setlength{\tabcolsep}{5pt}
\small
\begin{adjustbox}{max width=\textwidth,center}
\begin{tabular}{l l c c c c c}
\toprule
\textbf{Metric} & \textbf{Cohort} & \textbf{UNI} & \textbf{Virchow2} & \textbf{CHIEF} & \textbf{GigaPath} & \textbf{PulmoFoundation} \\
\midrule
Total cases (slides) & \multicolumn{6}{l}{Internal: 139 (201); H3: 258 (767)} \\
Adenocarcinoma in situ & \multicolumn{6}{l}{Internal: 30 (45; 21.6\%); H3: 13 (31; 5.0\%)} \\
Minimally invasive adenocarcinoma & \multicolumn{6}{l}{Internal: 24 (32; 17.3\%); H3: 100 (224; 38.8\%)} \\
Invasive adenocarcinoma & \multicolumn{6}{l}{Internal: 30 (37; 21.6\%); H3: 100 (324; 38.8\%)} \\
Neuroendocrine carcinoma & \multicolumn{6}{l}{Internal: 26 (29; 18.7\%); H3: 5 (19; 1.9\%)} \\
Adenosquamous carcinoma & \multicolumn{6}{l}{Internal: 4 (7; 2.9\%); H3: 3 (13; 1.2\%)} \\
Squamous cell carcinoma & \multicolumn{6}{l}{Internal: 25 (51; 18.0\%); H3: 37 (156; 14.3\%)} \\
\midrule
\multirow{2}{*}{\textbf{Macro AUC}} & Internal & 0.907 (0.868, 0.939) & 0.879 (0.831, 0.920) & 0.879 (0.839, 0.911) & 0.897 (0.858, 0.929) & \textbf{0.911 (0.872, 0.940)} \\
 & H3 & 0.882 (0.739, 0.922) & 0.853 (0.710, 0.893) & 0.889 (0.737, 0.923) & 0.835 (0.705, 0.873) & \textbf{0.890 (0.739, 0.923)} \\
\midrule
\multirow{2}{*}{\textbf{Macro Sensitivity}} & Internal & \textbf{0.570 (0.520, 0.625)} & 0.522 (0.473, 0.573) & 0.481 (0.438, 0.533) & 0.453 (0.396, 0.511) & 0.525 (0.484, 0.576) \\
 & H3 & 0.559 (0.499, 0.676) & 0.557 (0.498, 0.662) & 0.508 (0.460, 0.603) & 0.498 (0.463, 0.595) & \textbf{0.621 (0.552, 0.729)} \\
\midrule
\multirow{2}{*}{\textbf{Macro Specificity}} & Internal & \textbf{0.930 (0.914, 0.946)} & 0.915 (0.899, 0.931) & 0.907 (0.889, 0.923) & 0.900 (0.884, 0.917) & 0.918 (0.901, 0.934) \\
 & H3 & 0.898 (0.880, 0.911) & 0.886 (0.867, 0.900) & 0.875 (0.855, 0.886) & 0.888 (0.866, 0.904) & \textbf{0.922 (0.906, 0.936)} \\
\midrule
\multirow{2}{*}{\textbf{Macro PPV}} & Internal & \textbf{0.640 (0.585, 0.694)} & 0.447 (0.392, 0.500) & 0.537 (0.360, 0.631) & 0.538 (0.441, 0.646) & 0.540 (0.426, 0.655) \\
 & H3 & \textbf{0.563 (0.476, 0.690)} & 0.486 (0.426, 0.595) & 0.460 (0.312, 0.614) & 0.553 (0.519, 0.663) & 0.525 (0.466, 0.624) \\
\midrule
\multirow{2}{*}{\textbf{Macro NPV}} & Internal & \textbf{0.936 (0.920, 0.951)} & 0.925 (0.908, 0.940) & 0.918 (0.900, 0.935) & 0.912 (0.896, 0.929) & 0.927 (0.910, 0.943) \\
 & H3 & 0.906 (0.894, 0.918) & 0.885 (0.872, 0.899) & 0.875 (0.865, 0.884) & 0.911 (0.894, 0.928) & \textbf{0.920 (0.906, 0.934)} \\
\midrule
\multirow{2}{*}{\textbf{Adenocarcinoma in situ Sens. (Youden)}} & Internal & 0.867 (0.724, 1.000) & 0.733 (0.583, 0.952) & \textbf{1.000 (0.800, 1.000)} & 1.000 (0.633, 1.000) & 0.867 (0.735, 1.000) \\
 & H3 & \textbf{0.846 (0.470, 1.000)} & 0.692 (0.467, 1.000) & 0.769 (0.600, 1.000) & 0.615 (0.400, 1.000) & 0.615 (0.500, 1.000) \\
\midrule
\multirow{2}{*}{\textbf{Minimally invasive adenocarcinoma Sens. (Youden)}} & Internal & 0.708 (0.692, 1.000) & 0.625 (0.565, 1.000) & 0.917 (0.786, 1.000) & 0.917 (0.762, 1.000) & \textbf{0.958 (0.720, 1.000)} \\
 & H3 & 0.950 (0.806, 1.000) & 0.990 (0.789, 1.000) & \textbf{1.000 (0.699, 1.000)} & 0.970 (0.525, 1.000) & 0.900 (0.837, 0.969) \\
\midrule
\multirow{2}{*}{\textbf{Invasive adenocarcinoma Sens. (Youden)}} & Internal & 0.833 (0.636, 0.968) & 0.867 (0.742, 1.000) & 0.567 (0.487, 1.000) & 0.833 (0.727, 1.000) & \textbf{0.967 (0.643, 1.000)} \\
 & H3 & 0.730 (0.620, 0.933) & 0.500 (0.418, 0.898) & \textbf{0.810 (0.728, 0.908)} & 0.770 (0.527, 0.862) & 0.670 (0.582, 0.768) \\
\midrule
\multirow{2}{*}{\textbf{Neuroendocrine carcinoma Sens. (Youden)}} & Internal & 0.962 (0.897, 1.000) & 0.962 (0.905, 1.000) & 0.962 (0.905, 1.000) & \textbf{1.000 (0.913, 1.000)} & 0.962 (0.889, 1.000) \\
 & H3 & 1.000 (1.000, 1.000) & 1.000 (1.000, 1.000) & 1.000 (1.000, 1.000) & 1.000 (1.000, 1.000) & \textbf{1.000 (1.000, 1.000)} \\
\midrule
\multirow{2}{*}{\textbf{Adenosquamous carcinoma Sens. (Youden)}} & Internal & \textbf{1.000 (0.800, 1.000)} & 0.750 (0.400, 1.000) & 1.000 (1.000, 1.000) & 1.000 (1.000, 1.000) & 0.750 (0.600, 1.000) \\
 & H3 & 0.667 (0.333, 1.000) & \textbf{1.000 (1.000, 1.000)} & 1.000 (1.000, 1.000) & 1.000 (1.000, 1.000) & 0.667 (0.600, 1.000) \\
\midrule
\multirow{2}{*}{\textbf{Squamous cell carcinoma Sens. (Youden)}} & Internal & 0.960 (0.864, 1.000) & 0.960 (0.867, 1.000) & 0.920 (0.815, 1.000) & 0.960 (0.863, 1.000) & \textbf{0.960 (0.869, 1.000)} \\
 & H3 & \textbf{1.000 (0.970, 1.000)} & 1.000 (1.000, 1.000) & 0.973 (0.884, 1.000) & 0.973 (0.927, 1.000) & 0.919 (0.818, 1.000) \\
\bottomrule
\end{tabular}
\end{adjustbox}
\end{table}

\clearpage

\begin{table}[htbp]
\centering
\caption{\textbf{Vascular Invasion (Resection)}: Per-task performance of PulmoFoundation and four baseline foundation models (UNI, Virchow2, CHIEF, GigaPath). Performance metrics are reported with 95\% bootstrap percentile CIs. Macro AUC is threshold-free. Macro Sensitivity, Specificity, PPV, and NPV are macro-averaged one-versus-rest metrics computed from predicted class labels; per-class sensitivity rows are computed at class-specific Youden-optimal operating points. CIs are bootstrap percentiles over cases ($N{=}1{,}000$ replicates). Best-performing model per metric and cohort is shown in bold. Cohort-composition rows report case counts, with slide counts in parentheses; class percentages are case-level percentages. All metrics are computed on the same cohort.}
\label{tab:post_vascular_thrombus}
\renewcommand{\arraystretch}{1.15}
\setlength{\tabcolsep}{5pt}
\small
\begin{adjustbox}{max width=\textwidth,center}
\begin{tabular}{l l c c}
\toprule
\textbf{Metric} & \textbf{Model} & \textbf{Internal} & \textbf{External H3} \\
\midrule
Total cases (slides) & -- & 46 (116) & 249 (807) \\
Negative & -- & 30 (54; 65.2\%) & 0 (0; 0.0\%) \\
Positive & -- & 16 (62; 34.8\%) & 0 (0; 0.0\%) \\
\midrule
\multirow{5}{*}{\textbf{Macro AUC}} & UNI & 0.869 (0.754, 0.957) & 0.917 (0.880, 0.948) \\
 & Virchow2 & 0.885 (0.770, 0.967) & 0.931 (0.898, 0.959) \\
 & CHIEF & 0.823 (0.690, 0.940) & 0.917 (0.886, 0.947) \\
 & GigaPath & 0.808 (0.675, 0.926) & 0.788 (0.713, 0.854) \\
 & PulmoFoundation & \textbf{0.905 (0.798, 0.982)} & \textbf{0.933 (0.901, 0.960)} \\
\midrule
\multirow{5}{*}{\textbf{Macro Sensitivity}} & UNI & 0.710 (0.563, 0.852) & 0.749 (0.675, 0.819) \\
 & Virchow2 & 0.575 (0.484, 0.686) & 0.603 (0.541, 0.669) \\
 & CHIEF & 0.726 (0.580, 0.869) & \textbf{0.878 (0.831, 0.915)} \\
 & GigaPath & 0.743 (0.597, 0.881) & 0.730 (0.655, 0.802) \\
 & PulmoFoundation & \textbf{0.791 (0.646, 0.907)} & 0.672 (0.595, 0.750) \\
\midrule
\multirow{5}{*}{\textbf{Macro Specificity}} & UNI & 0.710 (0.563, 0.852) & 0.749 (0.675, 0.819) \\
 & Virchow2 & 0.575 (0.484, 0.686) & 0.603 (0.541, 0.669) \\
 & CHIEF & 0.726 (0.580, 0.869) & \textbf{0.878 (0.831, 0.915)} \\
 & GigaPath & 0.743 (0.597, 0.881) & 0.730 (0.655, 0.802) \\
 & PulmoFoundation & \textbf{0.791 (0.646, 0.907)} & 0.672 (0.595, 0.750) \\
\midrule
\multirow{5}{*}{\textbf{Macro PPV}} & UNI & 0.710 (0.558, 0.848) & \textbf{0.776 (0.694, 0.847)} \\
 & Virchow2 & 0.712 (0.355, 0.895) & 0.733 (0.610, 0.847) \\
 & CHIEF & 0.734 (0.589, 0.871) & 0.769 (0.716, 0.823) \\
 & GigaPath & 0.735 (0.590, 0.873) & 0.673 (0.611, 0.735) \\
 & PulmoFoundation & \textbf{0.783 (0.635, 0.899)} & 0.743 (0.650, 0.829) \\
\midrule
\multirow{5}{*}{\textbf{Macro NPV}} & UNI & 0.710 (0.558, 0.848) & \textbf{0.776 (0.694, 0.847)} \\
 & Virchow2 & 0.712 (0.355, 0.895) & 0.733 (0.610, 0.847) \\
 & CHIEF & 0.734 (0.589, 0.871) & 0.769 (0.716, 0.823) \\
 & GigaPath & 0.735 (0.590, 0.873) & 0.673 (0.611, 0.735) \\
 & PulmoFoundation & \textbf{0.783 (0.635, 0.899)} & 0.743 (0.650, 0.829) \\
\midrule
\multirow{5}{*}{\textbf{Negative Sensitivity (Youden)}} & UNI & 0.800 (0.560, 0.929) & \textbf{0.860 (0.731, 0.915)} \\
 & Virchow2 & 0.767 (0.600, 0.943) & 0.835 (0.789, 0.902) \\
 & CHIEF & 0.833 (0.444, 0.968) & 0.775 (0.736, 0.884) \\
 & GigaPath & 0.500 (0.407, 1.000) & 0.805 (0.508, 0.876) \\
 & PulmoFoundation & \textbf{0.833 (0.692, 0.960)} & 0.840 (0.797, 0.908) \\
\midrule
\multirow{5}{*}{\textbf{Positive Sensitivity (Youden)}} & UNI & 0.875 (0.809, 1.000) & 0.878 (0.827, 1.000) \\
 & Virchow2 & 0.938 (0.800, 1.000) & 1.000 (0.958, 1.000) \\
 & CHIEF & 0.750 (0.600, 1.000) & 1.000 (0.927, 1.000) \\
 & GigaPath & 1.000 (0.529, 1.000) & 0.673 (0.568, 0.943) \\
 & PulmoFoundation & \textbf{1.000 (1.000, 1.000)} & \textbf{1.000 (0.964, 1.000)} \\
\bottomrule
\end{tabular}
\end{adjustbox}
\end{table}

\clearpage

\begin{table}[htbp]
\centering
\caption{\textbf{Neural Invasion (Resection)}: Per-task performance of PulmoFoundation and four baseline foundation models (UNI, Virchow2, CHIEF, GigaPath). Performance metrics are reported with 95\% bootstrap percentile CIs. Macro AUC is threshold-free. Macro Sensitivity, Specificity, PPV, and NPV are macro-averaged one-versus-rest metrics computed from predicted class labels; per-class sensitivity rows are computed at class-specific Youden-optimal operating points. CIs are bootstrap percentiles over cases ($N{=}1{,}000$ replicates). Best-performing model per metric and cohort is shown in bold. Cohort-composition rows report case counts, with slide counts in parentheses; class percentages are case-level percentages. All metrics are computed on the same cohort.}
\label{tab:post_perineural_invasion}
\renewcommand{\arraystretch}{1.15}
\setlength{\tabcolsep}{5pt}
\small
\begin{adjustbox}{max width=\textwidth,center}
\begin{tabular}{l l c c}
\toprule
\textbf{Metric} & \textbf{Model} & \textbf{Internal} & \textbf{External H3} \\
\midrule
Total cases (slides) & -- & 38 (67) & 114 (366) \\
Negative & -- & 31 (57; 81.6\%) & 0 (0; 0.0\%) \\
Positive & -- & 7 (10; 18.4\%) & 0 (0; 0.0\%) \\
\midrule
\multirow{5}{*}{\textbf{Macro AUC}} & UNI & 0.911 (0.782, 0.993) & 0.902 (0.783, 0.982) \\
 & Virchow2 & \textbf{0.920 (0.794, 0.996)} & 0.968 (0.931, 0.992) \\
 & CHIEF & 0.833 (0.630, 0.971) & 0.944 (0.894, 0.979) \\
 & GigaPath & 0.913 (0.797, 0.995) & 0.876 (0.778, 0.954) \\
 & PulmoFoundation & 0.915 (0.782, 1.000) & \textbf{0.970 (0.934, 0.993)} \\
\midrule
\multirow{5}{*}{\textbf{Macro Sensitivity}} & UNI & 0.500 (0.500, 0.500) & 0.535 (0.500, 0.611) \\
 & Virchow2 & 0.635 (0.500, 0.833) & 0.774 (0.631, 0.909) \\
 & CHIEF & 0.500 (0.500, 0.500) & \textbf{0.920 (0.883, 0.953)} \\
 & GigaPath & 0.567 (0.500, 0.750) & 0.565 (0.490, 0.662) \\
 & PulmoFoundation & \textbf{0.635 (0.500, 0.833)} & 0.841 (0.709, 0.954) \\
\midrule
\multirow{5}{*}{\textbf{Macro Specificity}} & UNI & 0.500 (0.500, 0.500) & 0.535 (0.500, 0.611) \\
 & Virchow2 & 0.635 (0.500, 0.833) & 0.774 (0.631, 0.909) \\
 & CHIEF & 0.500 (0.500, 0.500) & \textbf{0.920 (0.883, 0.953)} \\
 & GigaPath & 0.567 (0.500, 0.750) & 0.565 (0.490, 0.662) \\
 & PulmoFoundation & \textbf{0.635 (0.500, 0.833)} & 0.841 (0.709, 0.954) \\
\midrule
\multirow{5}{*}{\textbf{Macro PPV}} & UNI & 0.408 (0.342, 0.461) & 0.765 (0.421, 0.965) \\
 & Virchow2 & 0.854 (0.408, 0.973) & \textbf{0.872 (0.716, 0.981)} \\
 & CHIEF & 0.408 (0.342, 0.461) & 0.732 (0.650, 0.826) \\
 & GigaPath & 0.731 (0.382, 0.972) & 0.759 (0.438, 0.965) \\
 & PulmoFoundation & \textbf{0.854 (0.408, 0.973)} & 0.866 (0.731, 0.980) \\
\midrule
\multirow{5}{*}{\textbf{Macro NPV}} & UNI & 0.408 (0.342, 0.461) & 0.765 (0.421, 0.965) \\
 & Virchow2 & 0.854 (0.408, 0.973) & \textbf{0.872 (0.716, 0.981)} \\
 & CHIEF & 0.408 (0.342, 0.461) & 0.732 (0.650, 0.826) \\
 & GigaPath & 0.731 (0.382, 0.972) & 0.759 (0.438, 0.965) \\
 & PulmoFoundation & \textbf{0.854 (0.408, 0.973)} & 0.866 (0.731, 0.980) \\
\midrule
\multirow{5}{*}{\textbf{Negative Sensitivity (Youden)}} & UNI & 0.871 (0.613, 1.000) & 0.900 (0.769, 0.960) \\
 & Virchow2 & 0.871 (0.618, 1.000) & 0.910 (0.859, 0.971) \\
 & CHIEF & 0.710 (0.414, 1.000) & 0.870 (0.804, 0.952) \\
 & GigaPath & 0.742 (0.625, 1.000) & 0.820 (0.660, 0.929) \\
 & PulmoFoundation & \textbf{0.871 (0.592, 1.000)} & \textbf{0.910 (0.851, 0.971)} \\
\midrule
\multirow{5}{*}{\textbf{Positive Sensitivity (Youden)}} & UNI & 0.857 (0.799, 1.000) & 0.857 (0.727, 1.000) \\
 & Virchow2 & 0.857 (0.714, 1.000) & 1.000 (1.000, 1.000) \\
 & CHIEF & 0.857 (0.500, 1.000) & 1.000 (0.950, 1.000) \\
 & GigaPath & \textbf{1.000 (0.778, 1.000)} & 0.857 (0.692, 1.000) \\
 & PulmoFoundation & 0.857 (0.667, 1.000) & \textbf{1.000 (1.000, 1.000)} \\
\bottomrule
\end{tabular}
\end{adjustbox}
\end{table}

\clearpage

\begin{table}[htbp]
\centering
\caption{\textbf{STAS (Resection)}: Per-task performance of PulmoFoundation and four baseline foundation models (UNI, Virchow2, CHIEF, GigaPath). Performance metrics are reported with 95\% bootstrap percentile CIs. Macro AUC is threshold-free. Macro Sensitivity, Specificity, PPV, and NPV are macro-averaged one-versus-rest metrics computed from predicted class labels; per-class sensitivity rows are computed at class-specific Youden-optimal operating points. CIs are bootstrap percentiles over cases ($N{=}1{,}000$ replicates). Best-performing model per metric and cohort is shown in bold. Cohort-composition rows report case counts, with slide counts in parentheses; class percentages are case-level percentages. All metrics are computed on the same cohort.}
\label{tab:post_STAS}
\renewcommand{\arraystretch}{1.15}
\setlength{\tabcolsep}{5pt}
\small
\begin{adjustbox}{max width=\textwidth,center}
\begin{tabular}{l l c}
\toprule
\textbf{Metric} & \textbf{Model} & \textbf{Internal} \\
\midrule
Total cases (slides) & -- & 55 (101) \\
Negative & -- & 25 (40; 45.5\%) \\
Positive & -- & 30 (61; 54.5\%) \\
\midrule
\multirow{5}{*}{\textbf{Macro AUC}} & UNI & 0.824 (0.703, 0.924) \\
 & Virchow2 & 0.887 (0.792, 0.964) \\
 & CHIEF & 0.832 (0.713, 0.932) \\
 & GigaPath & 0.802 (0.674, 0.910) \\
 & PulmoFoundation & \textbf{0.898 (0.811, 0.966)} \\
\midrule
\multirow{5}{*}{\textbf{Macro Sensitivity}} & UNI & 0.792 (0.682, 0.890) \\
 & Virchow2 & 0.791 (0.680, 0.891) \\
 & CHIEF & 0.737 (0.616, 0.847) \\
 & GigaPath & 0.715 (0.597, 0.832) \\
 & PulmoFoundation & \textbf{0.808 (0.698, 0.906)} \\
\midrule
\multirow{5}{*}{\textbf{Macro Specificity}} & UNI & 0.792 (0.682, 0.890) \\
 & Virchow2 & 0.791 (0.680, 0.891) \\
 & CHIEF & 0.737 (0.616, 0.847) \\
 & GigaPath & 0.715 (0.597, 0.832) \\
 & PulmoFoundation & \textbf{0.808 (0.698, 0.906)} \\
\midrule
\multirow{5}{*}{\textbf{Macro PPV}} & UNI & 0.793 (0.680, 0.892) \\
 & Virchow2 & 0.793 (0.686, 0.893) \\
 & CHIEF & 0.749 (0.624, 0.860) \\
 & GigaPath & 0.714 (0.597, 0.835) \\
 & PulmoFoundation & \textbf{0.808 (0.698, 0.906)} \\
\midrule
\multirow{5}{*}{\textbf{Macro NPV}} & UNI & 0.793 (0.680, 0.892) \\
 & Virchow2 & 0.793 (0.686, 0.893) \\
 & CHIEF & 0.749 (0.624, 0.860) \\
 & GigaPath & 0.714 (0.597, 0.835) \\
 & PulmoFoundation & \textbf{0.808 (0.698, 0.906)} \\
\midrule
\multirow{5}{*}{\textbf{Negative Sensitivity (Youden)}} & UNI & 0.880 (0.741, 1.000) \\
 & Virchow2 & 0.960 (0.679, 1.000) \\
 & CHIEF & 0.800 (0.666, 1.000) \\
 & GigaPath & 0.560 (0.440, 0.962) \\
 & PulmoFoundation & \textbf{0.960 (0.708, 1.000)} \\
\midrule
\multirow{5}{*}{\textbf{Positive Sensitivity (Youden)}} & UNI & 0.700 (0.467, 0.900) \\
 & Virchow2 & 0.700 (0.594, 1.000) \\
 & CHIEF & 0.833 (0.600, 0.960) \\
 & GigaPath & \textbf{0.933 (0.559, 1.000)} \\
 & PulmoFoundation & 0.733 (0.607, 0.974) \\
\bottomrule
\end{tabular}
\end{adjustbox}
\end{table}

\clearpage

\begin{table}[htbp]
\centering
\caption{\textbf{Pleural Invasion (Resection)}: Per-task performance of PulmoFoundation and four baseline foundation models (UNI, Virchow2, CHIEF, GigaPath). Performance metrics are reported with 95\% bootstrap percentile CIs. Macro AUC is threshold-free. Macro Sensitivity, Specificity, PPV, and NPV are macro-averaged one-versus-rest metrics computed from predicted class labels; per-class sensitivity rows are computed at class-specific Youden-optimal operating points. CIs are bootstrap percentiles over cases ($N{=}1{,}000$ replicates). Best-performing model per metric and cohort is shown in bold. Cohort-composition rows report case counts, with slide counts in parentheses; class percentages are case-level percentages. All metrics are computed on the same cohort.}
\label{tab:post_pleural_invasion}
\renewcommand{\arraystretch}{1.15}
\setlength{\tabcolsep}{5pt}
\small
\begin{adjustbox}{max width=\textwidth,center}
\begin{tabular}{l l c c}
\toprule
\textbf{Metric} & \textbf{Model} & \textbf{Internal} & \textbf{External H3} \\
\midrule
Total cases (slides) & -- & 72 (222) & 244 (789) \\
Negative & -- & 30 (39; 41.7\%) & 0 (0; 0.0\%) \\
Positive & -- & 42 (183; 58.3\%) & 0 (0; 0.0\%) \\
\midrule
\multirow{5}{*}{\textbf{Macro AUC}} & UNI & 0.811 (0.710, 0.904) & 0.887 (0.833, 0.932) \\
 & Virchow2 & 0.748 (0.636, 0.852) & 0.856 (0.797, 0.917) \\
 & CHIEF & 0.767 (0.653, 0.865) & 0.870 (0.808, 0.924) \\
 & GigaPath & 0.709 (0.589, 0.820) & 0.685 (0.609, 0.760) \\
 & PulmoFoundation & \textbf{0.827 (0.724, 0.915)} & \textbf{0.898 (0.849, 0.943)} \\
\midrule
\multirow{5}{*}{\textbf{Macro Sensitivity}} & UNI & \textbf{0.739 (0.644, 0.837)} & 0.700 (0.624, 0.771) \\
 & Virchow2 & 0.654 (0.550, 0.757) & 0.776 (0.709, 0.842) \\
 & CHIEF & 0.719 (0.620, 0.820) & 0.795 (0.735, 0.845) \\
 & GigaPath & 0.602 (0.530, 0.679) & 0.563 (0.512, 0.618) \\
 & PulmoFoundation & 0.717 (0.624, 0.815) & \textbf{0.837 (0.779, 0.890)} \\
\midrule
\multirow{5}{*}{\textbf{Macro Specificity}} & UNI & \textbf{0.739 (0.644, 0.837)} & 0.700 (0.624, 0.771) \\
 & Virchow2 & 0.654 (0.550, 0.757) & 0.776 (0.709, 0.842) \\
 & CHIEF & 0.719 (0.620, 0.820) & 0.795 (0.735, 0.845) \\
 & GigaPath & 0.602 (0.530, 0.679) & 0.563 (0.512, 0.618) \\
 & PulmoFoundation & 0.717 (0.624, 0.815) & \textbf{0.837 (0.779, 0.890)} \\
\midrule
\multirow{5}{*}{\textbf{Macro PPV}} & UNI & \textbf{0.745 (0.649, 0.845)} & \textbf{0.767 (0.671, 0.856)} \\
 & Virchow2 & 0.672 (0.556, 0.781) & 0.733 (0.664, 0.801) \\
 & CHIEF & 0.716 (0.617, 0.815) & 0.678 (0.629, 0.727) \\
 & GigaPath & 0.694 (0.570, 0.785) & 0.687 (0.540, 0.830) \\
 & PulmoFoundation & 0.733 (0.630, 0.834) & 0.735 (0.676, 0.788) \\
\midrule
\multirow{5}{*}{\textbf{Macro NPV}} & UNI & \textbf{0.745 (0.649, 0.845)} & \textbf{0.767 (0.671, 0.856)} \\
 & Virchow2 & 0.672 (0.556, 0.781) & 0.733 (0.664, 0.801) \\
 & CHIEF & 0.716 (0.617, 0.815) & 0.678 (0.629, 0.727) \\
 & GigaPath & 0.694 (0.570, 0.785) & 0.687 (0.540, 0.830) \\
 & PulmoFoundation & 0.733 (0.630, 0.834) & 0.735 (0.676, 0.788) \\
\midrule
\multirow{5}{*}{\textbf{Negative Sensitivity (Youden)}} & UNI & 0.833 (0.593, 0.971) & 0.835 (0.695, 0.912) \\
 & Virchow2 & 0.700 (0.433, 1.000) & \textbf{0.925 (0.648, 0.961)} \\
 & CHIEF & 0.700 (0.448, 0.966) & 0.845 (0.698, 0.895) \\
 & GigaPath & 0.800 (0.577, 0.968) & 0.450 (0.401, 0.848) \\
 & PulmoFoundation & \textbf{0.833 (0.529, 0.958)} & 0.790 (0.747, 0.922) \\
\midrule
\multirow{5}{*}{\textbf{Positive Sensitivity (Youden)}} & UNI & 0.690 (0.558, 0.925) & 0.841 (0.750, 0.976) \\
 & Virchow2 & 0.714 (0.412, 0.976) & 0.682 (0.564, 0.921) \\
 & CHIEF & 0.762 (0.458, 0.975) & 0.818 (0.729, 0.975) \\
 & GigaPath & 0.619 (0.400, 0.850) & \textbf{0.932 (0.571, 0.980)} \\
 & PulmoFoundation & \textbf{0.762 (0.651, 1.000)} & 0.909 (0.757, 0.977) \\
\bottomrule
\end{tabular}
\end{adjustbox}
\end{table}

\clearpage

\begin{table}[htbp]
\centering
\caption{\textbf{LN Metastasis (Resection)}: Per-task performance of PulmoFoundation and four baseline foundation models (UNI, Virchow2, CHIEF, GigaPath). Performance metrics are reported with 95\% bootstrap percentile CIs. Macro AUC is threshold-free. Macro Sensitivity, Specificity, PPV, and NPV are macro-averaged one-versus-rest metrics computed from predicted class labels; per-class sensitivity rows are computed at class-specific Youden-optimal operating points. CIs are bootstrap percentiles over cases ($N{=}1{,}000$ replicates). Best-performing model per metric and cohort is shown in bold. Cohort-composition rows report case counts, with slide counts in parentheses; class percentages are case-level percentages. All metrics are computed on the same cohort.}
\label{tab:post_lymph_node_metastasis}
\renewcommand{\arraystretch}{1.15}
\setlength{\tabcolsep}{5pt}
\small
\begin{adjustbox}{max width=\textwidth,center}
\begin{tabular}{l l c}
\toprule
\textbf{Metric} & \textbf{Model} & \textbf{Internal} \\
\midrule
Total cases (slides) & -- & 88 (152) \\
Negative & -- & 70 (110; 79.5\%) \\
Positive & -- & 18 (42; 20.5\%) \\
\midrule
\multirow{5}{*}{\textbf{Macro AUC}} & UNI & 0.920 (0.837, 0.986) \\
 & Virchow2 & 0.962 (0.917, 0.995) \\
 & CHIEF & 0.906 (0.792, 0.985) \\
 & GigaPath & 0.951 (0.901, 0.991) \\
 & PulmoFoundation & \textbf{0.975 (0.944, 0.997)} \\
\midrule
\multirow{5}{*}{\textbf{Macro Sensitivity}} & UNI & 0.854 (0.751, 0.948) \\
 & Virchow2 & 0.868 (0.749, 0.966) \\
 & CHIEF & 0.861 (0.752, 0.953) \\
 & GigaPath & 0.523 (0.480, 0.588) \\
 & PulmoFoundation & \textbf{0.904 (0.806, 0.993)} \\
\midrule
\multirow{5}{*}{\textbf{Macro Specificity}} & UNI & 0.854 (0.751, 0.948) \\
 & Virchow2 & 0.868 (0.749, 0.966) \\
 & CHIEF & 0.861 (0.752, 0.953) \\
 & GigaPath & 0.523 (0.480, 0.588) \\
 & PulmoFoundation & \textbf{0.904 (0.806, 0.993)} \\
\midrule
\multirow{5}{*}{\textbf{Macro PPV}} & UNI & 0.790 (0.689, 0.889) \\
 & Virchow2 & 0.886 (0.785, 0.975) \\
 & CHIEF & 0.862 (0.752, 0.958) \\
 & GigaPath & 0.636 (0.369, 0.930) \\
 & PulmoFoundation & \textbf{0.922 (0.828, 0.993)} \\
\midrule
\multirow{5}{*}{\textbf{Macro NPV}} & UNI & 0.790 (0.689, 0.889) \\
 & Virchow2 & 0.886 (0.785, 0.975) \\
 & CHIEF & 0.862 (0.752, 0.958) \\
 & GigaPath & 0.636 (0.369, 0.930) \\
 & PulmoFoundation & \textbf{0.922 (0.828, 0.993)} \\
\midrule
\multirow{5}{*}{\textbf{Negative Sensitivity (Youden)}} & UNI & 0.900 (0.611, 0.986) \\
 & Virchow2 & 0.814 (0.750, 1.000) \\
 & CHIEF & \textbf{0.943 (0.781, 0.987)} \\
 & GigaPath & 0.914 (0.750, 1.000) \\
 & PulmoFoundation & 0.843 (0.791, 1.000) \\
\midrule
\multirow{5}{*}{\textbf{Positive Sensitivity (Youden)}} & UNI & 0.833 (0.667, 1.000) \\
 & Virchow2 & 1.000 (0.810, 1.000) \\
 & CHIEF & 0.833 (0.667, 1.000) \\
 & GigaPath & 0.889 (0.765, 1.000) \\
 & PulmoFoundation & \textbf{1.000 (0.850, 1.000)} \\
\bottomrule
\end{tabular}
\end{adjustbox}
\end{table}

\clearpage

\begin{table}[htbp]
\centering
\caption{\textbf{IAC Differentiation (Resection)}: Per-task performance of PulmoFoundation and four baseline foundation models (UNI, Virchow2, CHIEF, GigaPath). Performance metrics are reported with 95\% bootstrap percentile CIs. Macro AUC is threshold-free. Macro Sensitivity, Specificity, PPV, and NPV are macro-averaged one-versus-rest metrics computed from predicted class labels; per-class sensitivity rows are computed at class-specific Youden-optimal operating points. CIs are bootstrap percentiles over cases ($N{=}1{,}000$ replicates). Best-performing model per metric and cohort is shown in bold. Cohort-composition rows report case counts, with slide counts in parentheses; class percentages are case-level percentages. All metrics are computed on the same cohort.}
\label{tab:post_iac_grading}
\renewcommand{\arraystretch}{1.15}
\setlength{\tabcolsep}{5pt}
\small
\begin{adjustbox}{max width=\textwidth,center}
\begin{tabular}{l l c c c c c}
\toprule
\textbf{Metric} & \textbf{Cohort} & \textbf{UNI} & \textbf{Virchow2} & \textbf{CHIEF} & \textbf{GigaPath} & \textbf{PulmoFoundation} \\
\midrule
Total cases (slides) & \multicolumn{6}{l}{Internal: 86 (292)} \\
High & \multicolumn{6}{l}{Internal: 9 (28; 10.5\%)} \\
Low & \multicolumn{6}{l}{Internal: 19 (75; 22.1\%)} \\
Medium & \multicolumn{6}{l}{Internal: 58 (189; 67.4\%)} \\
\midrule
\multirow{1}{*}{\textbf{Macro AUC}} & Internal & 0.901 (0.826, 0.966) & 0.814 (0.720, 0.898) & 0.893 (0.810, 0.964) & 0.879 (0.806, 0.940) & \textbf{0.902 (0.820, 0.965)} \\
\midrule
\multirow{1}{*}{\textbf{Macro Sensitivity}} & Internal & 0.539 (0.467, 0.613) & 0.656 (0.524, 0.783) & 0.774 (0.647, 0.895) & 0.620 (0.531, 0.719) & \textbf{0.795 (0.669, 0.911)} \\
\midrule
\multirow{1}{*}{\textbf{Macro Specificity}} & Internal & 0.805 (0.749, 0.867) & 0.823 (0.757, 0.884) & 0.897 (0.841, 0.948) & 0.841 (0.778, 0.903) & \textbf{0.907 (0.850, 0.954)} \\
\midrule
\multirow{1}{*}{\textbf{Macro PPV}} & Internal & 0.569 (0.506, 0.616) & 0.634 (0.515, 0.751) & \textbf{0.809 (0.666, 0.913)} & 0.683 (0.480, 0.909) & 0.803 (0.646, 0.923) \\
\midrule
\multirow{1}{*}{\textbf{Macro NPV}} & Internal & \textbf{0.909 (0.846, 0.953)} & 0.815 (0.751, 0.875) & 0.887 (0.830, 0.940) & 0.867 (0.800, 0.926) & 0.900 (0.838, 0.950) \\
\midrule
\multirow{1}{*}{\textbf{High Sens. (Youden)}} & Internal & 0.778 (0.555, 1.000) & 0.444 (0.250, 1.000) & 0.667 (0.500, 1.000) & 0.889 (0.625, 1.000) & \textbf{0.889 (0.545, 1.000)} \\
\midrule
\multirow{1}{*}{\textbf{Low Sens. (Youden)}} & Internal & 0.947 (0.842, 1.000) & 0.947 (0.737, 1.000) & 0.895 (0.750, 1.000) & 0.947 (0.800, 1.000) & \textbf{0.947 (0.786, 1.000)} \\
\midrule
\multirow{1}{*}{\textbf{Medium Sens. (Youden)}} & Internal & 0.862 (0.672, 0.984) & 0.879 (0.534, 1.000) & \textbf{0.931 (0.797, 0.984)} & 0.690 (0.593, 0.951) & 0.879 (0.688, 0.982) \\
\bottomrule
\end{tabular}
\end{adjustbox}
\end{table}

\clearpage

\begin{table}[htbp]
\centering
\caption{\textbf{TTF-1}: Per-task performance of PulmoFoundation and four baseline foundation models (UNI, Virchow2, CHIEF, GigaPath). Performance metrics are reported with 95\% bootstrap percentile CIs. Macro AUC is threshold-free. Macro Sensitivity, Specificity, PPV, and NPV are macro-averaged one-versus-rest metrics computed from predicted class labels; per-class sensitivity rows are computed at class-specific Youden-optimal operating points. CIs are bootstrap percentiles over cases ($N{=}1{,}000$ replicates). Best-performing model per metric and cohort is shown in bold. Cohort-composition rows report case counts, with slide counts in parentheses; class percentages are case-level percentages. All metrics are computed on the same cohort.}
\label{tab:post_TTF-1}
\renewcommand{\arraystretch}{1.15}
\setlength{\tabcolsep}{5pt}
\small
\begin{adjustbox}{max width=\textwidth,center}
\begin{tabular}{l l c c}
\toprule
\textbf{Metric} & \textbf{Model} & \textbf{Internal} & \textbf{External H3} \\
\midrule
Total cases (slides) & -- & 112 (138) & 251 (774) \\
Negative & -- & 65 (83; 58.0\%) & 51 (197; 20.3\%) \\
Positive & -- & 47 (55; 42.0\%) & 200 (577; 79.7\%) \\
\midrule
\multirow{5}{*}{\textbf{Macro AUC}} & UNI & 0.840 (0.759, 0.914) & 0.931 (0.871, 0.979) \\
 & Virchow2 & 0.916 (0.859, 0.965) & 0.939 (0.884, 0.983) \\
 & CHIEF & 0.889 (0.821, 0.946) & 0.908 (0.852, 0.951) \\
 & GigaPath & 0.869 (0.800, 0.929) & 0.932 (0.888, 0.968) \\
 & PulmoFoundation & \textbf{0.923 (0.868, 0.965)} & \textbf{0.974 (0.946, 0.994)} \\
\midrule
\multirow{5}{*}{\textbf{Macro Sensitivity}} & UNI & 0.805 (0.733, 0.870) & 0.890 (0.830, 0.937) \\
 & Virchow2 & 0.820 (0.747, 0.888) & 0.873 (0.831, 0.909) \\
 & CHIEF & \textbf{0.835 (0.759, 0.900)} & 0.771 (0.702, 0.841) \\
 & GigaPath & 0.793 (0.719, 0.862) & 0.840 (0.778, 0.896) \\
 & PulmoFoundation & 0.807 (0.733, 0.876) & \textbf{0.928 (0.876, 0.971)} \\
\midrule
\multirow{5}{*}{\textbf{Macro Specificity}} & UNI & 0.805 (0.733, 0.870) & 0.890 (0.830, 0.937) \\
 & Virchow2 & 0.820 (0.747, 0.888) & 0.873 (0.831, 0.909) \\
 & CHIEF & \textbf{0.835 (0.759, 0.900)} & 0.771 (0.702, 0.841) \\
 & GigaPath & 0.793 (0.719, 0.862) & 0.840 (0.778, 0.896) \\
 & PulmoFoundation & 0.807 (0.733, 0.876) & \textbf{0.928 (0.876, 0.971)} \\
\midrule
\multirow{5}{*}{\textbf{Macro PPV}} & UNI & 0.797 (0.724, 0.864) & 0.864 (0.804, 0.919) \\
 & Virchow2 & 0.833 (0.760, 0.900) & 0.760 (0.707, 0.809) \\
 & CHIEF & \textbf{0.839 (0.765, 0.903)} & 0.835 (0.764, 0.902) \\
 & GigaPath & 0.786 (0.715, 0.857) & 0.786 (0.724, 0.846) \\
 & PulmoFoundation & 0.829 (0.755, 0.893) & \textbf{0.934 (0.889, 0.974)} \\
\midrule
\multirow{5}{*}{\textbf{Macro NPV}} & UNI & 0.797 (0.724, 0.864) & 0.864 (0.804, 0.919) \\
 & Virchow2 & 0.833 (0.760, 0.900) & 0.760 (0.707, 0.809) \\
 & CHIEF & \textbf{0.839 (0.765, 0.903)} & 0.835 (0.764, 0.902) \\
 & GigaPath & 0.786 (0.715, 0.857) & 0.786 (0.724, 0.846) \\
 & PulmoFoundation & 0.829 (0.755, 0.893) & \textbf{0.934 (0.889, 0.974)} \\
\midrule
\multirow{5}{*}{\textbf{Negative Sensitivity (Youden)}} & UNI & 0.708 (0.629, 0.896) & 0.824 (0.736, 0.945) \\
 & Virchow2 & \textbf{0.877 (0.750, 0.944)} & \textbf{0.941 (0.860, 1.000)} \\
 & CHIEF & 0.846 (0.718, 0.931) & 0.784 (0.719, 0.960) \\
 & GigaPath & 0.846 (0.642, 0.925) & 0.784 (0.700, 0.941) \\
 & PulmoFoundation & 0.862 (0.754, 0.944) & 0.922 (0.841, 0.981) \\
\midrule
\multirow{5}{*}{\textbf{Positive Sensitivity (Youden)}} & UNI & \textbf{0.957 (0.807, 1.000)} & \textbf{0.975 (0.883, 1.000)} \\
 & Virchow2 & 0.872 (0.812, 1.000) & 0.900 (0.866, 0.974) \\
 & CHIEF & 0.872 (0.800, 0.981) & 0.915 (0.741, 0.955) \\
 & GigaPath & 0.809 (0.750, 1.000) & 0.935 (0.769, 0.985) \\
 & PulmoFoundation & 0.915 (0.826, 1.000) & 0.960 (0.922, 0.995) \\
\bottomrule
\end{tabular}
\end{adjustbox}
\end{table}

\clearpage

\begin{table}[htbp]
\centering
\caption{\textbf{CK7}: Per-task performance of PulmoFoundation and four baseline foundation models (UNI, Virchow2, CHIEF, GigaPath). Performance metrics are reported with 95\% bootstrap percentile CIs. Macro AUC is threshold-free. Macro Sensitivity, Specificity, PPV, and NPV are macro-averaged one-versus-rest metrics computed from predicted class labels; per-class sensitivity rows are computed at class-specific Youden-optimal operating points. CIs are bootstrap percentiles over cases ($N{=}1{,}000$ replicates). Best-performing model per metric and cohort is shown in bold. Cohort-composition rows report case counts, with slide counts in parentheses; class percentages are case-level percentages. All metrics are computed on the same cohort.}
\label{tab:post_CK-7}
\renewcommand{\arraystretch}{1.15}
\setlength{\tabcolsep}{5pt}
\small
\begin{adjustbox}{max width=\textwidth,center}
\begin{tabular}{l l c c}
\toprule
\textbf{Metric} & \textbf{Model} & \textbf{Internal} & \textbf{External H3} \\
\midrule
Total cases (slides) & -- & 92 (123) & 122 (377) \\
Negative & -- & 44 (63; 47.8\%) & 22 (87; 18.0\%) \\
Positive & -- & 48 (60; 52.2\%) & 100 (290; 82.0\%) \\
\midrule
\multirow{5}{*}{\textbf{Macro AUC}} & UNI & 0.873 (0.788, 0.936) & 0.958 (0.889, 1.000) \\
 & Virchow2 & 0.838 (0.747, 0.913) & 0.974 (0.930, 1.000) \\
 & CHIEF & 0.848 (0.761, 0.921) & 0.967 (0.921, 0.997) \\
 & GigaPath & 0.818 (0.726, 0.895) & 0.892 (0.797, 0.970) \\
 & PulmoFoundation & \textbf{0.899 (0.825, 0.956)} & \textbf{0.979 (0.948, 1.000)} \\
\midrule
\multirow{5}{*}{\textbf{Macro Sensitivity}} & UNI & 0.761 (0.672, 0.840) & \textbf{0.932 (0.857, 1.000)} \\
 & Virchow2 & 0.782 (0.689, 0.860) & 0.885 (0.790, 0.972) \\
 & CHIEF & 0.768 (0.673, 0.851) & 0.841 (0.739, 0.935) \\
 & GigaPath & 0.726 (0.627, 0.814) & 0.734 (0.623, 0.846) \\
 & PulmoFoundation & \textbf{0.796 (0.712, 0.872)} & 0.791 (0.688, 0.891) \\
\midrule
\multirow{5}{*}{\textbf{Macro Specificity}} & UNI & 0.761 (0.672, 0.840) & \textbf{0.932 (0.857, 1.000)} \\
 & Virchow2 & 0.782 (0.689, 0.860) & 0.885 (0.790, 0.972) \\
 & CHIEF & 0.768 (0.673, 0.851) & 0.841 (0.739, 0.935) \\
 & GigaPath & 0.726 (0.627, 0.814) & 0.734 (0.623, 0.846) \\
 & PulmoFoundation & \textbf{0.796 (0.712, 0.872)} & 0.791 (0.688, 0.891) \\
\midrule
\multirow{5}{*}{\textbf{Macro PPV}} & UNI & 0.761 (0.668, 0.844) & \textbf{0.985 (0.968, 1.000)} \\
 & Virchow2 & 0.781 (0.688, 0.860) & 0.976 (0.953, 0.995) \\
 & CHIEF & 0.769 (0.674, 0.853) & 0.968 (0.946, 0.987) \\
 & GigaPath & 0.727 (0.625, 0.816) & 0.842 (0.715, 0.950) \\
 & PulmoFoundation & \textbf{0.799 (0.712, 0.877)} & 0.923 (0.835, 0.977) \\
\midrule
\multirow{5}{*}{\textbf{Macro NPV}} & UNI & 0.761 (0.668, 0.844) & \textbf{0.985 (0.968, 1.000)} \\
 & Virchow2 & 0.781 (0.688, 0.860) & 0.976 (0.953, 0.995) \\
 & CHIEF & 0.769 (0.674, 0.853) & 0.968 (0.946, 0.987) \\
 & GigaPath & 0.727 (0.625, 0.816) & 0.842 (0.715, 0.950) \\
 & PulmoFoundation & \textbf{0.799 (0.712, 0.877)} & 0.923 (0.835, 0.977) \\
\midrule
\multirow{5}{*}{\textbf{Negative Sensitivity (Youden)}} & UNI & 0.750 (0.641, 0.978) & 0.909 (0.765, 1.000) \\
 & Virchow2 & 0.795 (0.590, 0.950) & 0.909 (0.773, 1.000) \\
 & CHIEF & 0.750 (0.643, 0.980) & \textbf{0.955 (0.821, 1.000)} \\
 & GigaPath & \textbf{0.886 (0.717, 0.977)} & 0.727 (0.607, 0.960) \\
 & PulmoFoundation & 0.795 (0.690, 0.978) & 0.909 (0.812, 1.000) \\
\midrule
\multirow{5}{*}{\textbf{Positive Sensitivity (Youden)}} & UNI & 0.896 (0.649, 0.977) & \textbf{1.000 (1.000, 1.000)} \\
 & Virchow2 & 0.792 (0.633, 0.975) & 0.990 (0.960, 1.000) \\
 & CHIEF & 0.854 (0.549, 0.950) & 0.900 (0.853, 1.000) \\
 & GigaPath & 0.688 (0.558, 0.854) & 0.960 (0.748, 1.000) \\
 & PulmoFoundation & \textbf{0.917 (0.706, 0.981)} & 0.980 (0.838, 1.000) \\
\bottomrule
\end{tabular}
\end{adjustbox}
\end{table}

\clearpage

\begin{table}[htbp]
\centering
\caption{\textbf{Napsin A}: Per-task performance of PulmoFoundation and four baseline foundation models (UNI, Virchow2, CHIEF, GigaPath). Performance metrics are reported with 95\% bootstrap percentile CIs. Macro AUC is threshold-free. Macro Sensitivity, Specificity, PPV, and NPV are macro-averaged one-versus-rest metrics computed from predicted class labels; per-class sensitivity rows are computed at class-specific Youden-optimal operating points. CIs are bootstrap percentiles over cases ($N{=}1{,}000$ replicates). Best-performing model per metric and cohort is shown in bold. Cohort-composition rows report case counts, with slide counts in parentheses; class percentages are case-level percentages. All metrics are computed on the same cohort.}
\label{tab:post_Napsin-A}
\renewcommand{\arraystretch}{1.15}
\setlength{\tabcolsep}{5pt}
\small
\begin{adjustbox}{max width=\textwidth,center}
\begin{tabular}{l l c c}
\toprule
\textbf{Metric} & \textbf{Model} & \textbf{Internal} & \textbf{External H3} \\
\midrule
Total cases (slides) & -- & 134 (165) & 138 (446) \\
Negative & -- & 63 (85; 47.0\%) & 38 (153; 27.5\%) \\
Positive & -- & 71 (80; 53.0\%) & 100 (293; 72.5\%) \\
\midrule
\multirow{5}{*}{\textbf{Macro AUC}} & UNI & 0.919 (0.875, 0.960) & 0.978 (0.953, 0.997) \\
 & Virchow2 & 0.922 (0.876, 0.959) & 0.966 (0.925, 0.994) \\
 & CHIEF & 0.859 (0.792, 0.916) & 0.887 (0.825, 0.940) \\
 & GigaPath & 0.897 (0.841, 0.948) & 0.914 (0.854, 0.963) \\
 & PulmoFoundation & \textbf{0.936 (0.893, 0.972)} & \textbf{0.980 (0.952, 0.999)} \\
\midrule
\multirow{5}{*}{\textbf{Macro Sensitivity}} & UNI & \textbf{0.824 (0.761, 0.882)} & 0.832 (0.750, 0.905) \\
 & Virchow2 & 0.814 (0.750, 0.873) & \textbf{0.927 (0.873, 0.973)} \\
 & CHIEF & 0.743 (0.665, 0.811) & 0.793 (0.712, 0.866) \\
 & GigaPath & 0.801 (0.736, 0.867) & 0.744 (0.676, 0.802) \\
 & PulmoFoundation & 0.796 (0.729, 0.857) & 0.903 (0.835, 0.963) \\
\midrule
\multirow{5}{*}{\textbf{Macro Specificity}} & UNI & \textbf{0.824 (0.761, 0.882)} & 0.832 (0.750, 0.905) \\
 & Virchow2 & 0.814 (0.750, 0.873) & \textbf{0.927 (0.873, 0.973)} \\
 & CHIEF & 0.743 (0.665, 0.811) & 0.793 (0.712, 0.866) \\
 & GigaPath & 0.801 (0.736, 0.867) & 0.744 (0.676, 0.802) \\
 & PulmoFoundation & 0.796 (0.729, 0.857) & 0.903 (0.835, 0.963) \\
\midrule
\multirow{5}{*}{\textbf{Macro PPV}} & UNI & \textbf{0.824 (0.763, 0.883)} & 0.910 (0.847, 0.958) \\
 & Virchow2 & 0.814 (0.748, 0.872) & 0.926 (0.868, 0.972) \\
 & CHIEF & 0.747 (0.670, 0.817) & 0.773 (0.693, 0.850) \\
 & GigaPath & 0.802 (0.736, 0.868) & 0.702 (0.639, 0.762) \\
 & PulmoFoundation & 0.802 (0.741, 0.861) & \textbf{0.951 (0.907, 0.985)} \\
\midrule
\multirow{5}{*}{\textbf{Macro NPV}} & UNI & \textbf{0.824 (0.763, 0.883)} & 0.910 (0.847, 0.958) \\
 & Virchow2 & 0.814 (0.748, 0.872) & 0.926 (0.868, 0.972) \\
 & CHIEF & 0.747 (0.670, 0.817) & 0.773 (0.693, 0.850) \\
 & GigaPath & 0.802 (0.736, 0.868) & 0.702 (0.639, 0.762) \\
 & PulmoFoundation & 0.802 (0.741, 0.861) & \textbf{0.951 (0.907, 0.985)} \\
\midrule
\multirow{5}{*}{\textbf{Negative Sensitivity (Youden)}} & UNI & 0.762 (0.667, 0.951) & 0.947 (0.868, 1.000) \\
 & Virchow2 & 0.730 (0.667, 0.913) & 0.947 (0.868, 1.000) \\
 & CHIEF & 0.762 (0.617, 0.875) & 0.816 (0.615, 1.000) \\
 & GigaPath & 0.794 (0.681, 0.915) & 0.737 (0.634, 0.973) \\
 & PulmoFoundation & \textbf{0.825 (0.741, 0.933)} & \textbf{0.974 (0.909, 1.000)} \\
\midrule
\multirow{5}{*}{\textbf{Positive Sensitivity (Youden)}} & UNI & 0.930 (0.758, 1.000) & \textbf{0.960 (0.922, 0.991)} \\
 & Virchow2 & \textbf{0.986 (0.853, 1.000)} & 0.940 (0.878, 0.990) \\
 & CHIEF & 0.859 (0.754, 0.971) & 0.800 (0.568, 0.979) \\
 & GigaPath & 0.915 (0.800, 0.987) & 0.940 (0.680, 1.000) \\
 & PulmoFoundation & 0.958 (0.892, 1.000) & 0.950 (0.917, 1.000) \\
\bottomrule
\end{tabular}
\end{adjustbox}
\end{table}

\clearpage

\begin{table}[htbp]
\centering
\caption{\textbf{P40}: Per-task performance of PulmoFoundation and four baseline foundation models (UNI, Virchow2, CHIEF, GigaPath). Performance metrics are reported with 95\% bootstrap percentile CIs. Macro AUC is threshold-free. Macro Sensitivity, Specificity, PPV, and NPV are macro-averaged one-versus-rest metrics computed from predicted class labels; per-class sensitivity rows are computed at class-specific Youden-optimal operating points. CIs are bootstrap percentiles over cases ($N{=}1{,}000$ replicates). Best-performing model per metric and cohort is shown in bold. Cohort-composition rows report case counts, with slide counts in parentheses; class percentages are case-level percentages. All metrics are computed on the same cohort.}
\label{tab:post_P40}
\renewcommand{\arraystretch}{1.15}
\setlength{\tabcolsep}{5pt}
\small
\begin{adjustbox}{max width=\textwidth,center}
\begin{tabular}{l l c c}
\toprule
\textbf{Metric} & \textbf{Model} & \textbf{Internal} & \textbf{External H3} \\
\midrule
Total cases (slides) & -- & 67 (86) & 243 (780) \\
Negative & -- & 47 (65; 70.1\%) & 200 (603; 82.3\%) \\
Positive & -- & 20 (21; 29.9\%) & 43 (177; 17.7\%) \\
\midrule
\multirow{5}{*}{\textbf{Macro AUC}} & UNI & 0.746 (0.596, 0.872) & 0.923 (0.863, 0.971) \\
 & Virchow2 & 0.758 (0.621, 0.871) & 0.886 (0.818, 0.942) \\
 & CHIEF & 0.782 (0.658, 0.896) & 0.882 (0.827, 0.931) \\
 & GigaPath & \textbf{0.803 (0.660, 0.923)} & 0.913 (0.859, 0.959) \\
 & PulmoFoundation & 0.802 (0.660, 0.917) & \textbf{0.952 (0.888, 0.995)} \\
\midrule
\multirow{5}{*}{\textbf{Macro Sensitivity}} & UNI & 0.684 (0.563, 0.795) & 0.852 (0.783, 0.917) \\
 & Virchow2 & 0.659 (0.534, 0.783) & 0.807 (0.736, 0.875) \\
 & CHIEF & 0.500 (0.500, 0.500) & 0.497 (0.490, 0.500) \\
 & GigaPath & 0.784 (0.674, 0.890) & 0.868 (0.805, 0.924) \\
 & PulmoFoundation & \textbf{0.786 (0.669, 0.889)} & \textbf{0.950 (0.903, 0.987)} \\
\midrule
\multirow{5}{*}{\textbf{Macro Specificity}} & UNI & 0.684 (0.563, 0.795) & 0.852 (0.783, 0.917) \\
 & Virchow2 & 0.659 (0.534, 0.783) & 0.807 (0.736, 0.875) \\
 & CHIEF & 0.500 (0.500, 0.500) & 0.497 (0.490, 0.500) \\
 & GigaPath & 0.784 (0.674, 0.890) & 0.868 (0.805, 0.924) \\
 & PulmoFoundation & \textbf{0.786 (0.669, 0.889)} & \textbf{0.950 (0.903, 0.987)} \\
\midrule
\multirow{5}{*}{\textbf{Macro PPV}} & UNI & 0.712 (0.572, 0.835) & 0.834 (0.760, 0.897) \\
 & Virchow2 & 0.670 (0.537, 0.803) & 0.720 (0.657, 0.784) \\
 & CHIEF & 0.351 (0.291, 0.403) & 0.412 (0.385, 0.434) \\
 & GigaPath & 0.785 (0.675, 0.890) & 0.802 (0.734, 0.867) \\
 & PulmoFoundation & \textbf{0.884 (0.777, 0.960)} & \textbf{0.926 (0.873, 0.973)} \\
\midrule
\multirow{5}{*}{\textbf{Macro NPV}} & UNI & 0.712 (0.572, 0.835) & 0.834 (0.760, 0.897) \\
 & Virchow2 & 0.670 (0.537, 0.803) & 0.720 (0.657, 0.784) \\
 & CHIEF & 0.351 (0.291, 0.403) & 0.412 (0.385, 0.434) \\
 & GigaPath & 0.785 (0.675, 0.890) & 0.802 (0.734, 0.867) \\
 & PulmoFoundation & \textbf{0.884 (0.777, 0.960)} & \textbf{0.926 (0.873, 0.973)} \\
\midrule
\multirow{5}{*}{\textbf{Negative Sensitivity (Youden)}} & UNI & 0.872 (0.511, 1.000) & 0.890 (0.867, 0.990) \\
 & Virchow2 & 0.787 (0.440, 1.000) & 0.785 (0.708, 0.975) \\
 & CHIEF & 0.617 (0.512, 0.857) & 0.940 (0.781, 0.969) \\
 & GigaPath & 0.872 (0.695, 1.000) & 0.910 (0.806, 0.950) \\
 & PulmoFoundation & \textbf{0.915 (0.804, 1.000)} & \textbf{0.970 (0.948, 0.995)} \\
\midrule
\multirow{5}{*}{\textbf{Positive Sensitivity (Youden)}} & UNI & 0.600 (0.389, 0.957) & 0.860 (0.704, 0.939) \\
 & Virchow2 & 0.650 (0.375, 1.000) & 0.860 (0.667, 0.971) \\
 & CHIEF & \textbf{0.950 (0.739, 1.000)} & 0.721 (0.636, 0.920) \\
 & GigaPath & 0.700 (0.500, 0.947) & 0.837 (0.743, 0.957) \\
 & PulmoFoundation & 0.700 (0.478, 0.900) & \textbf{0.930 (0.844, 1.000)} \\
\bottomrule
\end{tabular}
\end{adjustbox}
\end{table}

\clearpage

\begin{table}[htbp]
\centering
\caption{\textbf{P63}: Per-task performance of PulmoFoundation and four baseline foundation models (UNI, Virchow2, CHIEF, GigaPath). Performance metrics are reported with 95\% bootstrap percentile CIs. Macro AUC is threshold-free. Macro Sensitivity, Specificity, PPV, and NPV are macro-averaged one-versus-rest metrics computed from predicted class labels; per-class sensitivity rows are computed at class-specific Youden-optimal operating points. CIs are bootstrap percentiles over cases ($N{=}1{,}000$ replicates). Best-performing model per metric and cohort is shown in bold. Cohort-composition rows report case counts, with slide counts in parentheses; class percentages are case-level percentages. All metrics are computed on the same cohort.}
\label{tab:post_P63}
\renewcommand{\arraystretch}{1.15}
\setlength{\tabcolsep}{5pt}
\small
\begin{adjustbox}{max width=\textwidth,center}
\begin{tabular}{l l c}
\toprule
\textbf{Metric} & \textbf{Model} & \textbf{Internal} \\
\midrule
Total cases (slides) & -- & 58 (78) \\
Negative & -- & 40 (55; 69.0\%) \\
Positive & -- & 18 (23; 31.0\%) \\
\midrule
\multirow{5}{*}{\textbf{Macro AUC}} & UNI & 0.765 (0.630, 0.882) \\
 & Virchow2 & 0.738 (0.599, 0.854) \\
 & CHIEF & 0.714 (0.552, 0.853) \\
 & GigaPath & 0.625 (0.473, 0.770) \\
 & PulmoFoundation & \textbf{0.860 (0.746, 0.950)} \\
\midrule
\multirow{5}{*}{\textbf{Macro Sensitivity}} & UNI & 0.640 (0.529, 0.755) \\
 & Virchow2 & 0.579 (0.458, 0.700) \\
 & CHIEF & 0.670 (0.532, 0.796) \\
 & GigaPath & 0.618 (0.487, 0.754) \\
 & PulmoFoundation & \textbf{0.756 (0.634, 0.869)} \\
\midrule
\multirow{5}{*}{\textbf{Macro Specificity}} & UNI & 0.640 (0.529, 0.755) \\
 & Virchow2 & 0.579 (0.458, 0.700) \\
 & CHIEF & 0.670 (0.532, 0.796) \\
 & GigaPath & 0.618 (0.487, 0.754) \\
 & PulmoFoundation & \textbf{0.756 (0.634, 0.869)} \\
\midrule
\multirow{5}{*}{\textbf{Macro PPV}} & UNI & 0.753 (0.561, 0.908) \\
 & Virchow2 & 0.597 (0.445, 0.751) \\
 & CHIEF & 0.654 (0.528, 0.770) \\
 & GigaPath & 0.602 (0.489, 0.727) \\
 & PulmoFoundation & \textbf{0.787 (0.661, 0.897)} \\
\midrule
\multirow{5}{*}{\textbf{Macro NPV}} & UNI & 0.753 (0.561, 0.908) \\
 & Virchow2 & 0.597 (0.445, 0.751) \\
 & CHIEF & 0.654 (0.528, 0.770) \\
 & GigaPath & 0.602 (0.489, 0.727) \\
 & PulmoFoundation & \textbf{0.787 (0.661, 0.897)} \\
\midrule
\multirow{5}{*}{\textbf{Negative Sensitivity (Youden)}} & UNI & 0.625 (0.409, 1.000) \\
 & Virchow2 & 0.450 (0.333, 1.000) \\
 & CHIEF & 0.725 (0.477, 0.974) \\
 & GigaPath & 0.600 (0.279, 0.914) \\
 & PulmoFoundation & \textbf{0.975 (0.524, 1.000)} \\
\midrule
\multirow{5}{*}{\textbf{Positive Sensitivity (Youden)}} & UNI & 0.833 (0.400, 1.000) \\
 & Virchow2 & \textbf{0.944 (0.350, 1.000)} \\
 & CHIEF & 0.722 (0.364, 0.947) \\
 & GigaPath & 0.722 (0.421, 1.000) \\
 & PulmoFoundation & 0.611 (0.476, 1.000) \\
\bottomrule
\end{tabular}
\end{adjustbox}
\end{table}

\clearpage

\begin{table}[htbp]
\centering
\caption{\textbf{Ki-67}: Per-task performance of PulmoFoundation and four baseline foundation models (UNI, Virchow2, CHIEF, GigaPath). Performance metrics are reported with 95\% bootstrap percentile CIs. Macro AUC is threshold-free. Macro Sensitivity, Specificity, PPV, and NPV are macro-averaged one-versus-rest metrics computed from predicted class labels; per-class sensitivity rows are computed at class-specific Youden-optimal operating points. CIs are bootstrap percentiles over cases ($N{=}1{,}000$ replicates). Best-performing model per metric and cohort is shown in bold. Cohort-composition rows report case counts, with slide counts in parentheses; class percentages are case-level percentages. All metrics are computed on the same cohort.}
\label{tab:post_Ki-67}
\renewcommand{\arraystretch}{1.15}
\setlength{\tabcolsep}{5pt}
\small
\begin{adjustbox}{max width=\textwidth,center}
\begin{tabular}{l l c c c c c}
\toprule
\textbf{Metric} & \textbf{Cohort} & \textbf{UNI} & \textbf{Virchow2} & \textbf{CHIEF} & \textbf{GigaPath} & \textbf{PulmoFoundation} \\
\midrule
Total cases (slides) & \multicolumn{6}{l}{Internal: 48 (123)} \\
High & \multicolumn{6}{l}{Internal: 9 (19; 18.8\%)} \\
Low & \multicolumn{6}{l}{Internal: 22 (63; 45.8\%)} \\
Medium & \multicolumn{6}{l}{Internal: 17 (41; 35.4\%)} \\
\midrule
\multirow{1}{*}{\textbf{Macro AUC}} & Internal & 0.761 (0.650, 0.862) & 0.786 (0.680, 0.881) & 0.452 (0.325, 0.581) & 0.825 (0.710, 0.927) & \textbf{0.871 (0.786, 0.943)} \\
\midrule
\multirow{1}{*}{\textbf{Macro Sensitivity}} & Internal & 0.569 (0.415, 0.713) & 0.557 (0.423, 0.692) & 0.231 (0.104, 0.355) & 0.333 (0.333, 0.333) & \textbf{0.629 (0.475, 0.771)} \\
\midrule
\multirow{1}{*}{\textbf{Macro Specificity}} & Internal & 0.758 (0.686, 0.827) & 0.755 (0.692, 0.819) & 0.616 (0.563, 0.666) & 0.667 (0.667, 0.667) & \textbf{0.826 (0.751, 0.896)} \\
\midrule
\multirow{1}{*}{\textbf{Macro PPV}} & Internal & 0.579 (0.426, 0.731) & 0.590 (0.410, 0.732) & 0.133 (0.049, 0.240) & 0.119 (0.076, 0.167) & \textbf{0.700 (0.497, 0.847)} \\
\midrule
\multirow{1}{*}{\textbf{Macro NPV}} & Internal & 0.761 (0.681, 0.838) & 0.765 (0.690, 0.837) & 0.606 (0.532, 0.685) & 0.452 (0.410, 0.500) & \textbf{0.846 (0.771, 0.913)} \\
\midrule
\multirow{1}{*}{\textbf{High Sens. (Youden)}} & Internal & 0.667 (0.455, 1.000) & 0.667 (0.429, 1.000) & 0.333 (0.083, 1.000) & 0.778 (0.455, 1.000) & \textbf{1.000 (0.818, 1.000)} \\
\midrule
\multirow{1}{*}{\textbf{Low Sens. (Youden)}} & Internal & 0.773 (0.437, 1.000) & 0.636 (0.450, 0.955) & 0.955 (0.312, 1.000) & 0.682 (0.550, 0.957) & \textbf{0.955 (0.680, 1.000)} \\
\midrule
\multirow{1}{*}{\textbf{Medium Sens. (Youden)}} & Internal & 0.941 (0.368, 1.000) & 0.706 (0.476, 1.000) & \textbf{1.000 (0.384, 1.000)} & 0.824 (0.600, 1.000) & 0.882 (0.438, 1.000) \\
\bottomrule
\end{tabular}
\end{adjustbox}
\end{table}

\clearpage

\begin{table}[htbp]
\centering
\caption{\textbf{TMB}: Per-task performance of PulmoFoundation and four baseline foundation models (UNI, Virchow2, CHIEF, GigaPath). Performance metrics are reported with 95\% bootstrap percentile CIs. Macro AUC is threshold-free. Macro Sensitivity, Specificity, PPV, and NPV are macro-averaged one-versus-rest metrics computed from predicted class labels; per-class sensitivity rows are computed at class-specific Youden-optimal operating points. CIs are bootstrap percentiles over cases ($N{=}1{,}000$ replicates). Best-performing model per metric and cohort is shown in bold. Cohort-composition rows report case counts, with slide counts in parentheses; class percentages are case-level percentages. All metrics are computed on the same cohort.}
\label{tab:post_TMB}
\renewcommand{\arraystretch}{1.15}
\setlength{\tabcolsep}{5pt}
\small
\begin{adjustbox}{max width=\textwidth,center}
\begin{tabular}{l l c}
\toprule
\textbf{Metric} & \textbf{Model} & \textbf{Internal} \\
\midrule
Total cases (slides) & -- & 175 (213) \\
TMB-high & -- & 39 (59; 22.3\%) \\
TMB-low & -- & 136 (154; 77.7\%) \\
\midrule
\multirow{5}{*}{\textbf{Macro AUC}} & UNI & 0.584 (0.473, 0.684) \\
 & Virchow2 & 0.621 (0.505, 0.730) \\
 & CHIEF & 0.696 (0.594, 0.786) \\
 & GigaPath & 0.514 (0.417, 0.609) \\
 & PulmoFoundation & \textbf{0.712 (0.618, 0.800)} \\
\midrule
\multirow{5}{*}{\textbf{Macro Sensitivity}} & UNI & \textbf{0.565 (0.487, 0.641)} \\
 & Virchow2 & 0.500 (0.500, 0.500) \\
 & CHIEF & 0.500 (0.500, 0.500) \\
 & GigaPath & 0.500 (0.500, 0.500) \\
 & PulmoFoundation & 0.500 (0.500, 0.500) \\
\midrule
\multirow{5}{*}{\textbf{Macro Specificity}} & UNI & \textbf{0.565 (0.487, 0.641)} \\
 & Virchow2 & 0.500 (0.500, 0.500) \\
 & CHIEF & 0.500 (0.500, 0.500) \\
 & GigaPath & 0.500 (0.500, 0.500) \\
 & PulmoFoundation & 0.500 (0.500, 0.500) \\
\midrule
\multirow{5}{*}{\textbf{Macro PPV}} & UNI & \textbf{0.586 (0.483, 0.681)} \\
 & Virchow2 & 0.389 (0.357, 0.420) \\
 & CHIEF & 0.389 (0.357, 0.420) \\
 & GigaPath & 0.389 (0.357, 0.420) \\
 & PulmoFoundation & 0.389 (0.357, 0.420) \\
\midrule
\multirow{5}{*}{\textbf{Macro NPV}} & UNI & \textbf{0.586 (0.483, 0.681)} \\
 & Virchow2 & 0.389 (0.357, 0.420) \\
 & CHIEF & 0.389 (0.357, 0.420) \\
 & GigaPath & 0.389 (0.357, 0.420) \\
 & PulmoFoundation & 0.389 (0.357, 0.420) \\
\midrule
\multirow{5}{*}{\textbf{TMB-high Sensitivity (Youden)}} & UNI & 0.436 (0.238, 0.788) \\
 & Virchow2 & 0.564 (0.293, 0.727) \\
 & CHIEF & 0.564 (0.442, 0.875) \\
 & GigaPath & \textbf{0.974 (0.317, 1.000)} \\
 & PulmoFoundation & 0.667 (0.410, 0.952) \\
\midrule
\multirow{5}{*}{\textbf{TMB-low Sensitivity (Youden)}} & UNI & 0.772 (0.486, 0.928) \\
 & Virchow2 & 0.743 (0.669, 0.941) \\
 & CHIEF & \textbf{0.824 (0.479, 0.889)} \\
 & GigaPath & 0.147 (0.083, 0.827) \\
 & PulmoFoundation & 0.706 (0.377, 0.909) \\
\bottomrule
\end{tabular}
\end{adjustbox}
\end{table}

\clearpage

\begin{table}[htbp]
\centering
\caption{\textbf{EGFR}: Per-task performance of PulmoFoundation and four baseline foundation models (UNI, Virchow2, CHIEF, GigaPath). Performance metrics are reported with 95\% bootstrap percentile CIs. Macro AUC is threshold-free. Macro Sensitivity, Specificity, PPV, and NPV are macro-averaged one-versus-rest metrics computed from predicted class labels; per-class sensitivity rows are computed at class-specific Youden-optimal operating points. CIs are bootstrap percentiles over cases ($N{=}1{,}000$ replicates). Best-performing model per metric and cohort is shown in bold. Cohort-composition rows report case counts, with slide counts in parentheses; class percentages are case-level percentages. All metrics are computed on the same cohort.}
\label{tab:post_EGFR}
\renewcommand{\arraystretch}{1.15}
\setlength{\tabcolsep}{5pt}
\small
\begin{adjustbox}{max width=\textwidth,center}
\begin{tabular}{l l c c}
\toprule
\textbf{Metric} & \textbf{Model} & \textbf{Internal} & \textbf{External TCGA} \\
\midrule
Total cases (slides) & -- & 88 (146) & 414 (469) \\
Negative & -- & 43 (70; 48.9\%) & 357 (398; 86.2\%) \\
Positive & -- & 45 (76; 51.1\%) & 57 (71; 13.8\%) \\
\midrule
\multirow{5}{*}{\textbf{Macro AUC}} & UNI & 0.847 (0.753, 0.929) & 0.734 (0.660, 0.809) \\
 & Virchow2 & 0.835 (0.738, 0.914) & 0.720 (0.645, 0.793) \\
 & CHIEF & 0.862 (0.778, 0.931) & 0.697 (0.617, 0.767) \\
 & GigaPath & 0.832 (0.741, 0.914) & 0.729 (0.656, 0.800) \\
 & PulmoFoundation & \textbf{0.890 (0.813, 0.953)} & \textbf{0.751 (0.682, 0.822)} \\
\midrule
\multirow{5}{*}{\textbf{Macro Sensitivity}} & UNI & 0.746 (0.658, 0.826) & 0.603 (0.556, 0.647) \\
 & Virchow2 & 0.620 (0.530, 0.706) & 0.555 (0.514, 0.591) \\
 & CHIEF & \textbf{0.829 (0.749, 0.900)} & 0.656 (0.589, 0.718) \\
 & GigaPath & 0.500 (0.472, 0.533) & 0.524 (0.497, 0.554) \\
 & PulmoFoundation & 0.808 (0.721, 0.885) & \textbf{0.668 (0.605, 0.730)} \\
\midrule
\multirow{5}{*}{\textbf{Macro Specificity}} & UNI & 0.746 (0.658, 0.826) & 0.603 (0.556, 0.647) \\
 & Virchow2 & 0.620 (0.530, 0.706) & 0.555 (0.514, 0.591) \\
 & CHIEF & \textbf{0.829 (0.749, 0.900)} & 0.656 (0.589, 0.718) \\
 & GigaPath & 0.500 (0.472, 0.533) & 0.524 (0.497, 0.554) \\
 & PulmoFoundation & 0.808 (0.721, 0.885) & \textbf{0.668 (0.605, 0.730)} \\
\midrule
\multirow{5}{*}{\textbf{Macro PPV}} & UNI & 0.777 (0.691, 0.857) & 0.560 (0.530, 0.587) \\
 & Virchow2 & 0.668 (0.544, 0.780) & 0.548 (0.512, 0.579) \\
 & CHIEF & \textbf{0.832 (0.752, 0.905)} & 0.575 (0.540, 0.610) \\
 & GigaPath & 0.459 (0.201, 0.782) & \textbf{0.732 (0.437, 0.941)} \\
 & PulmoFoundation & 0.811 (0.718, 0.886) & 0.582 (0.548, 0.617) \\
\midrule
\multirow{5}{*}{\textbf{Macro NPV}} & UNI & 0.777 (0.691, 0.857) & 0.560 (0.530, 0.587) \\
 & Virchow2 & 0.668 (0.544, 0.780) & 0.548 (0.512, 0.579) \\
 & CHIEF & \textbf{0.832 (0.752, 0.905)} & 0.575 (0.540, 0.610) \\
 & GigaPath & 0.459 (0.201, 0.782) & \textbf{0.732 (0.437, 0.941)} \\
 & PulmoFoundation & 0.811 (0.718, 0.886) & 0.582 (0.548, 0.617) \\
\midrule
\multirow{5}{*}{\textbf{Negative Sensitivity (Youden)}} & UNI & 0.907 (0.689, 0.977) & \textbf{0.796 (0.695, 0.843)} \\
 & Virchow2 & 0.907 (0.755, 0.977) & 0.591 (0.457, 0.899) \\
 & CHIEF & 0.884 (0.778, 0.979) & 0.689 (0.542, 0.740) \\
 & GigaPath & 0.721 (0.595, 0.951) & 0.658 (0.570, 0.856) \\
 & PulmoFoundation & \textbf{0.953 (0.643, 1.000)} & 0.754 (0.457, 0.847) \\
\midrule
\multirow{5}{*}{\textbf{Positive Sensitivity (Youden)}} & UNI & 0.733 (0.651, 0.935) & 0.632 (0.531, 0.811) \\
 & Virchow2 & 0.733 (0.612, 0.892) & \textbf{0.772 (0.462, 0.925)} \\
 & CHIEF & 0.778 (0.625, 0.913) & 0.667 (0.564, 0.866) \\
 & GigaPath & \textbf{0.867 (0.612, 0.976)} & 0.737 (0.520, 0.867) \\
 & PulmoFoundation & 0.711 (0.625, 1.000) & 0.667 (0.547, 0.908) \\
\bottomrule
\end{tabular}
\end{adjustbox}
\end{table}

\clearpage

\begin{table}[htbp]
\centering
\caption{\textbf{STK11}: Per-task performance of PulmoFoundation and four baseline foundation models (UNI, Virchow2, CHIEF, GigaPath). Performance metrics are reported with 95\% bootstrap percentile CIs. Macro AUC is threshold-free. Macro Sensitivity, Specificity, PPV, and NPV are macro-averaged one-versus-rest metrics computed from predicted class labels; per-class sensitivity rows are computed at class-specific Youden-optimal operating points. CIs are bootstrap percentiles over cases ($N{=}1{,}000$ replicates). Best-performing model per metric and cohort is shown in bold. Cohort-composition rows report case counts, with slide counts in parentheses; class percentages are case-level percentages. All metrics are computed on the same cohort.}
\label{tab:post_STK11}
\renewcommand{\arraystretch}{1.15}
\setlength{\tabcolsep}{5pt}
\small
\begin{adjustbox}{max width=\textwidth,center}
\begin{tabular}{l l c}
\toprule
\textbf{Metric} & \textbf{Model} & \textbf{Internal} \\
\midrule
Total cases (slides) & -- & 83 (89) \\
Negative & -- & 67 (73; 80.7\%) \\
Positive & -- & 16 (16; 19.3\%) \\
\midrule
\multirow{5}{*}{\textbf{Macro AUC}} & UNI & 0.782 (0.649, 0.896) \\
 & Virchow2 & 0.803 (0.702, 0.892) \\
 & CHIEF & 0.768 (0.627, 0.902) \\
 & GigaPath & 0.813 (0.699, 0.911) \\
 & PulmoFoundation & \textbf{0.830 (0.723, 0.920)} \\
\midrule
\multirow{5}{*}{\textbf{Macro Sensitivity}} & UNI & 0.613 (0.483, 0.743) \\
 & Virchow2 & 0.588 (0.468, 0.723) \\
 & CHIEF & 0.541 (0.468, 0.643) \\
 & GigaPath & \textbf{0.698 (0.564, 0.827)} \\
 & PulmoFoundation & 0.554 (0.463, 0.664) \\
\midrule
\multirow{5}{*}{\textbf{Macro Specificity}} & UNI & 0.613 (0.483, 0.743) \\
 & Virchow2 & 0.588 (0.468, 0.723) \\
 & CHIEF & 0.541 (0.468, 0.643) \\
 & GigaPath & \textbf{0.698 (0.564, 0.827)} \\
 & PulmoFoundation & 0.554 (0.463, 0.664) \\
\midrule
\multirow{5}{*}{\textbf{Macro PPV}} & UNI & 0.614 (0.482, 0.743) \\
 & Virchow2 & 0.598 (0.467, 0.739) \\
 & CHIEF & 0.609 (0.394, 0.899) \\
 & GigaPath & \textbf{0.708 (0.565, 0.854)} \\
 & PulmoFoundation & 0.596 (0.415, 0.778) \\
\midrule
\multirow{5}{*}{\textbf{Macro NPV}} & UNI & 0.614 (0.482, 0.743) \\
 & Virchow2 & 0.598 (0.467, 0.739) \\
 & CHIEF & 0.609 (0.394, 0.899) \\
 & GigaPath & \textbf{0.708 (0.565, 0.854)} \\
 & PulmoFoundation & 0.596 (0.415, 0.778) \\
\midrule
\multirow{5}{*}{\textbf{Negative Sensitivity (Youden)}} & UNI & 0.776 (0.534, 0.905) \\
 & Virchow2 & 0.537 (0.463, 0.875) \\
 & CHIEF & \textbf{0.836 (0.746, 0.919)} \\
 & GigaPath & 0.612 (0.461, 0.923) \\
 & PulmoFoundation & 0.746 (0.479, 0.921) \\
\midrule
\multirow{5}{*}{\textbf{Positive Sensitivity (Youden)}} & UNI & 0.812 (0.600, 1.000) \\
 & Virchow2 & \textbf{1.000 (0.722, 1.000)} \\
 & CHIEF & 0.812 (0.600, 1.000) \\
 & GigaPath & 0.938 (0.643, 1.000) \\
 & PulmoFoundation & 0.812 (0.667, 1.000) \\
\bottomrule
\end{tabular}
\end{adjustbox}
\end{table}

\clearpage

\begin{table}[htbp]
\centering
\caption{\textbf{Lung adenocarcinoma overall survival}: Per-task performance of PulmoFoundation and four baseline foundation models (UNI, Virchow2, CHIEF, GigaPath). C-Index values are reported as mean with 95\% bootstrap CIs. Best-performing model per cohort is shown in bold. Event counts (deaths for overall survival; recurrences for disease-free survival) and median follow-up times in months with interquartile range (IQR) are reported at the top of the table.}
\label{tab:post_survival_luad}
\renewcommand{\arraystretch}{1.15}
\setlength{\tabcolsep}{5pt}
\begin{adjustbox}{max width=\textwidth,center}
\begin{tabular}{l l c c}
\toprule
\textbf{Metric} & \textbf{Model} & \textbf{Internal} & \textbf{External H1$^\dagger$} \\
\midrule
Events / Total $N$ & -- & 48 / 137 & 13 / 303 \\
Median follow-up, months (IQR) & -- & 23.5 (15.2, 37.7) & 61.0 (22.5, 73.0) \\
\midrule
\multirow{5}{*}{\textbf{C-Index}} & UNI & 0.613 (0.516, 0.709) & 0.695 (0.518, 0.850) \\
 & Virchow2 & 0.636 (0.548, 0.723) & 0.727 (0.592, 0.831) \\
 & CHIEF & 0.604 (0.511, 0.697) & 0.714 (0.583, 0.829) \\
 & GigaPath & 0.601 (0.489, 0.706) & 0.608 (0.478, 0.752) \\
 & PulmoFoundation & \textbf{0.651 (0.563, 0.739)} & \textbf{0.790 (0.691, 0.878)} \\
\bottomrule
\end{tabular}
\end{adjustbox}
\end{table}

\begin{table}[htbp]
\centering
\caption{\textbf{Lung squamous cell carcinoma overall survival}: Per-task performance of PulmoFoundation and four baseline foundation models (UNI, Virchow2, CHIEF, GigaPath). C-Index values are reported as mean with 95\% bootstrap CIs. Best-performing model per cohort is shown in bold. Event counts (deaths for overall survival; recurrences for disease-free survival) and median follow-up times in months with interquartile range (IQR) are reported at the top of the table.}
\label{tab:post_survival_lusc}
\renewcommand{\arraystretch}{1.15}
\setlength{\tabcolsep}{5pt}
\begin{adjustbox}{max width=\textwidth,center}
\begin{tabular}{l l c c}
\toprule
\textbf{Metric} & \textbf{Model} & \textbf{Internal} & \textbf{External H1$^\dagger$} \\
\midrule
Events / Total $N$ & -- & 57 / 136 & 2 / 73 \\
Median follow-up, months (IQR) & -- & 23.6 (12.5, 42.4) & 17.0 (8.0, 46.0) \\
\midrule
\multirow{5}{*}{\textbf{C-Index}} & UNI & \textbf{0.613 (0.536, 0.691)} & 0.380 (0.000, 0.857) \\
 & Virchow2 & 0.591 (0.499, 0.682) & 0.714 (0.470, 0.963) \\
 & CHIEF & 0.579 (0.490, 0.668) & 0.493 (0.267, 0.750) \\
 & GigaPath & 0.575 (0.495, 0.655) & 0.349 (0.162, 0.581) \\
 & PulmoFoundation & 0.612 (0.529, 0.695) & \textbf{0.766 (0.620, 0.897)} \\
\bottomrule
\end{tabular}
\end{adjustbox}
\end{table}

\begin{table}[htbp]
\centering
\caption{\textbf{Lung cancer disease-free survival (H1)}: Per-task performance of PulmoFoundation and four baseline foundation models (UNI, Virchow2, CHIEF, GigaPath). C-Index values are reported as mean with 95\% bootstrap CIs. Best-performing model per cohort is shown in bold. Event counts (deaths for overall survival; recurrences for disease-free survival) and median follow-up times in months with interquartile range (IQR) are reported at the top of the table.}
\label{tab:post_dfs}
\renewcommand{\arraystretch}{1.15}
\setlength{\tabcolsep}{5pt}
\begin{adjustbox}{max width=\textwidth,center}
\begin{tabular}{l l c}
\toprule
\textbf{Metric} & \textbf{Model} & \textbf{Internal} \\
\midrule
Events / Total $N$ & -- & 20 / 111 \\
Median follow-up, months (IQR) & -- & 46.0 (8.5, 63.0) \\
\midrule
\multirow{5}{*}{\textbf{C-Index}} & UNI & 0.749 (0.655, 0.844) \\
 & Virchow2 & 0.753 (0.656, 0.850) \\
 & CHIEF & 0.720 (0.611, 0.828) \\
 & GigaPath & 0.711 (0.569, 0.834) \\
 & PulmoFoundation & \textbf{0.755 (0.649, 0.860)} \\
\bottomrule
\end{tabular}
\end{adjustbox}
\end{table}

\begin{table}[htbp]
\centering
\caption{\textbf{Benign vs Malignant (Biopsy) -- Prospective observational cohort (PulmoFoundation).} Metrics were re-computed from predictions logged at the time of the observational prospective validation, using the unified evaluation pipeline applied throughout this revision (95\% bootstrap percentile CI, $N{=}1{,}000$ replicates over cases; operating-point metrics at the class-specific Youden-optimal threshold). Baselines were not deployed in the prospective workflow and are therefore reported only for the retrospective comparison (see companion baseline table). Cohort-composition rows report case counts, with slide counts in parentheses; class percentages are case-level percentages.}
\label{tab:cancer_vs_benign_biopsy_prospective}
\renewcommand{\arraystretch}{1.05}
\setlength{\tabcolsep}{5pt}
\begin{tabular}{l c}
\toprule
\textbf{Metric} & \textbf{PulmoFoundation} \\
\midrule
\multicolumn{2}{l}{\textit{Cohort composition}} \\
\midrule
Total cases (slides) & 250 (250) \\
Benign & 71 (71; 28.4\%) \\
Malignant & 179 (179; 71.6\%) \\
\midrule
\multicolumn{2}{l}{\textit{Discrimination and macro-averaged performance}} \\
\midrule
Macro AUC & 0.992 (0.978, 1.000) \\
Macro Sensitivity & 0.986 (0.972, 0.997) \\
Macro Specificity & 0.986 (0.972, 0.997) \\
Macro PPV & 0.967 (0.937, 0.993) \\
Macro NPV & 0.967 (0.937, 0.993) \\
\midrule
\multicolumn{2}{l}{\textit{Per-class sensitivity at Youden-optimal threshold}} \\
\midrule
Benign Sensitivity (Youden) & 1.000 (1.000, 1.000) \\
Malignant Sensitivity (Youden) & 0.972 (0.944, 0.995) \\
\bottomrule
\end{tabular}
\end{table}

\begin{table}[htbp]
\centering
\caption{\textbf{Benign vs Malignant (Frozen) -- Prospective observational cohort (PulmoFoundation).} Metrics were re-computed from predictions logged at the time of the observational prospective validation, using the unified evaluation pipeline applied throughout this revision (95\% bootstrap percentile CI, $N{=}1{,}000$ replicates over cases; operating-point metrics at the class-specific Youden-optimal threshold). Baselines were not deployed in the prospective workflow and are therefore reported only for the retrospective comparison (see companion baseline table). Cohort-composition rows report case counts, with slide counts in parentheses; class percentages are case-level percentages.}
\label{tab:cancer_vs_benign_frozen_prospective}
\renewcommand{\arraystretch}{1.05}
\setlength{\tabcolsep}{5pt}
\begin{tabular}{l c}
\toprule
\textbf{Metric} & \textbf{PulmoFoundation} \\
\midrule
\multicolumn{2}{l}{\textit{Cohort composition}} \\
\midrule
Total cases (slides) & 271 (380) \\
Benign & 31 (44; 11.4\%) \\
Malignant & 240 (336; 88.6\%) \\
\midrule
\multicolumn{2}{l}{\textit{Discrimination and macro-averaged performance}} \\
\midrule
Macro AUC & 0.970 (0.940, 0.993) \\
Macro Sensitivity & 0.939 (0.889, 0.981) \\
Macro Specificity & 0.939 (0.889, 0.981) \\
Macro PPV & 0.833 (0.768, 0.936) \\
Macro NPV & 0.833 (0.768, 0.936) \\
\midrule
\multicolumn{2}{l}{\textit{Per-class sensitivity at Youden-optimal threshold}} \\
\midrule
Benign Sensitivity (Youden) & 0.935 (0.839, 1.000) \\
Malignant Sensitivity (Youden) & 0.942 (0.916, 0.984) \\
\bottomrule
\end{tabular}
\end{table}

\begin{table}[htbp]
\centering
\caption{\textbf{Coarse-grained Subtyping (Resection) -- Prospective observational cohort (PulmoFoundation).} Metrics were re-computed from predictions logged at the time of the observational prospective validation, using the unified evaluation pipeline applied throughout this revision (95\% bootstrap percentile CI, $N{=}1{,}000$ replicates over cases; operating-point metrics at the class-specific Youden-optimal threshold). Baselines were not deployed in the prospective workflow and are therefore reported only for the retrospective comparison (see companion baseline table). Cohort-composition rows report case counts, with slide counts in parentheses; class percentages are case-level percentages.}
\label{tab:nanfang_lung_3class_prospective}
\renewcommand{\arraystretch}{1.05}
\setlength{\tabcolsep}{5pt}
\begin{tabular}{l c}
\toprule
\textbf{Metric} & \textbf{PulmoFoundation} \\
\midrule
\multicolumn{2}{l}{\textit{Cohort composition}} \\
\midrule
Total cases (slides) & 397 (1673) \\
Adenocarcinoma in situ & 127 (367; 32.0\%) \\
Invasive adenocarcinoma & 200 (906; 50.4\%) \\
Squamous cell carcinoma & 70 (400; 17.6\%) \\
\midrule
\multicolumn{2}{l}{\textit{Discrimination and macro-averaged performance}} \\
\midrule
Macro AUC & 0.883 (0.854, 0.908) \\
Macro Sensitivity & 0.832 (0.800, 0.891) \\
Macro Specificity & 0.842 (0.776, 0.886) \\
Macro PPV & 0.751 (0.683, 0.804) \\
Macro NPV & 0.881 (0.862, 0.913) \\
\midrule
\multicolumn{2}{l}{\textit{Per-class sensitivity at Youden-optimal threshold}} \\
\midrule
Adenocarcinoma in situ Sensitivity (Youden) & 0.882 (0.750, 0.953) \\
Invasive adenocarcinoma Sensitivity (Youden) & 0.685 (0.626, 0.819) \\
Squamous cell carcinoma Sensitivity (Youden) & 0.929 (0.892, 1.000) \\
\bottomrule
\end{tabular}
\end{table}

\begin{table}[htbp]
\centering
\caption{\textbf{NSCLC Subtyping (Resection) -- Prospective observational cohort (PulmoFoundation).} Metrics were re-computed from predictions logged at the time of the observational prospective validation, using the unified evaluation pipeline applied throughout this revision (95\% bootstrap percentile CI, $N{=}1{,}000$ replicates over cases; operating-point metrics at the class-specific Youden-optimal threshold). Baselines were not deployed in the prospective workflow and are therefore reported only for the retrospective comparison (see companion baseline table). Cohort-composition rows report case counts, with slide counts in parentheses; class percentages are case-level percentages.}
\label{tab:nanfang_lung_nsclc_prospective}
\renewcommand{\arraystretch}{1.05}
\setlength{\tabcolsep}{5pt}
\begin{tabular}{l c}
\toprule
\textbf{Metric} & \textbf{PulmoFoundation} \\
\midrule
\multicolumn{2}{l}{\textit{Cohort composition}} \\
\midrule
Total cases (slides) & 270 (1143) \\
LUAD & 200 (743; 74.1\%) \\
LUSC & 70 (400; 25.9\%) \\
\midrule
\multicolumn{2}{l}{\textit{Discrimination and macro-averaged performance}} \\
\midrule
Macro AUC & 0.993 (0.986, 0.998) \\
Macro Sensitivity & 0.968 (0.953, 0.988) \\
Macro Specificity & 0.968 (0.953, 0.988) \\
Macro PPV & 0.934 (0.896, 0.979) \\
Macro NPV & 0.934 (0.896, 0.979) \\
\midrule
\multicolumn{2}{l}{\textit{Per-class sensitivity at Youden-optimal threshold}} \\
\midrule
LUAD Sensitivity (Youden) & 0.950 (0.913, 0.986) \\
LUSC Sensitivity (Youden) & 0.986 (0.953, 1.000) \\
\bottomrule
\end{tabular}
\end{table}

\begin{table}[htbp]
\centering
\caption{\textbf{Fine-grained Subtyping (Resection) -- Prospective observational cohort (PulmoFoundation).} Metrics were re-computed from predictions logged at the time of the observational prospective validation, using the unified evaluation pipeline applied throughout this revision (95\% bootstrap percentile CI, $N{=}1{,}000$ replicates over cases; operating-point metrics at the class-specific Youden-optimal threshold). Baselines were not deployed in the prospective workflow and are therefore reported only for the retrospective comparison (see companion baseline table). Cohort-composition rows report case counts, with slide counts in parentheses; class percentages are case-level percentages.}
\label{tab:nanfang_lung_finegrained_prospective}
\renewcommand{\arraystretch}{1.05}
\setlength{\tabcolsep}{5pt}
\begin{tabular}{l c}
\toprule
\textbf{Metric} & \textbf{PulmoFoundation} \\
\midrule
\multicolumn{2}{l}{\textit{Cohort composition}} \\
\midrule
Total cases (slides) & 509 (1802) \\
Adenocarcinoma in situ & 127 (367; 25.0\%) \\
Minimally invasive adenocarcinoma & 150 (331; 29.5\%) \\
Invasive adenocarcinoma & 150 (647; 29.5\%) \\
Neuroendocrine carcinoma & 5 (23; 1.0\%) \\
Adenosquamous carcinoma & 7 (34; 1.4\%) \\
Squamous cell carcinoma & 70 (400; 13.8\%) \\
\midrule
\multicolumn{2}{l}{\textit{Discrimination and macro-averaged performance}} \\
\midrule
Macro AUC & 0.868 (0.843, 0.890) \\
Macro Sensitivity & 0.841 (0.831, 0.916) \\
Macro Specificity & 0.817 (0.756, 0.851) \\
Macro PPV & 0.455 (0.401, 0.627) \\
Macro NPV & 0.951 (0.939, 0.963) \\
\midrule
\multicolumn{2}{l}{\textit{Per-class sensitivity at Youden-optimal threshold}} \\
\midrule
Adenocarcinoma in situ Sensitivity (Youden) & 0.827 (0.694, 0.930) \\
Minimally invasive adenocarcinoma Sensitivity (Youden) & 0.920 (0.837, 0.967) \\
Invasive adenocarcinoma Sensitivity (Youden) & 0.687 (0.617, 0.780) \\
Neuroendocrine carcinoma Sensitivity (Youden) & 1.000 (1.000, 1.000) \\
Adenosquamous carcinoma Sensitivity (Youden) & 0.714 (0.714, 1.000) \\
Squamous cell carcinoma Sensitivity (Youden) & 0.900 (0.847, 0.977) \\
\bottomrule
\end{tabular}
\end{table}

\begin{table}[htbp]
\centering
\caption{\textbf{LN Metastasis Prediction from H\&E Slides (Resection) -- Prospective observational cohort (PulmoFoundation).} Metrics were re-computed from predictions logged at the time of the observational prospective validation, using the unified evaluation pipeline applied throughout this revision (95\% bootstrap percentile CI, $N{=}1{,}000$ replicates over cases; operating-point metrics at the class-specific Youden-optimal threshold). Baselines were not deployed in the prospective workflow and are therefore reported only for the retrospective comparison (see companion baseline table). Cohort-composition rows report case counts, with slide counts in parentheses; class percentages are case-level percentages.}
\label{tab:lymphnodemetastasis_merged_prospective}
\renewcommand{\arraystretch}{1.05}
\setlength{\tabcolsep}{5pt}
\begin{tabular}{l c}
\toprule
\textbf{Metric} & \textbf{PulmoFoundation} \\
\midrule
\multicolumn{2}{l}{\textit{Cohort composition}} \\
\midrule
Total cases (slides) & 386 (1770) \\
Negative & 300 (1269; 77.7\%) \\
Positive & 86 (501; 22.3\%) \\
\midrule
\multicolumn{2}{l}{\textit{Discrimination and macro-averaged performance}} \\
\midrule
Macro AUC & 0.849 (0.803, 0.892) \\
Macro Sensitivity & 0.795 (0.760, 0.841) \\
Macro Specificity & 0.795 (0.760, 0.841) \\
Macro PPV & 0.707 (0.676, 0.770) \\
Macro NPV & 0.707 (0.676, 0.770) \\
\midrule
\multicolumn{2}{l}{\textit{Per-class sensitivity at Youden-optimal threshold}} \\
\midrule
Negative Sensitivity (Youden) & 0.683 (0.646, 0.837) \\
Positive Sensitivity (Youden) & 0.907 (0.759, 0.955) \\
\bottomrule
\end{tabular}
\end{table}

\begin{table}[htbp]
\centering
\caption{\textbf{CK7 -- Prospective observational cohort (PulmoFoundation).} Metrics were re-computed from predictions logged at the time of the observational prospective validation, using the unified evaluation pipeline applied throughout this revision (95\% bootstrap percentile CI, $N{=}1{,}000$ replicates over cases; operating-point metrics at the class-specific Youden-optimal threshold). Baselines were not deployed in the prospective workflow and are therefore reported only for the retrospective comparison (see companion baseline table). Cohort-composition rows report case counts, with slide counts in parentheses; class percentages are case-level percentages.}
\label{tab:nanfang_lung_ck7_prospective}
\renewcommand{\arraystretch}{1.05}
\setlength{\tabcolsep}{5pt}
\begin{tabular}{l c}
\toprule
\textbf{Metric} & \textbf{PulmoFoundation} \\
\midrule
\multicolumn{2}{l}{\textit{Cohort composition}} \\
\midrule
Total cases (slides) & 128 (614) \\
Negative & 23 (131; 18.0\%) \\
Positive & 105 (483; 82.0\%) \\
\midrule
\multicolumn{2}{l}{\textit{Discrimination and macro-averaged performance}} \\
\midrule
Macro AUC & 0.921 (0.868, 0.963) \\
Macro Sensitivity & 0.885 (0.830, 0.950) \\
Macro Specificity & 0.885 (0.830, 0.950) \\
Macro PPV & 0.781 (0.692, 0.870) \\
Macro NPV & 0.781 (0.692, 0.870) \\
\midrule
\multicolumn{2}{l}{\textit{Per-class sensitivity at Youden-optimal threshold}} \\
\midrule
Negative Sensitivity (Youden) & 0.913 (0.833, 1.000) \\
Positive Sensitivity (Youden) & 0.857 (0.724, 0.926) \\
\bottomrule
\end{tabular}
\end{table}

\begin{table}[htbp]
\centering
\caption{\textbf{Napsin A -- Prospective observational cohort (PulmoFoundation).} Metrics were re-computed from predictions logged at the time of the observational prospective validation, using the unified evaluation pipeline applied throughout this revision (95\% bootstrap percentile CI, $N{=}1{,}000$ replicates over cases; operating-point metrics at the class-specific Youden-optimal threshold). Baselines were not deployed in the prospective workflow and are therefore reported only for the retrospective comparison (see companion baseline table). Cohort-composition rows report case counts, with slide counts in parentheses; class percentages are case-level percentages.}
\label{tab:nanfang_lung_napsina_prospective}
\renewcommand{\arraystretch}{1.05}
\setlength{\tabcolsep}{5pt}
\begin{tabular}{l c}
\toprule
\textbf{Metric} & \textbf{PulmoFoundation} \\
\midrule
\multicolumn{2}{l}{\textit{Cohort composition}} \\
\midrule
Total cases (slides) & 142 (700) \\
Negative & 68 (371; 47.9\%) \\
Positive & 74 (329; 52.1\%) \\
\midrule
\multicolumn{2}{l}{\textit{Discrimination and macro-averaged performance}} \\
\midrule
Macro AUC & 0.922 (0.876, 0.960) \\
Macro Sensitivity & 0.874 (0.826, 0.929) \\
Macro Specificity & 0.874 (0.829, 0.930) \\
Macro PPV & 0.873 (0.831, 0.930) \\
Macro NPV & 0.873 (0.831, 0.929) \\
\midrule
\multicolumn{2}{l}{\textit{Per-class sensitivity at Youden-optimal threshold}} \\
\midrule
Negative Sensitivity (Youden) & 0.882 (0.829, 0.985) \\
Positive Sensitivity (Youden) & 0.865 (0.734, 0.936) \\
\bottomrule
\end{tabular}
\end{table}

\begin{table}[htbp]
\centering
\caption{\textbf{P40 -- Prospective observational cohort (PulmoFoundation).} Metrics were re-computed from predictions logged at the time of the observational prospective validation, using the unified evaluation pipeline applied throughout this revision (95\% bootstrap percentile CI, $N{=}1{,}000$ replicates over cases; operating-point metrics at the class-specific Youden-optimal threshold). Baselines were not deployed in the prospective workflow and are therefore reported only for the retrospective comparison (see companion baseline table). Cohort-composition rows report case counts, with slide counts in parentheses; class percentages are case-level percentages.}
\label{tab:nanfang_lung_p40_prospective}
\renewcommand{\arraystretch}{1.05}
\setlength{\tabcolsep}{5pt}
\begin{tabular}{l c}
\toprule
\textbf{Metric} & \textbf{PulmoFoundation} \\
\midrule
\multicolumn{2}{l}{\textit{Cohort composition}} \\
\midrule
Total cases (slides) & 152 (737) \\
Negative & 106 (514; 69.7\%) \\
Positive & 46 (223; 30.3\%) \\
\midrule
\multicolumn{2}{l}{\textit{Discrimination and macro-averaged performance}} \\
\midrule
Macro AUC & 0.921 (0.867, 0.964) \\
Macro Sensitivity & 0.862 (0.812, 0.920) \\
Macro Specificity & 0.862 (0.813, 0.920) \\
Macro PPV & 0.816 (0.768, 0.911) \\
Macro NPV & 0.816 (0.768, 0.911) \\
\midrule
\multicolumn{2}{l}{\textit{Per-class sensitivity at Youden-optimal threshold}} \\
\midrule
Negative Sensitivity (Youden) & 0.811 (0.750, 0.963) \\
Positive Sensitivity (Youden) & 0.913 (0.740, 0.980) \\
\bottomrule
\end{tabular}
\end{table}

\begin{table}[htbp]
\centering
\caption{\textbf{P63 -- Prospective observational cohort (PulmoFoundation).} Metrics were re-computed from predictions logged at the time of the observational prospective validation, using the unified evaluation pipeline applied throughout this revision (95\% bootstrap percentile CI, $N{=}1{,}000$ replicates over cases; operating-point metrics at the class-specific Youden-optimal threshold). Baselines were not deployed in the prospective workflow and are therefore reported only for the retrospective comparison (see companion baseline table). Cohort-composition rows report case counts, with slide counts in parentheses; class percentages are case-level percentages.}
\label{tab:nanfang_lung_p63_prospective}
\renewcommand{\arraystretch}{1.05}
\setlength{\tabcolsep}{5pt}
\begin{tabular}{l c}
\toprule
\textbf{Metric} & \textbf{PulmoFoundation} \\
\midrule
\multicolumn{2}{l}{\textit{Cohort composition}} \\
\midrule
Total cases (slides) & 58 (297) \\
Negative & 24 (126; 41.4\%) \\
Positive & 34 (171; 58.6\%) \\
\midrule
\multicolumn{2}{l}{\textit{Discrimination and macro-averaged performance}} \\
\midrule
Macro AUC & 0.944 (0.871, 0.989) \\
Macro Sensitivity & 0.902 (0.824, 0.977) \\
Macro Specificity & 0.902 (0.825, 0.977) \\
Macro PPV & 0.922 (0.829, 0.980) \\
Macro NPV & 0.922 (0.829, 0.980) \\
\midrule
\multicolumn{2}{l}{\textit{Per-class sensitivity at Youden-optimal threshold}} \\
\midrule
Negative Sensitivity (Youden) & 0.833 (0.696, 1.000) \\
Positive Sensitivity (Youden) & 0.971 (0.700, 1.000) \\
\bottomrule
\end{tabular}
\end{table}

\begin{table}[htbp]
\centering
\caption{\textbf{TTF-1 -- Prospective observational cohort (PulmoFoundation).} Metrics were re-computed from predictions logged at the time of the observational prospective validation, using the unified evaluation pipeline applied throughout this revision (95\% bootstrap percentile CI, $N{=}1{,}000$ replicates over cases; operating-point metrics at the class-specific Youden-optimal threshold). Baselines were not deployed in the prospective workflow and are therefore reported only for the retrospective comparison (see companion baseline table). Cohort-composition rows report case counts, with slide counts in parentheses; class percentages are case-level percentages.}
\label{tab:nanfang_lung_ttf1_prospective}
\renewcommand{\arraystretch}{1.05}
\setlength{\tabcolsep}{5pt}
\begin{tabular}{l c}
\toprule
\textbf{Metric} & \textbf{PulmoFoundation} \\
\midrule
\multicolumn{2}{l}{\textit{Cohort composition}} \\
\midrule
Total cases (slides) & 246 (1056) \\
Negative & 60 (317; 24.4\%) \\
Positive & 186 (739; 75.6\%) \\
\midrule
\multicolumn{2}{l}{\textit{Discrimination and macro-averaged performance}} \\
\midrule
Macro AUC & 0.892 (0.853, 0.931) \\
Macro Sensitivity & 0.816 (0.780, 0.867) \\
Macro Specificity & 0.816 (0.780, 0.867) \\
Macro PPV & 0.737 (0.697, 0.836) \\
Macro NPV & 0.737 (0.697, 0.836) \\
\midrule
\multicolumn{2}{l}{\textit{Per-class sensitivity at Youden-optimal threshold}} \\
\midrule
Negative Sensitivity (Youden) & 0.917 (0.706, 0.982) \\
Positive Sensitivity (Youden) & 0.715 (0.648, 0.920) \\
\bottomrule
\end{tabular}
\end{table}

\begin{table}[htbp]
\centering
\caption{\textbf{Failure-mode analysis for prospective lymph-node-metastasis prediction from primary-tumor H\&E slides.} The analysis included 386 consecutive Center-H1 surgical-resection cases. N0 and N+ reference labels were established by pathological examination of separate regional lymph-node specimens. Case-level predictions were evaluated at the Youden-optimal operating point used in the prospective performance table. FP among N0 reports false-positive node-positive predictions among pathology-confirmed node-negative cases. FN among N+ reports false-negative node-negative predictions among node-positive cases. Histologic subtype was obtained from the structured subtype field or recovered from the final diagnostic text when the structured field was unavailable. Other specified carcinoma includes SMARCA4-deficient non-small-cell lung cancer, mucoepidermoid carcinoma, invasive carcinoma not otherwise specified, and pleomorphic or sarcomatoid carcinoma. Histologic subtype unavailable denotes cases without an assignable subtype in either report source. ``--'' indicates that the denominator was zero. Fractions for small subgroups are descriptive. N+ burden rows include node-positive cases with extractable positive-node counts in the pathology report. AIS, adenocarcinoma in situ; MIA, minimally invasive adenocarcinoma; IASLC, International Association for the Study of Lung Cancer; STAS, spread through air spaces; LN, lymph node.}
\label{tab:prospective_lnm_failure_modes}
\renewcommand{\arraystretch}{1.03}
\setlength{\tabcolsep}{3pt}
\small
\begin{tabular}{p{0.18\linewidth} p{0.27\linewidth} r r r r r}
\toprule
\textbf{Feature} & \textbf{Category} & \textbf{Cases} & \textbf{N0} & \textbf{N+} & \textbf{FP among N0} & \textbf{FN among N+} \\
\midrule
\multicolumn{7}{l}{\textit{Overall}} \\
\midrule
Overall & Overall & 386 & 300 & 86 & 95/300 & 8/86 \\
\multicolumn{7}{l}{\textit{Histologic subtype}} \\
\midrule
Histologic subtype & Adenocarcinoma in situ & 46 & 46 & 0 & 2/46 & -- \\
Histologic subtype & Minimally invasive adenocarcinoma & 62 & 61 & 1 & 2/61 & 0/1 \\
Histologic subtype & Invasive adenocarcinoma & 212 & 151 & 61 & 62/151 & 5/61 \\
Histologic subtype & Squamous cell carcinoma & 40 & 28 & 12 & 18/28 & 1/12 \\
Histologic subtype & Neuroendocrine neoplasm & 13 & 7 & 6 & 5/7 & 0/6 \\
Histologic subtype & Adenosquamous carcinoma & 2 & 0 & 2 & -- & 0/2 \\
Histologic subtype & Other specified carcinoma & 7 & 4 & 3 & 4/4 & 1/3 \\
Histologic subtype & Histologic subtype unavailable & 4 & 3 & 1 & 2/3 & 1/1 \\
\multicolumn{7}{l}{\textit{Node-positive burden}} \\
\midrule
Node-positive burden & 1 positive LN & 35 & 0 & 35 & -- & 4/35 \\
Node-positive burden & $\geq$2 positive LNs & 49 & 0 & 49 & -- & 4/49 \\
\multicolumn{7}{l}{\textit{Tumor size}} \\
\midrule
Tumor size & $\leq$1 cm & 120 & 111 & 9 & 11/111 & 2/9 \\
Tumor size & $>$1-2 cm & 162 & 130 & 32 & 46/130 & 3/32 \\
Tumor size & $>$2-3 cm & 56 & 34 & 22 & 20/34 & 1/22 \\
Tumor size & $>$3 cm & 44 & 22 & 22 & 17/22 & 2/22 \\
Tumor size & Unknown & 4 & 3 & 1 & 1/3 & 0/1 \\
\multicolumn{7}{l}{\textit{IASLC grade}} \\
\midrule
IASLC grade & G1 & 32 & 31 & 1 & 4/31 & 0/1 \\
IASLC grade & G2 & 81 & 75 & 6 & 24/75 & 3/6 \\
IASLC grade & G3 & 102 & 49 & 53 & 35/49 & 1/53 \\
IASLC grade & Unknown & 171 & 145 & 26 & 32/145 & 4/26 \\
\multicolumn{7}{l}{\textit{Lymphovascular invasion}} \\
\midrule
Lymphovascular invasion & Absent & 243 & 188 & 55 & 79/188 & 7/55 \\
Lymphovascular invasion & Present & 46 & 16 & 30 & 14/16 & 1/30 \\
Lymphovascular invasion & Unknown & 97 & 96 & 1 & 2/96 & 0/1 \\
\multicolumn{7}{l}{\textit{STAS}} \\
\midrule
STAS & Absent & 103 & 82 & 21 & 33/82 & 4/21 \\
STAS & Present & 184 & 120 & 64 & 59/120 & 4/64 \\
STAS & Unknown & 99 & 98 & 1 & 3/98 & 0/1 \\
\multicolumn{7}{l}{\textit{Pleural invasion}} \\
\midrule
Pleural invasion & Absent & 228 & 176 & 52 & 72/176 & 8/52 \\
Pleural invasion & Present & 57 & 26 & 31 & 21/26 & 0/31 \\
Pleural invasion & Unknown & 101 & 98 & 3 & 2/98 & 0/3 \\
\bottomrule
\end{tabular}
\end{table}

\begin{table}[htbp]
\centering
\caption{\textbf{Pre-specified triage operating points across three prospective workflows.} Locked operating points for the three triage workflows defined a priori on the prospective cohort: diagnostic biopsy (Workflow A), intra-operative frozen section (Workflow B), and the five-marker NSCLC IHC panel (Workflow C; per-marker rows plus pooled aggregate). Per-workflow safety bars: PPV $\geq$ 0.99 for binary diagnostic decisions (Workflows A, B); PPV $\geq$ 0.95 for each IHC marker before pooling (Workflow C). Threshold is the score cutoff at which a case or stain order is deferred under the one-tailed defer-positive scheme: cases or stain orders with predicted positive-class probability $\geq$ threshold are routed to the workflow-specific fast-track action; remaining cases or stain orders proceed under the conventional pathologist workup. Per-task thresholds are not hand-picked: for each task, the algorithm sweeps the score cutoff and selects the lowest cutoff at which the deferred set satisfies the pre-specified PPV floor, which maximizes the deferral rate subject to the safety bar. Per-task rows report cases deferred and PPV with 95\% bootstrap percentile CI (1{,}000 non-stratified resamples per task, paired by replicate index). The IHC pooled row reports point estimates computed from summed deferred-order and true-positive counts across the five markers; no pooled bootstrap CI is reported. Sensitivity in this table denotes the fraction of reference-positive cases included in the deferral-eligible subset, not the diagnostic sensitivity of PulmoFoundation or the combined workflow. A lower value therefore indicates lower triage coverage rather than poorer diagnostic performance, because all cases not selected for deferral remain in the conventional pathologist workflow. PPV is the safety metric used to constrain deferral. The IHC pooled row uses target-PPV parameterization: per-marker thresholds achieving PPV $\geq$ 0.95 are selected and the resulting deferred orders are pooled across markers, with the pooled PPV computed from the summed true-positive and deferred counts across the five markers. $N$, total cases evaluated; $N_{+}$, positive cases; $N_{-}$, negative cases.}
\label{tab:triage_supp}
\renewcommand{\arraystretch}{1.1}
\setlength{\tabcolsep}{4pt}
\small
\begin{tabular}{l c c c c c c c c}
\toprule
\textbf{Task} & \textbf{Thr.} & \textbf{\% Deferred (95\% CI)} & \textbf{PPV (95\% CI)} & \textbf{Sens.} & \textbf{Spec. retained} & \textbf{$N$} & \textbf{$N_{+}$} & \textbf{$N_{-}$} \\
\midrule
\multicolumn{9}{l}{\textit{Workflow A -- Diagnostic biopsy (PPV target $\geq$ 0.99)}} \\
\midrule
Diagnostic biopsy (B vs M) & 0.50 & 68.8 (62.4, 74.0) & 1.000 (1.000, 1.000) & 0.961 & 1.000 & 250 & 179 & 71 \\
\midrule
\multicolumn{9}{l}{\textit{Workflow B -- Intra-operative frozen section (PPV target $\geq$ 0.99)}} \\
\midrule
Frozen section (B vs M) & 0.50 & 83.0 (78.6, 87.1) & 0.991 (0.978, 1.000) & 0.929 & 0.935 & 271 & 240 & 31 \\
\midrule
\multicolumn{9}{l}{\textit{Workflow C -- IHC ancillary-test triage (PPV target $\geq$ 0.95)}} \\
\midrule
TTF-1 & 0.50 & 54.9 (48.8, 61.4) & 0.963 (0.930, 0.992) & 0.699 & 0.917 & 246 & 186 & 60 \\
Napsin-A & 0.90 & 36.6 (28.9, 44.4) & 0.962 (0.902, 1.000) & 0.676 & 0.971 & 142 & 74 & 68 \\
CK-7 & 0.54 & 75.0 (67.2, 82.8) & 0.958 (0.909, 0.990) & 0.876 & 0.826 & 128 & 105 & 23 \\
P40 & 0.90 & 15.1 (9.9, 21.1) & 1.000 (1.000, 1.000) & 0.500 & 1.000 & 152 & 46 & 106 \\
P63 & 0.50 & 29.3 (19.0, 41.4) & 1.000 (1.000, 1.000) & 0.500 & 1.000 & 58 & 34 & 24 \\
\midrule
IHC panel (pooled) & -- & 44.5 (323/726) & 0.966 & -- & -- & 726 & -- & -- \\
\bottomrule
\end{tabular}
\end{table}

\clearpage

\begin{table}[htbp]
\centering
\begin{threeparttable}
\caption{\textbf{Computational workflow runtime across biopsy, frozen-section, and resection WSI inference.} Runtime was measured from digitized WSI files to model prediction on one NVIDIA GeForce RTX 3090 GPU. Runtime entries are median (IQR). Total runtime excludes physical slide scanning. Native WSI formats were retained; because specimen type and file format were coupled in this run, file-format speed was not compared directly. The ``All WSIs'' row reports the pooled median across all WSIs and is not the average of specimen-type medians.}
\label{tab:runtime_summary}
\small
\setlength{\tabcolsep}{4pt}
\begin{tabular}{p{2.5cm}p{2.0cm}c p{4.3cm}p{4.5cm}}
\toprule
\textbf{Specimen type} & \textbf{Native WSI format} & \textbf{WSIs} & \textbf{Patches per WSI} & \textbf{Total computational workflow} \\
\midrule
Biopsy & .sdpc & 30 & 10,374.5 (6,313.3-14,138.3) & 394.8 s (284.8-515.8) \\
\hline
Frozen section & .svs & 30 & 5,711.0 (2,728.8-7,633.5) & 83.6 s (59.4-98.3) \\
\hline
Resection & .kfb & 30 & 9,145.5 (6,147.5-12,939.3) & 121.7 s (96.0-148.4) \\
\hline
All WSIs & Mixed & 90 & 8,024.0 (5,185.8-12,270.0) & 127.7 s (90.7-266.7) \\
\bottomrule
\end{tabular}
\end{threeparttable}
\end{table}

\begin{table}[htbp]
\centering
\begin{threeparttable}
\caption{\textbf{Runtime component breakdown for computational WSI inference.} Runtime entries are median (IQR). Classifier prediction includes feature loading, device transfer, and the classifier forward pass. The ``All WSIs'' row reports the pooled median across all WSIs and is not the average of specimen-type medians.}
\label{tab:runtime_components}
\small
\setlength{\tabcolsep}{5pt}
\begin{tabular}{p{3.0cm}p{4.0cm}p{3.5cm}p{3.5cm}}
\toprule
\textbf{Specimen type} & \textbf{Coordinate extraction} & \textbf{Feature extraction} & \textbf{Classifier prediction} \\
\midrule
Biopsy & 63.5 s (43.9-89.3) & 325.4 s (216.3-424.8) & 0.114 s (0.072-0.161) \\
\hline
Frozen section & 4.1 s (3.7-4.7) & 79.3 s (56.0-93.8) & 0.063 s (0.035-0.083) \\
\hline
Resection & 11.0 s (9.0-14.1) & 109.2 s (85.9-134.1) & 0.098 s (0.069-0.129) \\
\hline
All WSIs & 11.0 s (4.6-43.1) & 116.8 s (84.8-202.8) & 0.086 s (0.055-0.125) \\
\bottomrule
\end{tabular}
\end{threeparttable}
\end{table}

\clearpage

\begin{table}[htbp]
\centering
\caption{\textbf{Cohort-pair performance deltas with bootstrapped 95\% confidence intervals.} For each PulmoFoundation task evaluated on more than one cohort, the table reports the difference between cohort-marginal point estimates and a 95\% percentile bootstrap CI on that difference. The point delta is the arithmetic difference of the cohort-marginal AUC (classification, prospective; macro one-vs-rest) or C-Index (survival) values reported in the corresponding per-task Extended Data Tables. The 95\% CI is obtained by paired-replicate bootstrap on the underlying prediction probabilities (1{,}000 non-stratified resamples per cohort, paired by replicate index; cohorts are independent, so pairing reflects replicate alignment rather than unit-level pairing). Cohort-level class balance is reported in the per-task Extended Data Tables; prospective exclusion counts and exclusion reasons are reported in the prospective CONSORT diagram (Extended Data Fig.~\ref{ext_fig_consort_prospective}), and the crossover RCT flow is documented separately in Extended Data Fig.~\ref{ext_fig_consort_rct}. Center codes follow the manuscript convention; H1$^\dagger$ denotes the temporally-independent NF survival cohort. The prospective comparison pairs the prospective evaluation cohort against the internal validation set used during model development.}
\label{tab:perf_delta}
\renewcommand{\arraystretch}{1.1}
\setlength{\tabcolsep}{4pt}
\small
\begin{tabular}{l l c c l}
\toprule
\textbf{Task} & \textbf{Comparison} & \textbf{$N_{\text{int}}$} & \textbf{$N_{\text{ext}}$} & \textbf{$\Delta$ (95\% CI)} \\
\midrule
\multicolumn{5}{l}{\textit{A -- Retrospective classification (Macro AUC; External -- Internal)}} \\
\midrule
Benign vs Malignant (biopsy) & External (H4) & 159 & 400 & $-0.054$ ($-0.093$, $-0.015$) \\
Benign vs Malignant (frozen) & External (H10) & 113 & 169 & $-0.015$ ($-0.049$, $+0.018$) \\
Benign vs Malignant (frozen) & External (H2) & 113 & 514 & $+0.013$ ($-0.001$, $+0.042$) \\
CK-7 & External (H3) & 92 & 122 & $+0.080$ ($+0.016$, $+0.147$) \\
Coarse-grained Subtyping & External (H3) & 120 & 150 & $+0.136$ ($+0.068$, $+0.204$) \\
EGFR & External (TCGA) & 88 & 414 & $-0.139$ ($-0.240$, $-0.043$) \\
Fine-grained Subtyping & External (H3) & 139 & 258 & $-0.021$ ($-0.052$, $+0.023$) \\
NSCLC Subtyping & External (H5) & 120 & 294 & $+0.042$ ($+0.014$, $+0.078$) \\
NSCLC Subtyping & External (TCGA) & 120 & 1053 & $+0.022$ ($-0.008$, $+0.059$) \\
Napsin-A & External (H3) & 134 & 138 & $+0.044$ ($-0.001$, $+0.095$) \\
Neural Invasion & External (H3) & 38 & 114 & $+0.054$ ($-0.036$, $+0.190$) \\
P40 & External (H3) & 67 & 243 & $+0.150$ ($+0.014$, $+0.304$) \\
Pleural Invasion & External (H3) & 72 & 244 & $+0.072$ ($-0.032$, $+0.179$) \\
Primary Site & External (H5) & 139 & 525 & $+0.006$ ($-0.011$, $+0.027$) \\
Primary Site & External (H7) & 139 & 423 & $-0.026$ ($-0.050$, $-0.002$) \\
Primary vs Metastatic (resection) & External (H5) & 170 & 779 & $-0.087$ ($-0.108$, $-0.064$) \\
Primary vs Metastatic (resection) & External (H7) & 170 & 493 & $-0.120$ ($-0.153$, $-0.085$) \\
TTF-1 & External (H3) & 112 & 251 & $+0.051$ ($-0.001$, $+0.112$) \\
Vascular Invasion & External (H3) & 46 & 249 & $+0.028$ ($-0.057$, $+0.141$) \\
\midrule
\multicolumn{5}{l}{\textit{B -- Retrospective survival (C-Index; External -- Internal)}} \\
\midrule
LUAD OS & External (H1$^\dagger$) & 137 & 303 & $+0.139$ ($-0.006$, $+0.265$) \\
LUSC OS & External (H1$^\dagger$) & 136 & 73 & $+0.155$ ($-0.028$, $+0.319$) \\
\midrule
\multicolumn{5}{l}{\textit{C -- Prospective evaluation (Macro AUC; Prospective -- Internal)}} \\
\midrule
Benign vs Malignant (biopsy) & Prospective & 159 & 250 & $+0.022$ ($-0.004$, $+0.049$) \\
Benign vs Malignant (frozen) & Prospective & 113 & 271 & $-0.015$ ($-0.054$, $+0.022$) \\
CK-7 & Prospective & 92 & 128 & $+0.022$ ($-0.057$, $+0.095$) \\
Coarse-grained Subtyping & Prospective & 120 & 397 & $+0.051$ ($-0.009$, $+0.108$) \\
Fine-grained Subtyping & Prospective & 139 & 509 & $-0.043$ ($-0.077$, $+0.027$) \\
LN Metastasis (prospective) & Prospective & 88 & 386 & $-0.125$ ($-0.178$, $-0.070$) \\
NSCLC Subtyping & Prospective & 120 & 270 & $+0.038$ ($+0.009$, $+0.072$) \\
Napsin-A & Prospective & 134 & 142 & $-0.014$ ($-0.074$, $+0.042$) \\
P40 & Prospective & 67 & 152 & $+0.119$ ($-0.014$, $+0.257$) \\
P63 & Prospective & 58 & 58 & $+0.084$ ($-0.026$, $+0.208$) \\
TTF-1 & Prospective & 112 & 246 & $-0.031$ ($-0.094$, $+0.036$) \\
\bottomrule
\end{tabular}
\end{table}

\begin{table}[htbp]
  \centering
  \captionsetup{justification=centering, singlelinecheck=false}
  \caption{Characteristics of participating pathologists during the crossover RCT.}
  \label{tab:RCT_pathologist}
  \begin{tabular}{lr}
    \toprule
    \textbf{Characteristics} & \textbf{Num.} \\
    \midrule
    \textbf{Total pathologists} & 8 \\
    \textbf{Experience level} & \\
    \hspace{1em} Senior & 4 \\
    \hspace{1em} Junior & 4 \\
    \textbf{Sequence group allocation} & \\
    \hspace{1em} Group A (AI $\rightarrow$ No AI) & 4 \\
    \hspace{1em} Group B (No AI $\rightarrow$ AI) & 4 \\
    \textbf{Experience by sequence group} & \\
    \hspace{1em} Group A: Senior / Junior & 2 / 2 \\
    \hspace{1em} Group B: Senior / Junior & 2 / 2 \\
    \bottomrule
  \end{tabular}
\end{table}

\begin{table}[htbp]
  \centering
  \captionsetup{justification=centering, singlelinecheck=false}
  \caption{Characteristics of diagnostic cases across four pre-, intra-, and post-operative tasks involved in the crossover RCT. $\text{Total observations} = \text{Num. Pathologists} \times \text{Num. Cases} \times \text{Num. Periods}$.}
  \label{tab:RCT_cases}
  \begin{tabular}{lr}
    \toprule
      \textbf{Characteristics} & \textbf{Num.} \\
    \midrule
      \textbf{Total cases} & 658 \\
      \textbf{Total WSIs} & 1,803 \\
      \textbf{Total observations} & 10,528 \\
    \midrule
      \textbf{Task 1: Primary vs. Metastatic (Pre-operative)} & \\
      \hspace{1em} Cases & 129 \\
      \hspace{1em} WSIs & 136 \\
      \hspace{1em} Primary tumor & 95 \\
      \hspace{1em} Metastatic tumor & 34 \\
    \midrule
      \textbf{Task 2: Diagnostically Uncertain Benign vs. Malignant (Intra-operative)} & \\
      \hspace{1em} Cases & 99 \\
      \hspace{1em} WSIs & 235 \\
      \hspace{1em} Benign & 53 \\
      \hspace{1em} Malignant & 46 \\
    \midrule
      \textbf{Task 3: NSCLC Subtyping (Post-operative)} & \\
      \hspace{1em} Cases & 270 \\
      \hspace{1em} WSIs & 1,143 \\
      \hspace{1em} Adenocarcinoma & 200 \\
      \hspace{1em} Squamous cell carcinoma & 70 \\
    \midrule
      \textbf{Task 4: Metastasis Origin Prediction (Post-operative)} & \\
      \hspace{1em} Cases & 160 \\
      \hspace{1em} WSIs & 289 \\
      \hspace{1em} Lung primary & 87 \\
      \hspace{1em} Liver primary & 7 \\
      \hspace{1em} Breast primary & 11 \\
      \hspace{1em} Colorectal primary & 49 \\
      \hspace{1em} Kidney primary & 6 \\
    \bottomrule
  \end{tabular}
\end{table}

\begin{table}[t]
  \centering
  \captionsetup{justification=centering, singlelinecheck=false}
  \begin{adjustbox}{max width=\textwidth}
  \begin{threeparttable}
    \caption{Primary outcome from crossover RCT: diagnostic accuracy with and without AI assistance.}
    \label{tab:primary}
    \begin{tabular}{lccccc}
      \toprule
        & \textbf{N} & \textbf{w/o AI} & \textbf{w/ AI} & \textbf{OR (95\% CI)} & \textbf{P value} \\
      \midrule
      \textbf{Overall accuracy} & 5,264 & 83.2\% (82.2--84.2) & 91.7\% (90.9--92.4) & 2.31 (2.13--2.51) & $<$0.001 \\
      \midrule
      \textbf{By task} &  &  &  &  &  \\
      \hspace{1em} Pre-op: Primary vs. Metastatic & 1,032 & 79.7\% & 89.3\% & 2.18 (1.82--2.61) & $<$0.001 \\
      \hspace{1em} Intra-op: Benign vs. Malignant & 792 & 76.8\% & 88.0\% & 2.34 (2.11--2.61) & $<$0.001 \\
      \hspace{1em} Post-op: NSCLC subtyping & 2,160 & 90.1\% & 95.1\% & 2.18 (1.83--2.60) & $<$0.001 \\
      \hspace{1em} Post-op: Metastasis origin & 1,280 & 78.4\% & 90.1\% & 2.55 (2.08--3.12) & $<$0.001 \\
      \bottomrule
    \end{tabular}
    \begin{tablenotes}
      \small
      \item[1] N = number of diagnostic observations per condition (8 pathologists $\times$ cases per task).
      \item[2] Accuracy presented as percentage with 95\% confidence interval for overall outcome.
      \item[3] OR = odds ratio from generalized estimating equations (GEE) with exchangeable correlation structure, adjusting for period, task, and experience level. The overall model used pathologist-case pair as the clustering unit; task-specific models used pathologist as the clustering unit.
      \item[4] OR $>$ 1 indicates higher odds of correct diagnosis with AI assistance.
    \end{tablenotes}
  \end{threeparttable}
  \end{adjustbox}
\end{table}

\begin{table}[htbp]
  \centering
  \captionsetup{justification=centering, singlelinecheck=false}
  \begin{adjustbox}{max width=\linewidth}
  \begin{threeparttable}
    \caption{Carryover and repeated-exposure robustness analyses for the crossover RCT.}
    \label{tab:rct_carryover_robustness}
    \begin{tabular}{p{4.4cm}|p{5.8cm}|p{3.6cm}|p{2.0cm}}
      \toprule
      \textbf{Analysis} & \textbf{Comparison tested} & \textbf{Effect estimate} & \textbf{P value} \\
      \midrule
      Primary full crossover GEE & AI-assisted versus unassisted diagnosis across both periods & OR=2.31 (95\% CI: 2.13--2.51) & $<$0.001 \\
      \hline
      Period 1-only GEE & AI-assisted Group A versus unassisted Group B before crossover & OR=2.48 (95\% CI: 2.19--2.81) & $<$0.001 \\
      \hline
      Condition $\times$ Period interaction & Whether the AI effect differed between Period 1 and Period 2 & Not significant & 0.18 \\
      \bottomrule
    \end{tabular}
    \begin{tablenotes}
      \small
      \item[1] GEE = generalized estimating equation; OR = odds ratio; CI = confidence interval.
      \item[2] The Period 1-only analysis excludes all second-period reads and therefore removes within-trial repeat exposure to the same RCT cases.
      \item[3] The Condition $\times$ Period interaction tests whether the AI effect varied by evaluation period.
    \end{tablenotes}
  \end{threeparttable}
  \end{adjustbox}
\end{table}

\begin{table}[t]
  \centering
  \captionsetup{justification=centering, singlelinecheck=false}
  \begin{threeparttable}
    \caption{Secondary outcomes from crossover RCT: diagnostic time and confidence with and without AI assistance.}
    \label{tab:secondary}
    \begin{tabular}{lcccc}
      \toprule
       & \textbf{w/o AI} & \textbf{w/ AI} & \textbf{Effect (95\% CI)} & \textbf{P value} \\
      \midrule
      \textbf{Diagnostic time (seconds)} & & & & \\
      \hspace{1em} Median (IQR) & 95 (60 -- 141) & 76 (48 -- 116) & & \\
      \hspace{1em} Geometric mean & 90 & 74 & & \\
      \hspace{1em} Time ratio\tnote{1} & ref\tnote{2} & & 0.82 (0.81 -- 0.83) & $<$0.001 \\
      \hspace{1em} Percent change & & & $-$18.3\% & \\
      \midrule
      \textit{By task} & & & & \\
      \hspace{1em} Pre-Op: Primary vs. Metastasis & 100 (49 -- 156) & 82 (43 -- 125) & 0.82 (0.79 -- 0.85) & $<$0.001 \\
      \hspace{1em} Intra-Op: Benign vs. Malignant & 104 (68 -- 143) & 84 (58 -- 122) & 0.85 (0.83 -- 0.88) & $<$0.001 \\
      \hspace{1em} Post-Op: NSCLC subtyping & 96 (56 -- 129) & 75 (45 -- 108) & 0.81 (0.80 -- 0.82) & $<$0.001 \\
      \hspace{1em} Post-Op: Metastasis origin & 90 (62 -- 150) & 72 (50 -- 120) & 0.80 (0.78 -- 0.82) & $<$0.001 \\
      \midrule
      \textbf{Diagnostic confidence (1--10)} & & & & \\
      \hspace{1em} Mean (SD) & 8.3 (1.5) & 9.1 (1.0) & & \\
      \hspace{1em} Mean difference\tnote{3} & ref & & +0.75 (+0.71 -- +0.79) & $<$0.001 \\
      \midrule
      \textit{By task} & & & & \\
      \hspace{1em} Pre-Op: Primary vs. Metastasis & 8.3 (1.6) & 9.1 (1.1) & +0.82 (+0.72 -- +0.92) & $<$0.001 \\
      \hspace{1em} Intra-Op: Benign vs. Malignant & 8.0 (1.7) & 8.9 (1.1) & +0.85 (+0.74 -- +0.96) & $<$0.001 \\
      \hspace{1em} Post-Op: NSCLC subtyping & 8.6 (1.0) & 9.3 (0.7) & +0.66 (+0.61 -- +0.71) & $<$0.001 \\
      \hspace{1em} Post-Op: Metastasis origin & 8.1 (1.7) & 8.8 (1.3) & +0.79 (+0.70 -- +0.88) & $<$0.001 \\
      \bottomrule
    \end{tabular}
    \begin{tablenotes}
      \small
    \item[1] Time ratio from generalized estimating equations (GEE) on log-transformed time, adjusting for period, task, and experience level, with pathologist as the clustering unit. Ratio $<$ 1 indicates reduced time with AI.
    \item[2] ref = reference category; the without-AI condition served as the baseline for all effect estimates.
    \item[3] Mean difference from generalized estimating equations (GEE) adjusting for period, task, and experience level, with pathologist as the clustering unit. Cohen's d = 0.59 (medium effect size).
    \item[4] IQR = interquartile range; SD = standard deviation.
    \end{tablenotes}
  \end{threeparttable}
\end{table}

\begin{table}[htbp]
  \centering
  \captionsetup{justification=centering, singlelinecheck=false}
  \begin{adjustbox}{max width=\linewidth}
  \begin{threeparttable}
    \caption{Outcome-based AI impact analysis using displayed model predictions.}
    \label{tab:ai_utility}
    \begin{tabular}{lccccc}
      \toprule
      \textbf{Metric} & \textbf{Overall} & \textbf{Primary vs. Met} & \textbf{Benign vs. Mal} & \textbf{NSCLC} & \textbf{Origin} \\
      \midrule
      \textbf{Observations, n} & 5,264 & 1,032 & 792 & 2,160 & 1,280 \\
      \midrule
      \textbf{Outcome-based diagnostic change\tnote{1}, n (\%)} & & & & & \\
      \hspace{1em} Improved: incorrect to correct & 477 (9.1) & 107 (10.4) & 97 (12.2) & 119 (5.5) & 154 (12.0) \\
      \hspace{1em} Confirmed: AI correct and remained correct & 4,327 (82.2) & 804 (77.9) & 597 (75.4) & 1,933 (89.5) & 993 (77.6) \\
      \hspace{1em} Resilient: model prediction incorrect and remained correct & 22 (0.4) & 11 (1.1) & 3 (0.4) & 2 (0.1) & 6 (0.5) \\
      \hspace{1em} Failed: final diagnosis incorrect & 438 (8.3) & 110 (10.7) & 95 (12.0) & 106 (4.9) & 127 (9.9) \\
      \midrule
      \textbf{Mechanistic safety decomposition\tnote{2}, n (\%)} & & & & & \\
      \hspace{1em} Strict AI-mediated improvement\tnote{3} & 473 (9.0) & 105 (10.2) & 96 (12.1) & 119 (5.5) & 153 (12.0) \\
      \hspace{1em} Missed opportunity\tnote{4} & 324 (6.2) & 96 (9.3) & 75 (9.5) & 76 (3.5) & 77 (6.0) \\
      \hspace{1em} Accuracy loss after AI\tnote{5} & 32 (0.6) & 8 (0.8) & 8 (1.0) & 12 (0.6) & 4 (0.3) \\
      \hspace{1em} Strict erroneous-AI adoption harm\tnote{6} & 12 (0.2) & 5 (0.5) & 0 (0.0) & 4 (0.2) & 3 (0.2) \\
      \hspace{1em} Both failed\tnote{7} & 82 (1.6) & 6 (0.6) & 12 (1.5) & 18 (0.8) & 46 (3.6) \\
      \midrule
      \textbf{AI model performance} & & & & & \\
      \hspace{1em} Model accuracy, \% & 97.7 & 97.7 & 98.0 & 98.9 & 95.6 \\
      \hspace{1em} Adoption rate\tnote{8}, \% & 93.0 & 89.1 & 89.0 & 96.0 & 93.3 \\
      \bottomrule
    \end{tabular}
    \begin{tablenotes}
      \small
      \item[1] Displayed model predictions, ground truth, and pathologist diagnoses were defined at the observation level. The four diagnostic-change rows are mutually exclusive and sum to all AI-assisted observations.
      \item[2] Mechanistic rows decompose the diagnostic-change categories and are not all mutually exclusive. Strict erroneous-AI adoption harm is a subset of accuracy loss after AI.
      \item[3] Strict AI-mediated improvement: displayed model prediction was correct and the pathologist changed from an incorrect unassisted diagnosis to a correct assisted diagnosis.
      \item[4] Missed opportunity: displayed model prediction was correct, the unassisted diagnosis was incorrect, and the final assisted diagnosis remained incorrect.
      \item[5] Accuracy loss after AI: the unassisted diagnosis was correct and the final assisted diagnosis was incorrect.
      \item[6] Strict erroneous-AI adoption harm: the displayed model prediction was incorrect, the unassisted diagnosis was correct, and the pathologist adopted the incorrect model prediction.
      \item[7] Both failed: the displayed model prediction and the unassisted diagnosis were both incorrect, and the final assisted diagnosis was incorrect.
      \item[8] Adoption rate: proportion of observations where the final assisted diagnosis matched the displayed model prediction.
      \item[9] Met = metastasis; Mal = malignant; NSCLC = non-small cell lung cancer subtyping; Origin = metastasis origin prediction.
    \end{tablenotes}
  \end{threeparttable}
  \end{adjustbox}
\end{table}

\clearpage

\begin{table}[htbp]
  \centering
  \captionsetup{justification=centering, singlelinecheck=false}
  \begin{adjustbox}{max width=\textwidth}
  \begin{threeparttable}
    \caption{Individual pathologist accuracy (\%) by task with and without AI assistance during the crossover RCT.}
    \label{tab:individual_by_task}
    \begin{tabular}{lccc cccc cccc}
      \toprule
      & & & \multicolumn{4}{c}{\textbf{Without AI}} & \multicolumn{4}{c}{\textbf{With AI}} \\
      \cmidrule(lr){4-7} \cmidrule(lr){8-11}
      \textbf{ID} & \textbf{Exp.}\tnote{1} & \textbf{Seq.}\tnote{2} & \textbf{Task 1}\tnote{3} & \textbf{Task 2}\tnote{4} & \textbf{Task 3}\tnote{5} & \textbf{Task 4}\tnote{6} & \textbf{Task 1} & \textbf{Task 2} & \textbf{Task 3} & \textbf{Task 4} \\
      \midrule
      P1 & Junior & A & 73.6 & 63.6 & 86.3 & 73.1 & 83.7 & 79.8 & 91.1 & 86.2 \\
      P2 & Senior & B & 83.7 & 85.9 & 96.7 & 84.4 & 90.7 & 92.9 & 98.1 & 91.2 \\
      P3 & Senior & A & 90.7 & 93.9 & 95.9 & 90.0 & 91.5 & 97.0 & 98.1 & 93.1 \\
      P4 & Junior & B & 71.3 & 66.7 & 84.4 & 70.0 & 83.7 & 82.8 & 91.9 & 85.0 \\
      P5 & Senior & B & 86.8 & 79.8 & 95.9 & 83.1 & 94.6 & 92.9 & 97.4 & 91.2 \\
      P6 & Junior & B & 69.8 & 69.7 & 81.9 & 71.9 & 89.1 & 81.8 & 92.2 & 91.9 \\
      P7 & Senior & A & 89.9 & 96.0 & 95.9 & 85.0 & 95.3 & 99.0 & 98.1 & 93.1 \\
      P8 & Junior & A & 72.1 & 58.6 & 84.1 & 69.4 & 86.0 & 77.8 & 93.7 & 88.8 \\
      \midrule
      \textbf{Mean} &  &  & 79.7 & 76.8 & 90.1 & 78.4 & 89.3 & 88.0 & 95.1 & 90.1 \\
      \bottomrule
    \end{tabular}
    \begin{tablenotes}
      \small
      \item[1] Exp. = Experience level.
      \item[2] Seq. = Sequence group.
      \item[3] Task 1 = Primary vs. Metastatic for diagnostic biopsies.
      \item[4] Task 2 = Benign vs. Malignant for intra-operative frozen sections (uncertain cases from pathologists).
      \item[5] Task 3 = NSCLC subtyping for post-operative resections.
      \item[6] Task 4 = Metastasis origin prediction for post-operative resections.
    \end{tablenotes}
  \end{threeparttable}
  \end{adjustbox}
\end{table}

\begin{table}[htbp]
  \centering
  \captionsetup{justification=centering, singlelinecheck=false}
  \begin{threeparttable}
    \caption{Inter-rater agreement among pathologists with and without AI assistance.}
    \label{tab:interrater}
    \begin{tabular}{lccccc}
      \toprule
      & \multicolumn{2}{c}{\textbf{Fleiss' $\kappa$ (95\% CI)}} & & \\
      \cmidrule(lr){2-3}
      \textbf{Task} & \textbf{w/o AI} & \textbf{w/ AI} & \textbf{$\Delta\kappa$}\tnote{1} & \textbf{P value}\tnote{2} \\
      \midrule
      Overall & 0.55 (0.51--0.58) & 0.76 (0.73--0.79) & +0.21 & $<$0.001 \\
      \midrule
      Primary vs. Metastatic & 0.37 (0.28--0.45) & 0.61 (0.52--0.70) & +0.24 & $<$0.001 \\
      Benign vs. Malignant & 0.37 (0.29--0.44) & 0.62 (0.55--0.69) & +0.25 & $<$0.001 \\
      NSCLC Subtyping & 0.64 (0.58--0.69) & 0.83 (0.78--0.87) & +0.19 & $<$0.001 \\
      Metastasis Origin & 0.56 (0.50--0.62) & 0.81 (0.76--0.86) & +0.25 & $<$0.001 \\
      \bottomrule
    \end{tabular}
    \begin{tablenotes}
      \small
      \item[1] $\Delta\kappa$ = Fleiss' $\kappa$ with AI $-$ Fleiss' $\kappa$ without AI. Positive values indicate improved agreement.
      \item[2] P value from paired permutation test comparing $\kappa$ between conditions (10,000 iterations).
      \item[3] Interpretation: $\kappa < 0.20$ = poor; $0.21$--$0.40$ = fair; $0.41$--$0.60$ = moderate; $0.61$--$0.80$ = substantial; $>0.80$ = almost perfect.
    \end{tablenotes}
  \end{threeparttable}
\end{table}

\begin{table}[htbp]
  \centering
  \captionsetup{justification=centering, singlelinecheck=false}
  \begin{adjustbox}{max width=\linewidth}
  \begin{threeparttable}
    \caption{Pathologist behavior when the displayed model prediction was incorrect.}
    \label{tab:rct_wrong_model_behavior}
    \begin{tabular}{lccccc}
      \toprule
      \textbf{Metric} & \textbf{Overall} & \textbf{Primary vs. Met} & \textbf{Benign vs. Mal} & \textbf{NSCLC} & \textbf{Origin} \\
      \midrule
      \textbf{Model-incorrect observations\tnote{1}} & 120 (2.3) & 24 (2.3) & 16 (2.0) & 24 (1.1) & 56 (4.4) \\
      \midrule
      \hspace{1em} Final diagnosis correct despite incorrect model prediction & 26 (21.7) & 13 (54.2) & 4 (25.0) & 2 (8.3) & 7 (12.5) \\
      \hspace{1em} Maintained correct diagnosis despite incorrect model prediction & 22 (18.3) & 11 (45.8) & 3 (18.8) & 2 (8.3) & 6 (10.7) \\
      \hspace{1em} Corrected diagnosis despite incorrect model prediction & 4 (3.3) & 2 (8.3) & 1 (6.2) & 0 (0.0) & 1 (1.8) \\
      \hspace{1em} Adopted incorrect model prediction & 93 (77.5) & 11 (45.8) & 12 (75.0) & 22 (91.7) & 48 (85.7) \\
      \hspace{1em} Strict erroneous-AI adoption harm\tnote{2} & 12 (10.0) & 5 (20.8) & 0 (0.0) & 4 (16.7) & 3 (5.4) \\
      \hspace{1em} Final incorrect without adopting incorrect model prediction & 1 (0.8) & 0 (0.0) & 0 (0.0) & 0 (0.0) & 1 (1.8) \\
      \bottomrule
    \end{tabular}
    \begin{tablenotes}
      \small
      \item[1] The first row reports n (\%) using all AI-assisted observations as the denominator.
      \item[2] All subsequent rows report n (\%) using model-incorrect observations as the denominator. Strict erroneous-AI adoption harm requires an initially correct unassisted diagnosis, an incorrect displayed model prediction, final adoption of that incorrect model prediction, and an incorrect final assisted diagnosis.
      \item[3] Displayed model predictions refer to the task-specific prediction shown in the AI-assisted condition.
      \item[4] Met = metastasis; Mal = malignant; NSCLC = non-small cell lung cancer subtyping; Origin = metastasis origin prediction.
    \end{tablenotes}
  \end{threeparttable}
  \end{adjustbox}
\end{table}

\begin{table}[htbp]
  \centering
  \captionsetup{justification=centering, singlelinecheck=false}
  \begin{adjustbox}{max width=\linewidth}
  \begin{threeparttable}
    \caption{Model-confidence stratified sensitivity analysis.}
    \label{tab:rct_confidence_sensitivity}
    \begin{tabular}{lcccc}
      \toprule
      \textbf{Metric} & \textbf{Overall} & \textbf{Low} & \textbf{Middle} & \textbf{High} \\
      \midrule
      Observations, n & 5,264 & 1,936 & 1,672 & 1,656 \\
      Cases, n & 658 & 242 & 209 & 207 \\
      Mean max probability & 0.959 & 0.890 & 0.998 & 1.000 \\
      \midrule
      Model accuracy, \% & 97.7 & 93.8 & 100.0 & 100.0 \\
      Accuracy without AI, \% & 83.2 & 77.0 & 85.8 & 88.0 \\
      Accuracy with AI, \% & 91.7 & 87.0 & 93.3 & 95.5 \\
      Accuracy change, percentage points & +8.5 & +10.1 & +7.5 & +7.5 \\
      \midrule
      GEE OR (95\% CI)\tnote{2} & 2.31 (2.03--2.63) & 2.12 (1.82--2.48) & 2.42 (2.07--2.83) & 3.10 (2.40--3.98) \\
      \bottomrule
    \end{tabular}
    \begin{tablenotes}
      \small
      \item[1] The analysis included all 658 RCT cases and 5{,}264 case-reader observations. Model confidence was defined at the case level as the maximum predicted class probability. Cases were assigned within each task to low-, middle-, or high-confidence strata using task-specific confidence cutoffs. These probabilities were used only for post hoc stratification and were not displayed during RCT review.
      \item[2] Odds ratios were estimated with generalized estimating equations for diagnostic correctness, adjusting for AI condition, period, task, and experience level, with pathologist as the clustering unit. All GEE AI-effect P values were $<$0.001.
      \item[3] OR = odds ratio; CI = confidence interval.
    \end{tablenotes}
  \end{threeparttable}
  \end{adjustbox}
\end{table}

\begin{table}[htbp]
  \centering
  \captionsetup{justification=centering, singlelinecheck=false}
  \begin{adjustbox}{max width=\linewidth}
  \begin{threeparttable}
    \caption{Case-difficulty sensitivity analysis using unassisted pathologist accuracy as a post hoc proxy.}
    \label{tab:rct_difficulty_proxy_sensitivity}
    \begin{tabular}{lcccc}
      \toprule
      \textbf{Metric} & \textbf{Overall} & \textbf{Hard} & \textbf{Intermediate} & \textbf{Easy} \\
      \midrule
      Observations, n & 5,264 & 1,752 & 1,752 & 1,760 \\
      Cases, n & 658 & 219 & 219 & 220 \\
      Mean no-AI case accuracy, \% & 83.2 & 60.4 & 90.0 & 99.2 \\
      \midrule
      Model accuracy, \% & 97.7 & 94.1 & 99.1 & 100.0 \\
      Accuracy without AI, \% & 83.2 & 60.4 & 90.0 & 99.2 \\
      Accuracy with AI, \% & 91.7 & 80.1 & 95.8 & 99.1 \\
      Accuracy change, percentage points & +8.5 & +19.6 & +5.9 & $-$0.1 \\
      \midrule
      Improved\tnote{2}, n (\%) & 477 (9.1) & 356 (20.3) & 114 (6.5) & 7 (0.4) \\
      Accuracy loss\tnote{3}, n (\%) & 32 (0.6) & 12 (0.7) & 11 (0.6) & 9 (0.5) \\
      GEE OR (95\% CI)\tnote{4} & 2.31 (2.03--2.63) & 2.94 (2.29--3.76) & 2.79 (2.36--3.30) & NE \\
      \bottomrule
    \end{tabular}
    \begin{tablenotes}
      \small
      \item[1] Independent difficulty annotations were unavailable. Difficulty was approximated post hoc as 1 minus the unassisted accuracy across eight pathologists for each task-case pair. Strata were assigned within each task by ranked tertiles of this proxy.
      \item[2] Improved: unassisted diagnosis incorrect and assisted diagnosis correct.
      \item[3] Accuracy loss: unassisted diagnosis correct and assisted diagnosis incorrect. In the easy stratum, the displayed model prediction was correct in all nine accuracy-loss observations. None represented strict erroneous-AI adoption harm. Strict erroneous-AI adoption harm was 12 observations overall and is reported separately in Table~\ref{tab:rct_wrong_model_behavior}.
      \item[4] Odds ratios were estimated with generalized estimating equations for diagnostic correctness, adjusting for AI condition, period, task, and experience level, with pathologist as the clustering unit. GEE AI-effect P values were $<$0.001 for the overall, hard, and intermediate strata. NE indicates that sparse incorrect outcomes under near-ceiling accuracy prevented stable estimation of the adjusted OR and 95\% CI in the easy stratum.
      \item[5] OR = odds ratio; CI = confidence interval.
    \end{tablenotes}
  \end{threeparttable}
  \end{adjustbox}
\end{table}

\begin{table}[htbp]
  \centering
  \captionsetup{justification=centering, singlelinecheck=false}
  \begin{adjustbox}{max width=\linewidth}
  \begin{threeparttable}
    \caption{Full intra-pathologist diagnostic accuracy and AI-associated change across RCT tasks.}
    \label{tab:rct_individual_reader_performance}
    \begin{tabular}{lll lccc}
      \toprule
      \textbf{ID} & \textbf{Exp.} & \textbf{Seq.} & \textbf{Summary} & \textbf{Without AI, \%} & \textbf{With AI, \%} & \textbf{Change, pp} \\
      \midrule
      P1 & Junior & A & Overall & 77.2 & 86.8 & +9.6 \\
      P1 & Junior & A & Primary vs. Met & 73.6 & 83.7 & +10.1 \\
      P1 & Junior & A & Benign vs. Mal & 63.6 & 79.8 & +16.2 \\
      P1 & Junior & A & NSCLC & 86.3 & 91.1 & +4.8 \\
      P1 & Junior & A & Origin & 73.1 & 86.2 & +13.1 \\
      \addlinespace
      P2 & Senior & B & Overall & 89.5 & 94.2 & +4.7 \\
      P2 & Senior & B & Primary vs. Met & 83.7 & 90.7 & +7.0 \\
      P2 & Senior & B & Benign vs. Mal & 85.9 & 92.9 & +7.1 \\
      P2 & Senior & B & NSCLC & 96.7 & 98.1 & +1.5 \\
      P2 & Senior & B & Origin & 84.4 & 91.2 & +6.9 \\
      \addlinespace
      P3 & Senior & A & Overall & 93.2 & 95.4 & +2.3 \\
      P3 & Senior & A & Primary vs. Met & 90.7 & 91.5 & +0.8 \\
      P3 & Senior & A & Benign vs. Mal & 93.9 & 97.0 & +3.0 \\
      P3 & Senior & A & NSCLC & 95.9 & 98.1 & +2.2 \\
      P3 & Senior & A & Origin & 90.0 & 93.1 & +3.1 \\
      \addlinespace
      P4 & Junior & B & Overall & 75.7 & 87.2 & +11.6 \\
      P4 & Junior & B & Primary vs. Met & 71.3 & 83.7 & +12.4 \\
      P4 & Junior & B & Benign vs. Mal & 66.7 & 82.8 & +16.2 \\
      P4 & Junior & B & NSCLC & 84.4 & 91.9 & +7.4 \\
      P4 & Junior & B & Origin & 70.0 & 85.0 & +15.0 \\
      \addlinespace
      P5 & Senior & B & Overall & 88.6 & 94.7 & +6.1 \\
      P5 & Senior & B & Primary vs. Met & 86.8 & 94.6 & +7.8 \\
      P5 & Senior & B & Benign vs. Mal & 79.8 & 92.9 & +13.1 \\
      P5 & Senior & B & NSCLC & 95.9 & 97.4 & +1.5 \\
      P5 & Senior & B & Origin & 83.1 & 91.2 & +8.1 \\
      \addlinespace
      P6 & Junior & B & Overall & 75.2 & 90.0 & +14.7 \\
      P6 & Junior & B & Primary vs. Met & 69.8 & 89.1 & +19.4 \\
      P6 & Junior & B & Benign vs. Mal & 69.7 & 81.8 & +12.1 \\
      P6 & Junior & B & NSCLC & 81.9 & 92.2 & +10.4 \\
      P6 & Junior & B & Origin & 71.9 & 91.9 & +20.0 \\
      \addlinespace
      P7 & Senior & A & Overall & 92.1 & 96.5 & +4.4 \\
      P7 & Senior & A & Primary vs. Met & 89.9 & 95.3 & +5.4 \\
      P7 & Senior & A & Benign vs. Mal & 96.0 & 99.0 & +3.0 \\
      P7 & Senior & A & NSCLC & 95.9 & 98.1 & +2.2 \\
      P7 & Senior & A & Origin & 85.0 & 93.1 & +8.1 \\
      \addlinespace
      P8 & Junior & A & Overall & 74.3 & 88.6 & +14.3 \\
      P8 & Junior & A & Primary vs. Met & 72.1 & 86.0 & +14.0 \\
      P8 & Junior & A & Benign vs. Mal & 58.6 & 77.8 & +19.2 \\
      P8 & Junior & A & NSCLC & 84.1 & 93.7 & +9.6 \\
      P8 & Junior & A & Origin & 69.4 & 88.8 & +19.4 \\
      \bottomrule
    \end{tabular}
    \begin{tablenotes}
      \small
      \item[1] Overall rows pool all observations for a given pathologist across the four RCT tasks. Task rows report task-specific accuracy.
      \item[2] Change is the AI-assisted accuracy minus the unassisted accuracy in percentage points, computed from unrounded proportions.
      \item[3] Exp. = experience level; Seq. = randomized sequence group; Met = metastatic; Mal = malignant; NSCLC = non-small cell lung cancer subtyping; Origin = metastasis origin prediction; pp = percentage points.
    \end{tablenotes}
  \end{threeparttable}
  \end{adjustbox}
\end{table}

\clearpage

\end{document}